\documentclass[preprint,showpacs,preprintnumbers,amssymb,nofootinbib,aps,prd]{revtex4-1}
\usepackage{amssymb,amsmath,amsfonts,mathrsfs}
\usepackage[utf8]{inputenc}
\usepackage{amsmath}
\usepackage{slashed}
\usepackage{latexsym}
\usepackage{multirow}
\usepackage[usenames,dvipsnames,table]{xcolor}
\usepackage{graphicx}
\usepackage{epsfig}  
\usepackage{epsf}    
\usepackage{dcolumn}
\usepackage{bm}
\usepackage{dcolumn}
\usepackage{textcomp}
\usepackage{float}
\usepackage{subfig}
\usepackage{hyperref}
\usepackage{hypcap}
\usepackage{makecell}
\usepackage{color}
\usepackage{pifont}
\usepackage{appendix}
\usepackage{url}
\usepackage{cancel}
\allowdisplaybreaks[1]
\usepackage{import}
\usepackage{subfiles}
\usepackage{filecontents}
\usepackage{threeparttable, tablefootnote}
\usepackage{slashed}
\usepackage{lineno}
\usepackage{placeins}

\let\Oldsection\section
\renewcommand{\section}{\FloatBarrier\Oldsection}

\let\Oldsubsection\subsection
\renewcommand{\subsection}{\FloatBarrier\Oldsubsection}

\let\Oldsubsubsection\subsubsection
\renewcommand{\subsubsection}{\FloatBarrier\Oldsubsubsection}

\usepackage[normalem]{ulem}
\definecolor{darkgreen}{cmyk}{1,0,1,0.4}

\definecolor{pink}{cmyk}{0.4,1,0.3,0}

\def\com2#1{\textcolor{red}{\it{#1}}}

\makeatletter
\renewcommand{\fnum@table}{\textbf{\tablename~\thetable}}
\renewcommand{\fnum@figure}{\textbf{\figurename~\thefigure}}
\makeatother

\begin{document}
	
	\title
	{ Roadmap of left-right models based on GUTs}
	
	\author{Joydeep Chakrabortty}
	\email{joydeep@iitk.ac.in}
	\affiliation{Department of Physics, Indian Institute of Technology, Kanpur-208016, India}
	
	\author{Rinku Maji}
	\email{mrinku@iitk.ac.in}
	\affiliation{Department of Physics, Indian Institute of Technology, Kanpur-208016, India}
	
	\author{Subhendra Mohanty}
	\email{mohanty@prl.res.in}
	\affiliation{Theoretical Physics, Physical Research Laboratory, Ahmedabad-380009, India}
	
	\author{Sunando Kumar Patra}
	\email{sunando.patra@gmail.com}
	\affiliation{Department of Physics, Indian Institute of Technology, Kanpur-208016, India}
	
	\author{Tripurari Srivastava}
	\email{tripurar@iitk.ac.in}
	\affiliation{Department of Physics, Indian Institute of Technology, Kanpur-208016, India}
	
\begin{abstract}
	We perform a detailed study of the grand unified theories $SO(10)$ and $E(6)$ with left-right intermediate gauge symmetries of the form $SU(N)_L\otimes SU(N)_R \otimes \mathcal{G}$. 
	Proton decay lifetime constrains the unification scale to be  $ \gtrsim 10^{16}$ GeV and, as discussed in this paper, unwanted cosmological relics can be evaded if the intermediate symmetry scale is $ \gtrsim 10^{12}$ GeV. With these conditions, we study the renormalisation group  evolution of the gauge couplings and do a comparative analysis of all possible 
	left-right models where unification can occur. Both the D-parity conserved and broken scenarios as well as the supersymmetric (SUSY) and Non-supersymmetric (Non-SUSY) versions are considered. In addition to the fermion and scalar representations at each stage of the symmetry breaking, contributing to the $\beta$-functions, we list the intermediate left-right groups which successfully meet these requirements. We make use of the dimension-5 kinetic mixing effective operators for achieving unification and large intermediate scale. A significant result in the supersymmetric case is that to achieve successful unification for some breaking patterns, the scale of SUSY breaking needs to be at least a few TeV. In some of these cases, intermediate scale can be as low as $\sim 10^{12}$ GeV, for SUSY scale to be $\sim 30$ TeV. This has important consequences in the collider searches for SUSY particles and  phenomenology of the lightest neutralino as dark matter.
\end{abstract}

\maketitle
\flushbottom
\newpage
\section{Introduction}

Grand Unified Theories (GUTs) are the theories which attempt to discover a single gauge group for the unification of the strong, weak, and electromagnetic interactions and where the three couplings of the standard model (SM) $SU(3)_C\otimes SU(2)_L\otimes U(1)_Y \equiv \mathcal{G}_{2_L1_Y3_C}$ are unified at a high scale  (called the GUT scale, $M_X$) to a single coupling  $g_{U}$ of the GUT gauge group.
The grand unified gauge group must be in form of either $\mathcal{G}$ or $\mathcal{G} \otimes \mathcal{G}..$, as it must posseses an unified coupling $g_{U}$. The SM  is expected to emerge from the unified symmetry group, thus the minimum rank of the GUT group must be $\geq$ 4.  Some of the successful candidates for such theory are $SU(5)$, $SO(10)$, and $E(6)$. In this paper we focus on $SO(10)$ and $E(6)$, as we are interested in those unified groups which contain left-right gauge symmetries as the subgroup. The motivation behind left-right models as intermediate symmetry group is to raise P and CP violation to the same status as gauge symmetry breaking which take place via vacuum expectation values of specific scalar representations.  As the rank of these groups are 5 and 6 respectively, it is indeed possible to accommodate multiple intermediate symmetries  in the desert between  $M_X$ and $M_Z$.

In this paper we focus on the economical scenario of one intermediate symmetry at scale ($M_R$) below the GUT scale which further breaks to the SM directly. Among the numerous possibilities for the intermediate symmetry groups, we concentrate only on the left-right models, which are of the form $SU(N)_L\otimes SU(N)_R \otimes \mathcal{G}$, where $\mathcal{G}$ is any group or product of groups. These specific breaking patterns can be achieved by the suitable choice of representations and orientations of the vacuum expectation values of the GUT breaking scalars\cite{Fritzsch:1974nn,  Lazarides:1980nt,  Clark:1982ai, Aulakh:1982sw, Hewett:1985ss, Deshpande:1992au,  Amaldi:1991cn, Hung:2006kd, Howl:2007zi,  Chakrabortty:2008zk, Bertolini:2009es, Chakrabortty:2010az, DeRomeri:2011ie, Arbelaez:2013nga,Chakrabortty:2010xq}. Many phenomenological studies have been performed both in presence and absence of supersymmetry (SUSY) for $SO(10)$ \cite{Lazarides:1980nt, Chakrabortty:2010az, DeRomeri:2011ie, Patra:2015bga, Bandyopadhyay:2015fka, Babu:2016cri, Bandyopadhyay:2017uwc} and  $E(6)$ \cite{Gursey:1975ki,Achiman:1978vg,Hewett:1985ss, Hung:2006kd, Howl:2007zi, Chakrabortty:2013voa, Miller:2014jza, Chakrabortty:2015ika, Gogoladze:2014cha,Younkin:2012ui,Calmet:2011ic,Wang:2011eh,Biswas:2010yp,Atkins:2010re}. Successful generation of neutrino and fermion masses is one of finest achievements of GUT models \cite{Mohapatra:1979ia, Bajc:2002iw, Goh:2003hf, King:2003rf, Goh:2003sy, Mohapatra:2003tw, Bertolini:2004eq, Dev:2009aw, Joshipura:2009tg, Chakrabortty:2010az, Patel:2010hr, Blanchet:2010kw, BhupalDev:2011gi, Joshipura:2011nn, BhupalDev:2012nm, Meloni:2014rga, Babu:2016bmy, Meloni:2016rnt}.
Recently, different aspects of unification have been discussed in the context of dark matter \cite{Babu:2015bna, Nagata:2015dma, Garcia-Cely:2015quu, Brennan:2015psa, Boucenna:2015sdg, Mambrini:2015vna, Parida:2016hln, Nagata:2016knk, Sahoo:2017cqg, Arbelaez:2017ptu}. The implication of domain walls in presence of left-right symmetry in SUSY framework has been studied in \cite{Mishra:2009mk, Borah:2011qq, Borah:2012vk}. 
 
Our aim here is to check  all possible intermediate groups that arise from $SO(10)$ and $E(6)$ for both SUSY and Non-SUSY varieties, both in presence and absence of gravitational smearing at the unification scale. Moreover, we also want to check the viability of such scenarios which pass the scrutiny of proton decay limits and cosmological constraints, namely topological defects and baryon asymmetry of the universe. 
  
To be consistent with the observed limit \cite{Miura:2016krn} on proton lifetime\footnote{One can find the recent development in lattice computation for proton decay in Ref.~\cite{Aoki:2017puj}.}  ($\tau_p > 1.6 \times 10^{34}$), the unification scale\footnote{ In principle this bound needs to computed for individual model. But we have considered the conservative limit without loss of generality.} has to be raised above $10^{16}$ GeV. One way to achieve this is to include the contribution from the Planck mass-suppressed effective dimension-5 operators. These are expected  to arise by integrating out the full quantum gravity theory or string compactification leading to  an effective GUT theory at $M_X$. We study the effects of the Planck-suppressed effective dimension-5 operators along with  the RG evolution of the couplings and limit the Wilson coefficients of these operators from the requirement that $M_X  \gtrsim 10^{16}$ GeV. 
 
There are some critical constraints on the intermediate left-right symmetry models from cosmology, which are related to the existing D-parity in such models. It was shown by Shaposhnikov and Kuzmin \cite{Kuzmin:1980yp}, that the net baryon asymmetry must be zero in models with unbroken D-parity. Another cosmological problem that arises, is the formation of string-bounded stable domain walls, when D-parity is broken \cite{Kuzmin:1980yp}. A way out of both of these problems is if the inflation takes place after GUT symmetry breaking when one of the GUT scalars acts as the inflaton. Viable inflation from $SO(10)$ scalars as the inflaton, has been constructed  \cite{Garg:2015mra,Ellis:2016ipm} and it is seen that the reheat temperature at the end of inflation is $T_R \simeq 10^{12}$ GeV. If the scale of D-parity breaking is above the reheat temperature ($10^{12}$ GeV), there is no problem of stable domain walls and baryogenesis can be achieved via leptogenesis by heavy neutrino decay, in GUT models with left-right intermediate symmetries \cite{Fukugita:1986hr}. In this paper, we impose the criterion that the D-parity breaking of the intermediate scale must be above $10^{12}$ GeV and study the parameter space and gauge groups of the intermediate scale which satisfy this criterion. This ensures that after reheat the universe is in the SM phase and the harmful cosmological defects are not created. We do a detailed study of the role of the abelian mixing operators (which arise when there is a product of $U(1)$ groups in the intermediate scale) in raising the intermediate scale symmetry to above $10^{12}$  GeV and limit the  range of  couplings of the abelian mixing operators  using this criterion.
 
The rest of the paper is organized as follows: In section~\ref{sec:unification} we lay down some aspects of grand unified theories which are used for selecting the intermediate scale symmetries consistent with proton decay and cosmology. Here we have briefly discussed extended survival hypothesis, D-parity, renormalisation group evolutions (RGEs) of gauge couplings. We have also noted the modifications in the boundary conditions of RGEs at different scales due to threshold correction and Planck scale physics. We have concluded this section by introducing the homotopy structure of the vacuum manifolds and their respective topological defects. In section~\ref{sec:RGEs-intscales}, we have discussed all possible one-intermediate scale breaking patterns that carry explicit left-right gauge symmetry. We have computed the two loop beta functions for SUSY and Non-SUSY scenarios for each breaking chain. Then in section~\ref{sec:results} we have determined the values of the intermediate and unification scales and the unified gauge coupling, in accordance with the present experimental bounds of the low scale parameters, by simultaneously solving the RGEs and performing a goodness of fit test with a constructed $\Delta\chi^2$ statistic. This enables us to obtain the bounds with correlation among these high scale parameters including abelian mixing. The constraints due to topological defects and proton lifetime are implemented. We conclude by discussing their impacts on symmetry breaking scales and other free parameters of the theory.
\section{Some aspects of unification}
\label{sec:unification}

In this section we study some aspects of GUTs which have a bearing on fixing the unification and the intermediate symmetry scales.

%

\subsection{Extended Survival Hypothesis} 
\label{subsec:ESH}

 The direct breaking of GUT group to SM is not favoured as it does not predict the correct Weinberg angle ($\theta_W$) at low energy\footnote{This more specifically applicable for Non-supersymmetric scenario and also with minimal particle content. One can explain this by adding more particles and including their threshold corrections.}.  One or more intermediate scales are therefore necessary.  As the SM has rank $\geq$ 4, the GUT groups need to have large ranks ($\geq 5$) to posses one or more intermediate symmetry groups. We need extra scalars to break these  intermediate gauge groups. These scalars are usually embedded in large representations under the GUT group. But unlike the 
 GUT breaking scalars, they contribute in the RG between intermediate and unification scales. Due to their large dimensionality, their contribution to the beta coefficients may be large enough to spoil the unification picture. Also, the presence of such representations may require a significant fine tuning in the scalar potential to achieve correct vacuum structure. Thus to avoid the catastrophe due to the unnecessary sub-multiplets, a prescription named {\it Extended Survival Hypothesis} (ESH) has been proposed \cite{delAguila:1980qag}. According to this, {\it  at every stage of the symmetry breaking chain, only those scalars are light and relevant that develop a vacuum expectation value at that or the subsequent levels of the symmetry breaking}.  These sub-multiplets  play a crucial role in generating the fermion masses, specifically neutrino masses and $\sin^2 \theta_W$ without much fine tuning of the parameters of the scalar potential. We will use ESH to understand the symmetry breaking within a minimal fine tuned scenario.
 
%

\subsection{D-parity}
\label{subsec:Dparity}

D-parity is an important ingredient in the context of grand unified theories. Historically, D-parity was first introduced in \cite{Mohapatra:1974gc, Senjanovic:1975rk, Senjanovic:1978ev, Chang:1983fu, Chang:1984uy} in case of $SO(10)$, which contains $SU(2)_L\otimes SU(2)_R\otimes SU(4)_C$ as a maximal subgroup. D-parity, which plays a role analogous to charge conjugation, is defined as the product $\Gamma_{67} \Gamma_{23}$ where $\Gamma_{ij}$'s are the antisymmetric generators of $SO(10)$. As an example, a multiplet $(R_2,1,R_4)$ under $SU(2)_L\otimes SU(2)_R\otimes SU(4)_C$ is related to its conjugate representation $(1,\overline{R_2},\overline{R_4})$ by  D-parity.
D-parity is not realised in  all possible intermediate symmetries
. The characteristics of the vacuum orientation, in the wake of the breaking of GUT symmetry, decides whether the D-parity is broken or not. Though it is possible for the intermediate symmetry to have the form $SU(2)_L\otimes SU(2)_R\otimes ...$ in both cases, it is the D-parity which decides whether $g_{2L}$ and $g_{2R}$, the respective gauge couplings, will be the same or not at the intermediate scale. 

The minimum rank of the GUT group must be $\geq$ 5 to obtain the preferred form, mentioned in last paragraph, of the intermediate symmetry groups; thus $SO(10)$ is the minimal choice. As $E(6)$ is of rank 6 and it contains $SO(10)$ as a subgroup, we can realize D-parity through a few of its sub-groups. All these possibilities will be discussed in a later part of this paper.

%

D-parity and the scale at which it is broken has some significant implications for cosmology. If the intermediate symmetry is ${ SU(N)_L\otimes SU(N)_R\otimes \mathcal{G}}$, then the coupling constants of the two ${SU(N)}$ groups must be same ($g_{nL}=g_{nR}$) in the unbroken D-parity phase. In such a case, as pointed out by Kuzmin and Shaposhnikov  \cite{Kuzmin:1980yp}, baryon asymmetry cannot be generated by the decay of leptoquarks in the D-symmetric phase. To generate baryon asymmetry through leptoquarks \cite{Ignatiev:1979um}, the masses of these leptoquarks must be close to the unification scale. This implies that the D-parity  breaking must take place close to the unification scale in the left-right models, in the conventional GUT-baryogenesis scenario \cite{Ignatiev:1979um}.

A different cosmological problem associated with D-parity breaking is the formation of string-bounded domain walls which do not decay and would dominate the density of the late universe  \cite{Kuzmin:1980yp}. The formation of domain walls and monopoles is undesirable in the phase transition associated with the symmetry breaking, as it would dominate the energy density of the universe. On the other hand, textures harmlessly decay in the early universe while string networks are sub-dominant, can be accommodated in the energy density of the universe and may have observable signature in small angle anisotropy of the CMB \cite{Fraisse:2007nu}.
 
One scenario, that provides a way out of these cosmological problems associated with the string bounded domain walls (induced by D-parity breaking) and other harmful cosmological relics, is inflation \cite{Starobinsky:1980te, Guth:1980zm, Albrecht:1982wi}. Inflation can take place with the GUT scalars as inflaton and viable inflation models with $SO(10)$ scalars playing the role of the inflation have been constructed \cite{Aulakh:2012st, Garg:2015mra, Ellis:2016ipm, Gonzalo:2016gey, Leontaris:2016jty}. Any domain walls or other topological defects will be inflated away in these models, where inflation takes place following the GUT symmetry breaking scale. Following inflation, the reheat temperature from the decay of the inflation is around ${\rm 10^{12}\; GeV}$. If the intermediate symmetry and D-parity is broken at a scale above the reheat temperature of ${\rm 10^{12} \;GeV}$, then the dangerous walls bounded by strings \cite{Kibble:1982ae} will not form in the radiation era after inflation. The problem of baryogenesis can be solved through leptogenesis \cite{Fukugita:1986hr} in these models. This is possible with the decay of heavy right handed neutrinos with masses lower than the reheat temperature and the subsequent conversion of the lepton asymmetry to baryon asymmetry through sphalerons \cite{Kuzmin:1985mm} in the electroweak era. We will follow this cosmological scenario in this paper and will impose the criterion that only those GUT models are phenomenologically acceptable where the D-parity breaking scale is above the reheat temperature of ${\rm 10^{12}\; GeV}$. 
 
%

\subsection{RGEs of gauge couplings}
\label{subsec:RGEs}

The RGEs of the gauge couplings (upto two loop) can be written in terms of group theoretic invariants which encapsulate the contributions from the respective scalars and fermions of the theory 
\cite{Caswell:1974gg, Jones:1974mm, Jones:1981we, Machacek:1983tz, Machacek:1983fi, Machacek:1984zw}. These invariants depend solely on the representations of those scalars and fermions, under the gauge symmetries we are considering. Following Ref.~\cite{Jones:1981we}, the beta functions upto two-loop for gauge couplings for a product group $\mathcal{G}_i\otimes \mathcal{G}_j\otimes \mathcal{G}_k..$ can be written as\footnote{Here, we have not included the contributions of the Yukawa couplings.}:
\begin{align}
\mu \frac{dg_i}{d\mu}&= \frac{g_i^3}{(4\pi)^2} \left[ \frac{4\kappa}{3}T(F_i)D(F_j)
 +\frac{1}{3}T(S_i)D(S_j) - \frac{11}{3} C_2(G_i) \right] + \frac{1}{(4\pi)^4} g_i^5  \nonumber \\
 & \times \left[  \left(\frac{10}{3} C_2(G_i)+2C_2(F_i)\right) T(F_i)D(F_j)
 + \left(\frac{2}{3} C_2(G_i)+4C_2(S_i)\right) T(S_i)D(S_j)  \right. \nonumber  \\
 & \left. - \frac{34}{3} (C_2(G_i))^2 \right] 
  + \frac{1}{(4\pi)^4} g_i^3 g_j^2  \left[ 2C_2(F_j)T(F_i)D(F_j)
 + 4 C_2(S_j)T(S_i)D(S_j) \right].
 \end{align}
Here $S_i,F_i$ are the scalar and fermion representations transforming under group $\mathcal{G}_i$. $\kappa=1/2$ for the chiral fermions, otherwise it is 1. The $C_2(R)$ are the quadratic Casimir for scalar, fermion and adjoint representation for $R=S,F,G$ respectively.  $D(R)$ is the dimensionality of the representation and $T(R)$, the normalisation of the generators in $R-$dimensional representation. These group theoretic factors are related to each other by $C_2(R)=T(R)d/D(R)$, where $d$ is the number of generators of the group. 
These quantities have special values for abelian groups, e.g. $C_2(G)=0, T(R)=\sum_i q_i^2$ where $q_i$ are the normalised abelian charges.. 

In case of supersymmetry, the beta functions upto two-loop can be given as in Ref.~\cite{Jones:1981we}:
\begin{eqnarray}
\mu \frac{dg_i}{d\mu}&=& \frac{g_i^3}{(4\pi)^2} \left[ T(F_i)D(F_j) - 3 C_2(G_i) \right] \nonumber \\
 &+& \frac{1}{(4\pi)^4} g_i^5 \left[ \left(2 C_2(G_i)+ 4C_2(F_i)\right) T(F_i)D(F_j)
  - 6 (C_2(G_i))^2 \right] \nonumber \\
  &+& \frac{1}{(4\pi)^4} g_i^3 g_j^2  \left[ 4 C_2(F_j)T(F_i)D(F_j)  \right].
 \end{eqnarray}
Here the dimensions of the representations are assigned for the super-multiplets.

%

 \subsection{Abelian mixing}
\label{subsec:abelian-mixing}
In a theory when we have more than one abelian gauge group, the Lagrangian posses extra gauge invariant term in the gauge kinetic sector. Let us consider there are two abelian groups and  $F_{\mu \nu}, G_{\mu \nu}$ are their respective gauge invariant field strength tensors. Then, apart from their individual gauge kinetic terms there will be a term $\propto [F_{\mu \nu}G^{\mu \nu}]$ which leads to the abelian mixing.
As a result abelian gauge couplings start mixing with each other even at the one-loop level \cite{Holdom:1985ag,delAguila:1988jz,Lavoura:1993ut,delAguila:1995rb,Luo:2002iq,Fonseca:2013bua} and one needs to modify the structures of $\beta$ functions accordingly. 
In presence of multiple abelian gauge groups, e.g. $U(1)\otimes U(1) \otimes U(1)..$, the RGEs can be written as:
\begin{equation}
\mu \frac{dg_{kb}}{d\mu}=\beta_{ab} g_{ka},
\end{equation}
where
\begin{equation}
\beta_{ab}=\frac{1}{(4\pi)^2} g_{sa}\Sigma_{sr} g_{rb}.
\end{equation}
and $g_{ab}$ is the gauge coupling matrix, represented as 
\begin{equation}
g=
  \begin{bmatrix}
    g_{11} & g_{12} & g_{13} &... &g_{1n}\\
   g_{21}& g_{22} & g_{23} & ...& g_{2n} \\
   g_{31} & g_{32} & g_{33} & ...& g_{3n} \\
   ... & ......& ...& ...& ...\\
  ... & ...& ....& ...& .....\\  
  g_{n1} & g_{n2} & g_{n3} & .... & g_{nn}\\  
  \end{bmatrix},\label{eqn:gc_mat}
\end{equation}
where $\{a,b,k,s,r\} $ runs over number of  $U(1)$ groups. For example, for two $U(1)$ gauge symmetries, $\{a,b,k,s,r\} \in {1,2}$ and the above matrix will be of order 2.

The $\Sigma$'s are defined as \cite{Bertolini:2009qj, Chakrabortty:2009xm}:
\begin{equation}
\Sigma_{sr}=\sigma_{sr}^{(one-loop)}+\frac{1}{(4\pi)^2}\sigma_{sr}^{(two-loop)}.
\end{equation}  
 The beta coefficients $\sigma_{rs}$ can be written as  \cite{Bertolini:2009qj, Chakrabortty:2009xm}:
\begin{align}
\sigma_{sr}^{(one-loop)}& \equiv \tilde{b}_{sr} 
= \frac{2}{3}n_g \{y_s(F) y_r(F)  D(F)\}+\frac{1}{3} \{y_s(S) y_r(S)  D(S)\},
\end{align}
where  $y_s$ is the $s^{th}\;U(1)$'s normalized  charge  and $D(R)$ is the dimensionality of the non-singlet representations (fermion/scalar) that carry this charge. We would like to mention that for $s\neq r$, we get the abelian mixing terms. 

This mixing may lead to more complicated structures at two-loop level and the $\beta$ functions are given as  \cite{Bertolini:2009qj, Chakrabortty:2009xm}:
\begin{align}
\sigma_{ss}^{(2-loop)}&=\tilde{\tilde{b}}_{ss,ss} (g_{ss}^2+g_{sr}^2) +  \tilde{\tilde{b}}_{ss,sr}    (g_{ss} g_{rr}+g_{sr} g_{rr})
 + \tilde{\tilde{b}}_{ss,rr}   (g_{rr}^2+g_{rs}^2), \\
 \sigma_{sr}^{(2-loop)}&=\tilde{\tilde{b}}_{ss,sr} (g_{ss}^2+g_{sr}^2) +  \tilde{\tilde{b}}_{ss,rr}    (g_{ss} g_{rr}+g_{sr} g_{rr})
 + \tilde{\tilde{b}}_{sr,rr}   (g_{rr}^2+g_{rs}^2),\\
 \sigma_{rr}^{(2-loop)}&=\tilde{\tilde{b}}_{ss,rr} (g_{ss}^2+g_{sr}^2) +  \tilde{\tilde{b}}_{sr,rr}    (g_{ss} g_{rr}+g_{sr} g_{rr})
 + \tilde{\tilde{b}}_{rr,rr}   (g_{rr}^2+g_{rs}^2),
\end{align}
with $\beta$-coefficients given as
\begin{equation}
\tilde{\tilde{b}}_{ij,kl} 
= {2}n_g \Big \{\Big(y_i(F) y_j(F) y_k(F) y_l(F)\Big ) D(F)\Big  \}+{4} \Big \{ \Big(y_i(S) y_j(S) y_k(S) y_l(S)\Big ) D(S)\Big \}.
\end{equation}

At two-loop level, this abelian mixing gets entangled with non-abelian gauge couplings too. This affects their mutual running as follows \cite{Bertolini:2009qj, Chakrabortty:2009xm}:
  \begin{align}
\mu \frac{dg_k}{d\mu} \supset \frac{g_k^3g_{ij}^2}{(4\pi)^4} \tilde{\tilde{b}}_{ss,p} \;\;\;\; \text{and} &\;\;\;\;
\sigma_{sr}^{(two-loop)}\supset \tilde{\tilde{b}}_{ss,p} g_{p}^2\,,
\end{align}
  where
  $\tilde{\tilde{b}}_{rs,p} =[2 n_g \{y(F_r) y(F_s) T(F_k) D(F_l) \}+4 \{y(S_r) y(S_s) T(S_k) D(S_l)\}]$.
Here, $g_k$ is the non-abelian gauge coupling and $\tilde{\tilde{b}}_{rs,p}$ stands for abelian mixing with non-abelian gauge couplings, with $p$ as the non-abelian index.
The abelian mixing has been discussed in detail in Refs.~\cite{Bertolini:2009qj, Chakrabortty:2009xm, Fonseca:2013bua}, in the context of $SO(10)$ and $E(6)$ GUT groups. In the context of supersymmetric GUT, the effects of abelian mixing in SUSY spectrum, more precisely for gaugino masses, has been discussed in \cite{Fonseca:2011vn, Braam:2011xh,Hirsch:2012kv,Rizzo:2012rf,OLeary:2011vlq}.
    
%

\subsection{Matching conditions}
\label{subsec:matching}

In the instance of the breaking of a simple or a product gauge group into its subgroups, the gauge couplings of the broken groups are redistributed in terms of the unbroken symmetries. Thus the parent and the daughter gauge couplings need to be matched at the symmetry breaking scale which has been discussed in detail in \cite{Hall:1980kf, WEINBERG198051, Chang:1984qr, Binger:2003by}. If we neglect the heavy-mass-dependent logarithmic effects, we can write the matching condition of two gauge couplings as:
\begin{equation}
\frac{1}{\alpha_i} - \frac{C_2(\mathcal{G}_i)}{12\pi} = \frac{1}{\alpha_j} - \frac{C_2(\mathcal{G}_j)}{12\pi},
\end{equation}
where $\alpha_i=g_i^2/4\pi, C_2(\mathcal{G}_i)$ is the quadratic Casimir of group $\mathcal{G}_i$ in adjoint representation.  
This matching condition will get modified in presence of abelian gauge couplings. As an example, let us consider an abelian daughter group $U(1)_X$ and let the respective gauge coupling be $g_X$. The generator of this unbroken group ($I_X$) is an outcome of the spontaneous breaking of generators $I_m$, i.e. $I_X=w_m I_m$. Here $m$ indicates the number of broken generators and $w_m$ are the suitable weight factors leading to normalised $X$ charge and satisfy the following relation: $\sum_m w_m^2=1$. Now the the matching condition is given as \cite{Bertolini:2009qj, Chakrabortty:2009xm}:
\begin{equation}
\frac{1}{\alpha_X}  = \sum_m w_m^2 \Big[ \frac{1}{\alpha_m} - \frac{C_2(\mathcal{G}_m)}{12\pi} \Big].
\end{equation}
 $C_2(\mathcal{G}_m)=0$ for the abelian group. In the presence of more than one abelian groups, spontaneously broken at same scale and contributing to the $X-$charge, this matching condition is further modified. As we have discussed in the last section, the gauge couplings get mixed in the presence of two or more abelian gauge groups and we need to treat the full gauge coupling matrix together, in place of a single coupling (see Eqn.~\ref{eqn:gc_mat}). In this case, the matching condition reads as \cite{Bertolini:2009qj, Chakrabortty:2009xm}:
 \begin{equation}
\frac{1}{\alpha_X}  =  \Bigg[ Q.\frac{4\pi}{(g.g^T)}.Q^T + \sum_n w_n^2  \Big( \frac{1}{\alpha_n} - \frac{C_2(\mathcal{G}_n)}{12\pi} \Big) \Bigg],
\end{equation}
where the matrix $g$ is given in Eqn.~\ref{eqn:gc_mat}. $Q$ is a row vector in the above equation and satisfies the relation $Q.Q^T+\sum_n w_n^2=1$. In the absence of non-abelian groups in the parent sector, the above equation reduces to $1/\alpha_X  =   Q.[4\pi/(g.g^T)].Q^T$ with $Q.Q^T=1$.
  
%

\subsection{dimension-5 operators and unification boundary conditions}
\label{subsec:dim-5}

At the GUT scale, the unified renormalisable gauge kinetic term is written as:
\begin{equation}
\mathcal{L}^{kin}_{ren}=-\frac{1}{4C}\rm{Tr} (F^{\mu\nu} F_{\mu \nu}),
\label{Eq:dim-4}
\end{equation}
where the unified gauge field strength tensor $F_{\mu \nu}=\sum_iT^i F_{\mu \nu}^i$, and $T_i$'s are the generators of unified group and they are normalized as ${\rm Tr}(T_i T_j)=C\delta_{ij}$.  This $F_{\mu \nu}$ contains an unified gauge coupling $g_U=g_i(M_X)$.

In a typical unified theory, all the fundamental forces are included apart from gravity. Still, as the unification scale is fairly close to the Planck scale($M_{Pl}$), it is possible for string compactification or quantum gravity to have some impact on the unification boundary condition \cite{Hall:1980kf, Hill:1983xh, Shafi:1983gz, Hall:1992kq}.  These effects are expected to be through the higher dimensional operators suppressed by Planck scale and can be written as:
\begin{equation}
\mathcal{L}^{kin}_{non-ren}=-\frac{\eta}{M_{Pl}} \Big[\frac{1}{4C}{\rm Tr} (F^{\mu\nu} \Phi_D F_{\mu \nu})\Big ],
\label{Eq:dim-5}
\end{equation}
where $\eta$ is a dimensionless parameter.
\begin{table}[h!]
	\small
	\renewcommand*{\arraystretch}{1.2}
\begin{center}
    \begin{tabular}{| c | c | c | c | c | c | }
    \hline
     Group & Scalar Representation & $\delta_{3L}$  & $\delta_{3R}$  & $\delta_{3C}$  \\
     \hline
     $E(6)$ & 650 & $\frac{1}{2 \sqrt{2}}$  & $\frac{1}{2 \sqrt{2}}$  & $\frac{-1}{\sqrt{2}}$  \\
     \hline
     $E(6)$ & $650^{'}$ & $\frac{3}{2\sqrt{6}}$  & $\frac{-3}{2\sqrt{6}}$  & $0$  \\
    \hline
    \end{tabular}
     \caption{\small The group theoretic factors ($\delta_i$'s) arise from dimension-5 operators for the following breaking $E(6)\to \mathcal{G}_{3_L 3_R 3_C}$.  }
    \label{tab:dimension_5_I}
\end{center}
\end{table}
Here $F_{\mu \nu}$ transforms as the adjoint representation of the GUT group, and thus restricts the choice of $\Phi_D$ which can belong to only the
symmetric product of two adjoint representations. The GUT symmetry is spontaneously broken once the $\Phi_D$ acquires vacuum expectation value (VEV), $\textless \Phi_D \textgreater$, and the gauge couplings get additional contributions from the effective operator Eq.~\ref{Eq:dim-5}. These contributions are unequal due to the non-singlet nature of $\Phi_D$ and modify the unification boundary conditions as: $g_U^2=g_i^2(M_X)(1+\varepsilon \delta_i)$, where $\varepsilon =\eta \textless \Phi_D \textgreater/2M_{Pl} \sim \mathcal{O} (M_X/M_{Pl})$. It is worthy to mention that these effects could be important to evade the proton decay constraints. The extra free parameter $\varepsilon$ allows a range of solutions for the unification scale, and may help to revive certain breaking patterns which will be discussed in a later part of this paper.
\begin{table}[h!]
	\small
	\renewcommand*{\arraystretch}{1.2}
\begin{center}
    \begin{tabular}{| c | c | c | c | c | c | }
    \hline
     Group & Scalar Representation & $\delta_{2L}$  & $\delta_{2R}$  & $\delta_{4C}$  \\
     \hline
     $SO(10)$ & 54 & $\frac{3}{2 \sqrt{15}}$  & $\frac{3}{2 \sqrt{15}}$  & $\frac{-1}{ \sqrt{15}}$  \\
     \hline
     $SO(10)$ & 210 & $\frac{1}{\sqrt{2}}$  & $\frac{-1}{ \sqrt{2}}$  & $0$  \\
     \hline
     $SO(10)$ & 770 & $\frac{5}{3 \sqrt{5}}$  & $\frac{5}{3 \sqrt{5}}$  & $\frac{2}{3 \sqrt{5}}$  \\
     \hline
    \end{tabular}
     \caption{\small The group theoretic factors ($\delta_i$'s) arise from dimension-5 operators for the following breaking $SO(10)\rightarrow \mathcal{G}_{2_L2_R4_C}$. }
    \label{tab:dimension_5_II}
\end{center}
\end{table}
The relevant and  necessary  dimension-5 contributions are tabulated in Tables~\ref{tab:dimension_5_I}, \ref{tab:dimension_5_II}, and \ref{tab:dimension_5_III} (see \cite{Chakrabortty:2008zk,  Martin:2009ad, Chakrabortty:2010xq} for more). A set of new results has  been provided in Table~\ref{tab:dimension_5_IV} for breaking pattern $E(6)\to \mathcal{G}_{2_L2_R4_C1_X}$ for both D-parity conserved and broken cases. We would like to mention here that these dimension-5 operators may affect the unification scenario for $SO(10)$ and $E(6)$ GUT groups \cite{Patra:1991dy, Vayonakis:1993nn, Chakrabortty:2009xm, Calmet:2009hp} and these corrections lead to the non-universality of gaugino masses in the SUSY case \cite{Drees:1985bx, Ellis:1985jn, Anderson:1996bg, Chakrabortty:2008zk,  Martin:2009ad, Chakrabortty:2010xq, Bhattacharya:2009wv} leading to different phenomenology \cite{Bhattacharya:2009wv, Atkins:2010re, Biswas:2010yp, Calmet:2011ic,  Wang:2011eh, Younkin:2012ui, Chakrabortty:2013voa, Gogoladze:2014cha, Miller:2014jza, Chakrabortty:2015ika} compared to the usual minimal supersymmetric standard model.
\begin{table}[h!]
	\small
	\renewcommand*{\arraystretch}{1.2}
\begin{center}
    \begin{tabular}{| c | c | c | c | c | c | c | }
    \hline
     Group & Scalar Representation & $\delta_{2L}$  & $\delta_{2R}$  & $\delta_{3C}$ & $\delta_{1X}$  \\
     \hline
     $SO(10)$ & 54 + 210 & $\frac{3}{2\sqrt{15}}$  & $\frac{3}{2\sqrt{15}}$  & $\frac{-1}{\sqrt{15}}$  & $\frac{-1}{\sqrt{15}}$ \\
     \hline
     $SO(10)$ & 210 + 45 & $\frac{1}{\sqrt{2}}$  & $\frac{-1}{\sqrt{2}}$  & $0$  & $0$ \\
     \hline
      $SO(10)$ & 770 + 210 & $\frac{5}{3\sqrt{5}}$  & $\frac{5}{3\sqrt{5}}$  & $\frac{2}{3\sqrt{5}}$  & $\frac{2}{3\sqrt{5}}$ \\
     \hline
    \end{tabular}
 \caption{\small The group theoretic factors ($\delta_i$'s) arise from dimension-5 operators for the following breaking $SO(10)\rightarrow \mathcal{G}_{2_L2_R3_C 1_X}$. }
    \label{tab:dimension_5_III}
\end{center}
\end{table}
\begin{table}[h!]
	\small
	\renewcommand*{\arraystretch}{1.2}
	\begin{center}
		\begin{tabular}{| c | c | c | c | c | c | }
			\hline
			Group & Scalar Representation & $\delta_{2L}$  & $\delta_{2R}$  & $\delta_{4C}$ & $\delta_{1X}$ \\
			\hline
			$E(6)$ & $650\supset(54,0)$ & $\frac{9}{\sqrt{10}}$  & $\frac{9}{ \sqrt{10}}$  & $\frac{-6}{\sqrt{10}}$ & $0$  \\
			\hline
			$E(6)$ & $650\supset (210,0)$ & $\frac{6\sqrt{6}}{\sqrt{19}}$  & $-\frac{6\sqrt{6}}{\sqrt{19}}$  & $0$ & $0$ \\
			\hline
		\end{tabular}
		\caption{\small The group theoretic factors ($\delta_i$'s) arise from dimension-5 operators for the following breaking $E(6)\to \mathcal{G}_{2_L 2_R 4_C 1_X}$. D-parity is consereved and broken when $(54,0)$ and $(210,0)$ (under $SO(10)\otimes U(1)$) components of 650-dimensional scalars acquire VEVs respectively. }
		\label{tab:dimension_5_IV}
	\end{center}
\end{table}
 
%

\subsection{Topological defects associated with spontaneous symmetry breaking}
\label{subsec:TD}

It is worthwhile to properly analyse the topological structures of vacuum manifolds in spontaneously broken gauge field theories. Ref.s \cite{Kibble:1982dd, Vachaspati:1997rr} note that various types of topological defects, namely domain walls, cosmic strings, monopoles, and textures may appear. Investigating the homotopy groups of the respective vacuum manifolds can shed light on these structures. In this paper, we concentrate on those defects, which may appear from the subsequent breaking of GUT gauge groups to the SM \cite{Lazarides:1981fv, Kibble:1982ae, Pijushpani-Hill, Davis:1994py, Davis:1995bx, Jeannerot:2003qv}. During the breaking of a group $G$ down to its subgroup $H$, we can study the homotopy groups $\Pi_k(G/H) $ of the vacuum manifold ${\cal M}_n = G/H$ to see whether topological defects form during the phase transition associated with the said breaking. Topological defects are formed if  $\Pi_k(G/H) \neq \mathcal{I}$. Various types of topological defects which can form are : domain walls ($k = 0$),  cosmic strings ($k = 1$), monopoles ($k = 2$), and textures ($k = 3$).  Out of these, monopoles and domain walls are undesirable, as they dominate the energy density and would surpass that of the universe. Textures decay rapidly and leave no trace in the present universe. The energy density budget of the universe can accommodate cosmic strings and those may have observable signatures in the small angle anisotropy of the CMB \cite{Fraisse:2007nu}. We will later discuss whether these defects are isolated or hybrid ones.  

We list the homotopy of different groups below, which will appear in different stages of symmetry breaking using Bott periodicity theorem \cite{Bott}:\\
$~$\qquad(I) 
\begin{align}
\Pi_k(U(N))=\Pi_k(SU(N))&=\mathcal{I} {\rm \;for \;even}\; k\\
&= \mathbb{Z} {\rm \;for \;odd}\; k,
\end{align}
$~~~$\qquad with $k>1$ and $N\geq (k+1)/2$.\\
$~~~$\qquad $~\Pi_1(SU(N))=\mathcal{I}$  where $\Pi_1(U(N))=\mathbb{Z}$ $\forall$ $N$.\\
$~$\qquad(II)
\begin{eqnarray}
\Pi_k(O(N))=\Pi_k(SO(N))&=&\mathcal{I} {\rm \;for }\; k=2,4,5,6 (mod\; 8)\\
&=& \mathbb{Z}_2 {\rm \;for }\;  k=0,1(mod \;8) \\
&=& \mathbb{Z} {\rm \;for }\;  k=3,7 (mod\; 8),
\end{eqnarray}
$~~~$\qquad with $N\geq k+2$.

We would like to mention a few useful special cases \cite{Bott, Bruno, mimura1967, Tosa}:
\begin{table}[H]
	\small
	\renewcommand*{\arraystretch}{1.2}
\begin{center}
    \begin{tabular}{| c | c | c | c | c |  }
    \hline
   Lie   &  zeroth Homotopy  &  Fundamental &  2nd homotopy  &  3rd homotopy     \\
    Group       &  $(\Pi_0)$  &  group $(\Pi_1)$ &   group $(\Pi_2)$ &   group $(\Pi_3)$    \\
     \hline
       $U(1)$  & $\mathcal{I}$ & $\mathbb{Z}$ & $\mathcal{I}$  & $\mathcal{I}$    \\
     \hline
      $U(2)$  & $\mathcal{I}$ & $\mathbb{Z}$ & $\mathcal{I}$  & $\mathbb{Z}$    \\
     \hline
      $U(3)$  & $\mathcal{I}$ & $\mathbb{Z}$ & $\mathcal{I}$  & $\mathbb{Z}$    \\
     \hline
     $SO(2)$ & $\mathcal{I}$ & $\mathbb{Z}$  & $\mathcal{I}$ & $\mathcal{I}$   \\
     \hline
      $SO(3)$ & $\mathcal{I}$ & $\mathbb{Z}_2$  & $\mathcal{I}$ & $\mathbb{Z}$   \\
     \hline
     $SO(4)$ & $\mathcal{I}$ & $\mathbb{Z}_2$  & $\mathcal{I}$ & $(\mathbb{Z}\times \mathbb{Z})$   \\
     \hline
      $SO(6)$ & $\mathcal{I}$ & $\mathbb{Z}_2$  & $\mathcal{I}$ & $\mathbb{Z}$   \\
     \hline
       \end{tabular}
     \caption{Homotopy classification of Lie Groups.}
    \label{tab:homotopy}
\end{center}
\end{table}
We can define the homotopy as $\Pi_k(\mathcal{G}_i \otimes \mathcal{G}_j)=\Pi_k(\mathcal{G}_i) \otimes \Pi_k(\mathcal{G}_j)$ for a product group. The vacuum manifold is defined as $\mathcal{G} / (\mathcal{G}_i \otimes \mathcal{G}_j)$ for a given symmetry breaking chain $\mathcal{G} \to\mathcal{G}_i \otimes \mathcal{G}_j$. To investigate the topological structure, i.e., homotopy of the vacuum manifold we can write 
 $\Pi_k(\mathcal{G} / (\mathcal{G}_i \otimes \mathcal{G}_j))=\Pi_{k-1} (\mathcal{G}_i \otimes \mathcal{G}_j)$ when $\Pi_{k}(\mathcal{G})=\Pi_{k-1}(\mathcal{G})=\mathcal{I}$.
 
It becomes easy to classify the possible emergence of different topological defects, once we identify the homotopy of the vacuum manifold at every stages of symmetry breaking. We can assure the appearance of domain walls, cosmic strings, monopoles and textures for $k=0,1,2,3$ respectively, when the $k^{th}$ homotopy of vacuum manifold is non-trivial.
Here we demonstrate the generation of topological defects using two examples where we assume that $\Pi_{[2,1]}( \mathcal{G})=\Pi_{[1,0]}( \mathcal{G}_i)=\Pi_{[1,0]}( \mathcal{G}_j)=\mathcal{I}$:
\begin{itemize}
\item[I.]  Consider a symmetry breaking of the form \cite{Davis:1995bx}: $\mathcal{G} \to \mathcal{G}_i \otimes \mathcal{G}_j \otimes U(1)\to   \mathcal{G}_i \otimes \mathcal{G}_j $.
Analysing the vacuum manifold of the first stage of symmetry breaking, we note the following:
\begin{align}\label{eq:case-IA}
\Pi_1(\mathcal{G} /( \mathcal{G}_i \otimes \mathcal{G}_j \otimes U(1)))&=\Pi_0( \mathcal{G}_i \otimes \mathcal{G}_j \otimes U(1))=\mathcal{I}, \left(\text{No domain walls} \right. \nonumber \\
	& \left. \qquad\qquad\qquad\qquad\qquad\qquad \text{ and cosmic strings}\right) \nonumber \\
\Pi_2(\mathcal{G} /( \mathcal{G}_i \otimes \mathcal{G}_j \otimes U(1)))&=\Pi_1( \mathcal{G}_i \otimes \mathcal{G}_j \otimes U(1))=\mathbb{Z}, (\text{monopoles will be there}).
\end{align}
In the second stage of symmetry breaking, we find
\begin{eqnarray}
\Pi_1( \mathcal{G}_i \otimes \mathcal{G}_j \otimes U(1)/(\mathcal{G}_i \otimes \mathcal{G}_j ) ) &=& \mathbb{Z}, ({\rm presence\; of\; cosmic\; strings}) \nonumber \\
\Pi_2( \mathcal{G}_i \otimes \mathcal{G}_j \otimes U(1)/(\mathcal{G}_i \otimes \mathcal{G}_j ) )&=&\Pi_2( U(1) )=\mathcal{I}, ({\rm No\; monopole}).
\label{eq:case-IB}
\end{eqnarray}
\item[II.] Now consider another symmetry breaking of the form \cite{Davis:1995bx}: $\mathcal{G} \to \mathcal{G}_i \otimes \mathcal{G}_j \otimes \mathbb{Z}_2\to   \mathcal{G}_i \otimes \mathcal{G}_j $. 
If we analyse the vacuum manifold of first stage of symmetry breaking, we note the following:
\begin{eqnarray}
\Pi_1(\mathcal{G} /( \mathcal{G}_i \otimes \mathcal{G}_j \otimes \mathbb{Z}_2))&=&\Pi_0( \mathcal{G}_i \otimes \mathcal{G}_j \otimes \mathbb{Z}_2)=\mathbb{Z}_2, ({\rm cosmic\; strings})
\label{eq:case-IIA}
\end{eqnarray}
In the second stage of symmetry breaking, we find
\begin{eqnarray}
\Pi_0( \mathcal{G}_i \otimes \mathcal{G}_j \otimes \mathbb{Z}_2/(\mathcal{G}_i \otimes \mathcal{G}_j ) ) &=& \mathbb{Z}_2, ({\rm presence\; of\; domain\; walls}). 
\label{eq:case-IIB}
\end{eqnarray}
\end{itemize}
One can have hybrid topological defects \cite{Lazarides:1981fv, Kibble:1982dd, Vilenkin:1982hm} in case of a sequential symmetry breaking. For example, monopoles are produced in the first stage of symmetry breaking in case-I (see Eq.~\ref{eq:case-IA}) and we will have strings due the second stage of symmetry breaking  (see Eq.~\ref{eq:case-IB}). The monopole-antimonopole pair is connected by the strings in this scenario.
Unlike case-I, the strings, which are the outcomes of the first stage of symmetry breaking, are topologically unstable in the next type of symmetry breaking (see Eq.~\ref{eq:case-IIA}) but the domain walls, which are produced in the latter step, are stable (see Eq.~\ref{eq:case-IIB}).
All these discussions and conclusions regarding the topological defects are equally applicable for supersymmetry and Non-supersymmetric scenarios. The topological structures are based on the homotopy of the vacuum manifold corresponding to the spontaneous breaking of some Lie groups. Interestingly enough, the supersymmetry algebra is validated by the Lie algebra, and we can find the Lie algebra for SUSY by exponentiating the infinitesimal super-transformation. This has been discussed in detail in \cite{Albert:1948, Srivastava:1974zn, Santilli:1978pb, Davis:1995bx}.

Throughout this paper, we  will  impose the constraint that the scale of symmetry breaking (producing the harmful monopoles and domain walls) should be above the post-inflation reheat temperature of $10^{12} ~{\rm GeV}$, so that these defects do not form after inflation in the inflationary cosmology. This will restrict the symmetry breaking pattern which are acceptable {\it vis-\`{a}-vis} cosmology.

%


\section{RGEs of gauge couplings: $\beta$ coefficients}
\label{sec:RGEs-intscales}
\subsection{Breaking of $SO(10)$ to SM: $SO(10) \xrightarrow{M_X}  \mathcal{G}_{int} \xrightarrow{M_{R}}  \mathcal{G}_{2_L1_Y3_C}$}
\label{subsec:so10-SM}
\begin{figure}[h!]
	\begin{center}
		\includegraphics[scale=0.75]{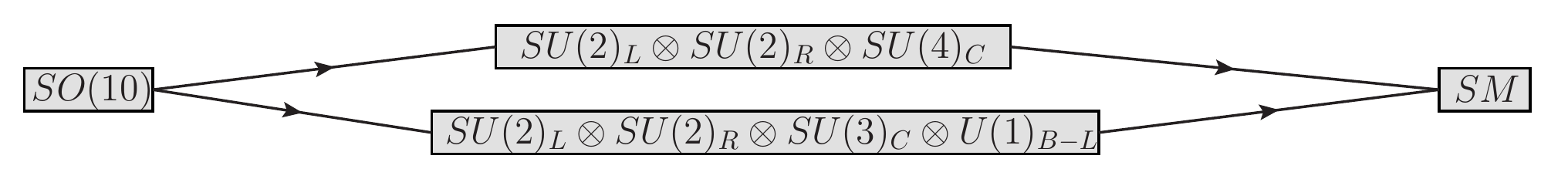}
	\end{center}\caption{Adopted one intermediate step breaking  of $SO(10)$ to the SM.}
	\label{SO10_breakings}
\end{figure}

$SO(10)$, whose rank is five and dimensionality of the adjoint representation is 45, is considered to be one of the favourite candidates for unification. Here we have considered all possible breaking patterns of $SO(10)$ to the SM through a single intermediate gauge group that includes $SU(N)_L\otimes SU(N)_R$ structure. These breaking patterns are all rank-conserving (see Fig.~\ref{SO10_breakings}). We stick to minimal field configurations, especially in scalar sectors. Using the novelty of extended survival hypothesis (ESH), we only make those sub-multiplets lighter, which participate in the process of symmetry breaking, including the electro-weak ones. Only these sub-multiplets participate in the evolution of the $\beta$-function. We have  illustrated the situation both in the presence and absence of D-parity. 
\subsubsection{ ${SO(10)\rightarrow \mathcal{G}_{2_L2_R4_C}\rightarrow  \mathcal{G}_{2_L1_Y3_C}}$}

$SO(10)$ spontaneously breaks to $SU(2)_L\otimes SU(2)_R\otimes SU(4)_C\equiv \mathcal{G}_{2_L 2_R 4_C}$ through the VEVs of  possible scalars $\Phi_{54}$, $\Phi_{210}$ and $\Phi_{770}$ which contain the sub-multiplet $(1,1,1)$ under $\mathcal{G}_{2_L 2_R 4_C}$. This ensures the presence of the desired intermediate symmetry. 
\begin{table}[h!]
	\small
	\renewcommand*{\arraystretch}{1.2}
	\begin{center}
		\begin{tabular}{ |c|c|c|c| }
			\hline
			\multirow{5}{4em}{Scalars} &
			$SO(10)$ & $\mathcal{G}_{224}$ & $\mathcal{G}_{213}$ \\
			\hline
			&$10$ & $(2,2,1)$ & $(2,\pm{\frac{1}{2}},1)$  \\
			&  $126$  & $(1,3,10)$ & -\\ 
			&  &${(3,1,\overline{10})}_D$ & - \\
			& $(54 , \ 770)_D$  & - & -\\
			& ${(210)}_\slashed{D}$ & - & -\\
			\hline
			\multirow{5}{4em}{Fermions}
			&${16}$ & $({2},{1},{4})$ & $(2,\frac{1}{6},3)$  \\
			&  &  & $(2,-\frac{1}{2},1)$\\ 
			& & $(1,2,\bar{4})$ & $(1,\frac{1}{3},\bar{3})$\\
			& & & $(1,-\frac{2}{3},\bar{3})$\\
			& & & $(1,1,1)$ \\
			& & & $(1,0,1)$ \\
			\hline
		\end{tabular}
		\caption{$SO(10)\rightarrow \mathcal{G}_{2_L2_R 4_C}\to \rm{SM}$}
		\label{SO10_G224_table}
	\end{center}
\end{table}
In Table~\ref{SO10_G224_table} we have listed the fermion and scalar representations which contribute to the  RGEs of the gauge couplings from $M_{R}$ to $M_X$.  The VEVs of $\Phi_{54}$ and $\Phi_{770}$ conserve D-parity, while  that of $\Phi_{210}$ does not. We have explicitly discussed both D-parity conserved and broken cases.

At the intermediate scale $M_{R}$, $SU(2)_R \otimes SU(4)_C$ is  spontaneously broken through the VEV of $(1,3,10)\subset 126$. Here $SU(2)_L$ remains unbroken, ensured by the singlet structure of $(1,3,10)$. The SM hyper-charge generator ($U(1)_Y$) is formed out of $SU(2)_R$ and $SU(4)_C$; $SU(3)_C$ pops out of the $SU(4)_C$ itself. This leads to the following matching conditions of the gauge couplings at the intermediate scale
\begin{align}
	\frac{1}{\alpha_{3C}(M_R)}&=\frac{1}{\alpha_{4C}(M_R)}-\frac{1}{12\pi},\\ 
	\frac{1}{\alpha_{1Y}(M_R)}&=\frac{3}{5}\left( \frac{1}{\alpha_{2R}(M_R)} - \frac{1}{6\pi} \right) + \frac{2}{5}\left( \frac{1}{\alpha_{4C}(M_R)} - \frac{1}{3\pi} \right),
\end{align}
where $\alpha_i=g_i^2/4\pi$.

We have computed the $\beta-$coefficients for the RGEs of the gauge couplings  from $M_{R}$ to $M_X$ scale upto two loop level  for Non-SUSY and SUSY cases respectively:

\underline{\textsl{D-parity not conserved}}\\
$ {\rm Non-SUSY}:\;\;\;\;\;\;
b_{2L} =-3, \; b_{4C} =-\frac{23}{3}, \; b_{2R} =\frac{11}{3}; \;\;\;
b_{ij}= 
\begin{pmatrix}
8 & \frac{45}{2} & 3 \\ 
\frac{9}{2} & \frac{643}{6} & \frac{153}{2} \\ 
3 & \frac{765}{2} & \frac{584}{3}
\end{pmatrix}.$
\\
${\rm SUSY}:\;\;\;\;\;\;
b_{2L} =1, \; b_{2R} =21, \; b_{4C} =3; \;\;\;
b_{ij}= 
\begin{pmatrix}
25 & 3 & 45 \\ 
3 & 265 & 405 \\ 
9 & 81 & 231
\end{pmatrix}.$

\underline{\textsl{D-parity conserved}}\\
$ {\rm Non-SUSY}:\;\;\;\;\;\; 
b_{2L} =\frac{11}{3}, \; b_{4C} =-\frac{14}{3}, \; b_{2R} =\frac{11}{3}; \;\;\;
b_{ij}= 
\begin{pmatrix}
\frac{584}{3} & \frac{765}{2} & 3 \\ 
\frac{153}{2} & \frac{1759}{6} & \frac{153}{2} \\ 
3 & \frac{765}{2} & \frac{584}{3}
\end{pmatrix}.
$
\\
${\rm SUSY}:\;\;\;\;\;\;
b_{2L} =21, \; b_{2R} =21, \; b_{4C} =12; \;\;\;
b_{ij}= 
\begin{pmatrix}
265 & 3 & 405 \\ 
3 & 265 & 405 \\ 
81 & 81 & 465
\end{pmatrix}.
$
%
\subsubsection{${SO(10)\rightarrow \mathcal{G}_{2_L2_R3_C1_{(B-L)}}\rightarrow  \mathcal{G}_{2_L1_Y3_C}}$}
$SO(10)$ can be spontaneously broken to $SU(2)_L\otimes SU(2)_R\otimes SU(3)_C \otimes U(1)_{B-L}$ $\equiv$ $\mathcal{G}_{2_L 2_R 3_C 1_{B-L}}$ through the VEVs of the possible scalars $\Phi_{45},\Phi_{210}$.  These two fields contain sub-multiplets $(1,1,1,0)$ under $\mathcal{G}_{2_L 2_R 3_C 1_{B-L}}$. One can also think of this possible breaking via $\mathcal{G}_{224}$, using the combined VEVs of these fields and the fields mentioned in earlier section. 
\begin{table}[h!]
	\small
	\renewcommand*{\arraystretch}{1.2}
	\begin{center}
		\begin{tabular}{ |c|c|c|c|}
			\hline
			&
			$SO(10)$ & $\mathcal{G}_{2231}$ & $\mathcal{G}_{213}$ \\
			\hline
			&$10$ & $(2,2,1,0)$ &$(2,\pm{\frac{1}{2}},1)$  \\
			&  $126$  & $(1,3,1,2)$  & -\\ 
			{{~Scalars}}& $(210)_D$ &-   & - \\
			& $(45)_\slashed{D}$ &-   & - \\
			\hline 
			&${16}$ & $(2,1,3,-\frac{1}{3})$  & $(2,\frac{1}{6},3)$  \\
			&  & $(2,1,1,1)$ & $(2,-\frac{1}{2},1)$\\ 
			& &$(1,2,\bar{3},\frac{1}{3})$ & $(1,\frac{1}{3},\bar{3})$\\
			& &$(1,2,1,-1)$  & $(1,-\frac{2}{3},\bar{3})$\\
			{{~Fermions}}&  & & $(1,1,1)$ \\
			&  &  & $(1,0,1)$ \\
			\hline
		\end{tabular}
		\caption{$SO(10)\rightarrow \mathcal{G}_{2_L2_R3_C1_{B-L}}\to \rm{SM}$}\label{SO10_G2231_table}
	\end{center}
\end{table}
All representations of fermions and scalars, which take part in the RG evolution of the gauge couplings from $M_{R}$ to $M_X$ scale and contribute to the respective $\beta-$coefficient computation, are tabulated in Table.~\ref{SO10_G2231_table}.  The VEVs of $\Phi_{210}$ and $\Phi_{45}$ conserve and break D-parity respectively. 

$SU(2)_R \otimes U(1)_{B-L}$ is broken spontaneously through the VEV of $(1,3,1,2)\subset 126$ at the intermediate scale $M_{R}$ and we find $ U(1)_Y$ as a remnant symmetry. Here, $SU(2)_L\otimes SU(3)_C$ remains unbroken, ensured by the singlet structure of $(1,3,1,2)$. The generator of $U(1)_Y$ is a linear combination of the generators of $U(1)_{B-L}$ and  $SU(2)_R$ at the intermediate scale $M_R$; this helps us write the matching condition at this scale: 
\begin{align}
	\frac{1}{\alpha_{1Y}(M_R)}=\frac{3}{5}\left( \frac{1}{\alpha_{2R}(M_R)} - \frac{1}{6\pi} \right) + \frac{2}{5}\left( \frac{1}{\alpha_{1(B-L)}(M_R)} \right)\,.
\end{align}

We have computed the $\beta-$coefficients  which are relevant for the running between $M_{R}$ and $M_X$ scales upto two loop level for both Non-SUSY and SUSY cases. These are listed below:

\underline{\textsl{D-parity not conserved}}\\
$ {\rm Non-SUSY}:\;\;\;
b_{2L} =-3, \; b_{3C} =-7, \; b_{1(B-L)} =\frac{11}{2}, \; b_{2R} =-\frac{7}{3}; \;\;\;
b_{ij}= 
\begin{pmatrix}
8 & 12 & \frac{3}{2} & 3 \\ 
\frac{9}{2} & -26 & \frac{1}{2} & \frac{9}{2} \\ 
\frac{9}{2} & 4 & \frac{61}{2} & \frac{81}{2} \\ 
3 & 12 & \frac{27}{2} & \frac{80}{3}
\end{pmatrix}.
$ 
\\
${\rm SUSY}:\;\;\;
b_{1(B-L)} =\frac{21}{2}, \; b_{2L} =1, \; b_{2R} =3, \; b_{3C} =-3; \;\;\;
b_{ij}= 
\begin{pmatrix}
34 & 9 & 45 & 8 \\ 
3 & 25 & 3 & 24 \\ 
15 & 3 & 49 & 24 \\ 
1 & 9 & 9 & 14
\end{pmatrix}.
$

\underline{\textsl{D-parity conserved}}\\
$ {\rm Non-SUSY}:\;\;\;
b_{2L} =-\frac{7}{3}, \; b_{3C} =-7, \; b_{1(B-L)} =7, \; b_{2R} =-\frac{7}{3}; \;\;\;
b_{ij}= 
\begin{pmatrix}
\frac{80}{3} & 12 & \frac{27}{2} & 3 \\ 
\frac{9}{2} & -26 & \frac{1}{2} & \frac{9}{2} \\ 
\frac{81}{2} & 4 & \frac{115}{2} & \frac{81}{2} \\ 
3 & 12 & \frac{27}{2} & \frac{80}{3}
\end{pmatrix}.
$
\\
${\rm SUSY}:\;\;\;
b_{1(B-L)} =15, \; b_{2L} =3, \; b_{2R} =3, \; b_{3C} =-3; \;\;\;
b_{ij}= 
\begin{pmatrix}
61 & 45 & 45 & 8 \\ 
15 & 49 & 3 & 24 \\ 
15 & 3 & 49 & 24 \\ 
1 & 9 & 9 & 14
\end{pmatrix}.
$

\subsection{Breaking of $E(6)$ to SM: $E(6) \xrightarrow{M_X}  \mathcal{G}_{int} \xrightarrow{M_{R}}  \mathcal{G}_{2_L1_Y3_C}$}
\label{subsec:e6-SM}
$E(6)$, one of the exceptional groups qualified to be a valid candidate for unification, is of rank six and has 78 as the dimensionality of adjoint representation. $E(6)$ contains $SO(10)$ as its subgroup. Thus it is expected that most of the features of $SO(10)$ can be realised within a more constrained system in $E(6)$ due to enhanced symmetry.
\begin{figure}[htb!]
	\begin{center}
		\includegraphics[scale=0.75]{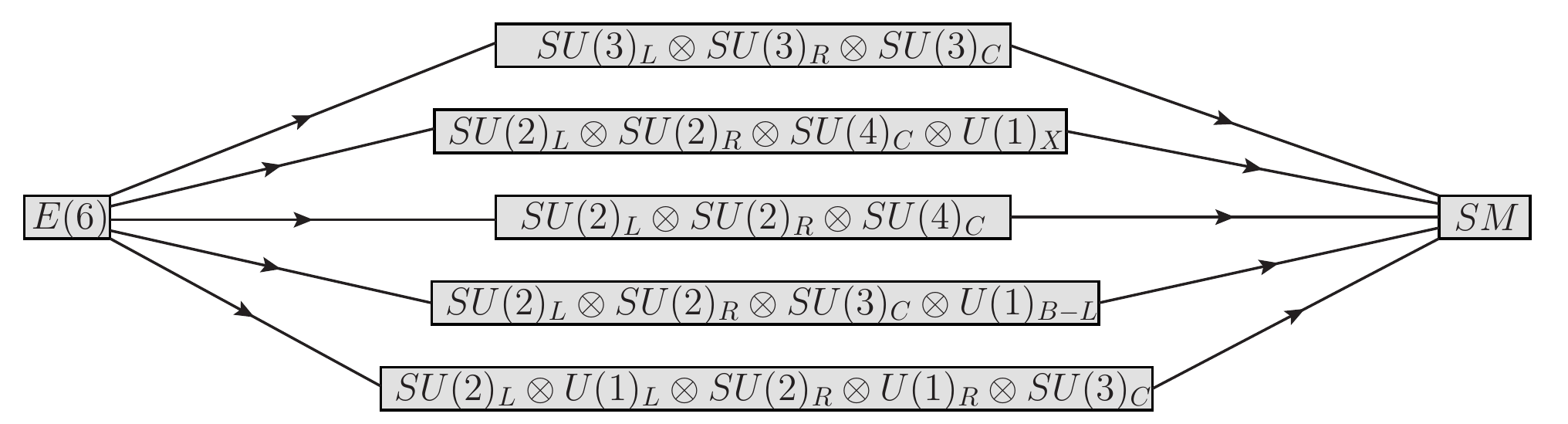}
	\end{center}\caption{Adopted one intermediate step breaking of $E(6)$ to the SM.}\label{E6_breakings}
\end{figure}
Here we have considered  all possible breaking patterns of $E(6)$ to the SM through one intermediate symmetry group that includes $SU(N)_L\otimes SU(N)_R$ structure. We have included rank-conserving as well as rank-reducing breaking in the process; see Fig.~\ref{E6_breakings}. We stick to the minimal field configurations, especially in scalar sectors. Using the virtue of ESH, we have incorporated only those sub-multiplets, which  participate in the process of symmetry breaking, including the electro-weak ones. We discuss the impact of D-parity and its breaking on the RGEs in this analysis. 
\subsubsection{$E(6) \xrightarrow{M_X}  \mathcal{G}_{3_L 3_R 3_c} \xrightarrow{M_{R}}   \mathcal{G}_{2_L1_Y3_C}$}
$E(6)$ spontaneously breaks to $SU(3)_L\otimes SU(3)_R\otimes SU(3)_C\equiv \mathcal{G}_{3_L 3_R 3_c}$  through the VEVs of three possible scalars $\Phi_{650}$, $\Phi_{2430}$, and $\Phi_{650^{'}}$ as they contain the $(1,1,1)$ sub-multiplet  under the intermediate symmetry  $\mathcal{G}_{3_L 3_R 3_c}$. 
\begin{table}[h!]
	\begin{center}\small
		\renewcommand*{\arraystretch}{1.2}
		\begin{tabular}{|c|c|c|c|}
			\hline
			& E(6) & $\mathcal{G}_{3_L 3_R 3_c}$  & $\rm{SM}$ \\
			\hline
			& 27 & $( 3,\bar{3},1)$ & $(2,\pm{\frac{1}{2}},1)$ \\
			{{~Scalars}}& $(650, \ 2430)_D$ & - & - \\ 
			& $(650^\prime)_\slashed{D}$ & - & - \\
			\hline
			& 27 & $(3,\bar{3},1)$ &$(2,\frac{1}{6},3)$ \\ 
			&  & $(3,1,3)$ &  $(2,-\frac{1}{2},1)$  \\ 
			& & $(1,\bar{3},\bar{3})$  & $(1,\frac{1}{3},\bar{3})$  \\
			& &   & $(1,-\frac{2}{3},\bar{3})$  \\ 
			{{~Fermions}} & &  & $(1,1,1)$ \\
			& & & $(1,0,1)$ \\
			\hline
		\end{tabular}
		\caption{$E(6)\to \mathcal{G}_{3_L 3_R 3_c} \to \rm{SM}$ }
		\label{chain_I_1}
	\end{center}
\end{table}
We have tabulated the  fermion and scalar fields that contribute in the $\beta-$coefficient computation for RGEs between $M_{R}$ and $M_X$ scale; see Table.~\ref{chain_I_1}.  VEVs of $\Phi_{650}$ and $\Phi_{2430}$ conserve D-parity and that of $\Phi_{650^{'}}$ does not. 

$SU(3)_L\otimes SU(3)_R$ is broken spontaneously through the VEV of $( 3,\bar{3},1)\subset27$ at the intermediate scale $M_{R}$. Here $SU(3)_C$ remains unbroken, which is ensured by the color singlet structure of $( 3,\bar{3},1)$. 
Following the convention of earlier sections, here we list the matching conditions at the intermediate scale:
\begin{align}
	\left( \frac{1}{\alpha_{2L}(M_R)} - \frac{1}{6\pi} \right)=& \left(\frac{1}{\alpha_{3L}(M_R)} - \frac{1}{4\pi} \right),
	\\
	\frac{1}{\alpha_{1Y}(M_R)}=& \frac{1}{5}\left( \frac{1}{\alpha_{3L}(M_R)} - \frac{1}{4\pi} \right) + \frac{4}{5}\left( \frac{1}{\alpha_{3R}(M_R)} - \frac{1}{4\pi} \right).
\end{align}
The relevant $\beta-$coefficients  for running of gauge coupling between $M_{R}$ and $M_X$ scale upto two loop level  for Non-SUSY and SUSY scenarios are given as: \\
$ {\rm Non-SUSY}:\;\;\;\;\;\;
b_{3C} =-5, \; b_{3R} =-\frac{9}{2}, \; b_{3L} =-\frac{9}{2}; \;\;\;\;
b_{ij}= 
\begin{pmatrix}
12 & 12 & 12 \\ 
12 & 23 & 20 \\ 
12 & 20 & 23
\end{pmatrix}.
$
\\
${\rm SUSY}:\;\;\;\;\;\;
b_{3L} =\frac{3}{2}, \; b_{3R} =\frac{3}{2}, \; b_{3C} =0; \;\;\;
b_{ij}= 
\begin{pmatrix}
65 & 32 & 24 \\ 
32 & 65 & 24 \\ 
24 & 24 & 48
\end{pmatrix}.
$

\subsubsection{$E(6)\to \mathcal{G}_{2_L 2_R  4_C 1_X} \to  \mathcal{G}_{2_L1_Y3_C}$}
$E(6)$ can also be broken to $SU(2)_L \otimes SU(2)_R \otimes SU(4)_C\otimes U(1)_X\equiv \mathcal{G}_{2_L 2_R  4_C 1_X}$ spontaneously through the VEV of scalar $\Phi_{650} \supset (1,1,1,0)$. 
\begin{table}[h!]
	\small
	\renewcommand*{\arraystretch}{1.2}
	\begin{center}
		\begin{tabular}{| c | c | c | c |}
			\hline
			& E(6) &$\mathcal{G}_{2241}$&  $\rm{SM}$ \\ \hline
			& 27 &  $(2,2,1,-2)$ & $(2,\pm{\frac{1}{2}},1)$ \\ 
			& $\overline{351^\prime}$  & $(1,3,10,2)$  & - \\  
			{{~Scalars}} & & $(3,1,\overline{10},-2)_D$  &- \\
			&  $(650)_{D,\slashed{D}}$  & - &- \\ 
			\hline
			& $27_F$ & $(2,2,1,-2)$ &$(2,\frac{1}{6},3)$  \\ 
			&   & $(2,1,\bar{4},1)$ & $(2,-\frac{1}{2},1)$ \\ 
			&   & $(1,1,1,4)$ &  $(1,\frac{1}{3},\bar{3})$ \\     
			{{~Fermions}} &  & $(1,2,4,1)$ & $(1,-\frac{2}{3},\bar{3})$ \\ 
			&   & $(1,1,6,-2)$ & $(1,1,1)$  \\   
			&&& $(1,0,1)$  \\
			\hline   
		\end{tabular}
		\caption{$E(6)\to \mathcal{G}_{2_L2_R4_C1_X}\to \rm{SM}$}
		\label{E6_G2241_table}
	\end{center}
\end{table}

The contributory scalar and fermion fields are given in  Table.~\ref{E6_G2241_table}. It is worthy to mention that $\Phi_{650}$ contains sub-multiplets which conserve as well as break D-parity and which can be realised in terms of the $SO(10)\otimes U(1)$ representations as $(54,0)$ and $(210,0)$ respectively.

At the intermediate scale $M_{R}$, $SU(2)_R \otimes SU(4)_C\otimes U(1)_X$ is broken spontaneously through the VEV of $( 1,3,10,2)\subset \overline{351'}$.
Here we have constructed the $U(1)_Y$ charges using normalised  $SU(4)_C$ and $U(1)_X$ quantum numbers.
Thus the necessary matching conditions at this scale read:
\begin{align}
	\frac{1}{\alpha_{3C}(M_R)}&=\frac{1}{\alpha_{4C}(M_R)}-\frac{1}{12\pi},
	\\
	\frac{1}{\alpha_{1Y}(M_R)}&=\frac{9}{10}\left( \frac{1}{\alpha_{4C}(M_R)} - \frac{1}{6\pi} \right)  + \frac{1}{10}\left( \frac{1}{\alpha_{1X}(M_R)} \right).
\end{align}	
Here, we would like to mention as a side-note that if $U(1)_X$ is just a mere spectator, then this case is very similar to $\mathcal{G}_{2_L2_R4_C}$, i.e.,
\begin{align*}
	\frac{1}{\alpha_{1Y}(M_R)}=\frac{3}{5}\left( \frac{1}{\alpha_{2R}(M_R)} - \frac{1}{6\pi} \right) + \frac{2}{5}\left( \frac{1}{\alpha_{4C}(M_R)} - \frac{1}{3\pi} \right)\,.
\end{align*}
The new set of  $\beta-$coefficients  upto two loop level  for Non-SUSY and SUSY cases are given as:\\
{\bf \underline{All 27 fermions present at $M_R$:-}}

\underline{\textsl{D-parity not conserved}}\\
$ {\rm NON-SUSY}:\;\;\;
b_{2L} =-1, \; b_{1X} =\frac{71}{9}, \; b_{4C} =-\frac{17}{3}, \; b_{2R} =\frac{17}{3}; \;\;\; 
b_{ij}= 
\begin{pmatrix}
\frac{65}{2} & \frac{13}{6} & \frac{45}{2} & \frac{15}{2} \\ 
\frac{13}{2} & \frac{149}{18} & \frac{225}{2} & \frac{93}{2} \\ 
\frac{9}{2} & \frac{15}{2} & \frac{973}{6} & \frac{153}{2} \\ 
\frac{15}{2} & \frac{31}{2} & \frac{765}{2} & \frac{1315}{6}
\end{pmatrix}.
$
\\
${\rm SUSY}:\;\;\;
b_{1X} =\frac{44}{3}, \; b_{2L} =4, \; b_{2R} =24, \; b_{4C} =6; \;\;\;
b_{ij}= 
\begin{pmatrix}
\frac{115}{9} & 11 & 51 & 135 \\ 
\frac{11}{3} & 46 & 12 & 45 \\ 
17 & 12 & 286 & 405 \\ 
9 & 9 & 81 & 285
\end{pmatrix}.
$

\underline{\textsl{D-parity conserved}}\\
${\rm NON-SUSY}:\;\;\;
b_{2L} =\frac{17}{3}, \; b_{1X} =\frac{86}{9}, \; b_{4C} =-\frac{8}{3}, \; b_{2R} =\frac{17}{3}; \;\;\;
b_{ij}= 
\begin{pmatrix}
\frac{1315}{6} & \frac{31}{2} & \frac{765}{2} & \frac{15}{2} \\ 
\frac{93}{2} & \frac{209}{18} & \frac{405}{2} & \frac{93}{2} \\ 
\frac{153}{2} & \frac{27}{2} & \frac{2089}{6} & \frac{153}{2} \\ 
\frac{15}{2} & \frac{31}{2} & \frac{765}{2} & \frac{1315}{6}
\end{pmatrix}.
$
\\
${\rm SUSY}:\;\;\;
b_{1X} =\frac{59}{3}, \; b_{2L} =24, \; b_{2R} =24, \; b_{4C} =15; \;\;\;
b_{ij}= 
\begin{pmatrix}
\frac{145}{9} & 51 & 51 & 225 \\ 
17 & 286 & 12 & 405 \\ 
17 & 12 & 286 & 405 \\ 
15 & 81 & 81 & 519
\end{pmatrix}.
$

{\bf \underline{Only 16 fermions present at $M_R$:-}}\\

\underline{\textsl{D-parity not conserved}}\\
${\rm NON-SUSY}:\;\;\;b_{2L} =-3, \; b_{1X} =\frac{29}{9}, \; b_{4C} =-\frac{23}{3}, \; b_{2R} =\frac{11}{3};  \;\;\; b_{ij}= 
\begin{pmatrix}
8 & \frac{7}{6} & \frac{45}{2} & 3 \\ 
\frac{7}{2} & \frac{71}{18} & \frac{195}{2} & \frac{87}{2} \\ 
\frac{9}{2} & \frac{13}{2} & \frac{643}{6} & \frac{153}{2} \\ 
3 & \frac{29}{2} & \frac{765}{2} & \frac{584}{3}
\end{pmatrix}.
$
\\
${\rm SUSY}:\;\;\;b_{1X} =\frac{23}{3}, \; b_{2L} =1, \; b_{2R} =21, \; b_{4C} =3; \; \; \; 
b_{ij}= 
\begin{pmatrix}
\frac{37}{9} & 5 & 45 & 105 \\ 
\frac{5}{3} & 25 & 3 & 45 \\ 
15 & 3 & 265 & 405 \\ 
7 & 9 & 81 & 231
\end{pmatrix}.
$

\underline{\textsl{D-parity conserved}}\\
${\rm NON-SUSY}:\;\;\;b_{2L} =\frac{11}{3}, \; b_{1X} =\frac{44}{9}, \; b_{4C} =-\frac{14}{3}, \; b_{2R} =\frac{11}{3};  \;\;\; b_{ij}= 
\begin{pmatrix}
\frac{584}{3} & \frac{29}{2} & \frac{765}{2} & 3 \\ 
\frac{87}{2} & \frac{131}{18} & \frac{375}{2} & \frac{87}{2} \\ 
\frac{153}{2} & \frac{25}{2} & \frac{1759}{6} & \frac{153}{2} \\ 
3 & \frac{29}{2} & \frac{765}{2} & \frac{584}{3}
\end{pmatrix}.
$
\\
${\rm SUSY}:\;\;\;b_{1X} =\frac{38}{3}, \; b_{2L} =21, \; b_{2R} =21, \; b_{4C} =12; \; \; \; b_{ij}= 
\begin{pmatrix}
\frac{67}{9} & 45 & 45 & 195 \\ 
15 & 265 & 3 & 405 \\ 
15 & 3 & 265 & 405 \\ 
13 & 81 & 81 & 465
\end{pmatrix}.
$
\subsubsection{$E(6)\to \mathcal{G}_{2_L 2_R 4_C } \to   \mathcal{G}_{2_L1_Y3_C}$}
$E(6)$ may posses a rank-reducing breaking, leading to $SU(2)_L\otimes SU(2)_R\otimes SU(4)_C\equiv \mathcal{G}_{2_L 2_R 4_C }$  through VEVs of 
$\Phi_{\overline{351^{\prime}}}$, and $\Phi_{1728}$, which contain sub-multiplet $(1,1,1)$ under $\mathcal{G}_{2_L 2_R 4_C}$.  
\begin{table}[htb!]
	\centering
	\small
	\renewcommand*{\arraystretch}{1.2}
	\begin{tabular}{| c | c | c | c |}
		\hline
		& E(6) & $ \mathcal{G}_{224}$ & $\mathcal{G}_{2_L1_Y3_C}$ \\ \hline
		& 27 & $(2,2,1)$ & $(2,\pm{\frac{1}{2}},1)$ \\ 
		& $(\overline{351^\prime})_D$ & $(1,3,10)$ & - \\ 
		{{~Scalars}}
		&   & ${(3,1,\overline{10})}_D$ &  \\
		&$(1728)_\slashed{D}$&-&-\\
		\hline
		& $27_F$  & $(2,1,\bar{4})$ &$(2,\frac{1}{6},3)$  \\ 
		&  &  $(1,2,4)$ & $(2,-\frac{1}{2},1)$  \\ 
		&  & $(2,\bar{2},1)$ & $(1,\frac{1}{3},\bar{3})$ \\     
		{{~Fermions}} &  & $(1,1,6)$ & $(1,-\frac{2}{3},\bar{3})$ \\ 
		&  & $(1,1,1)$  & $(1,1,1)$  \\   
		&&& $(1,0,1)$  \\
		\hline   
	\end{tabular}
	\caption{$E(6)\rightarrow \mathcal{G}_{2_L2_R 4_C}\to \rm{SM}$}
	\label{E6_G224_table}
\end{table}
We have listed the  fermion and scalar representations,  which participate in the RGEs of the gauge couplings from $M_{R}$ to $M_X$ scale in Table.~\ref{E6_G224_table}.  Here the VEV of $\Phi_{\overline{351^\prime}}$ conserves D-parity while that of $\Phi_{1728}$ does not.  

At $M_R$, $SU(2)_R \otimes SU(4)_C$ is broken spontaneously through the VEV of $(1,3,10)\subset \overline{351^{\prime}}$ to $ SU(3)_C \otimes U(1)_Y$. 
This leads to the  following matching conditions:
\begin{align}
	\frac{1}{\alpha_{3C}(M_R)}&=\frac{1}{\alpha_{4C}(M_R)}-\frac{1}{12\pi},
	\\ 
	\frac{1}{\alpha_{1Y}(M_R)}&=\frac{3}{5}\left( \frac{1}{\alpha_{2R}(M_R)} - \frac{1}{6\pi} \right) + \frac{2}{5}\left( \frac{1}{\alpha_{4C}(M_R)} - \frac{1}{3\pi} \right).
\end{align}
The relevant two loop $\beta-$coefficients for Non-SUSY and SUSY scenarios are given as:

\underline{\textsl{D-parity not conserved}}\\
${\rm Non-SUSY}:\;\;\;\;\;\;
b_{2L} =-1, \; b_{4C} =-\frac{17}{3}, \; b_{2R} =\frac{17}{3}; \;\;\;
b_{ij}= 
\begin{pmatrix}
\frac{65}{2} & \frac{45}{2} & \frac{15}{2} \\ 
\frac{9}{2} & \frac{973}{6} & \frac{153}{2} \\ 
\frac{15}{2} & \frac{765}{2} & \frac{1315}{6}
\end{pmatrix}.
$
\\
${\rm SUSY}:\;\;\;\;\;\;
b_{2L} =4, \; b_{2R} =24, \; b_{4C} =6; \;\;\;
b_{ij}= \begin{pmatrix}
46 & 12 & 45 \\ 
12 & 286 & 405 \\ 
9 & 81 & 285
\end{pmatrix}.
$

\underline{\textsl{D-parity conserved}}\\
${\rm Non-SUSY}:\;\;\;\;\;\;
b_{2L} =\frac{17}{3}, \; b_{4C} =-\frac{8}{3}, \; b_{2R} =\frac{17}{3}; \;\;\; 
b_{ij}= 
\begin{pmatrix}
\frac{1315}{6} & \frac{765}{2} & \frac{15}{2} \\ 
\frac{153}{2} & \frac{2089}{6} & \frac{153}{2} \\ 
\frac{15}{2} & \frac{765}{2} & \frac{1315}{6}
\end{pmatrix}.
$
\\
${\rm SUSY}:\;\;\;\;\;\;
b_{2L} =24, \; b_{2R} =24, \; b_{4C} =15; \;\;\;
b_{ij}= 
\begin{pmatrix}
286 & 12 & 405 \\ 
12 & 286 & 405 \\ 
81 & 81 & 519
\end{pmatrix}.
$
			
\subsubsection{$E(6)\to \mathcal{G}_{2_L 2_R 3_C 1_{B-L}} \to   \mathcal{G}_{2_L1_Y3_C}$}
$E(6)$ is spontaneously broken to $SU(2)_L\otimes SU(2)_R\otimes SU(3)_C \otimes U(1)_{B-L}\equiv \mathcal{G}_{2_L 2_R 3_C 1_{B-L}}$ through the VEVs of
$\Phi_{351}$ and $\Phi_{1728}$ that contain the sub-multiplet $(1,1,1,0)$ under $\mathcal{G}_{2_L 2_R 3_C 1_{B-L}}$. 
To have a clearer picture of this breaking, we have provided the detailed embedding of sub-multiplets under $SO(10)\otimes U(1)_X \supset G_{224}$ :
\begin{equation*}
	\Phi_{351} : \ (45,4) \supset (1,1,15) , \
	\Phi_{1728} : \ (210,4) \supset (1,1,15)  , \  \ (45,4) \supset (1,1,15) \,.
\end{equation*}

\begin{table}[h!]
	\small
	\renewcommand*{\arraystretch}{1.2}
	\begin{center}
		\begin{tabular}{| c | c | c | c |}
			\hline
			& E(6) & $ \mathcal{G}_{2231}$ & $\rm{SM}$ \\ \hline
			& 27 & $(2,2,1,0)$ & $(2,\pm{\frac{1}{2}},1)$ \\ 
			& $\overline{351^\prime}$ & $(1,3,1,2)$ & - \\ 
			{{~Scalars}}&& $(3,1,1,-2)_D$ &-\\
			& $(1728)_D$ & - & - \\
			& $(351 , \ 1728)_\slashed{D}$ & - & - \\
			\hline
			& $27_F$  & $(2,1,1,-1)$ &$(2,\frac{1}{6},3)$  \\ 
			&  & $(2,1,\bar{3},\frac{1}{3})$ & $(2,-\frac{1}{2},1)$  \\ 
			&  & $(1,2,1,1)$ & $(1,\frac{1}{3},\bar{3})$ \\     
			{{~Fermions}} &  & $(1,2,3,-\frac{1}{3})$ & $(1,-\frac{2}{3},\bar{3})$ \\ 
			&  & $(1, 1, 1,0)$  & $(1,1,1)$  \\   
			&& $(1, 1, 3,\frac{2}{3})$ & $(1,0,1)$  \\
			&& $(1, 1, \bar{3} , -\frac{2}{3})$ & \\
			&& $(2,\bar{2},1,0)$ & \\
			\hline   
		\end{tabular}
		\caption{$E(6)\rightarrow \mathcal{G}_{2_L2_R3_C1_X}\to \rm{SM}$}
		\label{E6_G2231_table}
	\end{center}
\end{table}
The scalar and fermion fields that are relevant for RG computation are listed in  Table.~\ref{E6_G2231_table}. Here the scalar  $\Phi_{1728}$ contains sub-multiplets whose VEVs  conserve as well as break D-parity. On the other hand, the VEV of $\Phi_{351}$ breaks D-parity.

At $M_{R}$, $SU(2)_R \otimes U(1)_{B-L}$ is broken spontaneously to $ U(1)_Y$ through the VEV of $(1,3,1,2)\subset \overline{351^{'}}$. The suitable matching condition is:
\begin{align}
	\frac{1}{\alpha_{1Y}(M_R)}=\frac{3}{5}\left( \frac{1}{\alpha_{2R}(M_R)} - \frac{1}{6\pi} \right) + \frac{2}{5}\left( \frac{1}{\alpha_{1(B-L)}(M_R)} \right).
\end{align}
The $\beta-$coefficients upto two loop level  for Non-SUSY and SUSY cases are:

\underline{\textsl{D-parity not conserved}}\\
$ {\rm Non-SUSY}:\;\;\;
b_{2L} =-1, \; b_{3C} =-5, \; b_{1(B-L)} =\frac{15}{2}, \; b_{2R} =-\frac{1}{3}; \;\; \;
b_{ij}= 
\begin{pmatrix}
\frac{65}{2} & 12 & \frac{3}{2} & \frac{15}{2} \\ 
\frac{9}{2} & 12 & \frac{3}{2} & \frac{9}{2} \\ 
\frac{9}{2} & 12 & \frac{63}{2} & \frac{81}{2} \\ 
\frac{15}{2} & 12 & \frac{27}{2} & \frac{307}{6}
\end{pmatrix}.
$
\\
${\rm SUSY}:\;\;\;
b_{1(B-L)} =\frac{27}{2}, \; b_{2L} =4, \; b_{2R} =6, \; b_{3C} =0; \;\;\;
b_{ij}= 
\begin{pmatrix}
36 & 9 & 45 & 24 \\ 
3 & 46 & 12 & 24 \\ 
15 & 12 & 70 & 24 \\ 
3 & 9 & 9 & 48
\end{pmatrix}.
$

\underline{\textsl{D-parity conserved}}\\
$ {\rm Non-SUSY}:\;\;\;
b_{2L} =-\frac{1}{3}, \; b_{3C} =-5, \; b_{1(B-L)} =9, \; b_{2R} =-\frac{1}{3}; \;\;\; 
b_{ij}= 
\begin{pmatrix}
\frac{307}{6} & 12 & \frac{27}{2} & \frac{15}{2} \\ 
\frac{9}{2} & 12 & \frac{3}{2} & \frac{9}{2} \\ 
\frac{81}{2} & 12 & \frac{117}{2} & \frac{81}{2} \\ 
\frac{15}{2} & 12 & \frac{27}{2} & \frac{307}{6}
\end{pmatrix}.
$ 
\\
${\rm SUSY}:\;\;\;
b_{1(B-L)} =18, \; b_{2L} =6, \; b_{2R} =6, \; b_{3C} =0; \;\;\;
b_{ij}= 
\begin{pmatrix}
63 & 45 & 45 & 24 \\ 
15 & 70 & 12 & 24 \\ 
15 & 12 & 70 & 24 \\ 
3 & 9 & 9 & 48
\end{pmatrix}.
$
\subsubsection{$E(6)\to \mathcal{G}_{2_L 1_L 2_R 1_R 3_c} \to  \mathcal{G}_{2_L1_Y3_C}$}
$E(6)$ can be broken to $SU(2)_L\otimes U(1)_L \otimes SU(2)_R \otimes U(1)_R\otimes SU(3)_C\equiv \mathcal{G}_{2_L 1_L 2_R 1_R 3_c}$  through the VEV of 
$\Phi_{650}$, which contains the sub-multiplet $(1,0,1,0,1)$ under $\mathcal{G}_{2_L 1_L 2_R 1_R 3_c}$. 
\begin{table}[htb!]
	\small
	\renewcommand*{\arraystretch}{1.2}
	\begin{center}
		\begin{tabular}{| c | c | c | c |}
			\hline
			& E(6) & $ \mathcal{G}_{2_L 1_L 2_R 1_R 3_c} $ & $\rm{SM}$ \\
			\hline
			& 27 & $ (2,-1,2,1,1) $ & $(2,\pm{\frac{1}{2}},1)$ \\ 
			&  & $ (1,2,2,1,1) $ & - \\ 
			{{~Scalars}} & &  ${(2,-2,1,-1,1)}_D $ &  \\ 
			& $(650)_D$ & - & - \\ 
			& $(650^\prime)_\slashed{D}$ & - & - \\
			\hline
			& 27 & $ (2,-1,2,1,1) $ & $(2,\frac{1}{6},3)$ \\ 
			&  & $ (1,2,2,1,1) $ & $(2,-\frac{1}{2},1)$  \\ 
			
			&  & $ (2,-1,1,-2,1) $ & $(1,\frac{1}{3},\bar{3})$  \\
			&  & $ (1,2,1,-2,1) $ & $(1,-\frac{2}{3},\bar{3})$  \\ 
			
			{{~Fermions}} &  & $ (2,-1,1,0,3) $ & $(1,1,1)$ \\
			& & $ (1,2,1,0,3) $ & $(1,0,1)$ \\
			& & $ (1,0,2,1,\bar{3}) $ & \\
			& & $ (1,0,1,-2,\bar{3}) $ & \\
			\hline
		\end{tabular}
		\caption{$E(6)\to \mathcal{G}_{2_L 1_L 2_R 1_R 3_c} \to \rm{SM}$ }
		\label{E6_G21213_table}
	\end{center}
\end{table}
We have provided the representations of fermions and scalars which are relevant for $\beta-$function computation in Table.~\ref{E6_G21213_table}. 
At  $M_{R}$, $U(1)_L \otimes SU(2)_R\otimes U(1)_R$ is broken to $U(1)_Y$  using  the VEV of $( 1,2,2,1,1)\subset27$.
We have the following matching condition for this breaking pattern, at $M_R$:
\begin{align}
	\frac{1}{\alpha_{1Y}(M_R)}=\frac{3}{5}\left( \frac{1}{\alpha_{2R}(M_R)} - \frac{1}{6\pi} \right) +4\pi Q(gg^T)^{-1}Q^T,
\end{align}
where
\begin{align*}
	Q=\left(\sqrt{\frac{1}{5}},\sqrt{\frac{1}{5}}\right) \  \text{and} \ g = \begin{pmatrix}
		g_{LL} & g_{LR} \\
		g_{RL} & g_{RR}
	\end{pmatrix}.
\end{align*}
As we have two abelian gauge groups here, we need to include the effects of abelian mixing while computing RGEs. The necessary $\beta-$coefficients employed between $M_{R}$ and $M_X$ scales are computed  upto two loop level  for Non-SUSY and SUSY cases respectively for different scenarios. Here we have provided only one-loop $\beta$-coefficients, and those for two loop are given in the appendix.\\
\\
{\bf \underline{All 27 fermions present at $M_R$:-}}\\
\\
{\bf D-parity not conserved:}\\
\underline{\textit{Non-SUSY:}}
{\small
	\begin{align*}
		(4\pi)^2\beta_{2L} = 
		&- \;g_{2L}^{3}\\
		(4\pi)^2\beta_{3C} = 
		&- 5 \;g_{3C}^{3}\\
		(4\pi)^2\beta_{2R} = 
		&- \frac{5 \;g_{2R}^{3}}{6}\\
		(4\pi)^2\beta_{LL} = 
		&\frac{19 }{3}(\;g_{LL}^{3} +  \;g_{LL} \;g_{LR}^{2}) + \frac{37}{6}( \;g_{RL}^{2}\;g_{LL} +  \;g_{RR} \;g_{LR} \;g_{RL})\\						(4\pi)^2\beta_{LR} = 
		&\frac{19}{3}( \;g_{LR}^{3}+ \;g_{LL}^{2} \;g_{LR})  + \frac{37}{6}( \;g_{RR}^{2} \;g_{LR} +  \;g_{LL} \;g_{RR} \;g_{RL}) \\						(4\pi)^2\beta_{RL} = 
		&\frac{19 }{3} (\;g_{LL}^{2} \;g_{RL} +  \;g_{LL} \;g_{RR} \;g_{LR})   + \frac{37}{6}( \;g_{RL}^{3}+ \;g_{RR}^{2} \;g_{RL})\\						(4\pi)^2\beta_{RR} = 
		&\frac{19}{3} (\;g_{LR}^{2} \;g_{RR} + \;g_{LL}{3} \;g_{LR} \;g_{RL}) + \frac{37}{6}( \;g_{RR}^{3} + \;g_{RL}^{2}\;g_{RR})
	\end{align*}
}
\underline{\textit{SUSY:}}
{\small
	\begin{align*} 
		(4\pi)^2{\beta_{2L}}& =4 \;g_{2 L}^3 \\ 
		(4\pi)^2{\beta_{2R}}& =\frac{9 \;g_{2 R}^3}{2} \\ 
		(4\pi)^2{\beta_{3C}}& =0 \\
		(4\pi)^2{\beta_{LL}}& =10( \;g_{LL}^3+ \;g_{LR}^2 \;g_{LL}) +\frac{19}{2}(  \;g_{RL}^2 \;g_{LL}+ \;g_{LR} \;g_{RL} \;g_{RR}) \\ 
		(4\pi)^2{\beta_{LR}}& =10( \;g_{LR}^3+ \;g_{LL}^2 \;g_{LR}) +\frac{19}{2}(  \;g_{RR}^2 \;g_{LR}+ \;g_{LL} \;g_{RL} \;g_{RR}) \\ 
		(4\pi)^2{\beta_{RL}}& = 10 (\;g_{LL}^2 \;g_{RL}+\;g_{LL} \;g_{LR} \;g_{RR})+\frac{19}{2}( \;g_{RL}^3+\;g_{RR}^2 \;g_{RL})  \\ 
		(4\pi)^2{\beta_{RR}}& =10( \;g_{LR}^2 \;g_{RR}+ \;g_{LL} \;g_{LR} \;g_{RL})+\frac{19}{2}(\;g_{RR}^3 +\;g_{RL}^2 \;g_{RR}) 
\end{align*}}

{\bf D-parity conserved}\\
\underline{\textit{Non-SUSY:}}
{\small
	\begin{align*}
		(4\pi)^2\beta_{2L} = 
		&- \frac{5 \;g_{2L}^{3}}{6}\\
		(4\pi)^2\beta_{3C} = 
		&- 5 \;g_{3C}^{3}\\
		(4\pi)^2\beta_{2R} = 
		&- \frac{5 \;g_{2R}^{3}}{6}\\
		(4\pi)^2\beta_{LL} = 
		&\frac{59}{9}( \;g_{LL}^{3} + \;g_{LR}^{2}\;g_{LL})					 + \frac{56}{9}( \;g_{LL} \;g_{RL}^{2} + \;g_{RR} \;g_{LR} \;g_{RL}) \\
		& +\frac{1}{9}(2 \;g_{LL}^{2} \;g_{RL}+\;g_{LR}^{2}\;g_{RL}+\;g_{LL} \;g_{RR} \;g_{LR})
		\\ 						(4\pi)^2\beta_{LR} = & \frac{59}{9}(\;g_{LR}^{3}+\;g_{LL}^{2}\;g_{LR}) + \frac{56}{9}(\;g_{RR}^{2}\;g_{LR}+\;g_{LL} \;g_{RR} \;g_{RL}) \\
		& + \frac{1}{9}(2 \;g_{LR}^{2} \;g_{RR} + \;g_{LL}^{2} \;g_{RR} +\;g_{LL} \;g_{LR} \;g_{RL} )
		\\
		(4\pi)^2\beta_{RL} = & \frac{59}{9}(\;g_{LL}^{2} \;g_{RL}+\;g_{LL} \;g_{RR} \;g_{LR}) + \frac{56}{9}(\;g_{RL}^{3}+\;g_{RR}^{2} \;g_{RL})\\
		& + \frac{1}{9}(2 \;g_{RL}^{2} \;g_{LL}+ \;g_{RR}^{2}\;g_{LL}+\;g_{RR} \;g_{LR} \;g_{RL})
		\\	(4\pi)^2\beta_{RR} = & \frac{59}{9}(\;g_{LR}^{2}\;g_{RR}+\;g_{LL} \;g_{LR} \;g_{RL})+\frac{56}{9}(\;g_{RR}^{3}+\;g_{RL}^{2}\;g_{RR}) \\
		&+\frac{1}{9}(2 \;g_{RR}^{2} \;g_{LR}+ \;g_{RL}^{2}\;g_{LR}+\;g_{LL} \;g_{RR} \;g_{RL}) 
\end{align*}}

\underline{\textit{SUSY:}}
{\small
	\begin{align*} 
		(4\pi)^2{\beta_{2L}}& =\frac{9 \;g_{2 L}^3}{2} \\ 
		(4\pi)^2{\beta_{2R}}& =\frac{9 \;g_{2 R}^3}{2} \\ 
		(4\pi)^2{\beta_{3C}}& =0 \\
		(4\pi)^2{\beta_{LL}}& =\frac{32}{3}(\;g_{LL}^3+ \;g_{LR}^2 \;g_{LL})+\frac{29}{3}( \;g_{RL}^2 \;g_{LL}+\;g_{LR} \;g_{RL} \;g_{RR})\\
		& \ +\frac{1}{3}(2 \;g_{LL}^2 \;g_{RL} + \;g_{LR}^2 \;g_{RL}+\;g_{LR} \;g_{RR} \;g_{LL}) \\ 
		(4\pi)^2{\beta_{LR}}& =\frac{32}{3} (\;g_{LR}^3 + \;g_{LL}^2 \;g_{LR}) +\frac{29}{3}( \;g_{RR}^2 \;g_{LR} + \;g_{LL} \;g_{RL} \;g_{RR})\\ & \
		+\frac{1}{3}(2 \;g_{LR}^2 \;g_{RR} + \;g_{LL}^2 \;g_{RR}+ \;g_{LL} \;g_{RL} \;g_{LR})  \\ 
		(4\pi)^2{\beta_{RL}}& = \frac{32}{3}( \;g_{LL}^2 \;g_{RL}+\;g_{LL} \;g_{LR} \;g_{RR})+\frac{29}{3}(\;g_{RL}^3+ \;g_{RR}^2 \;g_{RL})\\ & \
		+\frac{1}{3}(2 \;g_{RL}^2 \;g_{LL} + \;g_{RR}^2 \;g_{LL} + \;g_{LR} \;g_{RR} \;g_{RL} )  \\ 
		(4\pi)^2{\beta_{RR}}& =\frac{32}{3}( \;g_{LR}^2 \;g_{RR}+\;g_{LL} \;g_{LR} \;g_{RL})+\frac{29}{3}(\;g_{RR}^3 + \;g_{RL}^2 \;g_{RR})\\ & \
		+\frac{1}{3} (2 \;g_{RR}^2 \;g_{LR} + \;g_{RL}^2  \;g_{LR} + \;g_{LL} \;g_{RL} \;g_{RR}) 
\end{align*}}
{\bf \underline{Only 16 fermions present at $M_R$:-}}\\
\\
{\bf D-parity not conserved}\\
\underline{\textit{Non-SUSY:}}
{\small
	\begin{align*}
		(4\pi)^2\beta_{\;g_{2L}} = &
		- 3 \;g_{2L}^{3}\\
		(4\pi)^2\beta_{\;g_{3C}} = &
		- 7 \;g_{3C}^{3}\\
		(4\pi)^2\beta_{\;g_{2R}} = &
		- \frac{17 \;g_{2R}^{3}}{6}\\
		(4\pi)^2\beta_{LL} = 
		&3 (\;g_{LL}^{3} +  \;g_{LL} \;g_{LR}^{2})+ \frac{17 }{6}( \;g_{RL}^{2}\;g_{LL} + \;g_{RR} \;g_{LR} \;g_{RL})\\
		& + \frac{4}{3}( 2 \;g_{LL}^{2} \;g_{RL} +\;g_{LR}^{2} \;g_{RL} + \;g_{LL} \;g_{RR} \;g_{LR} ) \\
		(4\pi)^2\beta_{LR} = 
		& 3 (\;g_{LR}^{3} + \;g_{LL}^{2} \;g_{LR})
		+ \frac{17}{6} (\;g_{RR}^{2} \;g_{LR}+ \;g_{LL} \;g_{RR} \;g_{RL} )  \\
		&+\frac{4}{3}(2\;g_{LR}^{2}\;g_{RR} + \;g_{LL}^{2}\;g_{RR} +\;g_{LL} \;g_{LR} \;g_{RL}) \\
		(4\pi)^2\beta_{RL} = 
		&3 (\;g_{LL}^{2} \;g_{RL} + \;g_{LL} \;g_{RR} \;g_{LR}) + \frac{17}{6} (\;g_{RL}^{3}+\;g_{RR}^{2} \;g_{RL})   
		\\ & + \frac{4}{3} (2 \;g_{RL}^{2}\;g_{LL}+ \;g_{RR}^{2}\;g_{LL}+\;g_{RR}\;g_{LR} \;g_{RL})
		\\
		(4\pi)^2\beta_{RR} = 
		&  3( \;g_{RR} \;g_{LR}^{2} +\;g_{LL} \;g_{LR} \;g_{RL})+\frac{17}{6}( \;g_{RR}^{3}+\;g_{RL}^{2}\;g_{RR} )  \\
		&+\frac{4}{3} (2\;g_{RR}^{2}\;g_{LR} + \;g_{RL}^{2}\;g_{LR} + \;g_{LL}\;g_{RR} \;g_{RL})
\end{align*}}
\underline{\textit{SUSY:}}
{\small
	\begin{align*} 
		(4\pi)^2{\beta_{2L}}& =\;g_{2 L}^3 \\ 
		(4\pi)^2{\beta_{2R}}& =\frac{3 \;g_{2 R}^3}{2} \\ 
		(4\pi)^2{\beta_{3C}}& =-3 \;g_{3 C}^3 \\
		(4\pi)^2{\beta_{LL}}& =5( \;g_{LL}^3+\;g_{LR}^2 \;g_{LL})+\frac{9}{2} (\;g_{RL}^2 \;g_{LL}+\;g_{LR} \;g_{RL} \;g_{RR})\\&+2(2 \;g_{LL}^2\;g_{RL}+ \;g_{LR}^2 \;g_{RL}+ \;g_{LR} \;g_{RR} \;g_{LL})  \\ 
		(4\pi)^2{\beta_{LR}}& =5 (\;g_{LR}^3 + \;g_{LL}^2 \;g_{LR}) +\frac{9}{2} (\;g_{RR}^2 \;g_{LR}+ \;g_{LL} \;g_{RL} \;g_{RR} )\\&+2(2 \;g_{LR}^2\;g_{RR}+ \;g_{LL}^2 \;g_{RR}+ \;g_{LL} \;g_{RL} \;g_{LR})\\ 
		(4\pi)^2{\beta_{RL}}& = 5 (\;g_{LL}^2 \;g_{RL}+\;g_{LL} \;g_{LR} \;g_{RR})+\frac{9}{2}(\;g_{RL}^3+ \;g_{RR}^2 \;g_{RL})\\&+2 ( 2 \;g_{RL}^2\;g_{LL}+\;g_{RR}^2\;g_{LL}+\;g_{LR} \;g_{RR} \;g_{RL})  \\ 
		(4\pi)^2{\beta_{RR}}& =\frac{9 }{2}+4+5 (\;g_{LR}^2 \;g_{RR}+\;g_{LL} \;g_{LR} \;g_{RL})+\frac{9}{2}( \;g_{RR}^3 + \;g_{RL}^2 \;g_{RR})\\&+2(2 \;g_{RR}^2 \;g_{LR}+ \;g_{LL} \;g_{RL} \;g_{RR}+ \;g_{RL}^2 \;g_{LR})  
\end{align*}}
{\bf D-parity conserved}\\
\underline{\textit{Non-SUSY:}}
{\small
	\begin{align*}
		(4\pi)^2\beta_{2L} = 
		&- \frac{17 \;g_{2L}^{3}}{6}\\
		(4\pi)^2\beta_{3C} = 
		&- 7 \;g_{3C}^{3}\\
		(4\pi)^2\beta_{2R} = 
		&- \frac{17 \;g_{2R}^{3}}{6}\\
		(4\pi)^2\beta_{LL} = 
		&\frac{29 }{9}( \;g_{LL}^{3} + \;g_{LR}^{2}\;g_{LL}) + \frac{26 }{9}( \;g_{RL}^{2}\;g_{LL}+\;g_{RR} \;g_{LR} \;g_{RL} ) \\
		&+\frac{13 }{9}(2 \;g_{LL}^{2} \;g_{RL} + \;g_{LR}^{2} \;g_{RL} + \;g_{LL} \;g_{RR} \;g_{LR} )
		\\						(4\pi)^2\beta_{LR} = 
		&  \frac{29}{9}( \;g_{LR}^{3}+\;g_{LL}^{2} \;g_{LR})+\frac{26 }{9}(\;g_{RR}^{2} \;g_{LR}+\;g_{LL} \;g_{RR} \;g_{RL}) \\
		&+ \frac{13}{9}( 2 \;g_{LR}^{2}\;g_{RR}+ \;g_{LL}^{2} \;g_{RR} + \;g_{LL} \;g_{LR} \;g_{RL})\\						(4\pi)^2\beta_{RL} = 
		&\frac{29}{9}( \;g_{LL}^{2} \;g_{RL} +  \;g_{LL} \;g_{RR} \;g_{LR}) + \frac{26}{9}( \;g_{RL}^{3}+\;g_{RR}^{2} \;g_{RL}) \\
		&+ \frac{13}{9}(2 \;g_{RL}^{2}\;g_{LL} + \;g_{RR}^{2} \;g_{LL} + \;g_{RR} \;g_{LR} \;g_{RL})\\						(4\pi)^2\beta_{RR} = 
		& \frac{29}{9} (\;g_{LR}^{2} \;g_{RR} + \;g_{LL} \;g_{LR} \;g_{RL})+\frac{26 }{9}(\;g_{RR}^{3} + \;g_{RL}^{2}\;g_{RR})  \\
		&+\frac{13 }{9}(2 \;g_{RR}^{2} \;g_{LR} + \;g_{RL}^{2}\;g_{LR}+\;g_{LL}\;g_{RR} \;g_{RL} )
\end{align*}}
\underline{\textit{SUSY:}}
{\small
	\begin{align*} 
		(4\pi)^2{\beta_{2L}}& =\frac{3 \;g_{2 L}^3}{2} \\ 
		(4\pi)^2{\beta_{2R}}& =\frac{3 \;g_{2 R}^3}{2} \\ 
		(4\pi)^2{\beta_{3C}}& =-3 \;g_{3 C}^3 \\
		(4\pi)^2{\beta_{LL}}& =\frac{17 }{3}(\;g_{LL}^3+\;g_{LR}^2 \;g_{LL}) +\frac{14}{3}( \;g_{RL}^2 \;g_{LL}+\;g_{LR} \;g_{RL} \;g_{RR})\\& \ +\frac{7}{3}( 2 \;g_{LL}^2\;g_{RL}+\;g_{LR}^2 \;g_{RL} + \;g_{LR} \;g_{RR} \;g_{LL}) \\ 
		(4\pi)^2{\beta_{LR}}& =\frac{17}{3}( \;g_{LR}^3 + \;g_{LL}^2 \;g_{LR})+\frac{14}{3} (\;g_{RR}^2 \;g_{LR}+\;g_{LL} \;g_{RL} \;g_{RR}) \\ & \  +\frac{7}{3}(2 \;g_{LR}^2 \;g_{RR}+ \;g_{LL}^2 \;g_{RR}+ \;g_{LL} \;g_{RL} \;g_{LR}) \\ 
		(4\pi)^2{\beta_{RL}}& =\ \frac{17}{3} (\;g_{LL}^2 \;g_{RL}+\;g_{LL} \;g_{LR} \;g_{RR})+\frac{14}{3}(\;g_{RL}^3+ \;g_{RR}^2 \;g_{RL})\\ & \ +\frac{7}{3}(2\;g_{RL}^2\;g_{LL} + \;g_{RR}^2\;g_{LL} + \;g_{LR} \;g_{RR} \;g_{RL})  \\ 
		(4\pi)^2{\beta_{RR}}& =\frac{17}{3} (\;g_{LR}^2 \;g_{RR}+\;g_{LL} \;g_{LR} \;g_{RL})+\frac{14}{3}(\;g_{RR}^3 + \;g_{RL}^2 \;g_{RR}) \\ & \ +\frac{7}{3}(2 \;g_{RR}^2\;g_{LR}+ \;g_{RL}^2 \;g_{LR}+ \;g_{LL} \;g_{RL} \;g_{RR}). 
\end{align*}}

\section{Numerical Solutions of RGEs}
\label{sec:results}
In this section we have explicitly noted the gauge coupling unification for different breaking chains which include a single intermediate symmetry group emerging from $SO(10)$ and $E(6)$. Out of various possibilities, we have concentrated only on those intermediate gauge symmetries which contain the $SU(N)_L \otimes SU(N)_R$ structure. This special structure reflects the presence of D-parity. After enlisting all such possible breaking patterns in the previous section, we have categorically mentioned the participation of scalar and fermion representations in each individual chain. Then we have computed the $\beta$ coefficients for gauge coupling-running up to two loop, for Non-supersymmetric and supersymmetric scenarios for all the breaking patterns considered in this paper.  We have solved all such two loop renormalisation group equations numerically and found out the solutions in terms of the unified and intermediate scales and abelian gauge coupling mixing. 
\subsection{$\chi^2$ analysis: $M_X$, $M_R$, $g_U$}
\label{subsec:chisquare}
\begin{table}[h!]
\centering
\small
	\renewcommand*{\arraystretch}{1.2}
    \begin{tabular}{|c|c|}
    \hline
    Mass of $Z$-boson, $m_Z$& $91.1876(21)$ GeV \\
    \hline
    Strong coupling constant, $\alpha_s(m_Z)$ & $0.1185(6)$ \\
    \hline
    Fermi coupling constant, $G_F$ & $1.1663787(6)\times 10^{-5} $  $\text{GeV}^{-2}$\\
    \hline
     Weinberg angle, $\sin^2 \theta_W$ & $0.23126(5)$ \\
     \hline  
    \end{tabular}
    \caption{Values of different parameters to obtain the gauge couplings at the electroweak scale ($g_{a, Exp}$).} 
    \label{tab:parameters}
\end{table}

\begin{table}[h!]
\small
	\renewcommand*{\arraystretch}{1.2}
	\begin{center}
		\begin{tabular}{|c|c|c|c|c|c|}
			\hline
			\thead{GUT \\ Group} & \thead{Intermediate \\ gauge group} & \thead{D-parity} &$\log_{10} (\frac{M_R}{\rm GeV})$  & $ \log_{10} (\frac{M_X}{\rm GeV}) $ & \thead{Unified Coupling \\ $g_U (\times 10^{-2})$ }\\
			\hline
			$E(6)$&$\mathcal{G}_{2_L2_R4_C1_X}$& Conserved & $14.132(10)$ & $15.023(18)$ & $55.61(1)$ \\
			\cline{3-6}
			&($27$ Fermions)& Broken & $14.001(9)$ & $15.886(26)$ & $53.84(1)$ \\
			\cline{2-6}
			&$\mathcal{G}_{2_L2_R4_C1_X}$& Conserved & $14.154(10)$ & $15.029(18)$ & $55.121(6)$ \\
			\cline{3-6}
			&($16$ Fermions)& Broken & $14.052(9)$ & $15.865(22)$ &  $53.00(1)$ \\
			\cline{2-6}
			&$\mathcal{G}_{2_L2_R4_C}$& Conserved & $13.755(7)$ & $14.826(16)$ & $56.21(2)$ \\
			\cline{3-6}
			&& Broken & $11.513(20)$ & $15.769(26)$ & $55.088(6)$ \\
			\cline{2-6}
			&$\mathcal{G}_{2_L2_R3_C1_{B-L}}$& Conserved & $10.960(24)$ & $15.276(20)$ & $56.26(2)$ \\
			\cline{3-6}
			&& Broken & $9.959(34)$ & $16.057(28)$ &  $55.81(1)$ \\
				\cline{2-6}
			&$\mathcal{G}_{3_L3_R3_C}$& Conserved/Broken & NS & NS & NS \\
			\cline{3-6}
			\hline
			$SO(10)$&$\mathcal{G}_{2_L2_R4_C}$& Conserved & $13.755(7)$ & $14.820(16)$ & $55.64(1)$ \\
			\cline{3-6}
			&& Broken & $11.607(19)$ & $15.704(25)$ & $53.11(2)$ \\
			\cline{2-6}
			&$\mathcal{G}_{2_L2_R3_C1_{B-L}}$& Conserved & $10.964(24)$ & $15.257(20)$ & $54.088(7)$ \\
			\cline{3-6}
			&& Broken & $9.981(34)$ & $16.018(27)$ &  $52.895(17)$ \\
			\hline  
		\end{tabular}
		\caption{\small Best fit results of the unification and intermediate scales and  unified couplings for the Non-SUSY models, consistent with low energy experimental values showed in Table~\ref{tab:parameters}. ``NS" implies that we have not found any suitable solution for the breaking chain $E(6) \to \mathcal{G}_{3_L3_R3_C} \to$ SM. } 
		\label{tab:unific}
	\end{center}
\end{table}

\begin{figure}[h!]
        \centering
        \subfloat[$M_R$ - $M_X$]{
			\includegraphics[scale=0.45]{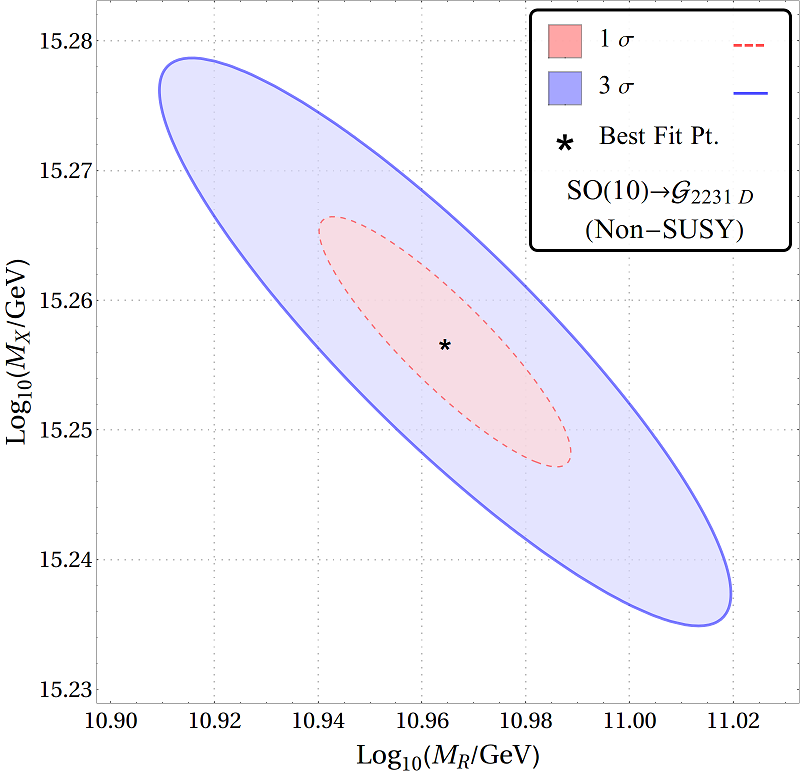}
        \label{fig:unichiso2231dNS_MRMX}}
		\subfloat[$M_R$ - $g_U$]{
			\includegraphics[scale=0.45]{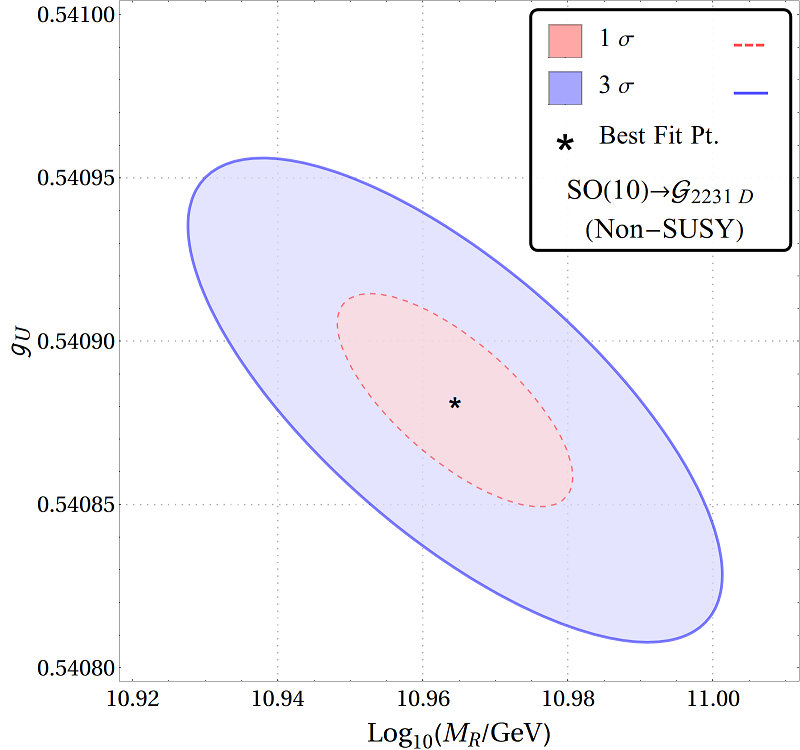}
		\label{fig:unichiso2231dNS_MRg}}
		\subfloat[$M_X$ - $g_U$]{
			\includegraphics[scale=0.45]{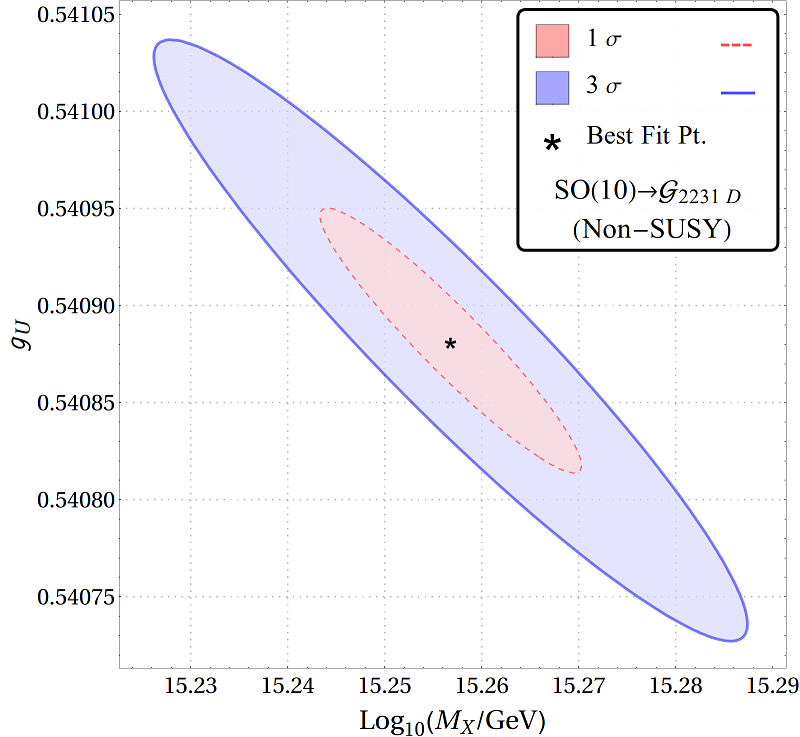}
		\label{fig:unichiso2231dNS_MXg}}\\
		\subfloat[$M_R$ - $M_X$]{
		\includegraphics[scale=0.45]{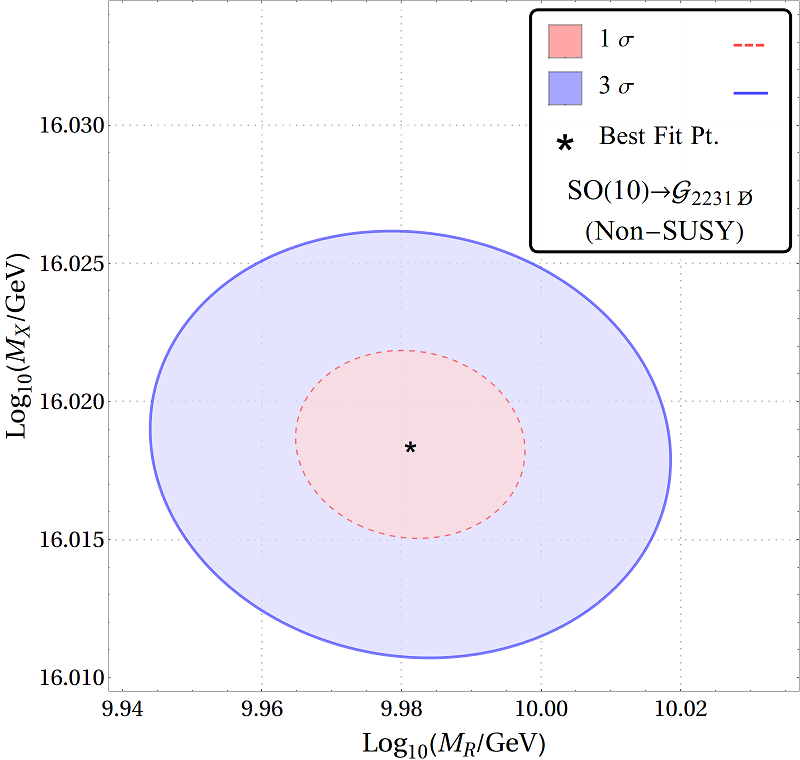}
	\label{fig:unichiso2231NS_MRMX}}
		\subfloat[$M_R$ - $g_U$]{
	\includegraphics[scale=0.45]{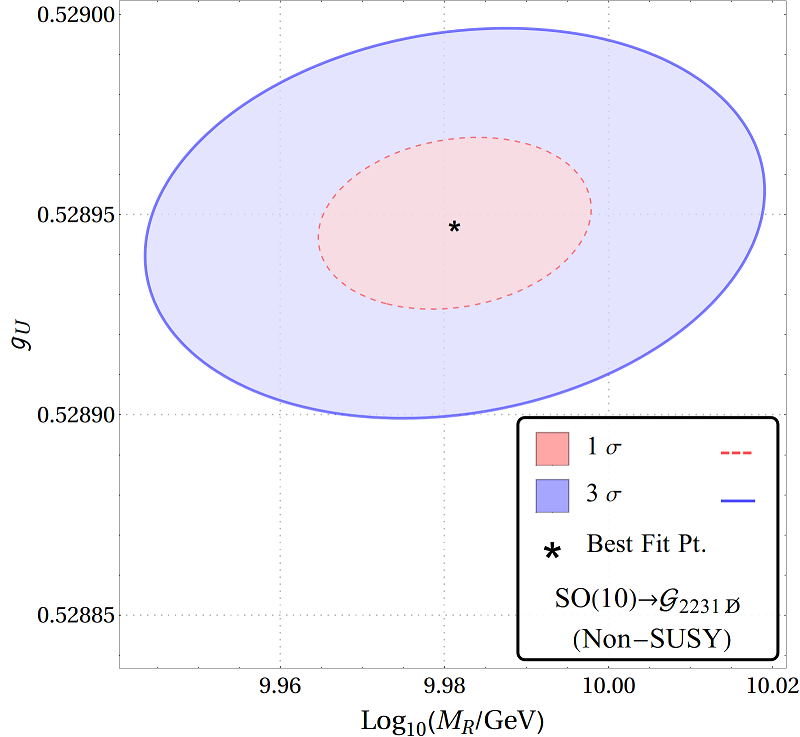}
	\label{fig:unichiso2231NS_MRg}}
		\subfloat[$M_X$ - $g_U$]{
	\includegraphics[scale=0.45]{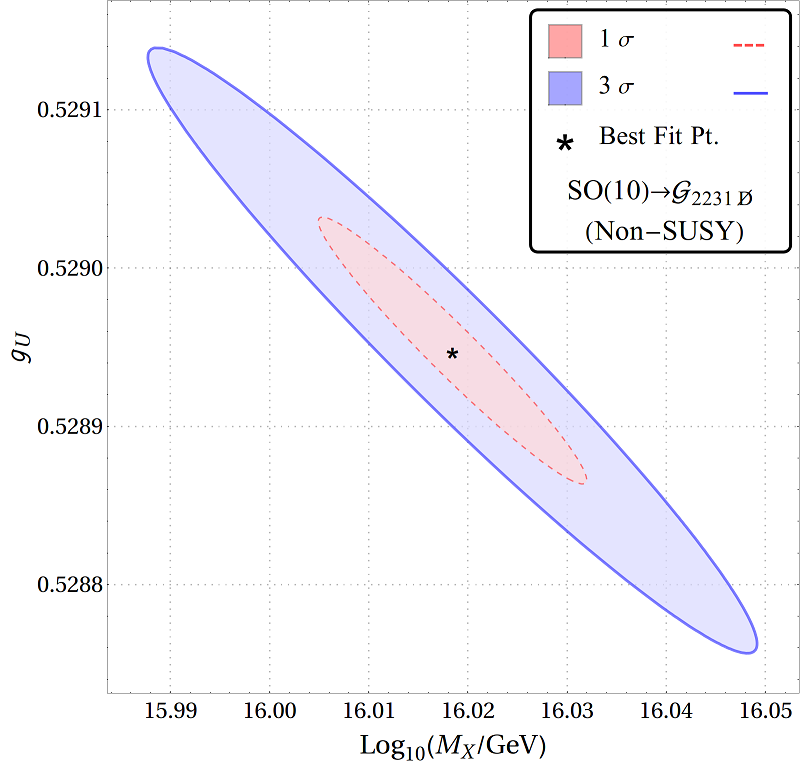}
	\label{fig:unichiso2231NS_MXg}}
	\caption{\small Correlations among intermediate ($M_R$) and unification ($M_X$) scales and the unified coupling ($g_{U}$) satisfying gauge coupling unification for breaking pattern $SO(10) \to \mathcal{G}_{2231}$ within Non-SUSY scenario for both D-parity conserved (Top-row) and broken (Bottom-row) cases.  The ``$\star$" denotes the best fit points, listed in Table \ref{tab:unific} and the blue and red regions depict the $3\sigma$ and $1\sigma$ contours respectively which satisfy the gauge coupling unification, consistent with low energy experimental values showed in Table~\ref{tab:parameters}.}
	\label{fig:unific-chisq-so-g2231-NS1}
\end{figure}
To estimate the level of unification quantitatively, we have performed a test of significance (goodness of fit) by defining a $\chi^2$ statistic, a function of the unification scale ($M_X$), the intermediate scale ($M_R$) and the unified coupling ($g_U$):
	\begin{align}
	\chi^2 = \sum_{a=1}^{3} \frac{(g^2_a - g^2_{a, Exp})}{\sigma^2(g^2_{a, Exp})}\,,
	\end{align}
where $g_a$ are the gauge couplings at the electroweak scale obtained by solving the two-loop RGEs under the assumption of unification, and $g_{a, Exp}$ are the experimental values of the corresponding gauge couplings, with $\sigma(g^2_{a,exp})$ signifying their uncertainties. The latter two are obtained from the input parameters listed in Table~\ref{tab:parameters}. A lack of a good fit will point to a possible absence of unification for a particular breaking scenario.

The summary of this analysis for Non-supersymmetric models are given in Table~\ref{tab:unific}. These results do not contain the correlations between parameters, for which we need to know the eigensystem of the covariance matrix, calculating which is improbable in the our present problem. We instead provide the correlations in graphical form, by showing the $1\sigma$ and $3\sigma$ confidence regions for each case.  Fig.~\ref{fig:unific-chisq-so-g2231-NS1} is such a representative scenario, containing the $1\sigma$ and $3\sigma$ contours in different parameter-planes for $SO(10) \to \mathcal{G}_{2231D}$ (and $\mathcal{G}_{2231\slashed{D}}$), that are equivalent to $p$-values of $0.3173$ and $0.0027$, corresponding to confidence levels of $68.27\%$ and $99.73\%$, respectively. For our purpose, each confidence interval corresponds to a particular value of $X = \Delta\chi^2$(i.e.\ $\chi^2 - \chi^2_{min}$) for $d.o.f = 2$ (no. of parameters), such that $p(X|{\rm d.o.f})$ is fixed. As an example, $\Delta\chi^2 = 2.30$ and $11.83$ for $1 \sigma$ and $3 \sigma$ regions respectively in 2 dimensions\footnote{Though $\Delta\chi^2 = 1$ gives $1 \sigma$ region for a single PDF and is needed for quoting uncertainties, it encloses a smaller region than the confidence level of $68.27\%$ for any higher dimensional PDF.}.
\begin{figure}[h!]\centering
	\subfloat[$\mathcal{G}_{224}$]{
		\includegraphics[scale=0.3]{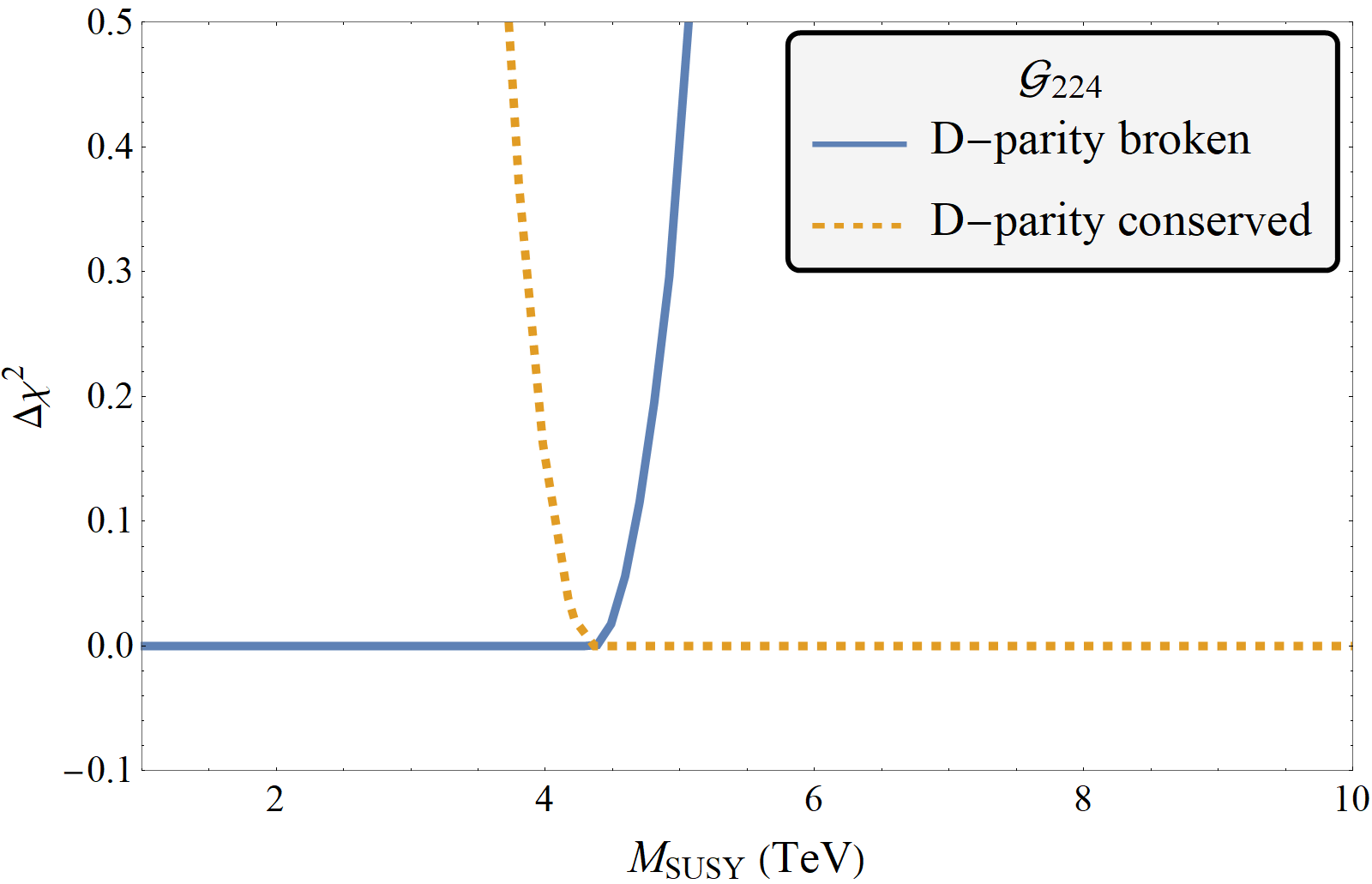}
		\label{fig:su-del1}}
	\subfloat[$\mathcal{G}_{2231}$]{
		\includegraphics[scale=0.3]{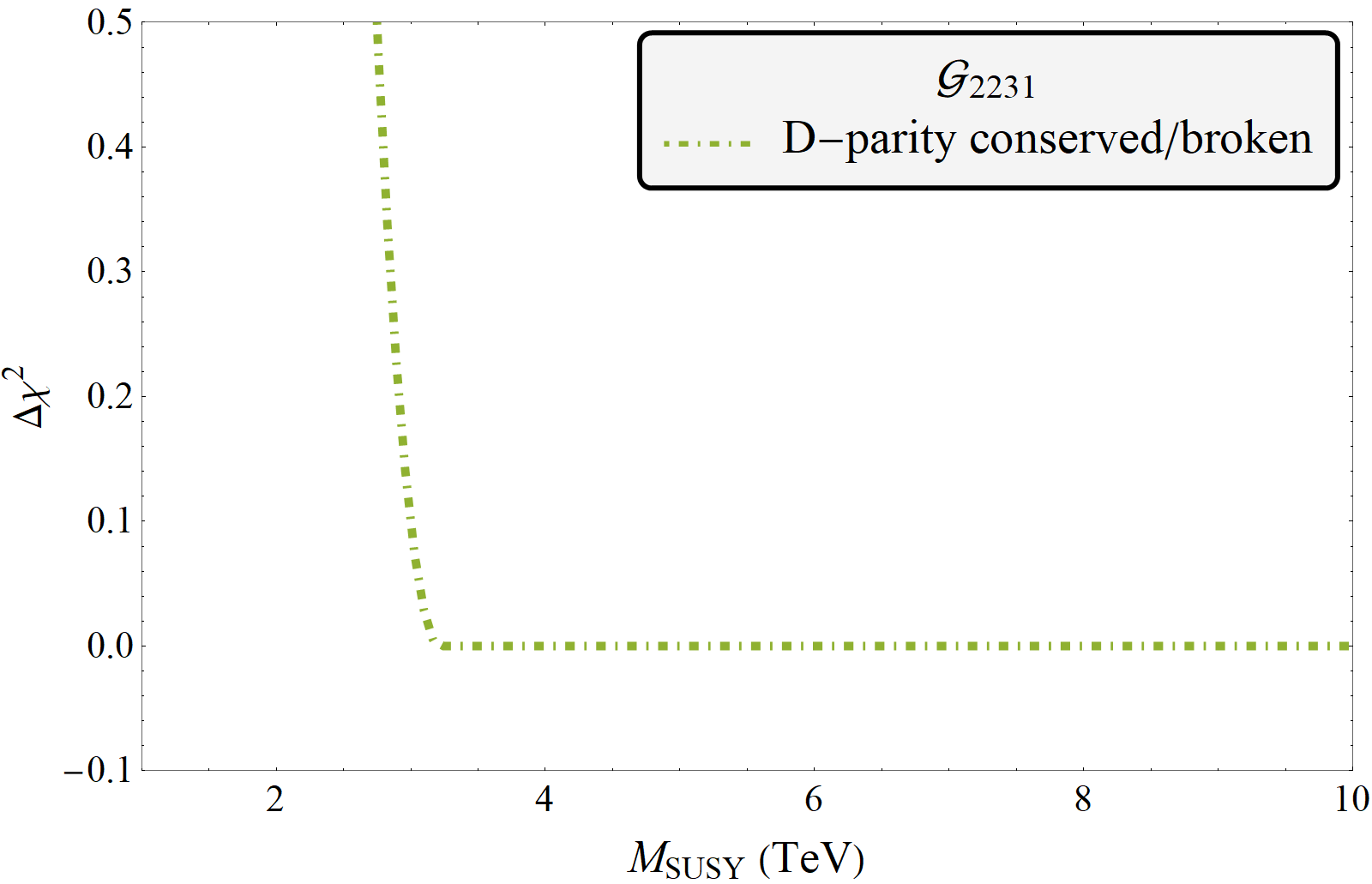}
		\label{fig:su-del2}}\\
	\subfloat[$\mathcal{G}_{333}$]{
		\includegraphics[scale=0.3]{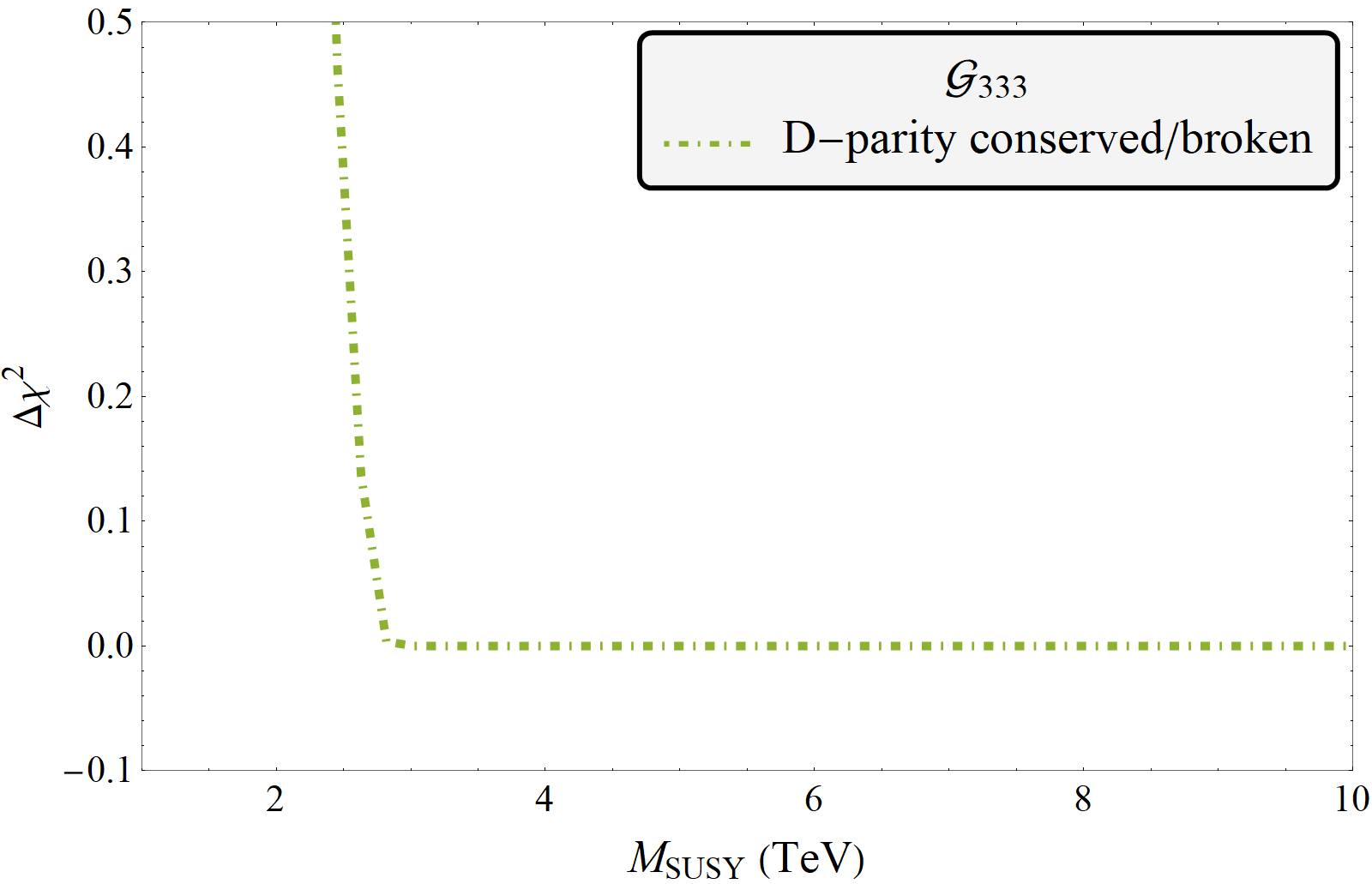}
		\label{fig:su-del3}}
	\subfloat[$\mathcal{G}_{2241}$]{
		\includegraphics[scale=0.3]{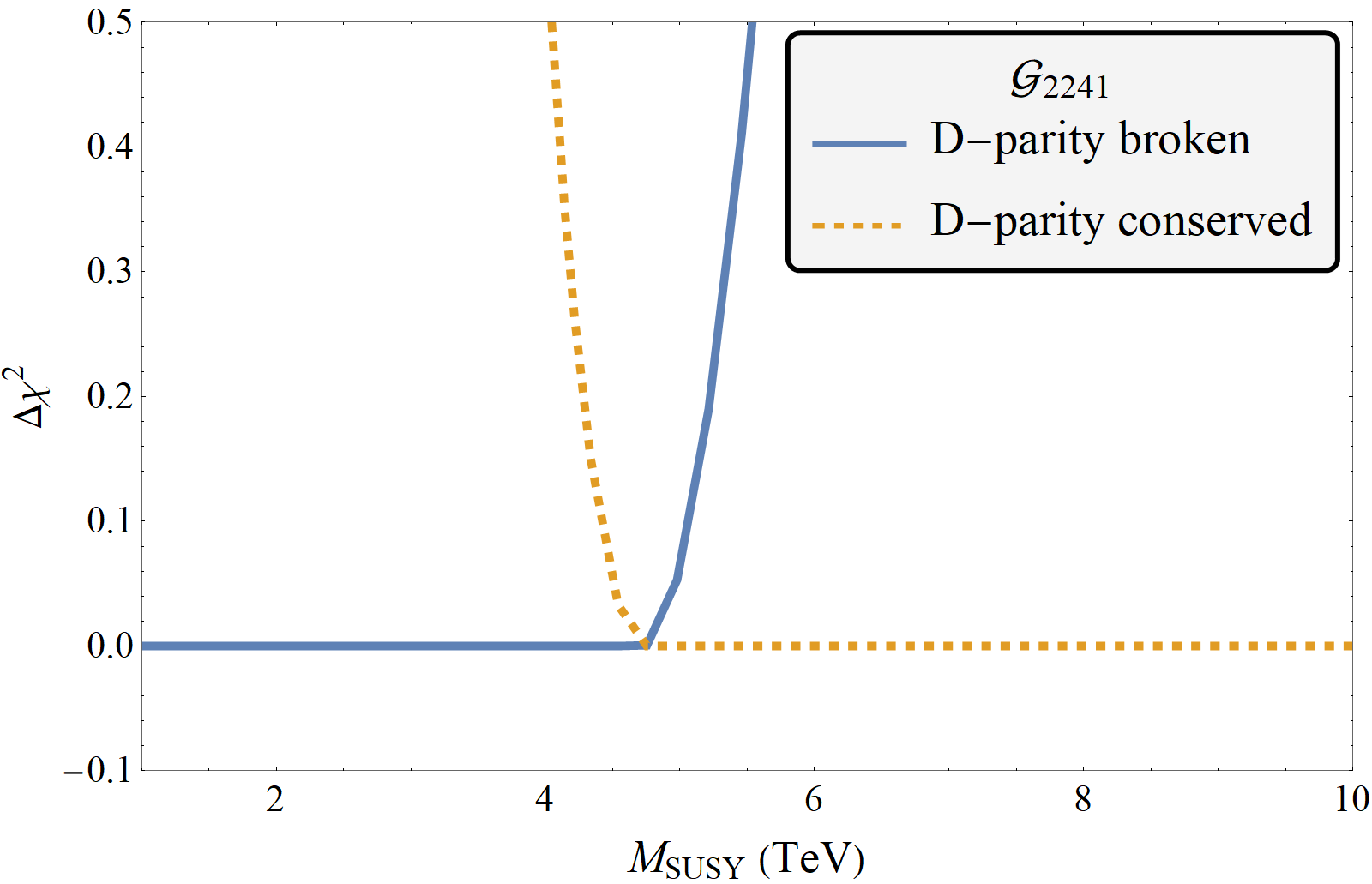}
		\label{fig:su-del4}}
	\caption{\small $\Delta \chi^2$ as a function of SUSY breaking  for intermediate gauge groups $\mathcal{G}_{224}$ (Fig.\ref{fig:su-del1}), $\mathcal{G}_{2231}$ (Fig.~\ref{fig:su-del2}), $\mathcal{G}_{333}$ (Fig.~\ref{fig:su-del3}), $\mathcal{G}_{2241}$ (Fig.~\ref{fig:su-del4}) for both D-parity conserved and broken cases.}
	\label{fig:susyscale-deltachisq}
\end{figure}
\begin{figure}[h!]
        \centering
            \subfloat[$E(6) \to \mathcal{G}_{333}$]{
	\includegraphics[scale=0.5]{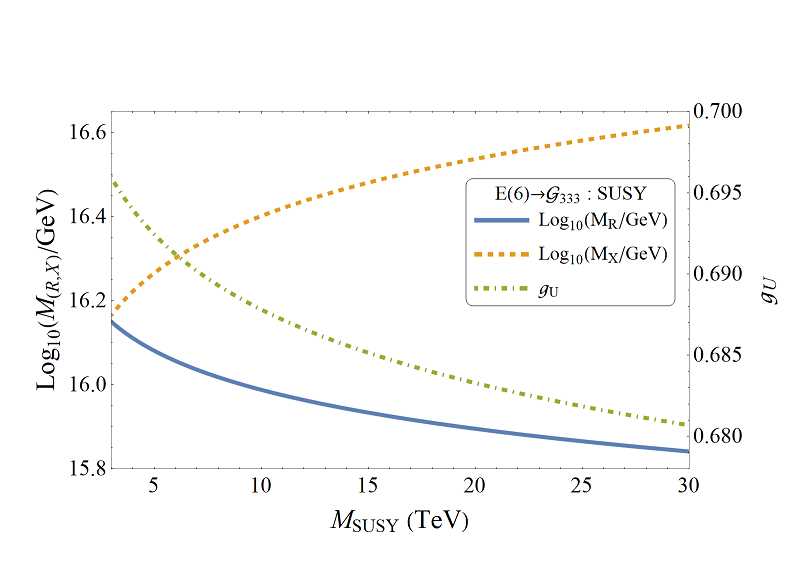}
    \label{fig:su-uni1}}
    \subfloat[$E(6) \to \mathcal{G}_{2241D} (27_F)$]{
    	\includegraphics[scale=0.5]{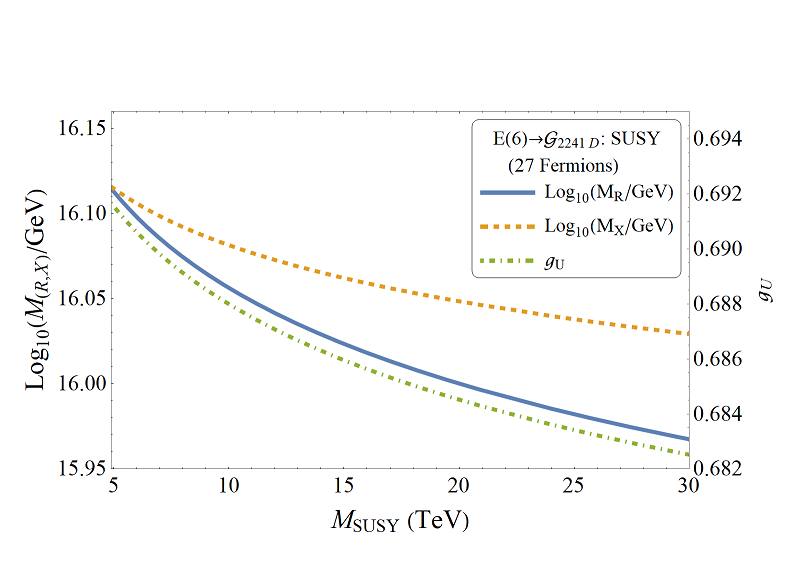}
    \label{fig:su-uni2}}
       \subfloat[$E(6) \to \mathcal{G}_{2231D}$]{
	\includegraphics[scale=0.5]{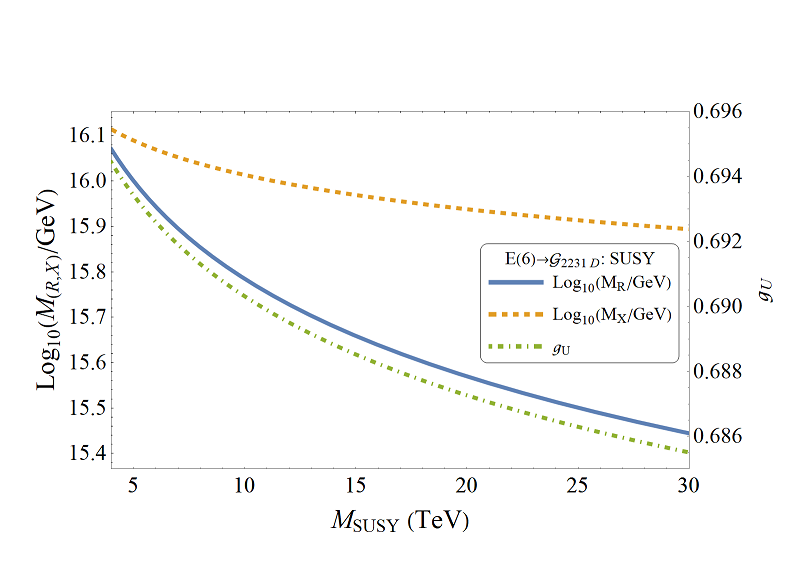}
	\label{fig:su-uni3}}\\
		 \subfloat[$E(6) \to \mathcal{G}_{224D} $]{
	\includegraphics[scale=0.5]{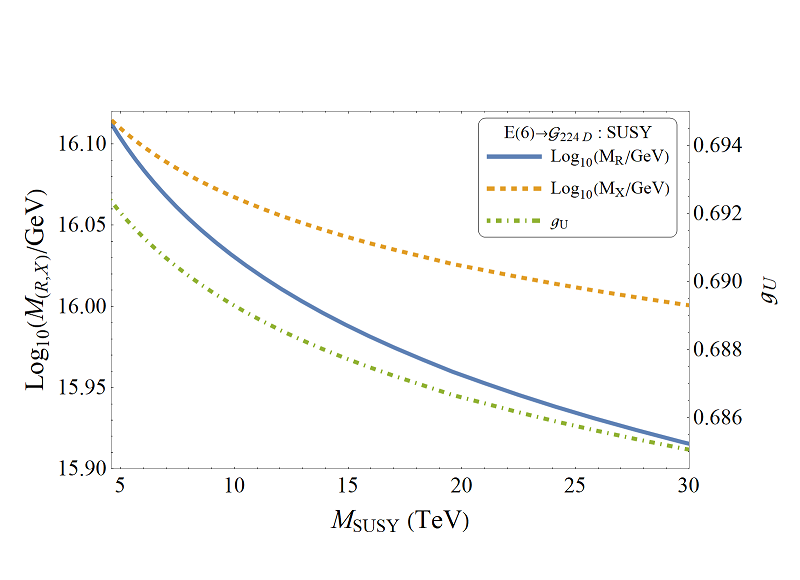}
	\label{fig:su-uni4}}
    \subfloat[$E(6) \to \mathcal{G}_{2241D} (16_F)$]{
	\includegraphics[scale=0.5]{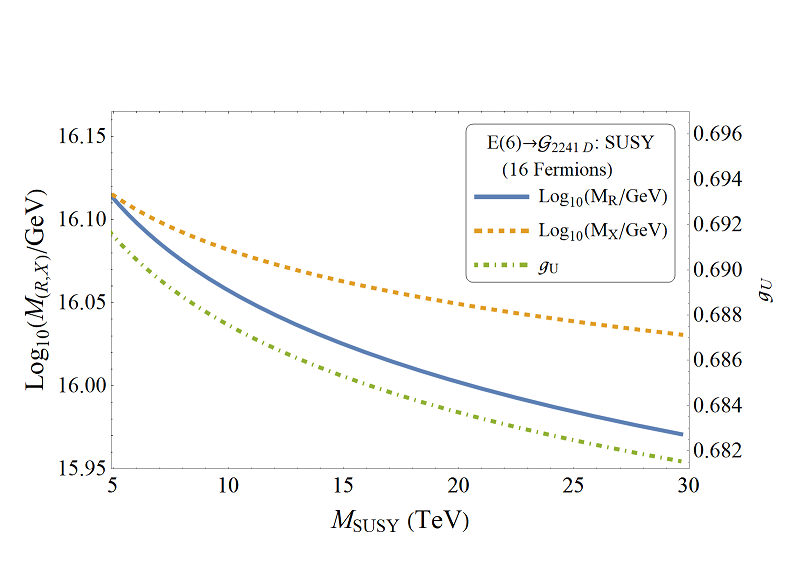}
	\label{fig:su-uni5}}
    \subfloat[$E(6) \to \mathcal{G}_{2231\slashed{D}}$]{
	\includegraphics[scale=0.5]{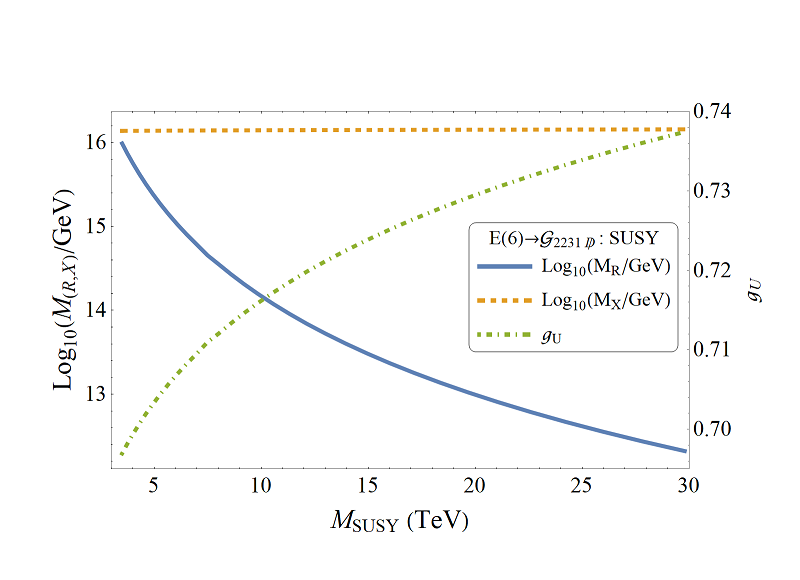}
	\label{fig:su-uni6}}\\
    \subfloat[$SO(10) \to \mathcal{G}_{224D}$]{
	\includegraphics[scale=0.5]{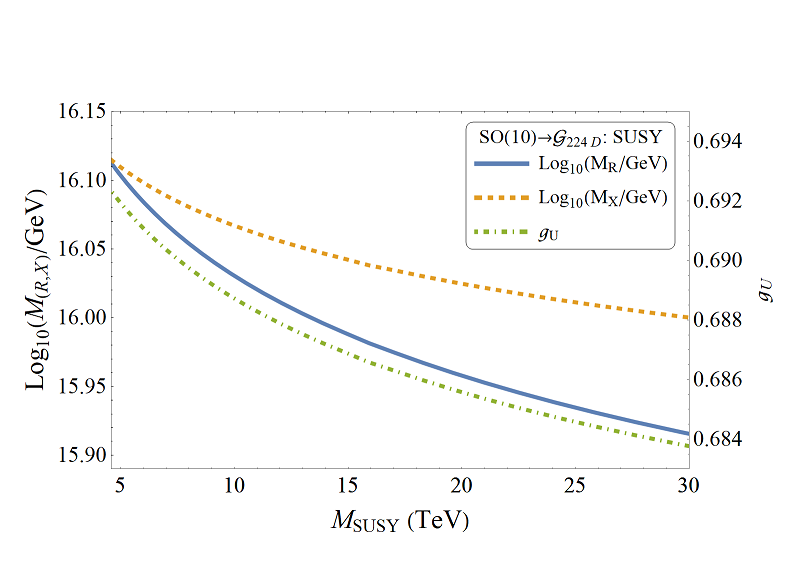}
    \label{fig:su-uni7}}
    \subfloat[$SO(10) \to \mathcal{G}_{2231D}$]{
	\includegraphics[scale=0.5]{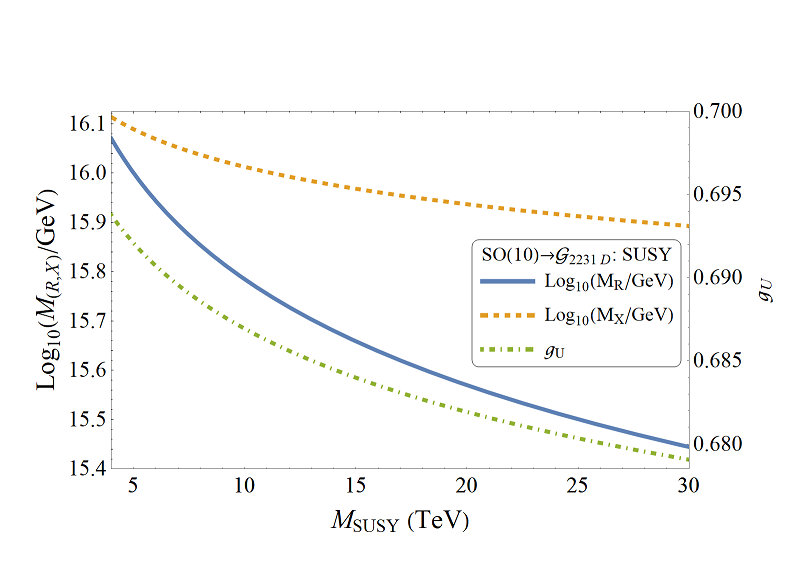}
	\label{fig:su-uni8}}
    \subfloat[$SO(10) \to \mathcal{G}_{2231\slashed{D}}$]{
	\includegraphics[scale=0.5]{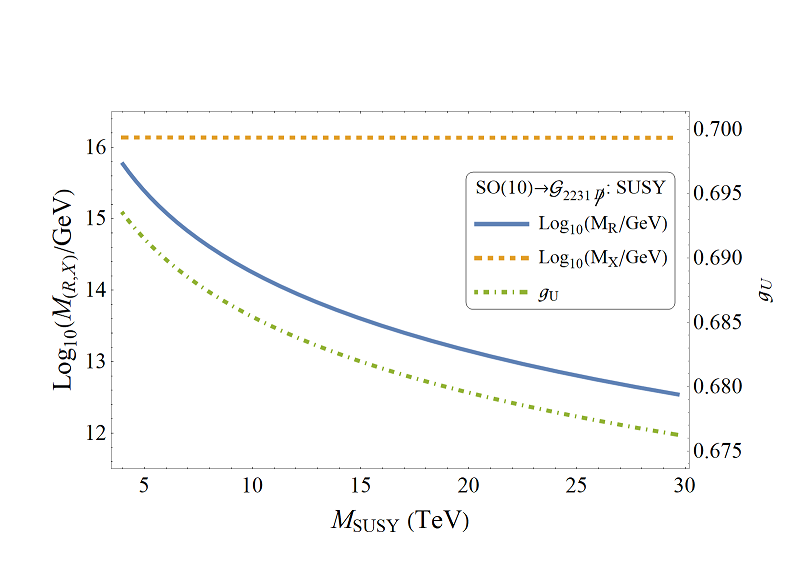}
    \label{fig:su-uni9}}
	\caption{\small Variation of intermediate ($M_R$) and unification ($M_X$) scales and the unified coupling ($g_{U}$) with the SUSY scale ($M_{SUSY}$) satisfying gauge coupling unification for various breaking patterns within supersymmetric scenario . }
	\label{fig:susy-unific}
\end{figure}
\begin{table}[h!]
\small
	\renewcommand*{\arraystretch}{1.1}
	\begin{center}
		\begin{tabular}{|c|c|c|c|c|c|}
			\hline
			\thead{GUT \\ Group} & \thead{Intermediate \\ gauge group} & \thead{D-parity} &$\log_{10} (\frac{M_R}{\rm GeV})$  & $ \log_{10} (\frac{M_X}{\rm GeV}) $ & \thead{Unified Coupling \\ $g_U \times 10^{-2}$ }\\
			\hline
			$E(6)$&$\mathcal{G}_{2_L2_R4_C1_X}$& Conserved* & $16.025(12)$ & $16.063(19)$ & $68.60(10)$ \\
			\cline{3-6}
			&($27$ Fermions)& Broken & $16.231(14)$ & $16.376(11)$ & $71.02(4)$ \\
			\cline{2-6}
			&$\mathcal{G}_{2_L2_R4_C1_X}$& Conserved* & $16.026(12)$ & $16.063(19)$ & $68.53(9)$ \\
			\cline{3-6}
			&($16$ Fermions)& Broken & $16.236(13)$ & $16.393(13)$ &  $70.798(8)$ \\
			\cline{2-6}
			&$\mathcal{G}_{2_L2_R4_C}$& Conserved* & $15.989 (3)$ & $16.043(16)$ & $68.78(13)$ \\
			\cline{3-6}
			&& Broken & $16.169(22)$ & $16.226(14)$ & $70.786(8)$ \\
			\cline{2-6}
			&$\mathcal{G}_{2_L2_R3_C1_{B-L}}$& Conserved* & $15.662(42)$ & $15.970(8)$ & $68.86(12)$ \\
			\cline{3-6}
			&& Broken* & $13.49 (36)$ & $16.149 (22)$ &  $72.38(49)$ \\
			\cline{2-6}
			&$\mathcal{G}_{3_L3_R3_C}$& Conserved/Broken* & $15.936 (7)$ & $16.478 (73)$ & $68.52 (7)$ \\
			\hline
			$SO(10)$&$\mathcal{G}_{2_L2_R4_C}$& Conserved* & $15.989 (8)$ & $16.043(16)$ & $68.69(11)$ \\
			\cline{3-6}
			&& Broken & $16.168 (22)$ & $16.226 (14)$ & $70.69(1)$ \\
			\cline{2-6}
			&$\mathcal{G}_{2_L2_R3_C1_{B-L}}$& Conserved* & $15.662 (42)$ & $15.969 (8)$ & $68.40(5)$ \\
			\cline{3-6}
			&& Broken* & $13.62 (33)$ & $16.134(29)$ &  $68.20 (2)$ \\
			\hline  
		\end{tabular}
		\caption{\small Best fit results of the unification and intermediate scale and  unified couplings for the SUSY models, consistent with low energy experimental values showed in Table~\ref{tab:parameters}.\\(* For these cases SUSY scale starts from $\sim 15$ TeV; for others, $M_{SUSY} \sim 1$ TeV)} 
		\label{tab:unific_SUSY}
	\end{center}
\end{table}
\begin{figure}[h!]
	\centering
	\subfloat[$M_R$ - $M_X$]{
		\includegraphics[scale=0.45]{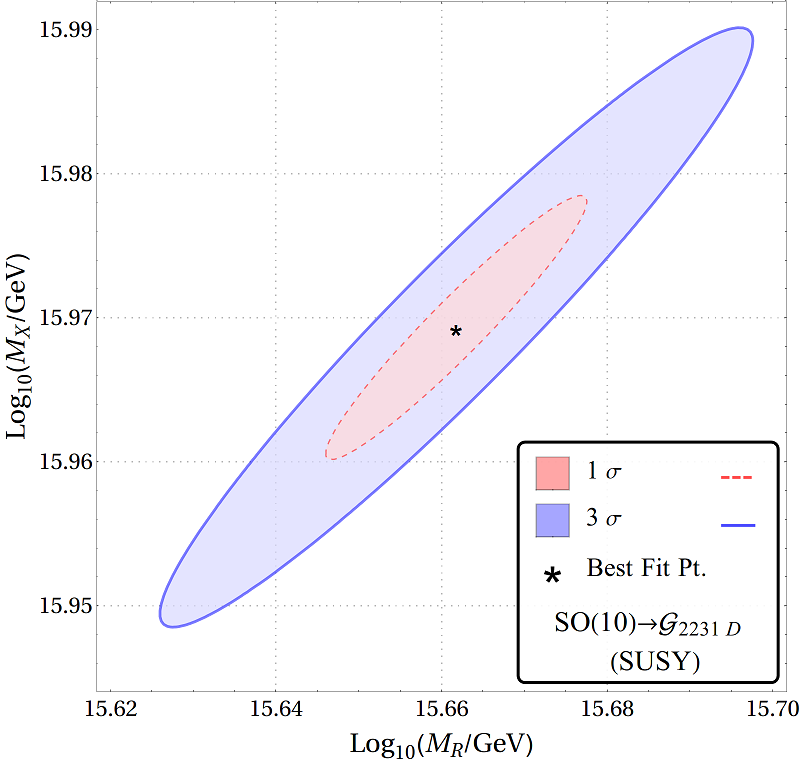}
		\label{fig:unichiso2231dS_MRMX}}
	\subfloat[$M_R$ - $g_U$]{
		\includegraphics[scale=0.45]{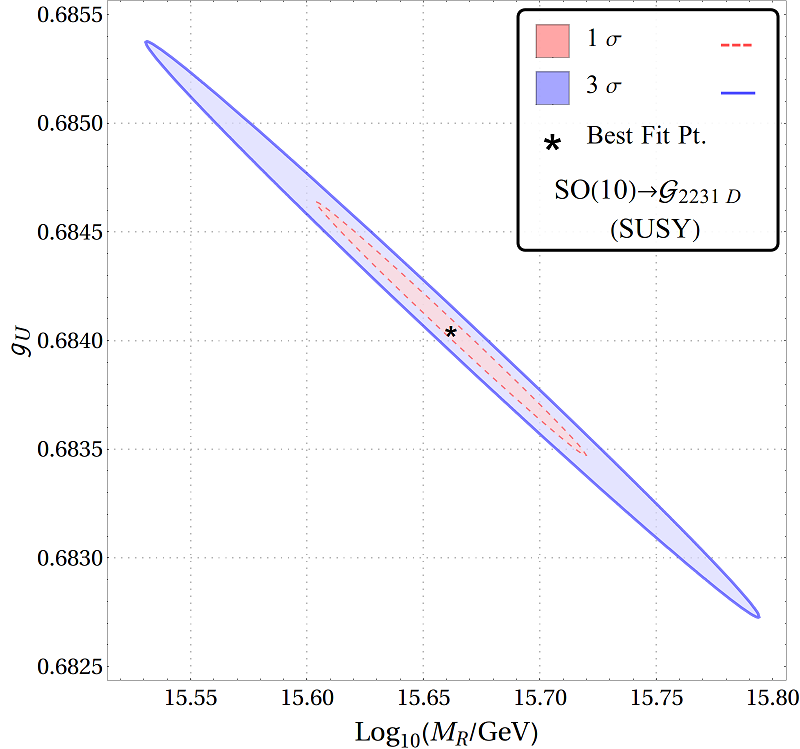}
		\label{fig:unichiso2231dS_MRg}}
	\subfloat[$M_X$ - $g_U$]{
		\includegraphics[scale=0.45]{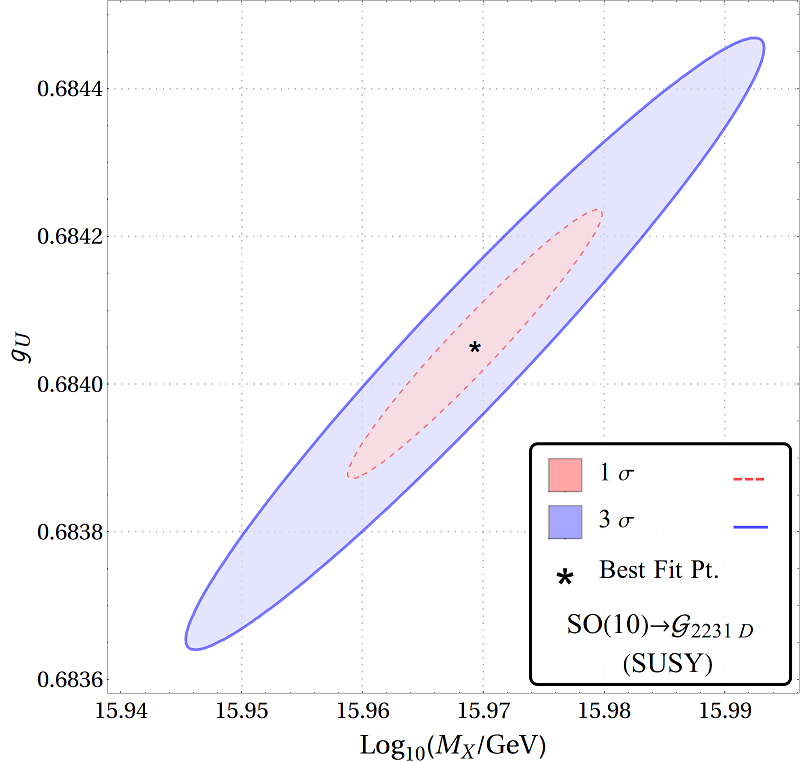}
		\label{fig:unichiso2231dS_MXg}}\\
	\subfloat[$M_R$ - $M_X$]{
		\includegraphics[scale=0.45]{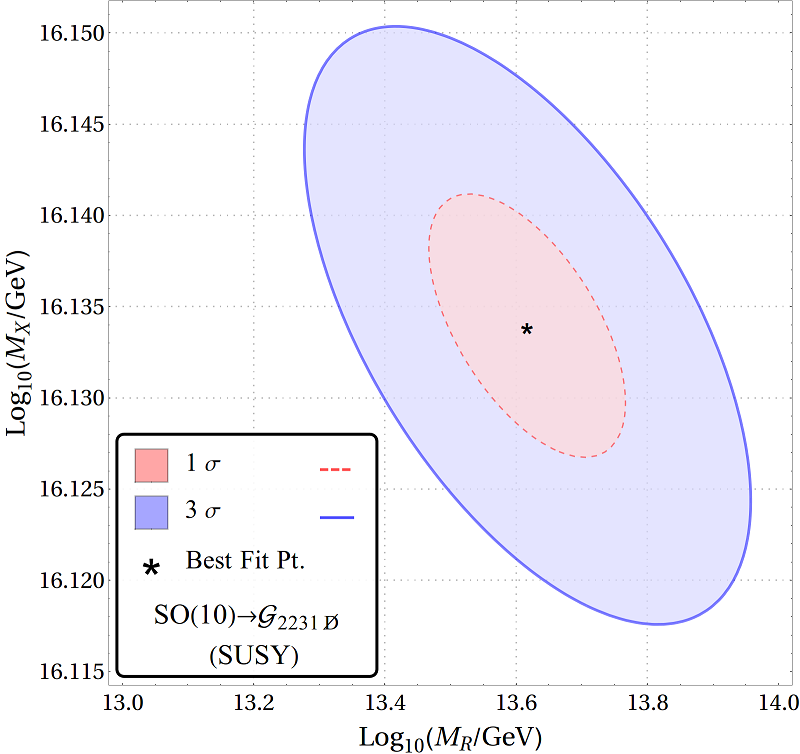}
		\label{fig:unichiso2231S_MRMX}}
	\subfloat[$M_R$ - $g_U$]{
		\includegraphics[scale=0.45]{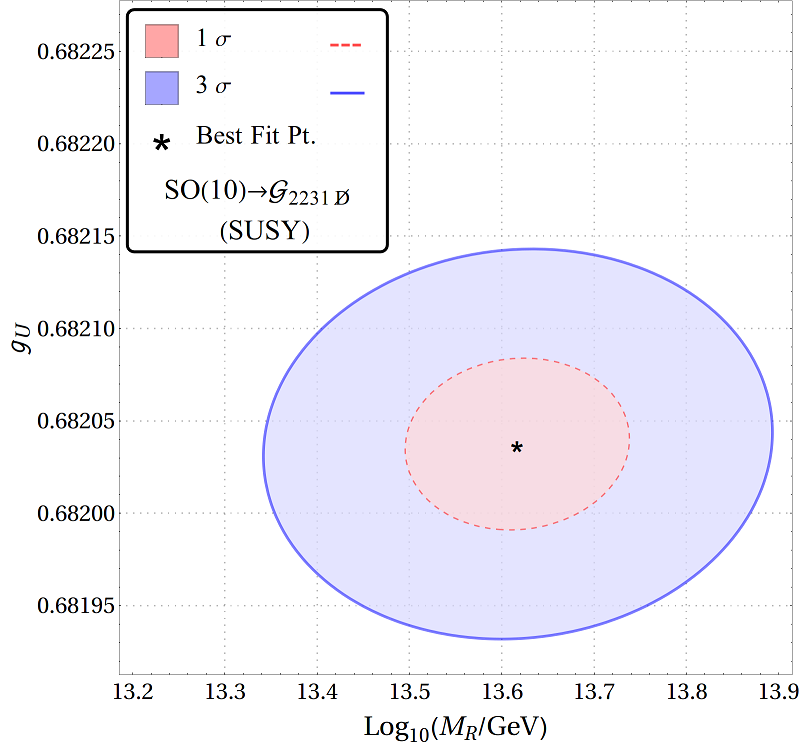}
		\label{fig:unichiso2231S_MRg}}
	\subfloat[$M_X$ - $g_U$]{
		\includegraphics[scale=0.45]{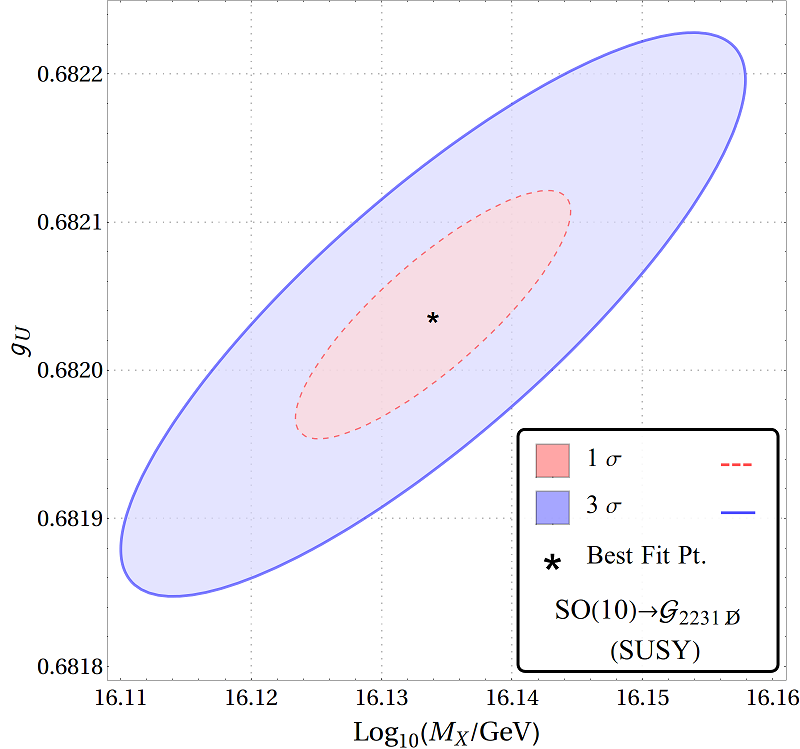}
		\label{fig:unichiso2231S_MXg}}
	\caption{\small Correlations among $M_R$ and $M_X$ and $g_{U}$ satisfying gauge coupling unification for breaking patter $SO(10) \to \mathcal{G}_{2231}$ within SUSY for both D-parity conserved (Top-row) and broken (Bottom-row) cases.  The SUSY scale is set at $M_{SUSY}=15$ TeV. The ``$\star$" implies the best fit point, and the blue and red regions depict the $3\sigma$ and $1\sigma$ contours respectively which satisfy the gauge coupling unification consistent with low energy experimental values showed in Table~\ref{tab:parameters}.}
	\label{fig:unific-chisq-so-g2231-S}
\end{figure}
To perform the similar analysis in presence of supersymmetry, there is a requirement of minor modification of the previous methodology as we do not know the precise value of the SUSY scale ($M_{SUSY}$). The general practice is to choose $M_{SUSY}\sim$ 1 TeV, buying the naturalness argument. But there is no such basic principle to do so. In fact, in some specific breaking chains, as will be shown later, this choice of $M_{SUSY}$ is not consistent with unification at all. For a given choice of $M_{SUSY}$, we have equal number of observables and parameters in context of a likelihood analysis, fixing our coveted $\chi^2_{min}$ at 0. We then vary the $M_{SUSY}$ between 1 - 30 TeV (accessible energy range by the Future Circular Collider (FCC) \cite{Golling:2016gvc,Contino:2016spe}) and find the range of $M_{SUSY}$ for which $\Delta \chi^2 = 0$ (and thus consistent with unification). 
The intermediate gauge groups for SUSY case are $\mathcal{G}_{333}, \mathcal{G}_{2241}, \mathcal{G}_{2231}, \mathcal{G}_{224}$ in presence and absence of D-parity.  Fig.~\ref{fig:susyscale-deltachisq} shows that the lowest possible value of $M_{SUSY}$, allowed by unification, is different for different intermediate groups and may be different for D-parity conserved-broken cases. It can be as low as 1 TeV for $\mathcal{G}_{224\slashed{D}}$ (see Fig.~\ref{fig:su-del1}) and $\mathcal{G}_{2241\slashed{D}}$ (see Fig.~\ref{fig:su-del4}) whereas for $\mathcal{G}_{224D}$ and $\mathcal{G}_{2241D}$, it is around 4.6 TeV. For both D-parity conserved and broken cases, lowest unification-allowed $M_{SUSY}$ is around 3.3 TeV for $\mathcal{G}_{2231}$ and is around 2.9 TeV for $\mathcal{G}_{333}$ (see Figs.~\ref{fig:su-del2} and \ref{fig:su-del3}). 

These plots provide a schematic understanding regarding the dependence of unification criteria on the choice of SUSY scale and encourage us to have a rigorous look into this.
Thus we have scrutinised the $M_{SUSY}$ dependence for each breaking chain we have considered in this paper.

In case of the breaking patterns satisfying unification, with $M_{SUSY}$ starting from 1 TeV, $M_R$ and $M_X$ start out being really close. With increasing $M_{SUSY}$, they get even closer up to the point of being indistinguishable from each other for $M_{SUSY} \approx 4.6$ TeV, after which they stop satisfying unification altogether. Thus, to create the correlation plots between parameters for these cases, we have taken $M_{SUSY} = 1$ TeV as the representative value. For other cases, where unification starts being satisfied from some higher $M_{SUSY}$, $M_R$ and $M_X$ start very close together and with increasing $M_{SUSY}$, get separated gradually. $M_{SUSY}$-dependence of the parameter values for these cases are showcased in Fig.~\ref{fig:susy-unific}, where we have noted the variations of $M_R$, $M_X$, and $g_{U}$ with $M_{SUSY}$, satisfying gauge coupling unification for breaking: (i) $E(6) \to \mathcal{G}_{333}$ (Fig.~\ref{fig:su-uni1}),  (ii) $E(6) \to \mathcal{G}_{2241D}$ with 27-fermions (Fig.~\ref{fig:su-uni2}), 
(iii) $E(6) \to \mathcal{G}_{2231D}$  (Fig.~\ref{fig:su-uni3}), (iv) $E(6) \to \mathcal{G}_{224D}$ (Fig.~\ref{fig:su-uni4}),
(v) $E(6) \to \mathcal{G}_{2241D}$ with 16-fermions (Fig.~\ref{fig:su-uni5}), (vi) $E(6) \to \mathcal{G}_{2231\slashed{D}}$ (Fig.~\ref{fig:su-uni6}), and (vii) $SO(10) \to \mathcal{G}_{224D}$ (Fig.~\ref{fig:su-uni7}), (viii)  $SO(10) \to \mathcal{G}_{2231D}$ (Fig.~\ref{fig:su-uni8}), (ix)  $SO(10) \to \mathcal{G}_{2231\slashed{D}}$ (Fig.~\ref{fig:su-uni9}).
 This $M_{SUSY}$-dependence of the parameters are not necessarily the same for the presence and absence of D-parity for a given intermediate symmetry group. 
To demonstrate the correlation between parameters for these cases, we needed to choose a value of $M_{SUSY}$ which is sufficiently large to ensure proper separation between $M_R$ and $M_X$, but considerably smaller still, than the scales attainable by colliders in near future; we chose $M_{SUSY} = 15$ TeV.
Table \ref{tab:unific_SUSY} summarises the results of our analysis for all the SUSY models. Correlation plots for $SO(10) \to \mathcal{G}_{2231}$ are listed in Fig.~\ref{fig:unific-chisq-so-g2231-S} as a representative case.

Variations of intermediate scale with the SUSY scale can have a large impact on supersymmetric phenomenology, e.g., for $M_{SUSY}\sim 20 $ TeV, $M_R$ is around $10^{13}$ GeV. Now within GUT-SUGRA scenario,  the boundary conditions will be provided in terms of the representations under the  intermediate gauge groups. And the low scale spectrum will be drastically affected by the non-negligible running of the spectrum from GUT to intermediate scale. This will certainly change the lightest neutralino composition which in turn will affect the conclusion based on constraints from dark matter, muon (g-2) and other low energy constraints. It will be worthy to include the impact of intermediate RGEs on SUSY phenomenology. It is also interesting to note that the intermediate scale is now related to the SUSY scale if we demand successful unification.
We leave this part for our future venture.

\subsection{Cosmological constraints and unification}
\label{subsec:TD-unific} 
In this section, we have solved the RGEs which are essentially coupled differential equations, and the solutions are given in terms of $M_X$, $M_R$, and other free couplings.
To start with we have adopted those breaking patterns, which predict unique intermediate scales, e.g. $\mathcal{G}_{2231}$, $\mathcal{G}_{2241}$,  $\mathcal{G}_{333}$, and $\mathcal{G}_{224}$. We have discussed both D-parity conserved and broken scenarios within  Non-SUSY and SUSY frameworks. We have listed the topological defects, that arise in the process of symmetry breaking involving the above-mentioned intermediate symmetry groups \cite{Jeannerot:2003qv}, in Table~\ref{tab:topological_defects}.
\begin{table}[h!]
	\begin{center}
		\begin{tabular}{|c|c|c|}
			\hline
			\multicolumn{2}{|c|}{\textbf{Intermediate Symmetry}}&\textbf{Topological defects}\\
			\hline
			\multirow{2}{2cm}{$~~~~\mathcal{G}_{224}$}&D-broken & monopoles\\
			\cline{2-3}
			&D-conserved & domain wall + monopoles + $Z_2$-strings \\
			\hline
			\multirow{2}{2cm}{$~~~~\mathcal{G}_{2231}$}&D-broken & monopoles + embedded strings\\
			\cline{2-3}
			&D-conserved & domain wall + monopoles + embedded strings \\
			\hline
			\multirow{2}{2cm}{$~~~~\mathcal{G}_{2241}$}&D-broken & monopoles + embedded strings\\
			\cline{2-3}
			&D-conserved & domain walls + monopoles + embedded strings \\
			\hline 
			\multirow{2}{2cm}{$~~~~\mathcal{G}_{333}$}&D-broken & textures\\
			\cline{2-3}
			&D-conserved & domain walls + textures \\
			\hline
		\end{tabular}
		\caption{\small Possible topological defects that can arise in the process of spontaneous breaking of GUT groups via different intermediate symmetries. The formation of topological defects does not get affected by the presence or absence of supersymmetry.} 
		\label{tab:topological_defects}
	\end{center}
\end{table}
We have further imposed the constraints arising from proton life time and cosmological non-observation of topological defects. The exclusion limit on proton decay ($\geq 10^{34}$ years) can be translated to constrain the lower limit of the unification scale which reads as $M_X  \gtrsim 10^{16}$ GeV. To be consistent with cosmological observations, there should not be any topological defects in nature, which are stable till date. Thus, if they are arising through the GUT or intermediate symmetry breaking, they must be inflated away which in turn implies that $M_R$ must be at pre-inflation era, i.e., $ \gtrsim 10^{12}$ GeV. 
\begin{figure}[h!]
        \centering
	\subfloat[Non-SUSY; D-Conserved]{
	\includegraphics[height=4.7cm]{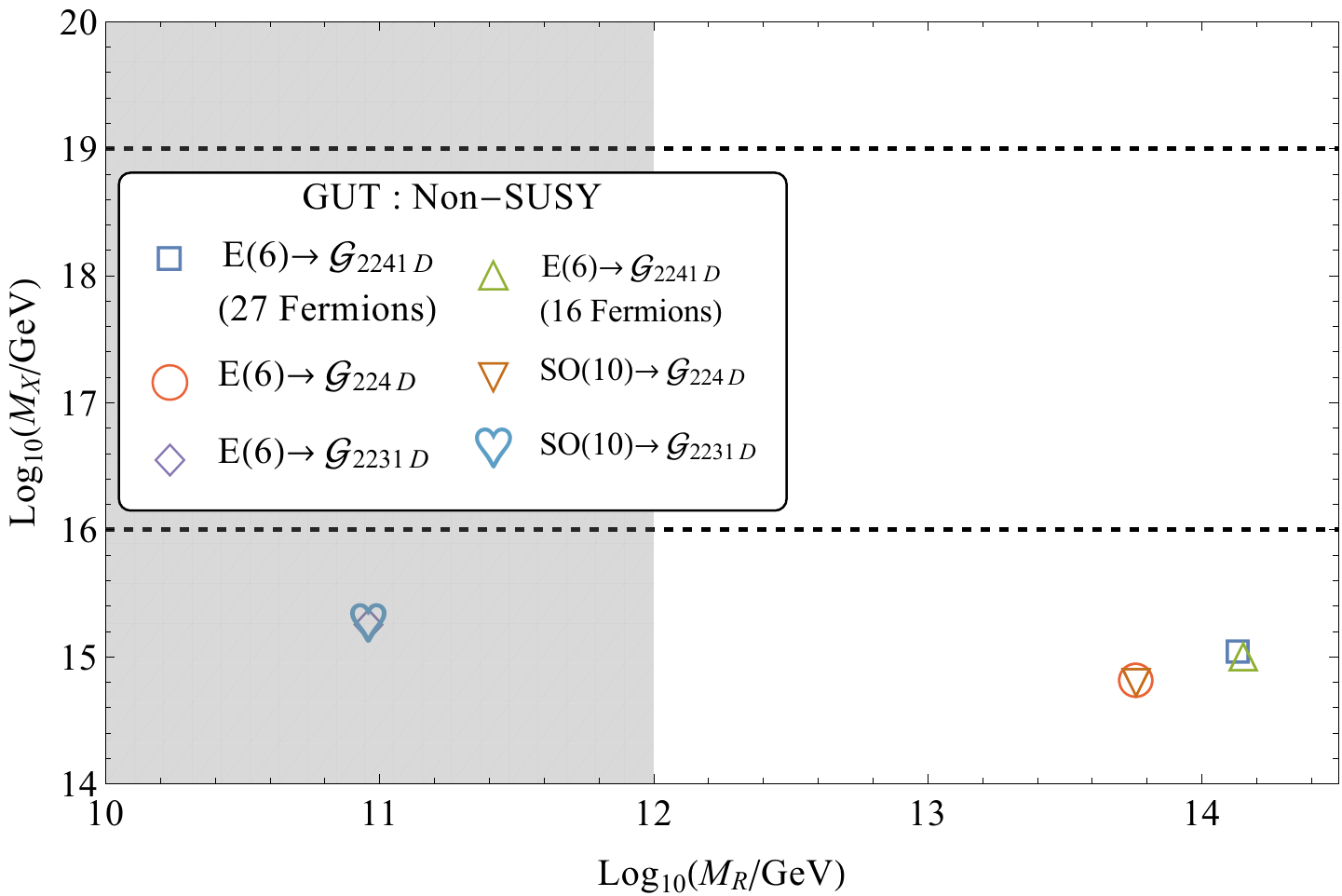}
	\label{fig:uniDpar_1}}
	\subfloat[SUSY; D-Conserved]{
	\includegraphics[height=4.7cm]{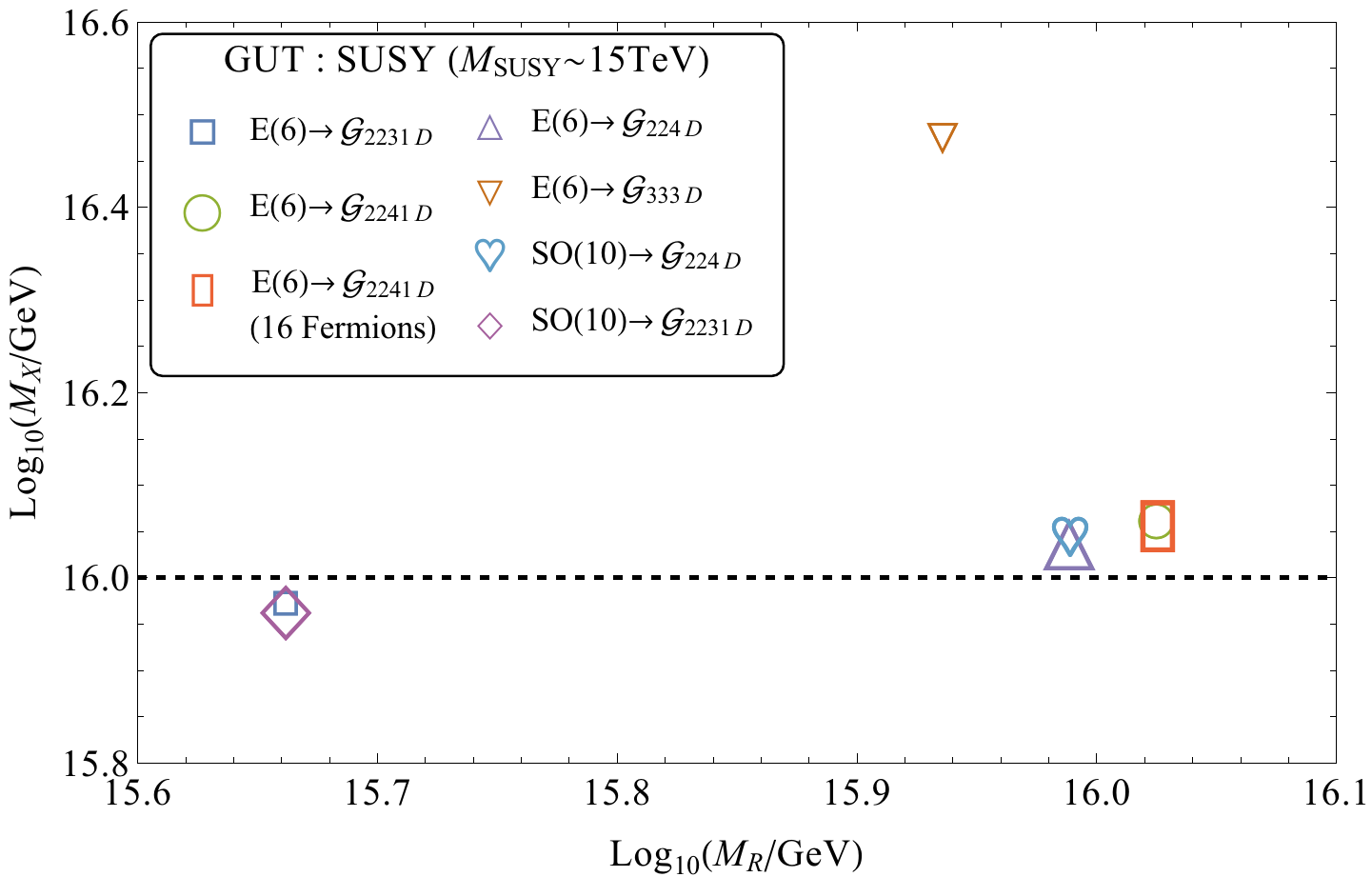}
	\label{fig:uniDpar_2}}\\
	\subfloat[Non-SUSY; D-Broken]{
	\includegraphics[height=4.7cm]{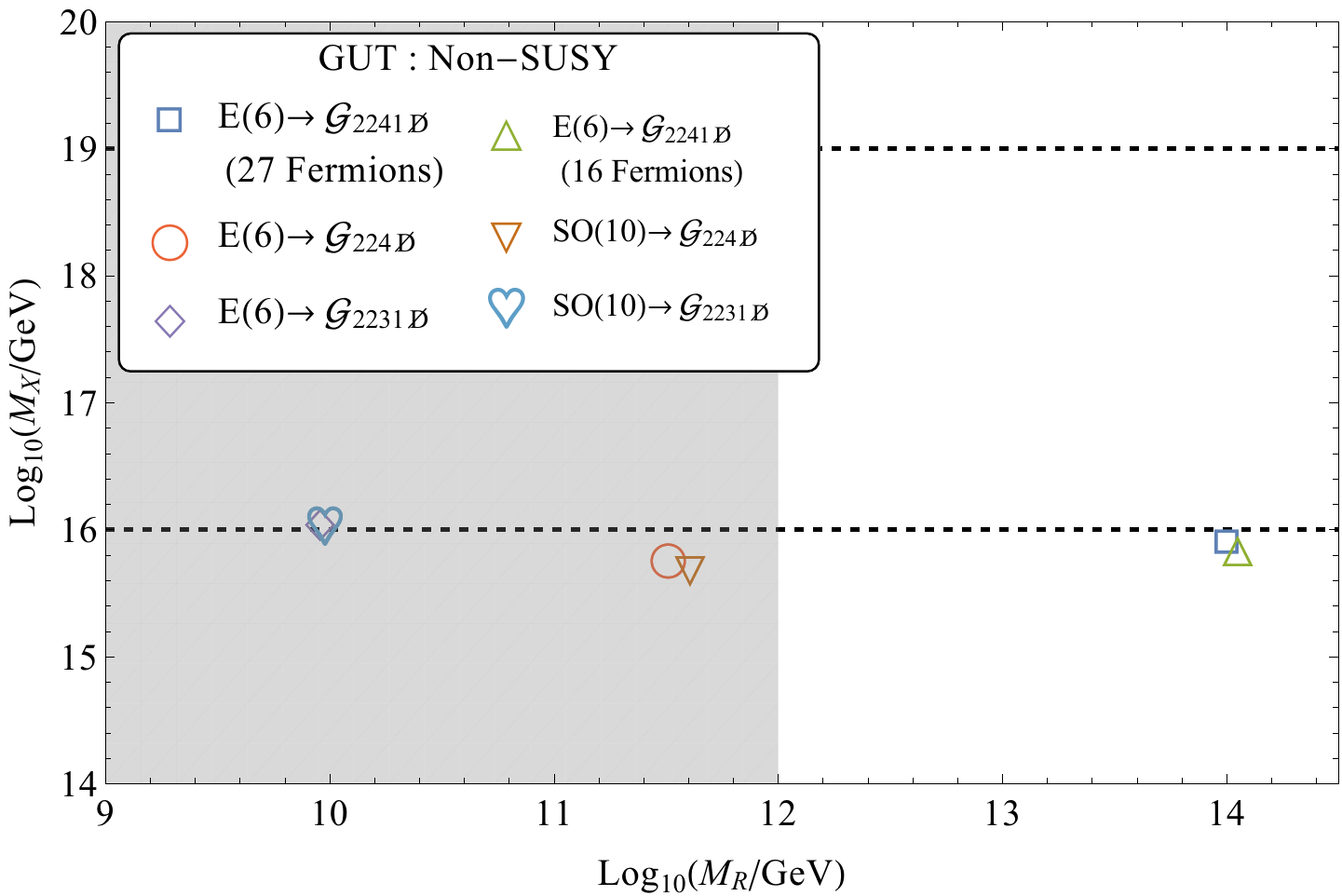}
	\label{fig:uniDpar_3}}
	\subfloat[SUSY; D-Broken]{
	\includegraphics[height=4.7cm]{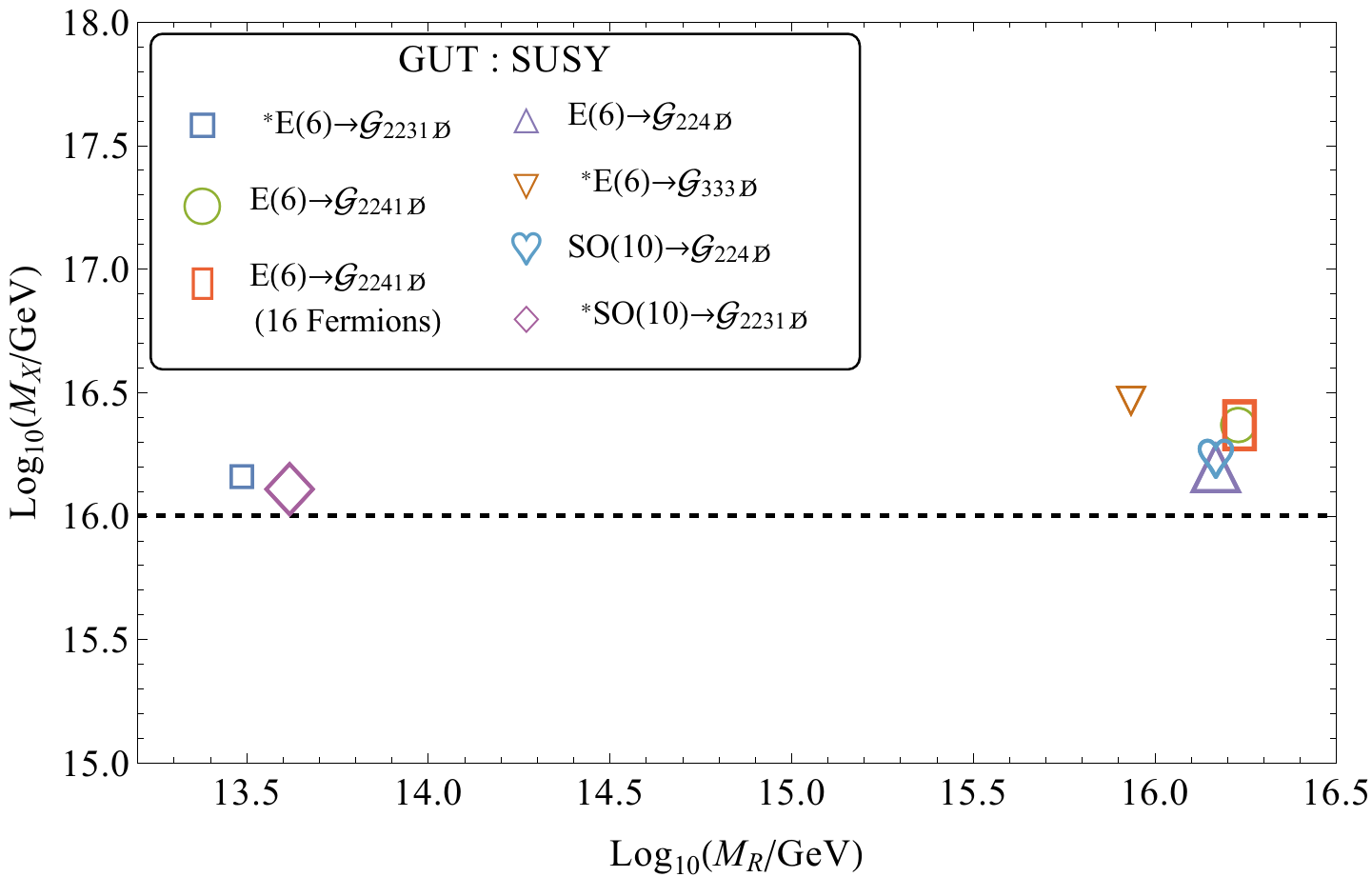}
	\label{fig:uniDpar_4}}
	\caption{\small Correlations among intermediate ($M_R$) and unification ($M_X$) scales: satisfying gauge coupling unification
	for conserved (Top row) and broken (Bottom row) D-parity within  Non-supersymmetric (Left column)) and supersymmetric (Right column) scenarios.  The grey shade depicts the exclusion limits on intermediate scale due to topological defects.}
	\label{fig:unific-D-parity}
\end{figure}
\begin{figure}[h!]
       \centering
	\subfloat[Non-SUSY; 16 fermions]{
	\includegraphics[scale=0.65]{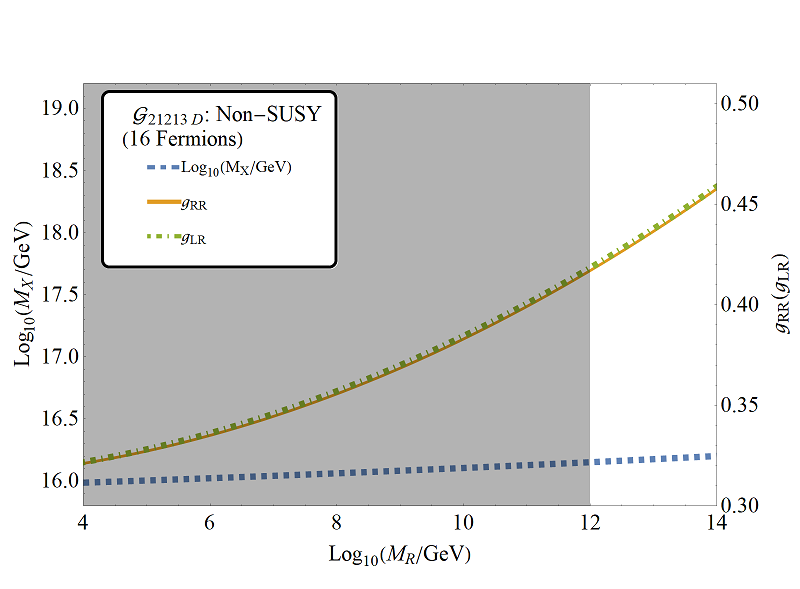}
	\label{fig:21213D-NS_1}}
	\subfloat[Non-SUSY; 27 fermions]{
	\includegraphics[scale=0.65]{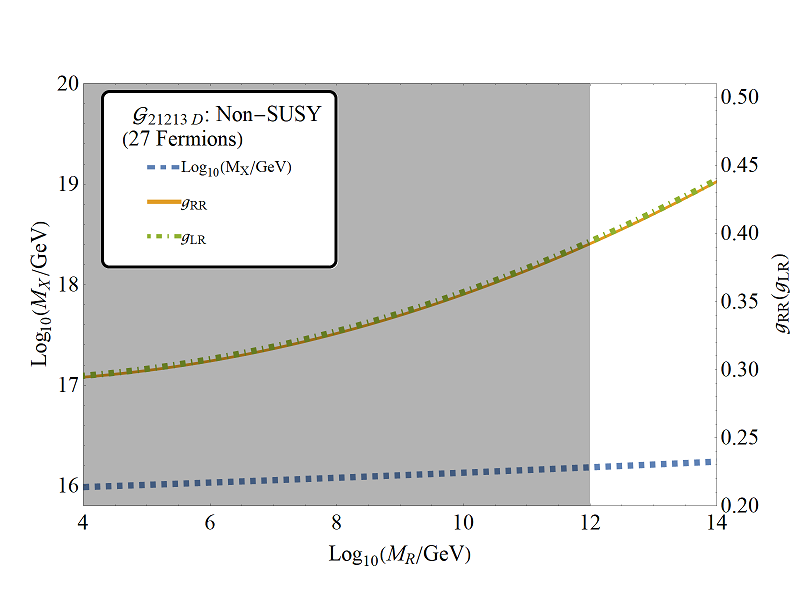}
	\label{fig:21213D-NS_2}}\\
	\subfloat[SUSY; 16 fermions]{
	\includegraphics[scale=0.65]{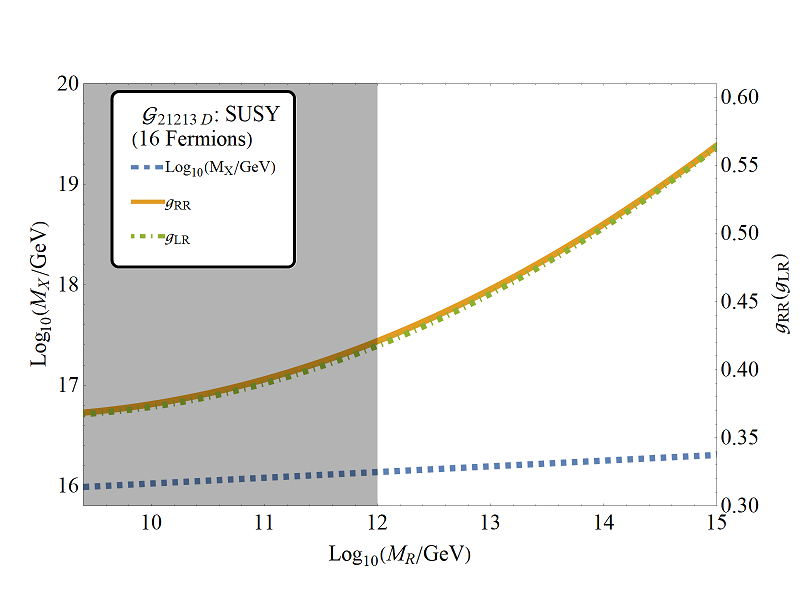}
	\label{fig:21213D-NS_3}}
	\subfloat[SUSY; 27 fermions]{
	\includegraphics[scale=0.65]{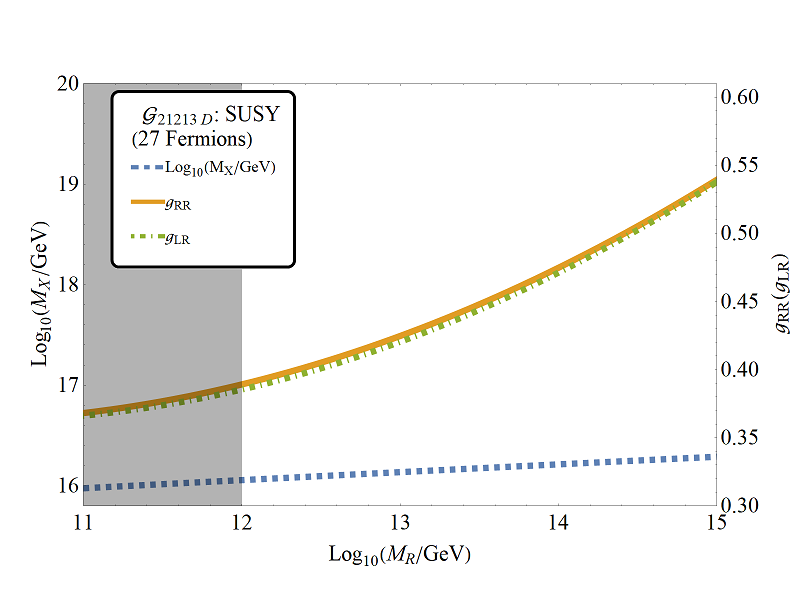}
	\label{fig:21213D-NS_4}}
	\caption{\small Gauge coupling unification for the following symmetry breaking pattern: $E(6) \to \mathcal{G}_{21213D}\to SM$. 
	Here D-parity is conserved, thus $g_{2R}= g_{2L}$. Due the presence of multiple $U(1)$ symmetry there will be abelian mixing in the gauge coupling running. There will be a $(2\times 2)$ abelian gauge coupling matrix where the elements are $g_{LL}=g_{RR}, g_{LR}=g_{RL}$. Variation of intermediate ($M_R$) scale with the unification ($M_X$) scale ($y_1$-axis), and abelian couplings ($y_2$-axis) after satisfying gauge coupling unification have been shown  with $16$-fermions  (a \& c) and 	$27$-fermions (b \& d) at the intermediate scale for Non-SUSY (a \& b) and SUSY (c \& d) scenarios. The grey shade depicts the exclusion limits on intermediate scale due to topological defects.}
	\label{fig:21213D-NS}
\end{figure}
\begin{figure}[h!]
        \centering
	\subfloat[Non-SUSY; 16 fermions]{
	\includegraphics[scale=0.65]{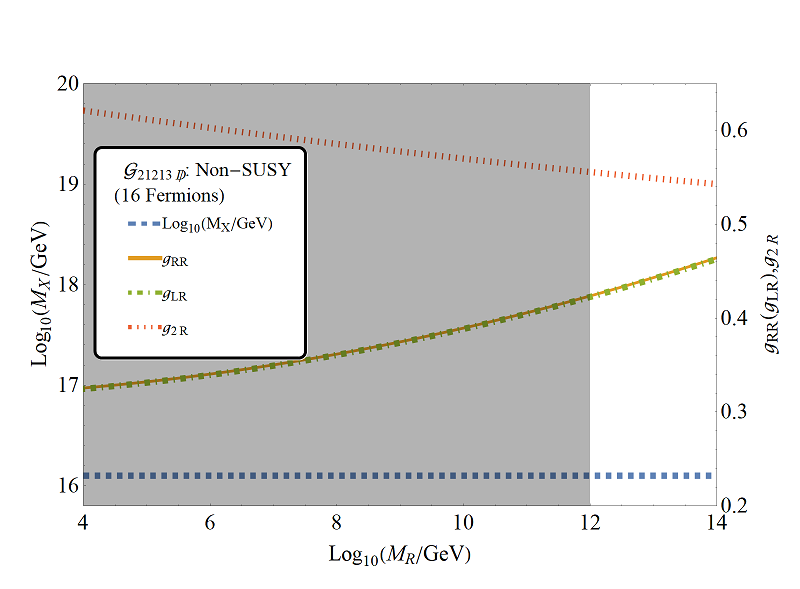}
	\label{fig:21213-NS_1}}
	\subfloat[Non-SUSY; 27 fermions]{
	\includegraphics[scale=0.65]{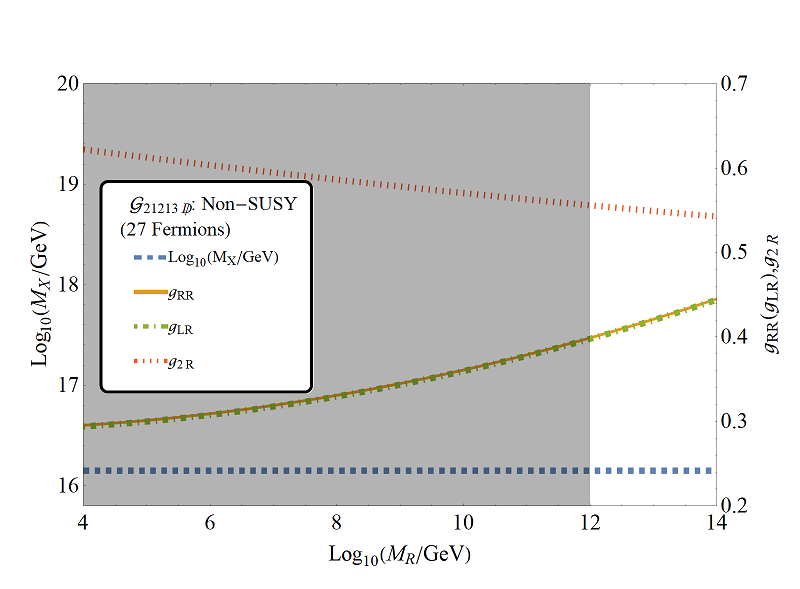}
	\label{fig:21213-NS_2}}\\
	\subfloat[SUSY; 16 fermions]{
	\includegraphics[scale=0.65]{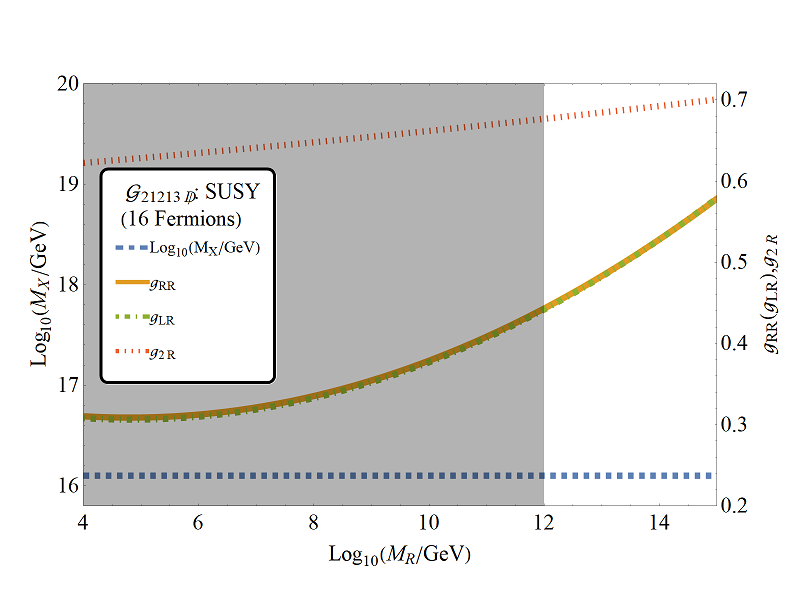}
	\label{fig:21213-NS_3}}
	\subfloat[SUSY; 27 fermions]{
	\includegraphics[scale=0.65]{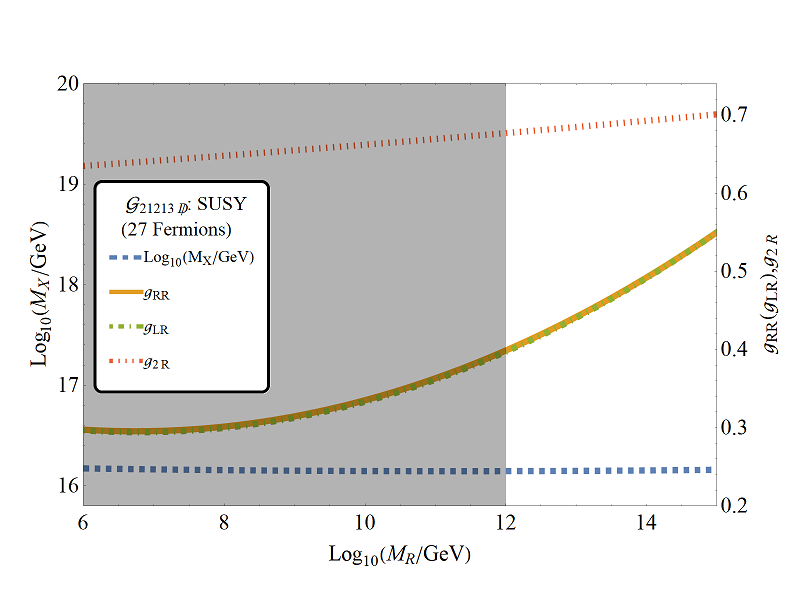}
	\label{fig:21213-NS_4}}
	\caption{\small Gauge coupling unification for the following symmetry breaking pattern: $E(6) \to \mathcal{G}_{21213\slashed{D}}\to SM$. 
	Here D-parity is broken, thus $g_{2R}\neq g_{2L}$, and treated as a free parameter. Due to the presence of multiple $U(1)$ symmetry there will be abelian mixing in the gauge coupling running. There will be a $(2\times 2)$ abelian gauge coupling matrix where the elements are $g_{LL}=g_{RR}, g_{LR}=g_{RL}$. Variation of intermediate ($M_R$) scale with the unification ($M_X$) scale ($y_1$-axis), and gauge couplings ($y_2$-axis) after satisfying gauge coupling unification have been shown  with $16$-fermions  (a \& c) and  
	$27$-fermions (b \& d) at the intermediate scale for Non-SUSY (a \& b) and SUSY (c \& d) scenarios. The grey shade depicts the exclusion limits on intermediate scale due to topological defects.}
	\label{fig:21213-NS}
\end{figure}
In Fig.~\ref{fig:unific-D-parity}, we have noted the solutions of the two loop RGEs in terms of the $M_R$ and $M_X$ scales for D-parity conserving (Top row) and broken (Bottom row)  cases. 
 We have shown the correlations between $M_R$ and $M_X$ scales  for Non-supersymmetric case in Fig.s~\ref{fig:uniDpar_1} \& \ref{fig:uniDpar_3},   and for supersymmetric scenario in Fig.s~\ref{fig:uniDpar_2} \& \ref{fig:uniDpar_4}.  
The grey shade depicts the exclusion limits due to topological defects, i.e., $M_R \gtrsim 10^{12}$ GeV. It is worth mentioning that apart from intermediate symmetry $\mathcal{G}_{2231D}$, which arise from the breaking of $SO(10)$ and $E(6)$, all other breaking chains are consistent with cosmological constraints; see Fig.~\ref{fig:uniDpar_1}. The horizontal dotted lines depict the minimum ($10^{16}$ GeV) and maximum ($10^{19}$ GeV) values of GUT scales consistent with proton decay non-observation. It is easy to realize that in Non-SUSY scenarios almost all of the breaking patterns are troubled by either  proton life time or cosmological constraints; see Fig.s~\ref{fig:uniDpar_1} \& \ref{fig:uniDpar_3}.  But in SUSY case, the picture is different. All the intermediate scales are way beyond $10^{12}$ GeV and unification scales are also around $10^{16}$ GeV; see Fig.s~\ref{fig:uniDpar_2} \& \ref{fig:uniDpar_4}. Thus all the breaking chains are safe from constraints due to the stable topological defects and proton life time.

It is interesting to note that apart from the intermediate groups $ \mathcal{G}_{2231}$ and $\mathcal{G}_{224}$, which arise either from $E(6)$ or $SO(10)$, other breaking chains are compatible with the constraints  due to stable topological defects; see Fig.~\ref{fig:uniDpar_3}. These scenarios are also lying at the edge of the lower bound on $M_X$. The group $\mathcal {G}_{2241\slashed{D}}$ is also living dangerously at the edge of this bound. These observations are only for Non-SUSY models. 
The status of the solutions for SUSY scenario Fig.~\ref{fig:uniDpar_4} are completely different.  As the intermediate scale for all models are $\geq 10^{15}$ GeV, they are safe from stable topological defects. For these models unification scales are also above $10^{16}$ GeV, thus consistent with proton decay non-observation data Fig.~\ref{fig:uniDpar_4}.
So far we have discussed the breaking patterns which predict unique intermediate scales. Now we will discuss the intermediate gauge groups that contain more than one abelian symmetries. This implies that there will be abelian mixing even at the one-loop level in the RGEs from $M_R$ to $M_X$ scale. 
We have considered the breaking of $E(6)$ to $ \mathcal{G}_{21213D}$ which is the only breaking chain under consideration. For this particular intermediate gauge group, we have discussed the D-parity conserved and broken scenarios within both  Non-SUSY and SUSY frameworks. We have further adopted two varieties, (i) only 16-fermions surviving at the intermediate scale (see  Fig.s~\ref{fig:21213D-NS_1} \& \ref{fig:21213D-NS_3}), and (ii) all the 27 fermions are light enough to be present till the $M_R$ scale (see Figs.~\ref{fig:21213D-NS_2} \& \ref{fig:21213D-NS_4}). In Fig.s~\ref{fig:21213D-NS_3} \& \ref{fig:21213D-NS_4} we have adopted the similar scenario as in Fig.s~\ref{fig:21213D-NS_1} \& \ref{fig:21213D-NS_2} within SUSY frameworks with similar two situations.
As D-parity is conserved, we have $g_{2L}=g_{2R}$. We have also considered $g_{LL}=g_{RR}$ and $g_{LR}=g_{RL}$. 
Similar to the earlier analysis, the grey shade depicts the exclusion limit $M_R \geq 10^{12}$ GeV due to stable topological defects. In these plots, the blue (dashed) lines stand for the allowed ranges of $M_X$, where the same for $g_{RR}$ and $g_{LR}$ are shown by the orange (solid) and green (dot-dashed) lines. These indicate that the ranges of both $g_{RR}$ and $g_{LR}$, allowed by cosmological constraints, are [0.411 : 0.469], [0.385 : 0.453], [0.412 : 0.526], and [0.386 : 0.492], for Non-SUSY (16 fermions), Non-SUSY (27 fermions), SUSY (16 fermions), and SUSY (16 fermions) cases respectively.

We have considered the breaking of $E(6)$ to $ \mathcal{G}_{21213\slashed{D}}$ where D-parity is not conserved within Non-SUSY (a \& b) and SUSY (c \& d) frameworks. We have assumed two varieties, (i) when only 16-fermions survive at the intermediate scale, see  Fig.s~\ref{fig:21213-NS_1} \& \ref{fig:21213-NS_3}, and (ii) all the 27 fermions are light enough to be present till the $M_R$ scale, see Fig.s~\ref{fig:21213-NS_2} \& \ref{fig:21213-NS_4}. In Fig.s~\ref{fig:21213-NS_3} \& \ref{fig:21213-NS_4} we have performed the similar analysis but in presence of SUSY. Similar to the previous cases, we cannot ignore the abelian mixing in the presence of two $U(1)$ at the intermediate scale. As the  D-parity is broken at the intermediate scale, we have $g_{2L} \neq g_{2R}$, and thus $g_{2R}$ is a free parameter unlike the earlier D-conserved scenario. Here we have further considered $g_{LR}=g_{RL}$, and $g_{RR}$ is a free parameter while $g_{LL}$ is obtained from the suitable matching conditions. Here too, the grey shade depicts the exclusion limit $M_R \geq 10^{12}$ GeV due to topological defects. In these plots, the blue (dashed) lines stand for the allowed ranges of $M_X$, where the same for $g_{RR}$ and $g_{LR}$ are shown by the orange (solid) and green (dot-dashed) lines. Unlike the previous case, here we have an allowed range of solutions for $g_{2R}$ as well, shown by the red (dotted) lines. These indicate that the ranges of both $g_{RR}$ and $g_{LR}$, allowed by cosmological constraints, are [0.419 : 0.470], [0.391 : 0.454], [0.431 : 0.530], and [0.388 : 0.497], for Non-SUSY (16 fermions), Non-SUSY (27 fermions), SUSY (16 fermions), and SUSY (16 fermions) cases respectively. The ranges allowed for $g_{2R}$ for these cases in the same order are [0.540 : 0.558], [0.539 : 0.560], [0.672 : 0.697], and [0.670 : 0.701].

\subsection{Unification in presence of dimension-5 operator}
\label{subsec:dim-5-unific}

So far we have discussed the gauge coupling unification in the light of the renormalisable unified gauge kinetic term. We have found in the previous section that for some models, either the unification scale is not compatible with proton life time or the gauge couplings are not unifying within Planck scale. These have been a motivation for us to pursue the unification program by smearing the unified boundary condition, through  the higher dimensional operator in the gauge kinetic sector. Incorporation of such non-renormalisable terms in the lagrangian is not unrealistic. In fact, GUT is often considered as an effective theory, as it does not contain gravity. It is indeed possible to have footprints of Planck scale physics through some higher dimensional terms. As discussed in section~\ref{subsec:dim-5}, gravitational effects can smear the gauge coupling unification criteria at the GUT scale itself. 

We can add a non-renormalisable dimension-5 operator in the gauge kinetic sector as in Eq.~\ref{Eq:dim-5}. The corrections due to this operator are weighted by the group theoretic factors ($\delta_i$) which are listed in Tables~\ref{tab:dimension_5_I}, \ref{tab:dimension_5_II}, and \ref{tab:dimension_5_III}. These $\delta_i$'s  depend on the choice of GUT breaking scalars and their vacuum orientations which decide the breaking patterns. Here we have used the modified unification boundary conditions and taken the impact of the dimension-5 contribution ($\varepsilon$) into consideration. 

\begin{figure}[ht]
	\centering
	\subfloat[Non-SUSY]{
		\includegraphics[height=5cm]{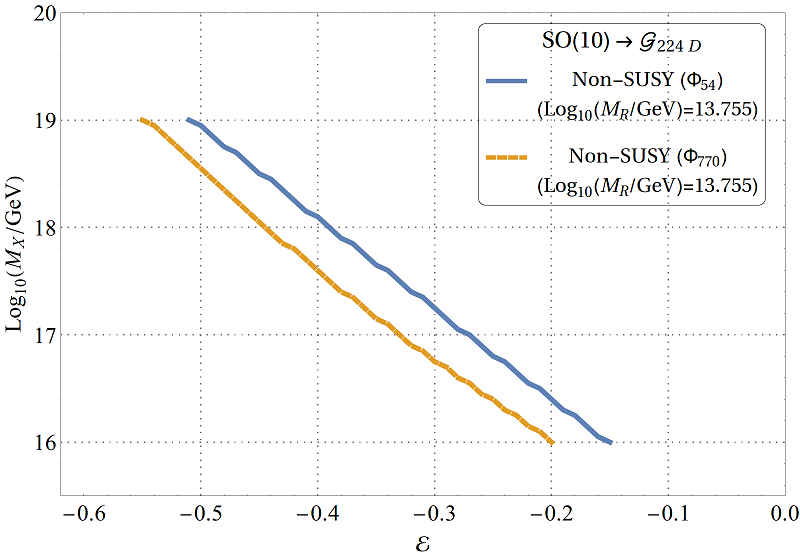}
		\label{fig:224D-dim5_1}}
	\subfloat[SUSY]{
		\includegraphics[height=5cm]{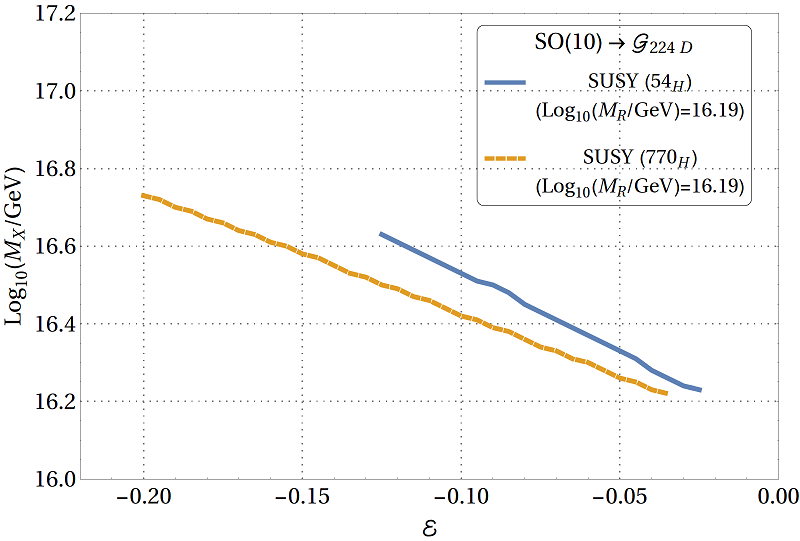}
		\label{fig:224D-dim5_2}}
	\caption{\small Gauge coupling unification for the following symmetry breaking pattern: $SO(10) \to \mathcal{G}_{224D}\to SM$.
		$M_X$ as a function of  strength of dimension-5 operator $(\varepsilon$) for fixed $M_R$ is noted for Non-SUSY (a) and SUSY (b) scenarios for GUT breaking scalars of 54 and 770-dimensional.  All of the $M_R$ values are allowed by cosmological constraints ($M_R \gtrsim 10^{12}$ GeV).}
	\label{fig:224D-dim5}
\end{figure}

The impact of dimension-5 operators can only be realised for rank-preserving breaking.
First, we have considered the breaking-chain $SO(10) \to \mathcal{G}_{224D}$ using the VEVs of 54 and 770 dimensional scalars. As we have mentioned, the intermediate scale is not affected by this new operator and thus also not by the choice of GUT breaking scalar. 
For this particular case, the intermediate scales are fixed at $M_R\simeq 10^{13.76}$ GeV, and $10^{16.19}$ GeV for Non-SUSY and SUSY scenarios respectively\footnote{We have set $M_{SUSY} \sim 1$ TeV in this section.}.
We have determined  the allowed range of $M_X$ depending on the range of values of $\varepsilon$.  
In Fig.~\ref{fig:224D-dim5_1}, we have found that for Non-SUSY case, unification scale can vary in between $[10^{16}:10^{19}]$ GeV with 
$\varepsilon \in [-0.15:-0.51]$ and $\in [-0.20:-0.55]$ for 54- and 770-dimensional scalars respectively. In Fig.~\ref{fig:224D-dim5_2}, we have performed the similar analysis in presence of  SUSY. Here, we have noted the solutions as $M_X \in [10^{16.23}:10^{16.63}]$ GeV with $\varepsilon \in [-0.025:-0.125]$ for $\Phi_{54}$ and $M_X \in [10^{16.22}:10^{16.73}]$ GeV with $\varepsilon \in [-0.035:-0.2]$ for $\Phi_{770}$. 
\begin{figure}[h!]
        \centering
	\subfloat[Non-SUSY]{
	\includegraphics[scale=0.75]{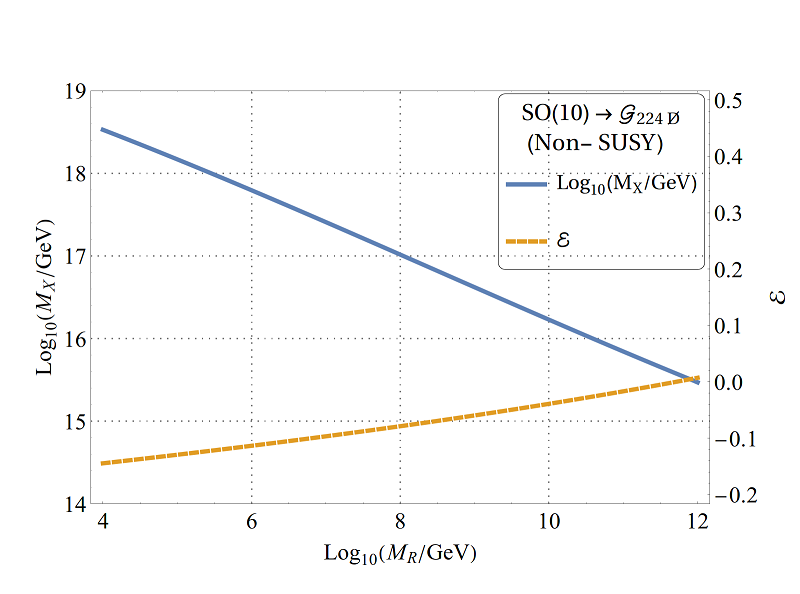}
	\label{fig:224-dim5-dodd_1}}
	\subfloat[SUSY]{
	\includegraphics[scale=0.75]{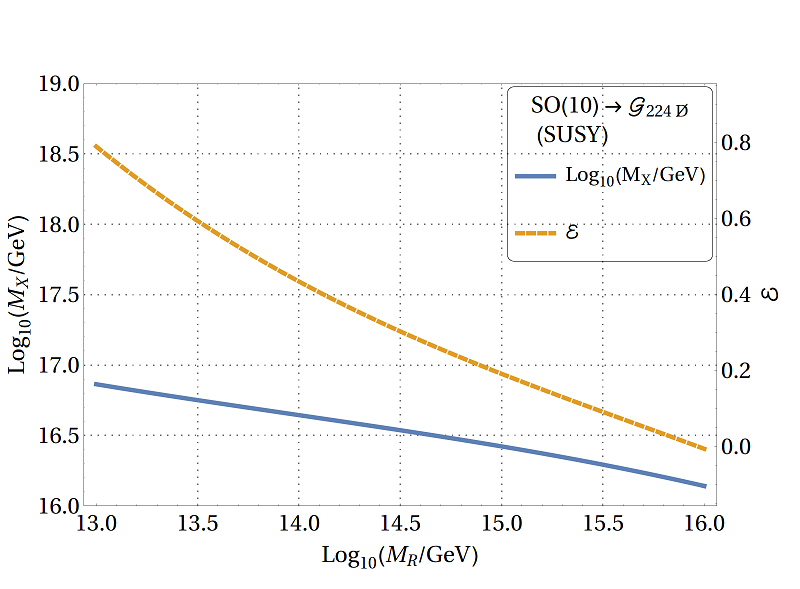}
	\label{fig:224-dim5-dodd_2}}
	\caption{\small Gauge coupling unification for the following symmetry breaking pattern: $SO(10) \to \mathcal{G}_{224\slashed{D}}\to SM$, in presence of the GUT breaking scalar $\Phi_{210}$.
	$M_X$ and strength of dimension-5 operator $(\varepsilon$) as a function of $M_R$ is noted 
	for Non-SUSY (a) and SUSY (b) scenarios.  Whereas the case depicted in Fig. (a) is in trouble from cosmological constraints, the whole range of $M_R$ is allowed in Fig. (b).}
	\label{fig:224-dim5-dodd}
\end{figure}
Next we have considered the similar breaking pattern, $SO(10) \to \mathcal{G}_{224\slashed{D}}$,  but now D-parity is broken. This is achieved through the VEV of $\Phi_{210}$. As the D-parity is not a good symmetry,  we do not have $g_{2L} \neq g_{2R}$. Thus, unlike the previous cases, $M_R$ is  a free parameter here and we can get a range of solutions for $M_R$ consistent with unification. In Fig.~\ref{fig:224-dim5-dodd_1}, we have analysed the unification for Non-SUSY scenario. The blue and yellow lines show the allowed range of $M_X\in [10^{16}:10^{18.5}]$ GeV (read $y_1$ axis for label), and $\varepsilon \in [-0.026:-0.145]$ (read $y_2$ axis for label) compatible with the range of $M_R\in[10^{4}:10^{10.6}]$ GeV respectively. It is clear from this plot that the $M_R$ could marginally satisfy the cosmological constraints at $10^{12}$ GeV at the cost of proton life time constraint as the respective unification scale is $\sim 10^{15.5}$ GeV.
In Fig.~\ref{fig:224-dim5-dodd_2}, we have performed a similar analysis in presence of supersymmetry. Here, blue and yellow lines show the allowed range of $M_X\in [10^{16.14}:10^{16.86}]$ GeV ($y_1$ axis), and $\varepsilon \in [0.006:0.79]$ ($y_2$ axis) compatible with the range of $M_R\in[10^{13}:10^{16}]$ GeV respectively.
\begin{figure}[h!]
        \centering
	\subfloat[Non-SUSY]{
	\includegraphics[scale=0.75]{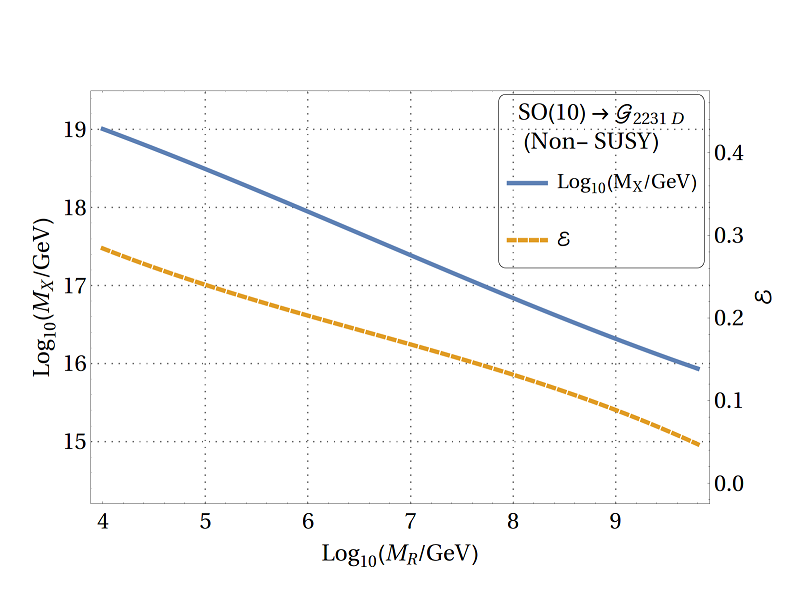}
	\label{fig:2231-dim5_1}}
	\subfloat[SUSY]{
	\includegraphics[scale=0.75]{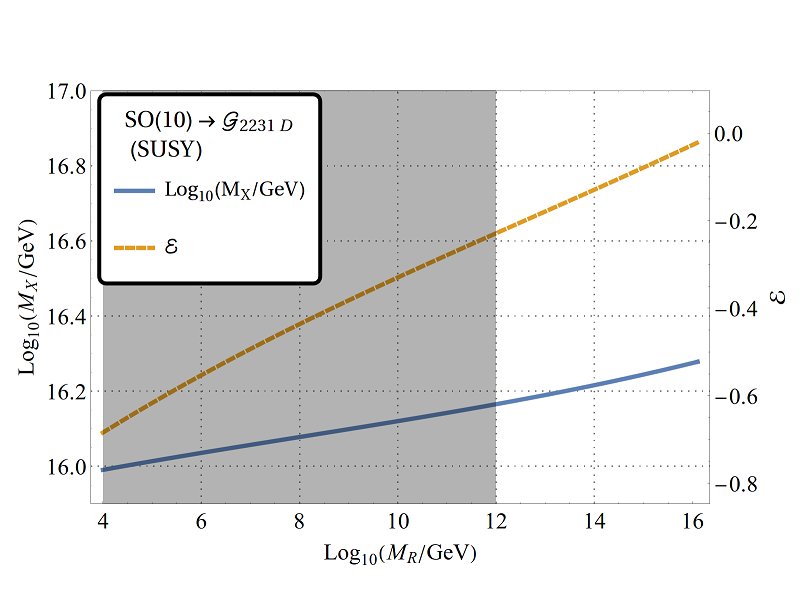}
	\label{fig:2231-dim5_2}}
	\caption{\small Gauge coupling unification for the following symmetry breaking pattern: $SO(10) \to \mathcal{G}_{2231D}\to SM$, in presence of the GUT breaking scalar $\Phi_{(54 + 210)}$.
	$M_X$ and $\varepsilon$ as a function of $M_R$ is noted 
	for Non-SUSY (a) and SUSY (b) scenarios.  Whereas the case depicted in Fig. (a) is in trouble from cosmological constraints, the range of $M_R$ values incompatible with them is shaded grey in Fig. (b).}
	\label{fig:2231-dim5}
\end{figure}
\begin{figure}[h!]
	\centering
	\subfloat[Non-SUSY]{
		\includegraphics[scale=0.75]{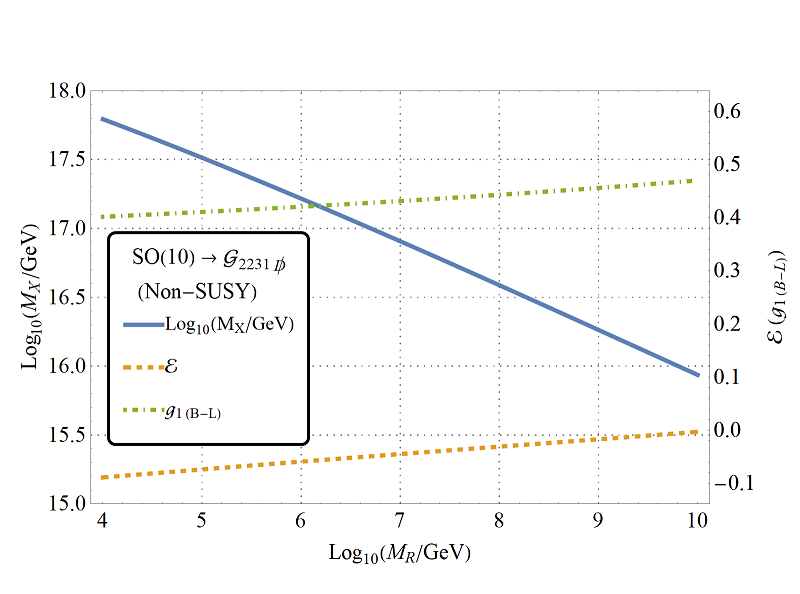}
		\label{fig:2231do-dim5_1}}
	\subfloat[SUSY]{
		\includegraphics[scale=0.75]{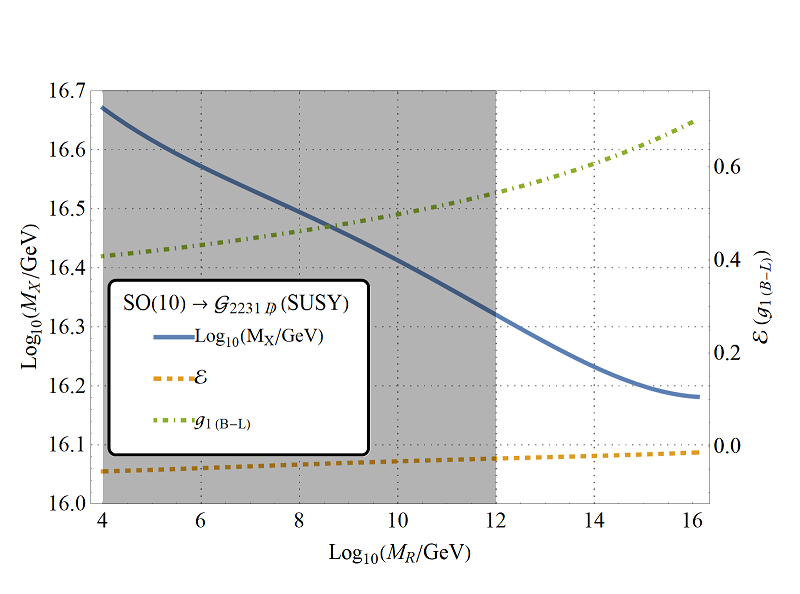}
		\label{fig:2231do-dim5_2}}
	\caption{\small Gauge coupling unification for the following symmetry breaking pattern: $SO(10) \to \mathcal{G}_{2231\slashed{D}}\to SM$, in presence of the GUT breaking scalar $\Phi_{(210+45)}$.
		$M_X$ and $\varepsilon$ as a function of $M_R$ is noted 
		for Non-SUSY (a) and SUSY (b) scenarios.  Whereas the case depicted in Fig. (a) is in trouble from cosmological constraints, the range of $M_R$ values incompatible with them is shaded grey in Fig. (b).}
	\label{fig:2231do-dim5}
\end{figure}
Next we have considered breaking of $SO(10)$ to $ \mathcal{G}_{2231D}$ using the VEVs of a (54+210)-dimensional scalars. As the D-parity is intact,  we have $g_{2L} = g_{2R}$ at the intermediate scale, but unlike the previous scenario (Fig.~\ref{fig:224D-dim5}),  $M_R$ is a free parameter. Thus the unification solutions can be given in terms of $M_X,M_R$, and $\varepsilon$, similar to Fig.~\ref{fig:224-dim5-dodd}. 
In Fig.~\ref{fig:2231-dim5_1}, we have analysed the unification for Non-SUSY scenario. The blue and yellow lines show the allowed range of $M_X\in [10^{16}:10^{19}]$ GeV (read $y_1$ axis for label), and $\varepsilon \in [0.045:0.284]$ (read $y_2$ axis for label) compatible with the range of $M_R\in[10^{4}:10^{9.8}]$ GeV respectively. For this particular scenario, the whole range of intermediate scale is incompatible with the constraints due to topological defects.
In Fig.~\ref{fig:2231-dim5_2}, we have performed a similar analysis in presence of supersymmetry. Here, blue and yellow lines show the allowed range of $M_X\in [10^{16}:10^{16.3}]$ GeV ($y_1$ axis), and $\varepsilon \in [-0.68:-0.03]$ ($y_2$ axis) compatible with the range of $M_R\in[10^{4}:10^{16}]$ GeV respectively. The grey-shaded region in this plot depicts the exclusion limits due to the cosmological constraints.

In Fig.~\ref{fig:2231do-dim5}, we have performed an analysis similar to Fig.~\ref{fig:2231-dim5}, but with the D-parity broken case. Here $SO(10)$ is broken to $\mathcal{G}_{2231\slashed{D}}$ using the VEVs of (210+45)-dimensional scalars. In Fig.~\ref{fig:2231do-dim5_1}, we have analysed the unification for the Non-SUSY scenario. The blue and yellow lines show the allowed range of $M_X\in [10^{16}:10^{17.8}]$ GeV (read $y_1$ axis for label), and $\varepsilon \in [-0.006:-0.089]$, $g_{1(B-L)} \in [0.40:0.47]$ (read $y_2$ axis for label) compatible with the range of $M_R\in[10^{4}:10^{9.8}]$ GeV respectively.  For this scenario too, the full range of intermediate scale is incompatible with the constraints due to topological defects.
In Fig.~\ref{fig:2231do-dim5_2}, we have performed a similar analysis in presence of SUSY. Here, blue and yellow lines show the allowed range of $M_X\in [10^{16.18}:10^{16.67}]$ GeV ($y_1$ axis), and $\varepsilon \in [-0.014:-0.054]$, $g_{1(B-L)} \in [0.41:0.70]$ ($y_2$ axis) compatible with the range of $M_R\in[10^{4}:10^{16}]$ GeV respectively.  The grey-shaded region in this plot depicts the exclusion limits due to the cosmological constraints.
\begin{figure}[h!]
        \centering
	\includegraphics[scale=0.750]{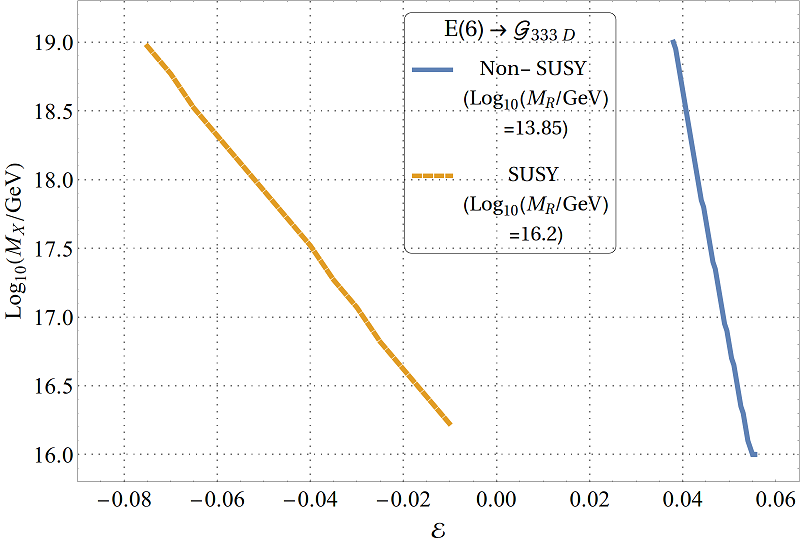}
	\caption{\small Gauge coupling unification for the following symmetry breaking pattern: $E(6) \to \mathcal{G}_{333D}\to SM$.
	$M_X$ as a function of $\varepsilon$ for fixed $M_R$ is noted 
	for Non-SUSY and SUSY scenarios for GUT breaking scalar $\Phi_{650}$.  All of the $M_R$ values are allowed by cosmological constraints ($M_R \gtrsim 10^{12}$ GeV).}
	\label{fig:333D-dim5}
\end{figure}

$\mathcal{G}_{333D}$ is the maximal subgroup of $E(6)$. We have noted in the last section that D-parity conserved and broken cases cannot be discriminated from each other by looking into their respective RGEs, as they are exactly the same. But the contributions from the dimension-5 operator are different based on whether D-parity is broken or not.  
\begin{figure}[h!]
        \centering
	\subfloat[Non-SUSY]{
	\includegraphics[scale=0.75]{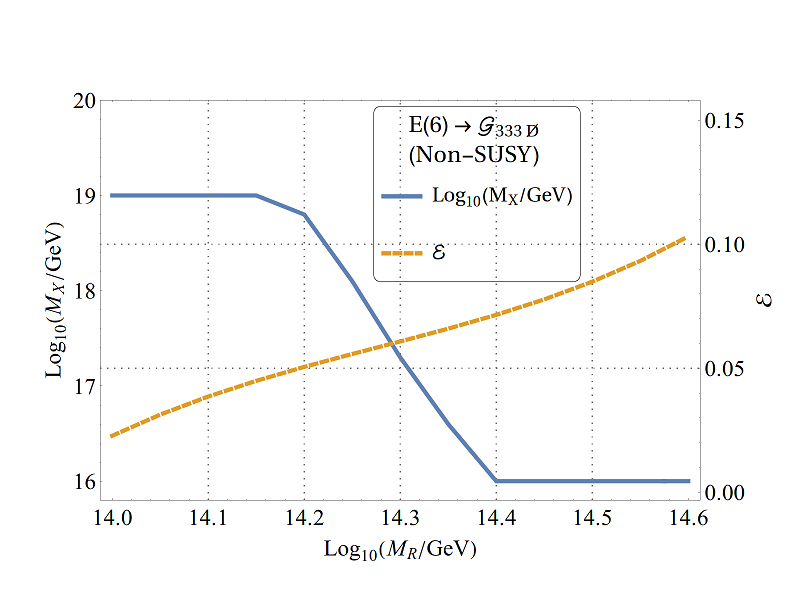}
	\label{fig:333-dim5-dodd_1}}
	\subfloat[SUSY]{
	\includegraphics[scale=0.75]{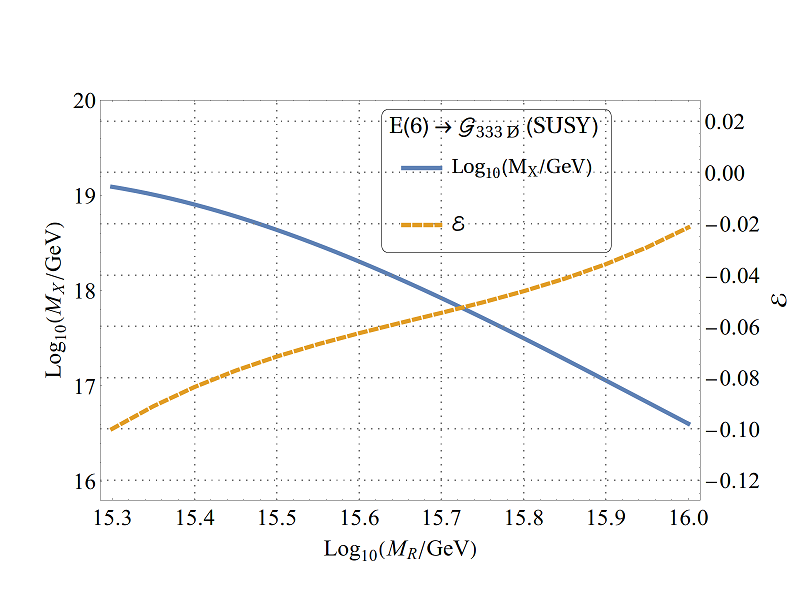}
	\label{fig:333-dim5-dodd_2}}
	\caption{\small Gauge coupling unification for the following symmetry breaking pattern: $E(6) \to \mathcal{G}_{333\slashed{D}}\to SM$.
	$M_X$ and $\varepsilon$ as a function of $M_R$ is noted 
	for Non-SUSY (a) and SUSY (b) scenarios.  The whole range of $M_R$ values are allowed by cosmological constraints ($M_R \gtrsim 10^{12}$ GeV).}
	\label{fig:333-dim5-dodd}
\end{figure}
At the intermediate scale, unbroken D-parity is ensured by the VEVs of $\Phi_{650}$ and $\Phi_{2430}$. The $\delta_i$'s computed using 
$\Phi_{2430}$ are all equal and they do not alter the unification what we have achieved in earlier sections. This is true for both Non-SUSY and SUSY cases. 
In this analysis we have included the impact of $\Phi_{650}$ only.
Here the intermediate scales are fixed at $M_R\simeq 10^{13.85}$ GeV and $10^{16.20}$ GeV in Non-SUSY and SUSY scenarios respectively. 
In Fig.~\ref{fig:333D-dim5} , we have adjudged the Non-SUSY (blue-solid line), and SUSY (yellow-dotted) scenarios. Here the unification scale can vary in between $[10^{16}:10^{19}]$ with $\varepsilon \in [0.038:0.055]$ for Non-SUSY and $[10^{16.22}:10^{19}]$ GeV with $\varepsilon \in [-0.01:-0.075]$ for SUSY cases respectively.

In Fig.~\ref{fig:333-dim5-dodd}, we have considered the similar breaking pattern $E(6) \to \mathcal{G}_{333\slashed{D}}$ using the VEV of a $650^{'}$-dimensional scalar. Unlike the previous scenario, here D-parity is not a good symmetry which implies $g_{3L} \neq g_{3R}$. Thus $M_R$ is  not constrained and a range of solutions for $M_R$ can be consistent with unification. 
In Fig.~\ref{fig:333-dim5-dodd_1}, the blue and yellow lines show the allowed range of unification scale, $M_X\in [10^{16}:10^{19}]$ GeV ($y_1$ axis), 
and $\varepsilon \in [0.023:0.103]$ ($y_2$ axis) compatible with the range of $M_R$ [$10^{14.0}:10^{14.6}$] GeV respectively. This analysis is within Non-SUSY framework. 
In Fig.~\ref{fig:333-dim5-dodd_2}, the blue and yellow lines depict the allowed range of unification scale, $M_X\in [10^{16.6}:10^{19.0}]$ GeV ($y_1$ axis), 
and $\varepsilon \in [-0.02:-0.10]$ ($y_2$ axis) compatible with the range of $M_R\in [10^{15.3}:10^{16.0}$] GeV respectively. These solutions are in the presence of supersymmetry.

\begin{figure}[h!]
	\centering
	\subfloat[Non-SUSY]{
		\includegraphics[scale=0.65]{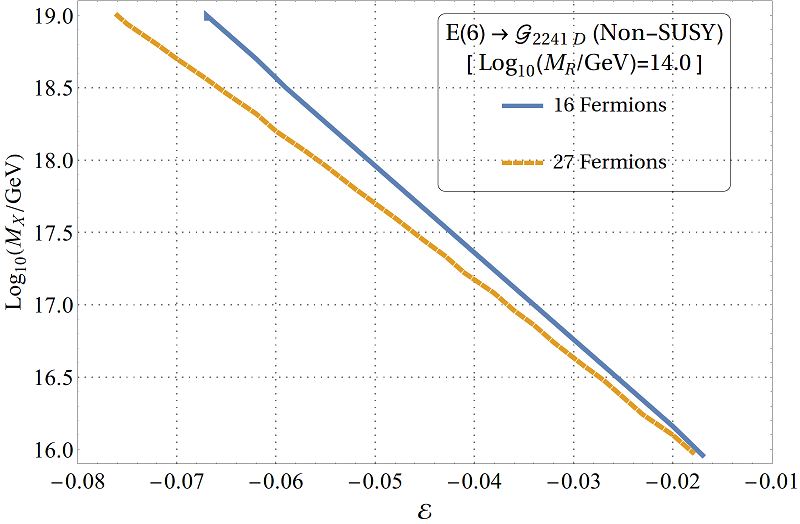}
		\label{fig:2241-dim5-deven_nsusy}}
	\subfloat[SUSY]{
		\includegraphics[scale=0.65]{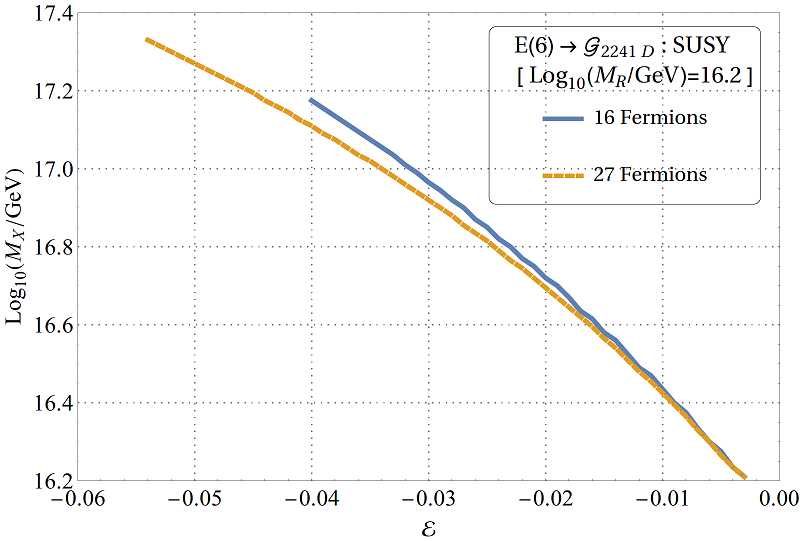}
		\label{fig:2241-dim5-deven_susy}}
	\caption{\small Gauge coupling unification for the following symmetry breaking pattern: $E(6) \to \mathcal{G}_{2241D}\to SM$.
		$M_X$  as a function of $\varepsilon$ is noted 
		for Non-SUSY (a) and SUSY (b) scenarios.  These scenarios are compatible with  cosmological constraint ($M_R \gtrsim 10^{12}$ GeV).}
	\label{fig:2241-dim5-deven}
\end{figure}
Next, we have considered the breaking-chain $E(6) \to \mathcal{G}_{2241D}$ using the VEV of 650-dimensional scalar which contains D-parity preserving $(54,0)$ component  under $SO(10)\otimes U(1)$. 
Here, the intermediate scales are fixed at $M_R\simeq 10^{14}$ GeV, and $10^{16.2}$ GeV for Non-SUSY and SUSY cases respectively.
We have found that the allowed range of $M_X$ depends on the range of values of $\varepsilon$.  
In Fig.~\ref{fig:2241-dim5-deven_nsusy}, we have found that for Non-SUSY case, unification scale can vary in between $[10^{16}:10^{19}]$ GeV with 
$\varepsilon \in [-0.076:-0.018]$ and $ [-0.067:-0.017]$ when 27 and 16 fermions are contributing in the RGEs respectively. In Fig.~\ref{fig:2241-dim5-deven_susy},  we have noted that for SUSY scenario,  the unification scale ($M_X$) can vary in between $[10^{16.21}:10^{17.33}]$ and $[10^{16.21}:10^{17.18}]$ GeV with 
$\varepsilon \in [-0.054:-0.003]$ and $ [-0.040:-0.003]$  when  27 and 16 fermions are contributing in the RGEs, respectively.

In Fig.~\ref{fig:2241-dim5-dodd}, we have considered the breaking pattern $E(6) \to \mathcal{G}_{2241\slashed{D}}$ using the VEV of a $650$-dimensional scalar whose $(210,0)$ component under $SO(10)\otimes U(1)_X$ breaks D-parity. As D-parity is broken, $g_{2L} \neq g_{2R}$. 	Here, the intermediate scale is fixed at $M_R\simeq 10^{14}$ GeV, and $10^{16.1}$ GeV for Non-SUSY and SUSY cases respectively for both 27 and 16-fermion scenarios. 
In Fig.s~\ref{fig:2241-dim5-dodd_nsusy_1} (\ref{fig:2241-dim5-dodd_nsusy_2}), the blue and yellow lines show the allowed ranges of unification scale, $M_X\in [10^{16}:10^{19}]\; \Big([10^{16}:10^{19}]\Big)$ GeV ($y_1$ axis), 
and $g_{1X} \in [0.43:0.51]\; \Big([0.43:0.51]\Big)$ ($y_2$ axis) compatible with the range of $\varepsilon$ [$-0.035:-0.002$] \Big([$-0.032:-0.001$]\Big) GeV when 27 (16) fermions are contributing to the RGEs respectively, within the Non-SUSY framework. 
A similar analysis has been performed within the SUSY scenario. In Fig.s~\ref{fig:2241-dim5-dodd_susy_1} (\ref{fig:2241-dim5-dodd_susy_2}),  the blue and yellow lines show the allowed range of unification scale, $M_X\in [10^{16.14}:10^{18.94}] \Big([10^{16.14}:10^{19}]\Big)$ GeV ($y_1$ axis), 
and $g_{1X} \in [0.54:0.70] \Big([0.54:0.70]\Big)$ ($y_2$ axis) compatible with the range of $\varepsilon$ [$-0.006:+0.060$] \Big([$-0.008:+0.044$]\Big) GeV when 27 (16) fermions are contributing to the RGEs respectively. 

\begin{figure}[h!]
	\centering
	\subfloat[Non-SUSY]{
		\includegraphics[scale=0.65]{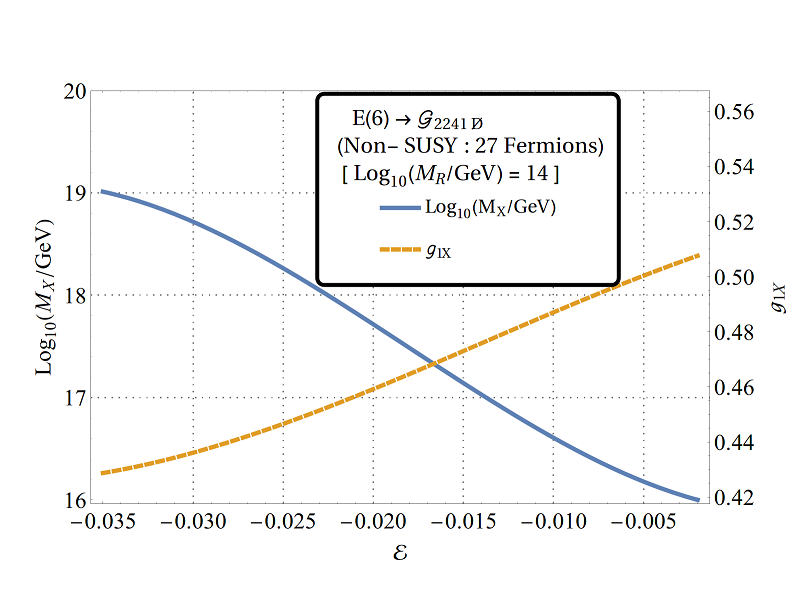}
		\label{fig:2241-dim5-dodd_nsusy_1}}
	\subfloat[Non-SUSY]{
		\includegraphics[scale=0.65]{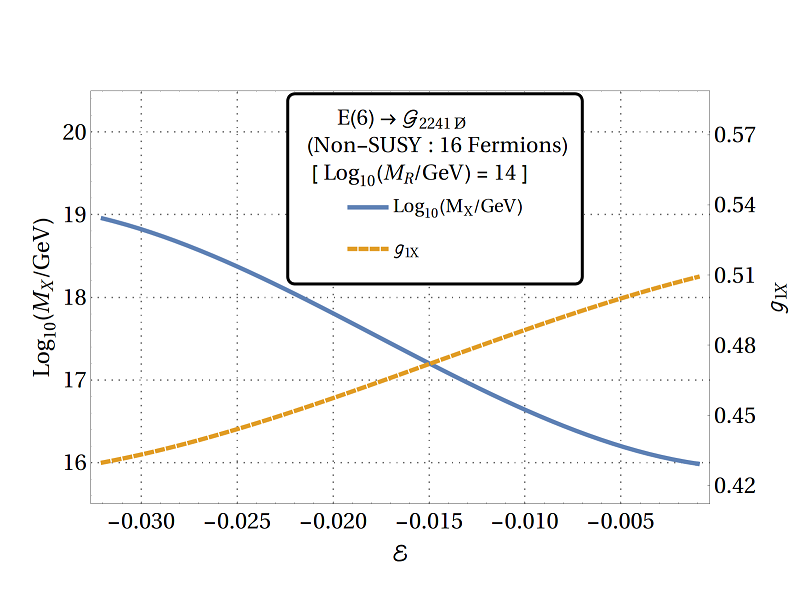}
		\label{fig:2241-dim5-dodd_nsusy_2}}
\\
	\subfloat[SUSY]{
	\includegraphics[scale=0.65]{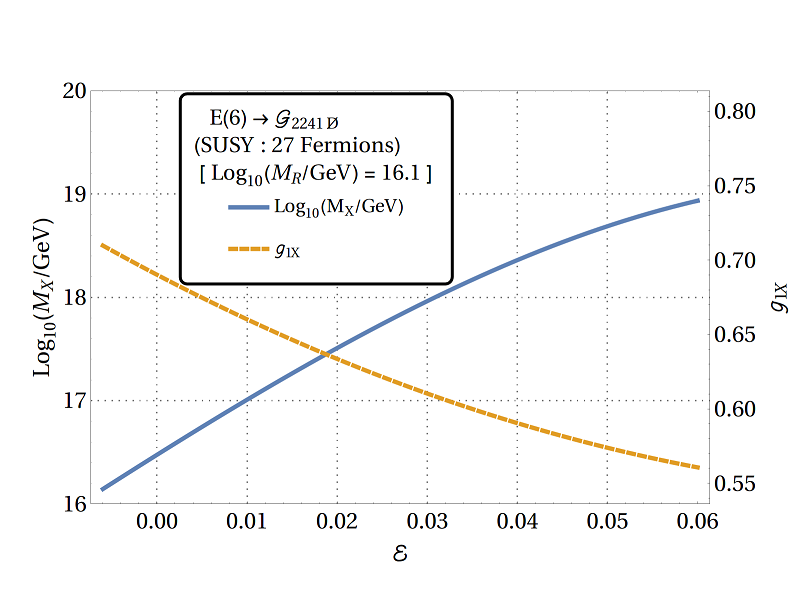}
	\label{fig:2241-dim5-dodd_susy_1}}
\subfloat[SUSY]{
	\includegraphics[scale=0.65]{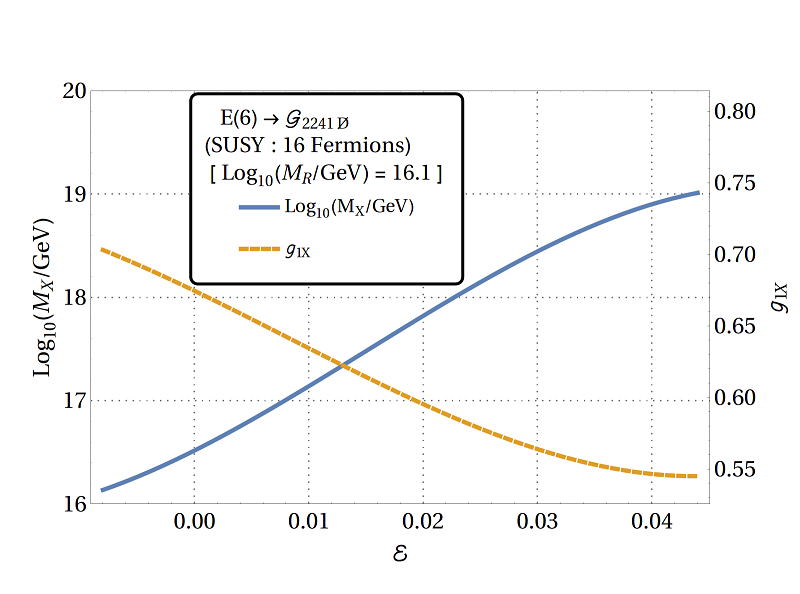}
	\label{fig:2241-dim5-dodd_susy_2}}
	\caption{\small Gauge coupling unification for the following symmetry breaking pattern: $E(6) \to \mathcal{G}_{2241\slashed{D}}\to SM$.
		$M_X$  as a function of $\varepsilon$ is noted 
		for Non-SUSY (a) and SUSY (b) scenarios.  These scenarios are compatible with  cosmological constraint ($M_R \gtrsim 10^{12}$ GeV).}
	\label{fig:2241-dim5-dodd}
\end{figure}

In Tables \ref{tab:model-status-non-susy} and \ref{tab:model-status-susy}, we have provided a qualitative comparison among different scenarios considered in this paper in the light of the implemented constraints due to proton life time and topological defects.

\begin{table}[h!]
	\begin{center}
		\scriptsize
		\begin{tabular}{|c|c|c|c|c|c|}
			\hline
			\multicolumn{2}{|c|}{\textbf{Intermediate Symmetry}}&\multicolumn{2}{c|}{\textbf{Topological defects}} &\multicolumn{2}{c|}{\textbf{Proton life time}}\\
			\multicolumn{2}{|c|}{(Non-SUSY)}	& \multicolumn{2}{c|}{$M_R\gtrsim 10^{12}$ GeV} & \multicolumn{2}{c|}{$M_X\gtrsim 10^{16}$ GeV} \\
			\cline{3-6}
			\multicolumn{2}{|c|}{$~$}	& No dim-5 & dim-5 & No dim-5 & dim-5 \\
			\hline
			\multirow{2}{2cm}{$~~~~\mathcal{G}_{224}$}& D-conserved & \checkmark & \checkmark & $\times$ & \checkmark\\
			\cline{2-6}
			& D-broken & $\times$ & $\times$ & $\times$ & \checkmark \\
			\hline
			\multirow{2}{2cm}{$~~~~\mathcal{G}_{2231}$}& D-conserved & $\times$ & $\times$ & $\times$ & \checkmark \\
			\cline{2-6}
			& D-broken & $\times$ & $\times$ & \checkmark & \checkmark\\
			\hline
			\multirow{2}{2cm}{$~~~~\mathcal{G}_{2241}$}& D-conserved& \checkmark & \checkmark & $\times$ & \checkmark\\
			\cline{2-6}
			& D-broken & \checkmark & \checkmark &  \checkmark & \checkmark\\
			\hline 
			\multirow{2}{2cm}{$~~~~\mathcal{G}_{333}$}& D-conserved & NS &  \checkmark & NS  & \checkmark\\
			\cline{2-6}
			& D-broken & NS & \checkmark  & NS & \checkmark\\
			\hline
		\end{tabular}
		\caption{Status of the Non-SUSY models in the light of proton decay and cosmological constraints considered in this paper.
	 ``NS" implies no solution for unification exists for $\mathcal{G}_{333}$ intermediate symmetry.} 
		\label{tab:model-status-non-susy}
	\end{center}
\end{table}
\begin{table}[h!]
	\begin{center}
		\scriptsize
		\begin{tabular}{|c|c|c|c|c|c|}
			\hline
			\multicolumn{2}{|c|}{\textbf{Intermediate Symmetry}}&\multicolumn{2}{c|}{\textbf{Topological defects}} &\multicolumn{2}{c|}{\textbf{Proton life time}}\\
			\multicolumn{2}{|c|}{(SUSY)}	& \multicolumn{2}{c|}{$M_R\gtrsim 10^{12}$ GeV} & \multicolumn{2}{c|}{$M_X\gtrsim 10^{16}$ GeV} \\
			\cline{3-6}
			\multicolumn{2}{|c|}{$~$}	& No dim-5 & dim-5 & No dim-5 & dim-5 \\
			\hline
			\multirow{2}{2cm}{$~~~~\mathcal{G}_{224}$}& D-conserved & \checkmark & \checkmark &  \checkmark & \checkmark\\
			\cline{2-6}
			& D-broken &  \checkmark &  \checkmark &  \checkmark & \checkmark \\
			\hline
			\multirow{2}{2cm}{$~~~~\mathcal{G}_{2231}$}& D-conserved &  \checkmark &  \checkmark &  \checkmark & \checkmark \\
			\cline{2-6}
			& D-broken & \checkmark & \checkmark & \checkmark & \checkmark\\
			\hline
			\multirow{2}{2cm}{$~~~~\mathcal{G}_{2241}$}& D-conserved& \checkmark & \checkmark &  \checkmark & \checkmark\\
			\cline{2-6}
			& D-broken & \checkmark & \checkmark &  \checkmark & \checkmark \\
			\hline 
			\multirow{2}{2cm}{$~~~~\mathcal{G}_{333}$}& D-conserved &  \checkmark &  \checkmark & \checkmark & \checkmark\\
			\cline{2-6}
			& D-broken &  \checkmark& \checkmark & \checkmark & \checkmark\\
			\hline
		\end{tabular}
		\caption{Status of the SUSY models in the light of proton decay and cosmological constraints considered in this paper The choice of $M_{SUSY}$ for individual cases are mentioned in Table~\ref{fig:unific-chisq-so-g2231-S}.} 
		\label{tab:model-status-susy}
	\end{center}
\end{table}

\section{Conclusions}

In this work, we have enumerated all possible intermediate symmetry groups, which are of the form $SU(N)_L\otimes SU(N)_R \otimes \mathcal{G}$ and can be embedded in the GUT gauge groups $SO(10)$ and $E(6)$. We have further assumed that the GUT symmetry is broken to the SM through only one intermediate symmetry, i.e.  $\mathcal{G}_{\rm GUT} \to SU(N)_L\otimes SU(N)_R \otimes \mathcal{G }\to SM$. We have listed the scalar representations and the direction of VEVs for each stage of symmetry breaking. We have also listed the representations of scalar and fermion fields, which contribute to the beta functions at each stage of the symmetry breaking chain. Though we have started with the full multiplets under the unified group, we have assumed that only those sub-multiplets are light, which induce the symmetry breaking and contain SM particles. This assumption is in accordance with the extended survival hypothesis (ESH).  

We have computed the RGEs for all the breaking chains up to two loop order for both SUSY and Non-SUSY scenarios\footnote{We have checked some of them with \cite{Lyonnet:2016xiz} for Non-SUSY and \cite{Fonseca:2011sy} for SUSY cases, and they are in well agreement with our results.}. We have also included the effect of the abelian mixing, whenever it arises. We have solved the RGEs numerically and noted the correlated solutions in terms of $M_X$, $M_R$, and $g_U$, within the present experimental bounds on the low scale parameters and under a few confidence levels. This has been achieved by doing a goodness of fit test with the construction and optimisation of a $\Delta\chi^2$ statistic.  We would like to mention that in this analysis we have not included the effect of Yukawa couplings in two loop RGEs. The extra contributions due to the Yukawa couplings are of the order of two loop corrections. As the two loop results are not much departed from the one loop ones, we would expect our conclusions to remain unchanged. We leave the effect of the Yukawa couplings for our upcoming paper where the fermion mass generation is the primary goal.

The theory is less constrained in the case of abelian mixing. We have noted the allowed ranges of the mixing parameters along with other gauge couplings, which are not constrained by the matching conditions. Different types of topological defects may arise in the process of symmetry breaking and we have discussed such possibilities in detail. Such defects are in conflict with the cosmological observation, thus they should be inflated away. This implies that inflation must occur after the $\mathcal{G}_{\rm GUT}$ symmetry breaking and before the breaking of the intermediate symmetry. Also, those intermediate symmetries, which could give rise to harmful topological defects, must occur above the reheat temperature. This ensures that following inflation and reheating, the universe is in the SM phase and these undesirable topological defects are not produced. This puts a severe constraint on the intermediate scale, $M_R\gtrsim  10^{12}$ GeV. We have applied this constraint in all possible breaking patterns and studied the allowed parameter space. We have also implemented the bound from proton decay life time, which allows $M_X \gtrsim 10^{16}$ GeV. We have listed all possible left-right symmetric intermediate groups starting from $SO(10)$ and $E(6)$ which survive these two constraints.

There are further implications on the other phenomenological observations related to neutrino masses and inflationary dynamics from these intermediate scales. For example \cite{Mohapatra:1979ia}, the right handed neutrino mass is related to the $\mathcal{G}_{224}$ breaking scale, and thus affects the low scale masses and mixing parameters in the neutrino sector. As these intermediate scales are very high, the Yukawa coupling could be large and may have significant impact on low energy observables through RGEs. Phenomenologically favoured plateau inflation models, constructed within the GUT framework, work for specific intermediate symmetries only \cite{Garg:2015mra, Ellis:2016ipm}. Models which clear the constraint of proton decay and cosmological relics, which have been pointed out in this paper, can further be investigated in the light of neutrino mass models and inflation phenomenology.

The mSUGRA assumption of universal gaugino masses at the unification scale is no longer true when the SUSY breaking is induced by the $\mathcal{F}$-term of the non-singlet field under the GUT gauge group \cite{Chakrabortty:2008zk,Martin:2009ad, Bhattacharya:2009wv, Chakrabortty:2010xq}. The low energy gaugino masses will depend not only on the GUT scale mass-ratios but also on the intermediate symmetry group and the scale of intermediate symmetry through the RG running. This will have important phenomenological implications in the context of dark matter and muon (g-2) \cite{Younkin:2012ui, Chakrabortty:2013voa, Miller:2014jza, Gogoladze:2014cha, Chakrabortty:2015ika}. In order to evade negative results from super-partner searches at LHC  \cite{Aad:2015iea}, the  particle masses need to have either split \cite{Dine:2004dv, Giudice:2004tc} or compressed spectrum \cite{LeCompte:2011cn}. These spectra may be achieved using non-universal gauginos at the GUT scale along with suitable choice of intermediate symmetry and breaking scale \cite{Deppisch:2017xhv}. We have found in our work that in order to have the intermediate symmetry breaking above $10^{12}$ GeV, as favoured by cosmology, the SUSY breaking scale cannot be pushed beyond $\sim 30$ TeV and it depends on both the intermediate symmetry group and scale. This will have implications in both dark matter and collider searches and we leave the detailed analysis of the SUSY spectra in left-right models for future studies.

\section*{Acknowledgements}
The authors acknowledge the useful discussions with Suneel Chandra Reddy Vemula on homotopy.
We would like to thank Anindya Datta and  Amitava Raychaudhuri for useful suggestions. 
J.C., R.M., S.P., and T.S.  are supported by the Department of Science and Technology,
Government of India, under the Grant IFA12-PH-34 (INSPIRE Faculty Award); and the Science and Engineering
Research Board, Government of India, under the agreement SERB/PHY/2016348 (Early Career Research Award). 

\appendix
\numberwithin{equation}{section}
\numberwithin{table}{section}

\section*{APPENDIX}
\label{sec:appendix}


\subsection*{Two loop RGEs for the breaking $E(6)\to\mathcal{G}_{21213}\to\rm{SM}$}
{\bf \underline{When all the 27-fermions are present at the $M_R$ scale:}}\\

{\bf D-parity not conserved}\\

\underline{\textsl{Non-SUSY}}
			{\scriptsize		
	\begin{align*}
		(4\pi)^4\beta^{\rm 2loop}_{2L} = 
		& \frac{11 \;g_{LL}^{2}}{6} \;g_{2L}^{3} + \frac{65 \;g_{2L}^{5}}{2} + 12 \;g_{2L}^{3} \;g_{3C}^{2} + \frac{11 \;g_{2L}^{3}}{6} ~\;g_{RR}^{2} + \frac{11 \;g_{2L}^{3}}{6} \;g_{LR}^{2}\\
		&- \frac{2 \;g_{LL}}{3} \;g_{2L}^{3} \;g_{RL} + \frac{15 \;g_{2L}^{3}}{2} \;g_{2R}^{2} - \frac{2 ~\;g_{RR}}{3} \;g_{2L}^{3} \;g_{LR} + \frac{11 \;g_{2L}^{3}}{6} \;g_{RL}^{2}
		\\
		(4\pi)^4\beta^{\rm 2loop}_{3C} = 
		& \frac{3 \;g_{LL}^{2}}{2} \;g_{3C}^{3} + \frac{9 \;g_{2L}^{2}}{2} \;g_{3C}^{3} + 12 \;g_{3C}^{5} + \frac{3 \;g_{3C}^{3}}{2} ~\;g_{RR}^{2} + \frac{3 \;g_{3C}^{3}}{2} \;g_{LR}^{2} + \frac{3 \;g_{3C}^{3}}{2} \;g_{RL}^{2}\\
		&+\frac{9 \;g_{2R}^{2}}{2} \;g_{3C}^{3}
		\\
		(4\pi)^4\beta^{\rm 2loop}_{2R} = 
		& \frac{5 \;g_{LL}^{2}}{2} \;g_{2R}^{3} + \frac{104 \;g_{2R}^{5}}{3} + 12 \;g_{2R}^{3} \;g_{3C}^{2} + 2 \;g_{2R}^{3} ~\;g_{RR}^{2} + \frac{5 \;g_{2R}^{3}}{2} \;g_{LR}^{2} + 2 \;g_{2R}^{3} \;g_{RL}^{2}\\
		&+\frac{15 \;g_{2L}^{2}}{2} \;g_{2R}^{3}
		\\
		(4\pi)^4\beta^{\rm 2loop}_{LL} = 
		& \frac{11 \;g_{LL}^{5}}{2} + \frac{4 \;g_{LL}^{4}}{3} \;g_{RL} + 12 \;g_{LL}^{3} \;g_{3C}^{2} + 11 \;g_{LL}^{3} \;g_{LR}^{2} + 11 \;g_{LL}^{3} \;g_{RL}^{2} + \frac{\;g_{RL}}{3} \;g_{LR}^{4}\\
		&+\frac{11 \;g_{LL}^{3}}{2} \;g_{2L}^{2} + \frac{15 \;g_{LL}^{3}}{2} \;g_{2R}^{2} + \frac{11 \;g_{LL}^{3}}{6} ~\;g_{RR}^{2} + \frac{11 \;g_{LL}}{2} \;g_{LR}^{4} + \frac{14 \;g_{LL}}{3} \;g_{RL}^{4}+\;g_{LL}^{3} ~\;g_{RR} \;g_{LR}\\
		& - 2 \;g_{LL}^{2} \;g_{2L}^{2} \;g_{RL} + 6 \;g_{LL} \;g_{2R}^{2} \;g_{RL}^{2} + 12 \;g_{LL} \;g_{3C}^{2} \;g_{LR}^{2} + \;g_{LL} ~\;g_{RR} \;g_{LR}^{3} - \;g_{2L}^{2} \;g_{LR}^{2} \;g_{RL}\\
		&+\frac{5 \;g_{LL}^{2}}{3} \;g_{LR}^{2} \;g_{RL} + \frac{11 \;g_{LL}}{2} \;g_{2L}^{2} \;g_{LR}^{2} + \frac{11 \;g_{LL}}{2} \;g_{2L}^{2} \;g_{RL}^{2} + 12 \;g_{LL} \;g_{3C}^{2} \;g_{RL}^{2}\\
		&+\frac{11 \;g_{LL}}{2} ~\;g_{RR}^{2} \;g_{LR}^{2} + \frac{14 \;g_{LL}}{3} ~\;g_{RR}^{2} \;g_{RL}^{2} + \frac{11 ~\;g_{RR}}{2} \;g_{LR}^{3} \;g_{RL} + \frac{14 ~\;g_{RR}}{3} \;g_{LR} \;g_{RL}^{3}\\
		&- \;g_{LL} \;g_{2L}^{2} ~\;g_{RR} \;g_{LR} + \frac{15 \;g_{LL}}{2} \;g_{2R}^{2} \;g_{LR}^{2} + \frac{22 \;g_{LL}}{3} \;g_{LR}^{2} \;g_{RL}^{2} + \frac{14 ~\;g_{RR}^{3}}{3} \;g_{LR} \;g_{RL}\\
		&+\frac{77 ~\;g_{RR}}{6} \;g_{LL}^{2} \;g_{LR} \;g_{RL} + \frac{11 ~\;g_{RR}}{2} \;g_{2L}^{2} \;g_{LR} \;g_{RL} + 6 \;g_{2R}^{2} ~\;g_{RR} \;g_{LR} \;g_{RL}\\
		& + 12 \;g_{3C}^{2} ~\;g_{RR} \;g_{LR} \;g_{RL}
		\\
		(4\pi)^4\beta^{\rm 2loop}_{LR} = 
		& \frac{\;g_{LL}^{4} ~\;g_{RR}}{3} + 11 \;g_{LL}^{2} \;g_{LR}^{3} + 12 \;g_{3C}^{2} \;g_{LR}^{3} + 11 ~\;g_{RR}^{2} \;g_{LR}^{3} + \frac{4 ~\;g_{RR}}{3} \;g_{LR}^{4} + \frac{11}{2} \;g_{LR}^{5}\\
		&+\frac{11 \;g_{LL}^{4}}{2} \;g_{LR} + \frac{11 \;g_{2L}^{2}}{2} \;g_{LR}^{3} + \frac{15 \;g_{2R}^{2}}{2} \;g_{LR}^{3} + \frac{14 ~\;g_{RR}^{4}}{3} \;g_{LR} + \frac{11}{6} \;g_{LR}^{3} \;g_{RL}^{2}\\
		& - \;g_{LL}^{2} \;g_{2L}^{2} ~\;g_{RR} + 12 \;g_{LL}^{2} \;g_{3C}^{2} \;g_{LR} + \;g_{LL} \;g_{LR}^{3} \;g_{RL} - 2 \;g_{2L}^{2} ~\;g_{RR} \;g_{LR}^{2} + 6 \;g_{2R}^{2} ~\;g_{RR}^{2} \;g_{LR}\\
		&+\;g_{LL}^{3} \;g_{LR} \;g_{RL}+\frac{11 ~\;g_{RR}}{2} \;g_{LL}^{3} \;g_{RL} + \frac{11 \;g_{LL}^{2}}{2} \;g_{2L}^{2} \;g_{LR} + \frac{5 ~\;g_{RR}}{3} \;g_{LL}^{2} \;g_{LR}^{2} + 12 \;g_{3C}^{2} ~\;g_{RR}^{2} \;g_{LR}\\
		&+\frac{11 \;g_{LL}^{2}}{2} \;g_{LR} \;g_{RL}^{2} + \frac{14 \;g_{LL}}{3} ~\;g_{RR}^{3} \;g_{RL} + \frac{14 \;g_{LL}}{3} ~\;g_{RR} \;g_{RL}^{3} + \frac{11 \;g_{2L}^{2}}{2} ~\;g_{RR}^{2} \;g_{LR}\\
		&+\frac{15 \;g_{LL}^{2}}{2} \;g_{2R}^{2} \;g_{LR} + \frac{22 \;g_{LL}^{2}}{3} ~\;g_{RR}^{2} \;g_{LR} - \;g_{LL} \;g_{2L}^{2} \;g_{LR} \;g_{RL} + \frac{14 ~\;g_{RR}^{2}}{3} \;g_{LR} \;g_{RL}^{2}\\
		&+\frac{11 \;g_{LL}}{2} \;g_{2L}^{2} ~\;g_{RR} \;g_{RL} + 6 \;g_{LL} \;g_{2R}^{2} ~\;g_{RR} \;g_{RL} + 12 \;g_{LL} \;g_{3C}^{2} ~\;g_{RR} \;g_{RL} + \frac{77 \;g_{LL}}{6} ~\;g_{RR} \;g_{LR}^{2} \;g_{RL}
		\\
		(4\pi)^4\beta^{\rm 2loop}_{RL} = 
		& \frac{\;g_{LL}^{3} ~\;g_{RR}^{2}}{3} + \frac{4 \;g_{LL}^{3}}{3} \;g_{RL}^{2} + 11 \;g_{LL}^{2} \;g_{RL}^{3} + 6 \;g_{2R}^{2} \;g_{RL}^{3} + 12 \;g_{3C}^{2} \;g_{RL}^{3} + \frac{14}{3} \;g_{RL}^{5}\\
		&+\frac{11 \;g_{LL}^{4}}{2} \;g_{RL} + \frac{11 \;g_{2L}^{2}}{2} \;g_{RL}^{3} + \frac{14 ~\;g_{RR}^{4}}{3} \;g_{RL} + \frac{28 ~\;g_{RR}^{2}}{3} \;g_{RL}^{3} + \frac{11}{6} \;g_{LR}^{2} \;g_{RL}^{3}\\
		&+12 \;g_{LL}^{2} \;g_{3C}^{2} \;g_{RL} - \;g_{LL} \;g_{2L}^{2} ~\;g_{RR}^{2} - 2 \;g_{LL} \;g_{2L}^{2} \;g_{RL}^{2} + \;g_{LL} ~\;g_{RR}^{2} \;g_{LR}^{2} + 6 \;g_{2R}^{2} ~\;g_{RR}^{2} \;g_{RL} \\
		& + 12 \;g_{3C}^{2} ~\;g_{RR}^{2} \;g_{RL}+\frac{11 \;g_{LL}}{2} ~\;g_{RR}^{3} \;g_{LR} + \frac{11 \;g_{LL}}{2} ~\;g_{RR} \;g_{LR}^{3} + \frac{2 \;g_{LL}}{3} \;g_{LR}^{2} \;g_{RL}^{2}\\
		&  + \frac{~\;g_{RR}}{3} \;g_{LR}^{3} \;g_{RL}+\frac{11 ~\;g_{RR}}{2} \;g_{LL}^{3} \;g_{LR} + \frac{11 \;g_{LL}^{2}}{2} \;g_{2L}^{2} \;g_{RL} + \frac{11 \;g_{LL}^{2}}{2} \;g_{LR}^{2} \;g_{RL}\\
		& + \frac{11 \;g_{2L}^{2}}{2} ~\;g_{RR}^{2} \;g_{RL}+\frac{15 \;g_{LL}^{2}}{2} \;g_{2R}^{2} \;g_{RL} + \frac{22 \;g_{LL}^{2}}{3} ~\;g_{RR}^{2} \;g_{RL} - \;g_{2L}^{2} ~\;g_{RR} \;g_{LR} \;g_{RL} \\
		&+\frac{5 ~\;g_{RR}}{3} \;g_{LL}^{2} \;g_{LR} \;g_{RL} + \frac{11 \;g_{LL}}{2} \;g_{2L}^{2} ~\;g_{RR} \;g_{LR} + \frac{15 \;g_{LL}}{2} \;g_{2R}^{2} ~\;g_{RR} \;g_{LR} + 12 \;g_{LL} \;g_{3C}^{2} ~\;g_{RR} \;g_{LR}\\
		&+ \frac{11 ~\;g_{RR}^{2}}{2} \;g_{LR}^{2} \;g_{RL}+\frac{77 \;g_{LL}}{6} ~\;g_{RR} \;g_{LR} \;g_{RL}^{2}
		\\
		(4\pi)^4\beta^{\rm 2loop}_{RR} = 
		& 6 \;g_{2R}^{2} ~\;g_{RR}^{3} + 12 \;g_{3C}^{2} ~\;g_{RR}^{3} + \frac{14 ~\;g_{RR}^{5}}{3} + 11 ~\;g_{RR}^{3} \;g_{LR}^{2} + \frac{4 ~\;g_{RR}^{2}}{3} \;g_{LR}^{3} + \frac{\;g_{LR}^{3}}{3} \;g_{RL}^{2}\\
		&+\frac{11 \;g_{LL}^{2}}{6} ~\;g_{RR}^{3} + \frac{11 \;g_{2L}^{2}}{2} ~\;g_{RR}^{3} + \frac{28 ~\;g_{RR}^{3}}{3} \;g_{RL}^{2} + \frac{11 ~\;g_{RR}}{2} \;g_{LR}^{4} + \frac{14 ~\;g_{RR}}{3} \;g_{RL}^{4}\\
		&+\;g_{LL}^{2} \;g_{LR} \;g_{RL}^{2} - 2 \;g_{2L}^{2} ~\;g_{RR}^{2} \;g_{LR} - \;g_{2L}^{2} \;g_{LR} \;g_{RL}^{2} + 6 \;g_{2R}^{2} ~\;g_{RR} \;g_{RL}^{2} + 12 \;g_{3C}^{2} ~\;g_{RR} \;g_{LR}^{2}\\
		&+\frac{\;g_{LL}^{3} ~\;g_{RR}}{3} \;g_{RL} + \frac{2 \;g_{LL}^{2}}{3} ~\;g_{RR}^{2} \;g_{LR} + \frac{11 \;g_{LL}}{2} \;g_{LR} \;g_{RL}^{3} + 12 \;g_{3C}^{2} ~\;g_{RR} \;g_{RL}^{2}\\
		&+\frac{11 ~\;g_{RR}}{2} \;g_{LL}^{2} \;g_{LR}^{2} + \frac{11 ~\;g_{RR}}{2} \;g_{LL}^{2} \;g_{RL}^{2} + \frac{11 \;g_{LL}}{2} \;g_{LR}^{3} \;g_{RL} + \frac{11 ~\;g_{RR}}{2} \;g_{2L}^{2} \;g_{LR}^{2}\\
		&+\frac{11 \;g_{LL}^{3}}{2} \;g_{LR} \;g_{RL} + \frac{11 ~\;g_{RR}}{2} \;g_{2L}^{2} \;g_{RL}^{2} + \frac{15 ~\;g_{RR}}{2} \;g_{2R}^{2} \;g_{LR}^{2} + \frac{22 ~\;g_{RR}}{3} \;g_{LR}^{2} \;g_{RL}^{2}\\
		&- \;g_{LL} \;g_{2L}^{2} ~\;g_{RR} \;g_{RL} + \frac{11 \;g_{LL}}{2} \;g_{2L}^{2} \;g_{LR} \;g_{RL} + 12 \;g_{LL} \;g_{3C}^{2} \;g_{LR} \;g_{RL} + \frac{5 \;g_{LL}}{3} ~\;g_{RR} \;g_{LR}^{2} \;g_{RL}\\
		&+\frac{15 \;g_{LL}}{2} \;g_{2R}^{2} \;g_{LR} \;g_{RL} + \frac{77 \;g_{LL}}{6} ~\;g_{RR}^{2} \;g_{LR} \;g_{RL}
	\end{align*}
}
\underline{\textsl{SUSY}}
{\scriptsize
	\begin{align*} 
		(4\pi)^4{\beta^{\rm 2loop}_{LL}}& =24 \;g_{LL}^3 \;g_{3 C}^2+24 \;g_{LR}^2 \;g_{LL} \;g_{3 C}^2+24 \;g_{RL}^2 \;g_{LL} \;g_{3 C}^2+24 \;g_{LR} \;g_{RL} \;g_{RR} \;g_{3 C}^2+10 \;g_{LL}^3 \;g_{2 L}^2-2 \;g_{RL} \;g_{LL}^2 \;g_{2 L}^2\\&+10 \;g_{LR}^2 \;g_{LL} \;g_{2 L}^2+10 \;g_{RL}^2 \;g_{LL} \;g_{2 L}^2-\;g_{LR} \;g_{RR} \;g_{LL} \;g_{2 L}^2-\;g_{LR}^2 \;g_{RL} \;g_{2 L}^2+10 \;g_{LR} \;g_{RL} \;g_{RR} \;g_{2 L}^2+12 \;g_{LL}^3 \;g_{2 R}^2\\&+12 \;g_{LR}^2 \;g_{LL} \;g_{2 R}^2+\frac{21}{2} \;g_{RL}^2 \;g_{LL} \;g_{2 R}^2+\frac{21}{2} \;g_{LR} \;g_{RL} \;g_{RR} \;g_{2 R}^2+10 \;g_{LL}^5+\frac{4}{3} \;g_{RL} \;g_{LL}^4+20 \;g_{LR}^2 \;g_{LL}^3 \\& +20 \;g_{RL}^2 \;g_{LL}^3+\frac{10}{3} \;g_{RR}^2 \;g_{LL}^3+\;g_{LR} \;g_{RR} \;g_{LL}^3+\frac{5}{3} \;g_{LR}^2 \;g_{RL} \;g_{LL}^2+\frac{70}{3} \;g_{LR} \;g_{RL} \;g_{RR} \;g_{LL}^2+10 \;g_{LR}^4 \;g_{LL}\\& +\frac{55}{6} \;g_{RL}^4 \;g_{LL}+\frac{40}{3} \;g_{LR}^2 \;g_{RL}^2 \;g_{LL}+10 \;g_{LR}^2 \;g_{RR}^2 \;g_{LL}+\frac{55}{6} \;g_{RL}^2 \;g_{RR}^2 \;g_{LL}+\;g_{LR}^3 \;g_{RR} \;g_{LL}+\frac{55}{6} \;g_{LR} \;g_{RL} \;g_{RR}^3\\&+\frac{1}{3} \;g_{LR}^4 \;g_{RL}+\frac{55}{6} \;g_{LR} \;g_{RL}^3 \;g_{RR}+10 \;g_{LR}^3 \;g_{RL} \;g_{RR} \\ 
		(4\pi)^4{\beta^{\rm 2loop}_{LR}}& =24 \;g_{LR}^3 \;g_{3 C}^2+24 \;g_{LL}^2 \;g_{LR} \;g_{3 C}^2+24 \;g_{RR}^2 \;g_{LR} \;g_{3 C}^2+24 \;g_{LL} \;g_{RL} \;g_{RR} \;g_{3 C}^2+10 \;g_{LR}^3 \;g_{2 L}^2-2 \;g_{RR} \;g_{LR}^2 \;g_{2 L}^2\\& +10 \;g_{LL}^2 \;g_{LR} \;g_{2 L}^2+10 \;g_{RR}^2 \;g_{LR} \;g_{2 L}^2-\;g_{LL} \;g_{RL} \;g_{LR} \;g_{2 L}^2-\;g_{LL}^2 \;g_{RR} \;g_{2 L}^2+10 \;g_{LL} \;g_{RL} \;g_{RR} \;g_{2 L}^2\\&+12 \;g_{LR}^3 \;g_{2 R}^2+12 \;g_{LL}^2 \;g_{LR} \;g_{2 R}^2+\frac{21}{2} \;g_{RR}^2 \;g_{LR} \;g_{2 R}^2+\frac{21}{2} \;g_{LL} \;g_{RL} \;g_{RR} \;g_{2 R}^2+10 \;g_{LR}^5+\frac{4}{3} \;g_{RR} \;g_{LR}^4\\&+20 \;g_{LL}^2 \;g_{LR}^3+\frac{10}{3} \;g_{RL}^2 \;g_{LR}^3+20 \;g_{RR}^2 \;g_{LR}^3+\;g_{LL} \;g_{RL} \;g_{LR}^3+\frac{5}{3} \;g_{LL}^2 \;g_{RR} \;g_{LR}^2+\frac{70}{3} \;g_{LL} \;g_{RL} \;g_{RR} \;g_{LR}^2\\&+10 \;g_{LL}^4 \;g_{LR}+\frac{55}{6} \;g_{RR}^4 \;g_{LR}+10 \;g_{LL}^2 \;g_{RL}^2 \;g_{LR}+\frac{40}{3} \;g_{LL}^2 \;g_{RR}^2 \;g_{LR}+\frac{55}{6} \;g_{RL}^2 \;g_{RR}^2 \;g_{LR}\\&+\;g_{LL}^3 \;g_{RL} \;g_{LR}+\frac{55}{6} \;g_{LL} \;g_{RL} \;g_{RR}^3+\frac{1}{3} \;g_{LL}^4 \;g_{RR}+\frac{55}{6} \;g_{LL} \;g_{RL}^3 \;g_{RR}+10 \;g_{LL}^3 \;g_{RL} \;g_{RR} \\ 
		(4\pi)^4{\beta^{\rm 2loop}_{RL}}& =24 \;g_{RL}^3 \;g_{3 C}^2+24 \;g_{LL}^2 \;g_{RL} \;g_{3 C}^2+24 \;g_{RR}^2 \;g_{RL} \;g_{3 C}^2+24 \;g_{LL} \;g_{LR} \;g_{RR} \;g_{3 C}^2+10 \;g_{RL}^3 \;g_{2 L}^2-2 \;g_{LL} \;g_{RL}^2 \;g_{2 L}^2\\&+10 \;g_{LL}^2 \;g_{RL} \;g_{2 L}^2+10 \;g_{RR}^2 \;g_{RL} \;g_{2 L}^2-\;g_{LR} \;g_{RR} \;g_{RL} \;g_{2 L}^2-\;g_{LL} \;g_{RR}^2 \;g_{2 L}^2+10 \;g_{LL} \;g_{LR} \;g_{RR} \;g_{2 L}^2\\&+\frac{21}{2} \;g_{RL}^3 \;g_{2 R}^2+12 \;g_{LL}^2 \;g_{RL} \;g_{2 R}^2+\frac{21}{2} \;g_{RR}^2 \;g_{RL} \;g_{2 R}^2+12 \;g_{LL} \;g_{LR} \;g_{RR} \;g_{2 R}^2+\frac{55 \;g_{RL}^5}{6}+20 \;g_{LL}^2 \;g_{RL}^3\\&+\frac{10}{3} \;g_{LR}^2 \;g_{RL}^3+\frac{55}{3} \;g_{RR}^2 \;g_{RL}^3+\frac{4}{3} \;g_{LL}^3 \;g_{RL}^2+\frac{2}{3} \;g_{LL} \;g_{LR}^2 \;g_{RL}^2+\frac{70}{3} \;g_{LL} \;g_{LR} \;g_{RR} \;g_{RL}^2+10 \;g_{LL}^4 \;g_{RL}\\&+\frac{55}{6} \;g_{RR}^4 \;g_{RL}+10 \;g_{LL}^2 \;g_{LR}^2 \;g_{RL}+\frac{40}{3} \;g_{LL}^2 \;g_{RR}^2 \;g_{RL}+10 \;g_{LR}^2 \;g_{RR}^2 \;g_{RL}+\frac{1}{3} \;g_{LR}^3 \;g_{RR} \;g_{RL}\\&+\frac{5}{3} \;g_{LL}^2 \;g_{LR} \;g_{RR} \;g_{RL}+10 \;g_{LL} \;g_{LR} \;g_{RR}^3+\frac{1}{3} \;g_{LL}^3 \;g_{RR}^2+\;g_{LL} \;g_{LR}^2 \;g_{RR}^2+10 \;g_{LL} \;g_{LR}^3 \;g_{RR}+10 \;g_{LL}^3 \;g_{LR} \;g_{RR} \\ 
		(4\pi)^4{\beta^{\rm 2loop}_{RR}}& =24 \;g_{RR}^3 \;g_{3 C}^2+24 \;g_{LR}^2 \;g_{RR} \;g_{3 C}^2+24 \;g_{RL}^2 \;g_{RR} \;g_{3 C}^2+24 \;g_{LL} \;g_{LR} \;g_{RL} \;g_{3 C}^2+10 \;g_{RR}^3 \;g_{2 L}^2-2 \;g_{LR} \;g_{RR}^2 \;g_{2 L}^2\\&+10 \;g_{RL}^2 \;g_{RR} \;g_{2 L}^2-\;g_{LL} \;g_{RL} \;g_{RR} \;g_{2 L}^2-\;g_{LR} \;g_{RL}^2 \;g_{2 L}^2+10 \;g_{LL} \;g_{LR} \;g_{RL} \;g_{2 L}^2+\frac{21}{2} \;g_{RR}^3 \;g_{2 R}^2+12 \;g_{LR}^2 \;g_{RR} \;g_{2 R}^2\\&+\frac{21}{2} \;g_{RL}^2 \;g_{RR} \;g_{2 R}^2+12 \;g_{LL} \;g_{LR} \;g_{RL} \;g_{2 R}^2+\frac{55 \;g_{RR}^5}{6}+\frac{10}{3} \;g_{LL}^2 \;g_{RR}^3+20 \;g_{LR}^2 \;g_{RR}^3+\frac{55}{3} \;g_{RL}^2 \;g_{RR}^3+\frac{4}{3} \;g_{LR}^3 \;g_{RR}^2\\&+\frac{2}{3} \;g_{LL}^2 \;g_{LR} \;g_{RR}^2+\frac{70}{3} \;g_{LL} \;g_{LR} \;g_{RL} \;g_{RR}^2+10 \;g_{LR}^4 \;g_{RR}+\frac{55}{6} \;g_{RL}^4 \;g_{RR}+10 \;g_{LL}^2 \;g_{LR}^2 \;g_{RR}+10 \;g_{LL}^2 \;g_{RL}^2 \;g_{RR}\\&+\frac{1}{3} \;g_{LL}^3 \;g_{RL} \;g_{RR}+\frac{5}{3} \;g_{LL} \;g_{LR}^2 \;g_{RL} \;g_{RR}+10 \;g_{LL} \;g_{LR} \;g_{RL}^3+\frac{1}{3} \;g_{LR}^3 \;g_{RL}^2+\;g_{LL}^2 \;g_{LR} \;g_{RL}^2+10 \;g_{LL} \;g_{LR}^3 \;g_{RL}\\&+10 \;g_{LL}^3 \;g_{LR} \;g_{RL}+10 \;g_{LR}^2 \;g_{RR} \;g_{2 L}^2+\frac{40}{3} \;g_{LR}^2 \;g_{RL}^2 \;g_{RR} \\ 
		(4\pi)^4{\beta^{\rm 2loop}_{2L}}& =24 \;g_{3 C}^2 \;g_{2 L}^3+12 \;g_{2 L}^3 \;g_{2 R}^2+46 \;g_{2 L}^5+\frac{10}{3} \;g_{LL}^2 \;g_{2 L}^3+\frac{10}{3} \;g_{LR}^2 \;g_{2 L}^3+\frac{10}{3} \;g_{RL}^2 \;g_{2 L}^3+\frac{10}{3} \;g_{RR}^2 \;g_{2 L}^3\\&-\frac{2}{3} \;g_{LL} \;g_{RL} \;g_{2 L}^3-\frac{2}{3} \;g_{LR} \;g_{RR} \;g_{2 L}^3 \\ 
		(4\pi)^4{\beta^{\rm 2loop}_{2R}}& =24 \;g_{3 C}^2 \;g_{2 R}^3+12 \;g_{2 L}^2 \;g_{2 R}^3+\frac{99 \;g_{2 R}^5}{2}+4 \;g_{LL}^2 \;g_{2 R}^3+4 \;g_{LR}^2 \;g_{2 R}^3+\frac{7}{2} \;g_{RL}^2 \;g_{2 R}^3+\frac{7}{2} \;g_{RR}^2 \;g_{2 R}^3 \\ 
		(4\pi)^4{\beta^{\rm 2loop}_{3C}}& =9 \;g_{3 C}^3 \;g_{2 L}^2+9 \;g_{3 C}^3 \;g_{2 R}^2+48 \;g_{3 C}^5+3 \;g_{LL}^2 \;g_{3 C}^3+3 \;g_{LR}^2 \;g_{3 C}^3+3 \;g_{RL}^2 \;g_{3 C}^3+3 \;g_{RR}^2 \;g_{3 C}^3 \\
	\end{align*}
}
{\bf D-parity conserved}\\

\underline{\textsl{Non-SUSY}}
{\scriptsize
	\begin{align*}
		(4\pi)^4\beta^{\rm 2loop}_{2L} = 
		& \frac{5 \;g_{LL}^{2}}{2} \;g_{2L}^{3} + \frac{104 \;g_{2L}^{5}}{3} + 12 \;g_{2L}^{3} \;g_{3C}^{2} + 2 \;g_{2L}^{3} \;g_{RR}^{2} + \frac{5 \;g_{2L}^{3}}{2} \;g_{LR}^{2} + 2 \;g_{2L}^{3} \;g_{RL}^{2}\\
		&+\frac{15 \;g_{2L}^{3}}{2} \;g_{2R}^{2}
		\\
		(4\pi)^4\beta^{\rm 2loop}_{3C} = 
		& \frac{3 \;g_{LL}^{2}}{2} \;g_{3C}^{3} + \frac{9 \;g_{2L}^{2}}{2} \;g_{3C}^{3} + 12 \;g_{3C}^{5} + \frac{3 \;g_{3C}^{3}}{2} \;g_{RR}^{2} + \frac{3 \;g_{3C}^{3}}{2} \;g_{LR}^{2} + \frac{3 \;g_{3C}^{3}}{2} \;g_{RL}^{2}\\
		&+\frac{9 \;g_{2R}^{2}}{2} \;g_{3C}^{3}
		\\
		(4\pi)^4\beta^{\rm 2loop}_{2R} = 
		& \frac{5 \;g_{LL}^{2}}{2} \;g_{2R}^{3} + \frac{104 \;g_{2R}^{5}}{3} + 12 \;g_{2R}^{3} \;g_{3C}^{2} + 2 \;g_{2R}^{3} \;g_{RR}^{2} + \frac{5 \;g_{2R}^{3}}{2} \;g_{LR}^{2} + 2 \;g_{2R}^{3} \;g_{RL}^{2}\\
		&+\frac{15 \;g_{2L}^{2}}{2} \;g_{2R}^{3}
		\\
		(4\pi)^4\beta^{\rm 2loop}_{LL} = 
		& \frac{115 \;g_{LL}^{5}}{18} + \frac{15 \;g_{LL}^{3}}{2} \;g_{2L}^{2} + 12 \;g_{LL}^{3} \;g_{3C}^{2} + \frac{4 \;g_{LL}^{2}}{9} \;g_{RL}^{3} + \frac{7}{9} \;g_{LR}^{4} \;g_{RL} + \frac{\;g_{LR}^{2}}{9} \;g_{RL}^{3}\\
		&+\frac{28 \;g_{LL}^{4}}{9} \;g_{RL} + \frac{15 \;g_{LL}^{3}}{2} \;g_{2R}^{2} + \frac{37 \;g_{LL}^{3}}{18} \;g_{RR}^{2} + \frac{37 \;g_{LL}^{3}}{3} \;g_{RL}^{2} + \frac{85 \;g_{LL}}{18} \;g_{RL}^{4}\\
		&+\frac{115 \;g_{LL}^{3}}{9} \;g_{LR}^{2} + 6 \;g_{LL} \;g_{2L}^{2} \;g_{RL}^{2} + 6 \;g_{LL} \;g_{2R}^{2} \;g_{RL}^{2} + 12 \;g_{LL} \;g_{3C}^{2} \;g_{LR}^{2} + \frac{115 \;g_{LL}}{18} \;g_{LR}^{4}\\
		&+\frac{2 \;g_{LL}^{2}}{9} \;g_{RR}^{2} \;g_{RL} + 12 \;g_{LL} \;g_{3C}^{2} \;g_{RL}^{2} + \frac{\;g_{LL} \;g_{RR}^{3}}{9} \;g_{LR} + \frac{7 \;g_{LL}}{3} \;g_{RR} \;g_{LR}^{3} + \frac{\;g_{RL}}{3} \;g_{RR}^{2} \;g_{LR}^{2}\\
		&+\frac{7 \;g_{RR}}{3} \;g_{LL}^{3} \;g_{LR} + \frac{35 \;g_{LL}^{2}}{9} \;g_{LR}^{2} \;g_{RL} + \frac{15 \;g_{LL}}{2} \;g_{2L}^{2} \;g_{LR}^{2} + \frac{15 \;g_{LL}}{2} \;g_{2R}^{2} \;g_{LR}^{2}\\
		&+\frac{37 \;g_{LL}}{6} \;g_{RR}^{2} \;g_{LR}^{2} + \frac{85 \;g_{LL}}{18} \;g_{RR}^{2} \;g_{RL}^{2} + \frac{74 \;g_{LL}}{9} \;g_{LR}^{2} \;g_{RL}^{2} + \frac{37 \;g_{RR}}{6} \;g_{LR}^{3} \;g_{RL}\\
		&+6 \;g_{2L}^{2} \;g_{RR} \;g_{LR} \;g_{RL} + 6 \;g_{2R}^{2} \;g_{RR} \;g_{LR} \;g_{RL} + \frac{85 \;g_{RR}^{3}}{18} \;g_{LR} \;g_{RL} + \frac{85 \;g_{RR}}{18} \;g_{LR} \;g_{RL}^{3}\\
		&+\frac{259 \;g_{RR}}{18} \;g_{LL}^{2} \;g_{LR} \;g_{RL} + \frac{5 \;g_{LL}}{9} \;g_{RR} \;g_{LR} \;g_{RL}^{2} + 12 \;g_{3C}^{2} \;g_{RR} \;g_{LR} \;g_{RL}
		\\
		(4\pi)^4\beta^{\rm 2loop}_{LR} = 
		& \frac{7 \;g_{RR}}{9} \;g_{LL}^{4} + \frac{\;g_{LL}^{2} \;g_{RR}^{3}}{9} + \frac{15 \;g_{2L}^{2}}{2} \;g_{LR}^{3} + 12 \;g_{3C}^{2} \;g_{LR}^{3} + \frac{4 \;g_{RR}^{3}}{9} \;g_{LR}^{2} + \frac{115}{18} \;g_{LR}^{5}\\
		&+\frac{15 \;g_{2R}^{2}}{2} \;g_{LR}^{3} + \frac{85 \;g_{RR}^{4}}{18} \;g_{LR} + \frac{37 \;g_{RR}^{2}}{3} \;g_{LR}^{3} + \frac{28 \;g_{RR}}{9} \;g_{LR}^{4} + \frac{37}{18} \;g_{LR}^{3} \;g_{RL}^{2}\\
		&+\frac{115 \;g_{LL}^{4}}{18} \;g_{LR} + 12 \;g_{LL}^{2} \;g_{3C}^{2} \;g_{LR} + \frac{115 \;g_{LL}^{2}}{9} \;g_{LR}^{3} + 6 \;g_{2L}^{2} \;g_{RR}^{2} \;g_{LR} + 6 \;g_{2R}^{2} \;g_{RR}^{2} \;g_{LR}\\
		&+\frac{\;g_{LL}^{2} \;g_{RR}}{3} \;g_{RL}^{2} + \frac{\;g_{LL}}{9} \;g_{LR} \;g_{RL}^{3} + 12 \;g_{3C}^{2} \;g_{RR}^{2} \;g_{LR} + \frac{2 \;g_{RR}}{9} \;g_{LR}^{2} \;g_{RL}^{2}\\
		&+\frac{7 \;g_{LL}^{3}}{3} \;g_{LR} \;g_{RL} + \frac{15 \;g_{LL}^{2}}{2} \;g_{2L}^{2} \;g_{LR} + \frac{15 \;g_{LL}^{2}}{2} \;g_{2R}^{2} \;g_{LR} + \frac{7 \;g_{LL}}{3} \;g_{LR}^{3} \;g_{RL}\\
		&+\frac{37 \;g_{RR}}{6} \;g_{LL}^{3} \;g_{RL} + \frac{74 \;g_{LL}^{2}}{9} \;g_{RR}^{2} \;g_{LR} + \frac{35 \;g_{RR}}{9} \;g_{LL}^{2} \;g_{LR}^{2} + \frac{37 \;g_{LL}^{2}}{6} \;g_{LR} \;g_{RL}^{2}\\
		&+6 \;g_{LL} \;g_{2L}^{2} \;g_{RR} \;g_{RL} + \frac{85 \;g_{LL}}{18} \;g_{RR}^{3} \;g_{RL} + \frac{85 \;g_{LL}}{18} \;g_{RR} \;g_{RL}^{3} + \frac{85 \;g_{RR}^{2}}{18} \;g_{LR} \;g_{RL}^{2}\\
		&+6 \;g_{LL} \;g_{2R}^{2} \;g_{RR} \;g_{RL} + 12 \;g_{LL} \;g_{3C}^{2} \;g_{RR} \;g_{RL} + \frac{5 \;g_{LL}}{9} \;g_{RR}^{2} \;g_{LR} \;g_{RL} + \frac{259 \;g_{LL}}{18} \;g_{RR} \;g_{LR}^{2} \;g_{RL}
		\\
		(4\pi)^4\beta^{\rm 2loop}_{RL} = 
		& \frac{\;g_{LL} \;g_{RR}^{4}}{9} + \frac{4 \;g_{LL}}{9} \;g_{RL}^{4} + 6 \;g_{2L}^{2} \;g_{RL}^{3} + 6 \;g_{2R}^{2} \;g_{RL}^{3} + 12 \;g_{3C}^{2} \;g_{RL}^{3} + \frac{85}{18} \;g_{RL}^{5}\\
		&+\frac{7 \;g_{LL}^{3}}{9} \;g_{RR}^{2} + \frac{28 \;g_{LL}^{3}}{9} \;g_{RL}^{2} + \frac{37 \;g_{LL}^{2}}{3} \;g_{RL}^{3} + \frac{85 \;g_{RR}^{2}}{9} \;g_{RL}^{3} + \frac{37}{18} \;g_{LR}^{2} \;g_{RL}^{3}\\
		&+\frac{115 \;g_{LL}^{4}}{18} \;g_{RL} + 12 \;g_{LL}^{2} \;g_{3C}^{2} \;g_{RL} + 6 \;g_{2L}^{2} \;g_{RR}^{2} \;g_{RL} + 6 \;g_{2R}^{2} \;g_{RR}^{2} \;g_{RL} + \frac{85 \;g_{RR}^{4}}{18} \;g_{RL}\\
		&+\frac{5 \;g_{LL}}{9} \;g_{RR}^{2} \;g_{RL}^{2} + 12 \;g_{3C}^{2} \;g_{RR}^{2} \;g_{RL} + \frac{\;g_{LR}}{3} \;g_{RR}^{3} \;g_{RL} + \frac{\;g_{RR}}{3} \;g_{LR} \;g_{RL}^{3}\\
		&+\frac{15 \;g_{LL}^{2}}{2} \;g_{2L}^{2} \;g_{RL} + \frac{7 \;g_{LL}}{3} \;g_{RR}^{2} \;g_{LR}^{2} + \frac{14 \;g_{LL}}{9} \;g_{LR}^{2} \;g_{RL}^{2} + \frac{7 \;g_{RR}}{9} \;g_{LR}^{3} \;g_{RL}\\
		&+\frac{15 \;g_{LL}^{2}}{2} \;g_{2R}^{2} \;g_{RL} + \frac{74 \;g_{LL}^{2}}{9} \;g_{RR}^{2} \;g_{RL} + \frac{37 \;g_{LL}}{6} \;g_{RR}^{3} \;g_{LR} + \frac{37 \;g_{RR}^{2}}{6} \;g_{LR}^{2} \;g_{RL}\\
		&+\frac{115 \;g_{RR}}{18} \;g_{LL}^{3} \;g_{LR} + \frac{115 \;g_{LL}^{2}}{18} \;g_{LR}^{2} \;g_{RL} + 12 \;g_{LL} \;g_{3C}^{2} \;g_{RR} \;g_{LR} + \frac{115 \;g_{LL}}{18} \;g_{RR} \;g_{LR}^{3}\\
		&+\frac{35 \;g_{RR}}{9} \;g_{LL}^{2} \;g_{LR} \;g_{RL} + \frac{15 \;g_{LL}}{2} \;g_{2L}^{2} \;g_{RR} \;g_{LR} + \frac{15 \;g_{LL}}{2} \;g_{2R}^{2} \;g_{RR} \;g_{LR} + \frac{259 \;g_{LL}}{18} \;g_{RR} \;g_{LR} \;g_{RL}^{2}
		\\
		(4\pi)^4\beta^{\rm 2loop}_{RR} = 
		& 6 \;g_{2L}^{2} \;g_{RR}^{3} + 6 \;g_{2R}^{2} \;g_{RR}^{3} + 12 \;g_{3C}^{2} \;g_{RR}^{3} + \frac{85 \;g_{RR}^{5}}{18} + \frac{4 \;g_{RR}^{4}}{9} \;g_{LR} + \frac{\;g_{LR}}{9} \;g_{RL}^{4}\\
		&+\frac{37 \;g_{LL}^{2}}{18} \;g_{RR}^{3} + \frac{37 \;g_{RR}^{3}}{3} \;g_{LR}^{2} + \frac{85 \;g_{RR}^{3}}{9} \;g_{RL}^{2} + \frac{28 \;g_{RR}^{2}}{9} \;g_{LR}^{3} + \frac{7}{9} \;g_{LR}^{3} \;g_{RL}^{2}\\
		&+6 \;g_{2L}^{2} \;g_{RR} \;g_{RL}^{2} + 6 \;g_{2R}^{2} \;g_{RR} \;g_{RL}^{2} + 12 \;g_{3C}^{2} \;g_{RR} \;g_{LR}^{2} + \frac{115 \;g_{RR}}{18} \;g_{LR}^{4} + \frac{85 \;g_{RR}}{18} \;g_{RL}^{4}\\
		&+\frac{\;g_{LL} \;g_{RR}^{3}}{3} \;g_{RL} + \frac{\;g_{LL} \;g_{RR}}{3} \;g_{RL}^{3} + 12 \;g_{3C}^{2} \;g_{RR} \;g_{RL}^{2} + \frac{5 \;g_{RR}^{2}}{9} \;g_{LR} \;g_{RL}^{2}\\
		&+\frac{7 \;g_{RR}}{9} \;g_{LL}^{3} \;g_{RL} + \frac{14 \;g_{LL}^{2}}{9} \;g_{RR}^{2} \;g_{LR} + \frac{7 \;g_{LL}^{2}}{3} \;g_{LR} \;g_{RL}^{2} + \frac{15 \;g_{RR}}{2} \;g_{2L}^{2} \;g_{LR}^{2}\\
		&+\frac{37 \;g_{RR}}{6} \;g_{LL}^{2} \;g_{RL}^{2} + \frac{37 \;g_{LL}}{6} \;g_{LR} \;g_{RL}^{3} + \frac{15 \;g_{RR}}{2} \;g_{2R}^{2} \;g_{LR}^{2} + \frac{74 \;g_{RR}}{9} \;g_{LR}^{2} \;g_{RL}^{2}\\
		&+\frac{115 \;g_{LL}^{3}}{18} \;g_{LR} \;g_{RL} + \frac{115 \;g_{RR}}{18} \;g_{LL}^{2} \;g_{LR}^{2} + 12 \;g_{LL} \;g_{3C}^{2} \;g_{LR} \;g_{RL} + \frac{115 \;g_{LL}}{18} \;g_{LR}^{3} \;g_{RL}\\
		&+\frac{15 \;g_{LL}}{2} \;g_{2L}^{2} \;g_{LR} \;g_{RL} + \frac{15 \;g_{LL}}{2} \;g_{2R}^{2} \;g_{LR} \;g_{RL} + \frac{259 \;g_{LL}}{18} \;g_{RR}^{2} \;g_{LR} \;g_{RL} + \frac{35 \;g_{LL}}{9} \;g_{RR} \;g_{LR}^{2} \;g_{RL}
	\end{align*}
}
\underline{\textsl{SUSY}}
{\scriptsize
	\begin{align*} 
		(4\pi)^4{\beta^{\rm 2loop}_{LL}}& =24 \;g_{LL}^3 \;g_{3 C}^2+24 \;g_{LR}^2 \;g_{LL} \;g_{3 C}^2+24 \;g_{RL}^2 \;g_{LL} \;g_{3 C}^2+24 \;g_{LR} \;g_{RL} \;g_{RR} \;g_{3 C}^2+12 \;g_{LL}^3 \;g_{2 L}^2 \\ &+12 \;g_{LR}^2 \;g_{LL} \;g_{2 L}^2 +\frac{21}{2} \;g_{RL}^2 \;g_{LL} \;g_{2 L}^2+\frac{21}{2} \;g_{LR} \;g_{RL} \;g_{RR} \;g_{2 L}^2+12 \;g_{LL}^3 \;g_{2 R}^2+12 \;g_{LR}^2 \;g_{LL} \;g_{2 R}^2 \\ & +\frac{21}{2} \;g_{RL}^2 \;g_{LL} \;g_{2 R}^2+\frac{21}{2} \;g_{LR} \;g_{RL} \;g_{RR} \;g_{2 R}^2  +\frac{98 \;g_{LL}^5}{9}+\frac{28}{9} \;g_{RL} \;g_{LL}^4+\frac{196}{9} \;g_{LR}^2 \;g_{LL}^3+\frac{64}{3} \;g_{RL}^2 \;g_{LL}^3 \\ & +\frac{32}{9} \;g_{RR}^2 \;g_{LL}^3+\frac{7}{3} \;g_{LR} \;g_{RR} \;g_{LL}^3+\frac{4}{9} \;g_{RL}^3 \;g_{LL}^2  +\frac{2}{9} \;g_{RL} \;g_{RR}^2 \;g_{LL}^2+\frac{35}{9} \;g_{LR}^2 \;g_{RL} \;g_{LL}^2 \\ & +\frac{224}{9} \;g_{LR} \;g_{RL} \;g_{RR} \;g_{LL}^2+\frac{98}{9} \;g_{LR}^4 \;g_{LL}+\frac{83}{9} \;g_{RL}^4 \;g_{LL}+\frac{1}{9} \;g_{LR} \;g_{RR}^3 \;g_{LL}  +\frac{128}{9} \;g_{LR}^2 \;g_{RL}^2 \;g_{LL} \\ & +\frac{32}{3} \;g_{LR}^2 \;g_{RR}^2 \;g_{LL}+\frac{83}{9} \;g_{RL}^2 \;g_{RR}^2 \;g_{LL}+\frac{7}{3} \;g_{LR}^3 \;g_{RR} \;g_{LL}+\frac{5}{9} \;g_{LR} \;g_{RL}^2 \;g_{RR} \;g_{LL}  +\frac{1}{9} \;g_{LR}^2 \;g_{RL}^3 \\ & +\frac{83}{9} \;g_{LR} \;g_{RL} \;g_{RR}^3+\frac{1}{3} \;g_{LR}^2 \;g_{RL} \;g_{RR}^2+\frac{7}{9} \;g_{LR}^4 \;g_{RL}+\frac{83}{9} \;g_{LR} \;g_{RL}^3 \;g_{RR}+\frac{32}{3} \;g_{LR}^3 \;g_{RL} \;g_{RR} \\ 
		(4\pi)^4{\beta^{\rm 2loop}_{LR}}& =24 \;g_{LR}^3 \;g_{3 C}^2+24 \;g_{LL}^2 \;g_{LR} \;g_{3 C}^2+24 \;g_{RR}^2 \;g_{LR} \;g_{3 C}^2+24 \;g_{LL} \;g_{RL} \;g_{RR} \;g_{3 C}^2+12 \;g_{LR}^3 \;g_{2 L}^2+12 \;g_{LL}^2 \;g_{LR} \;g_{2 L}^2 \\ & +\frac{21}{2} \;g_{RR}^2 \;g_{LR} \;g_{2 L}^2+\frac{21}{2} \;g_{LL} \;g_{RL} \;g_{RR} \;g_{2 L}^2+12 \;g_{LR}^3 \;g_{2 R}^2+12 \;g_{LL}^2 \;g_{LR} \;g_{2 R}^2+\frac{21}{2} \;g_{RR}^2 \;g_{LR} \;g_{2 R}^2 \\& +\frac{21}{2} \;g_{LL} \;g_{RL} \;g_{RR} \;g_{2 R}^2  +\frac{98 \;g_{LR}^5}{9}+\frac{28}{9} \;g_{RR} \;g_{LR}^4+\frac{196}{9} \;g_{LL}^2 \;g_{LR}^3+\frac{32}{9} \;g_{RL}^2 \;g_{LR}^3+\frac{64}{3} \;g_{RR}^2 \;g_{LR}^3 \\& +\frac{7}{3} \;g_{LL} \;g_{RL} \;g_{LR}^3+\frac{4}{9} \;g_{RR}^3 \;g_{LR}^2 +\frac{35}{9} \;g_{LL}^2 \;g_{RR} \;g_{LR}^2+\frac{2}{9} \;g_{RL}^2 \;g_{RR} \;g_{LR}^2+\frac{224}{9} \;g_{LL} \;g_{RL} \;g_{RR} \;g_{LR}^2 \\& +\frac{98}{9} \;g_{LL}^4 \;g_{LR}+\frac{83}{9} \;g_{RR}^4 \;g_{LR}+\frac{1}{9} \;g_{LL} \;g_{RL}^3 \;g_{LR}  +\frac{32}{3} \;g_{LL}^2 \;g_{RL}^2 \;g_{LR}+\frac{128}{9} \;g_{LL}^2 \;g_{RR}^2 \;g_{LR} \\& +\frac{83}{9} \;g_{RL}^2 \;g_{RR}^2 \;g_{LR}+\frac{5}{9} \;g_{LL} \;g_{RL} \;g_{RR}^2 \;g_{LR}+\frac{7}{3} \;g_{LL}^3 \;g_{RL} \;g_{LR}+\frac{1}{9} \;g_{LL}^2 \;g_{RR}^3  +\frac{83}{9} \;g_{LL} \;g_{RL} \;g_{RR}^3 \\& +\frac{7}{9} \;g_{LL}^4 \;g_{RR}+\frac{83}{9} \;g_{LL} \;g_{RL}^3 \;g_{RR}+\frac{1}{3} \;g_{LL}^2 \;g_{RL}^2 \;g_{RR}+\frac{32}{3} \;g_{LL}^3 \;g_{RL} \;g_{RR} \\ 
		(4\pi)^4{\beta^{\rm 2loop}_{RL}}& =24 \;g_{RL}^3 \;g_{3 C}^2+24 \;g_{LL}^2 \;g_{RL} \;g_{3 C}^2+24 \;g_{RR}^2 \;g_{RL} \;g_{3 C}^2+24 \;g_{LL} \;g_{LR} \;g_{RR} \;g_{3 C}^2+\frac{21}{2} \;g_{RL}^3 \;g_{2 L}^2+12 \;g_{LL}^2 \;g_{RL} \;g_{2 L}^2 \\& +\frac{21}{2} \;g_{RR}^2 \;g_{RL} \;g_{2 L}^2+12 \;g_{LL} \;g_{LR} \;g_{RR} \;g_{2 L}^2+\frac{21}{2} \;g_{RL}^3 \;g_{2 R}^2+12 \;g_{LL}^2 \;g_{RL} \;g_{2 R}^2+\frac{21}{2} \;g_{RR}^2 \;g_{RL} \;g_{2 R}^2 \\& +12 \;g_{LL} \;g_{LR} \;g_{RR} \;g_{2 R}^2 +\frac{83 \;g_{RL}^5}{9}+\frac{4}{9} \;g_{LL} \;g_{RL}^4+\frac{64}{3} \;g_{LL}^2 \;g_{RL}^3+\frac{32}{9} \;g_{LR}^2 \;g_{RL}^3+\frac{166}{9} \;g_{RR}^2 \;g_{RL}^3+\frac{1}{3} \;g_{LR} \;g_{RR} \;g_{RL}^3 \\& +\frac{28}{9} \;g_{LL}^3 \;g_{RL}^2 +\frac{14}{9} \;g_{LL} \;g_{LR}^2 \;g_{RL}^2 +\frac{5}{9} \;g_{LL} \;g_{RR}^2 \;g_{RL}^2+\frac{224}{9} \;g_{LL} \;g_{LR} \;g_{RR} \;g_{RL}^2+\frac{98}{9} \;g_{LL}^4 \;g_{RL} \\& +\frac{83}{9} \;g_{RR}^4 \;g_{RL}+\frac{1}{3} \;g_{LR} \;g_{RR}^3 \;g_{RL}+\frac{98}{9} \;g_{LL}^2 \;g_{LR}^2 \;g_{RL}  +\frac{128}{9} \;g_{LL}^2 \;g_{RR}^2 \;g_{RL}+\frac{32}{3} \;g_{LR}^2 \;g_{RR}^2 \;g_{RL}\\& +\frac{7}{9} \;g_{LR}^3 \;g_{RR} \;g_{RL}+\frac{35}{9} \;g_{LL}^2 \;g_{LR} \;g_{RR} \;g_{RL}+\frac{1}{9} \;g_{LL} \;g_{RR}^4+\frac{32}{3} \;g_{LL} \;g_{LR} \;g_{RR}^3 \\& +\frac{7}{9} \;g_{LL}^3 \;g_{RR}^2+\frac{7}{3} \;g_{LL} \;g_{LR}^2 \;g_{RR}^2+\frac{98}{9} \;g_{LL} \;g_{LR}^3 \;g_{RR}+\frac{98}{9} \;g_{LL}^3 \;g_{LR} \;g_{RR} \\ 
		(4\pi)^4{\beta^{\rm 2loop}_{RR}}& =24 \;g_{RR}^3 \;g_{3 C}^2+24 \;g_{LR}^2 \;g_{RR} \;g_{3 C}^2+24 \;g_{RL}^2 \;g_{RR} \;g_{3 C}^2+24 \;g_{LL} \;g_{LR} \;g_{RL} \;g_{3 C}^2+\frac{21}{2} \;g_{RR}^3 \;g_{2 L}^2+12 \;g_{LR}^2 \;g_{RR} \;g_{2 L}^2 \\& +\frac{21}{2} \;g_{RL}^2 \;g_{RR} \;g_{2 L}^2+12 \;g_{LL} \;g_{LR} \;g_{RL} \;g_{2 L}^2+\frac{21}{2} \;g_{RR}^3 \;g_{2 R}^2+12 \;g_{LR}^2 \;g_{RR} \;g_{2 R}^2+\frac{21}{2} \;g_{RL}^2 \;g_{RR} \;g_{2 R}^2\\&+12 \;g_{LL} \;g_{LR} \;g_{RL} \;g_{2 R}^2+\frac{83 \;g_{RR}^5}{9} +\frac{4}{9} \;g_{LR} \;g_{RR}^4+\frac{32}{9} \;g_{LL}^2 \;g_{RR}^3+\frac{64}{3} \;g_{LR}^2 \;g_{RR}^3+\frac{166}{9} \;g_{RL}^2 \;g_{RR}^3  \\& +\frac{28}{9} \;g_{LR}^3 \;g_{RR}^2+\frac{5}{9} \;g_{LR} \;g_{RL}^2 \;g_{RR}^2  +\frac{14}{9} \;g_{LL}^2 \;g_{LR} \;g_{RR}^2 +\frac{224}{9} \;g_{LL} \;g_{LR} \;g_{RL} \;g_{RR}^2+\frac{98}{9} \;g_{LR}^4 \;g_{RR} \\& +\frac{83}{9} \;g_{RL}^4 \;g_{RR}+\frac{1}{3} \;g_{LL} \;g_{RL}^3 \;g_{RR}+\frac{98}{9} \;g_{LL}^2 \;g_{LR}^2 \;g_{RR}  +\frac{32}{3} \;g_{LL}^2 \;g_{RL}^2 \;g_{RR}+\frac{128}{9} \;g_{LR}^2 \;g_{RL}^2 \;g_{RR}\\& +\frac{7}{9} \;g_{LL}^3 \;g_{RL} \;g_{RR}+\frac{35}{9} \;g_{LL} \;g_{LR}^2 \;g_{RL} \;g_{RR}+\frac{1}{9} \;g_{LR} \;g_{RL}^4 +\frac{32}{3} \;g_{LL} \;g_{LR} \;g_{RL}^3  +\frac{7}{9} \;g_{LR}^3 \;g_{RL}^2 \\& +\frac{7}{3} \;g_{LL}^2 \;g_{LR} \;g_{RL}^2+\frac{98}{9} \;g_{LL} \;g_{LR}^3 \;g_{RL}+\frac{98}{9} \;g_{LL}^3 \;g_{LR} \;g_{RL} +\frac{1}{3} \;g_{LL} \;g_{RL} \;g_{RR}^3\\ 
		(4\pi)^4{\beta^{\rm 2loop}_{2L}}& =24 \;g_{3 C}^2 \;g_{2 L}^3+12 \;g_{2 L}^3 \;g_{2 R}^2+\frac{99 \;g_{2 L}^5}{2}+4 \;g_{LL}^2 \;g_{2 L}^3+4 \;g_{LR}^2 \;g_{2 L}^3+\frac{7}{2} \;g_{RL}^2 \;g_{2 L}^3+\frac{7}{2} \;g_{RR}^2 \;g_{2 L}^3 \\ 
		(4\pi)^4{\beta^{\rm 2loop}_{2R}}& =24 \;g_{3 C}^2 \;g_{2 R}^3+12 \;g_{2 L}^2 \;g_{2 R}^3+\frac{99 \;g_{2 R}^5}{2}+4 \;g_{LL}^2 \;g_{2 R}^3+4 \;g_{LR}^2 \;g_{2 R}^3+\frac{7}{2} \;g_{RL}^2 \;g_{2 R}^3+\frac{7}{2} \;g_{RR}^2 \;g_{2 R}^3 \\ 
		(4\pi)^4{\beta^{\rm 2loop}_{3C}}& =9 \;g_{3 C}^3 \;g_{2 L}^2+9 \;g_{3 C}^3 \;g_{2 R}^2+48 \;g_{3 C}^5+3 \;g_{LL}^2 \;g_{3 C}^3+3 \;g_{LR}^2 \;g_{3 C}^3+3 \;g_{RL}^2 \;g_{3 C}^3+3 \;g_{RR}^2 \;g_{3 C}^3 \\
	\end{align*}
}

{\bf \underline{When only 16-fermions are present at the $M_R$ scale:}}\\

{\bf D-parity not conserved}\\
\underline{\textsl{Non-SUSY}}


{\scriptsize
	\begin{align*}
		(4\pi)^4\beta^{\rm 2loop}_{\;g_{2L}} = 
		& \frac{4 \;g_{LL}^{2}}{3} \;g_{2L}^{3} + 8 \;g_{2L}^{5} + 3 \;g_{2L}^{3} \;g_{2R}^{2} + 12 \;g_{2L}^{3} \;g_{3C}^{2} + \frac{4 \;g_{2L}^{3}}{3} \;g_{RR}^{2} + \frac{4 \;g_{2L}^{3}}{3} \;g_{LR}^{2}\\
		&+\frac{\;g_{LL} \;g_{2L}^{3}}{3} \;g_{RL} + \frac{\;g_{2L}^{3} \;g_{RR}}{3} \;g_{LR} + \frac{4 \;g_{2L}^{3}}{3} \;g_{RL}^{2}
		\\
		(4\pi)^4\beta^{\rm 2loop}_{\;g_{3C}} = 
		&\frac{\;g_{LL}^{2} \;g_{3C}^{3}}{2} + \frac{9 \;g_{2L}^{2}}{2} \;g_{3C}^{3} - 26 \;g_{3C}^{5} + \frac{\;g_{3C}^{3} \;g_{RR}^{2}}{2} + \frac{\;g_{3C}^{3}}{2} \;g_{LR}^{2} + \frac{\;g_{3C}^{3}}{2} \;g_{RL}^{2}\\
		&+\frac{9 \;g_{2R}^{2}}{2} \;g_{3C}^{3}
		\\
		(4\pi)^4\beta^{\rm 2loop}_{\;g_{2R}} = 
		&2 \;g_{LL}^{2} \;g_{2R}^{3} + 3 \;g_{2L}^{2} \;g_{2R}^{3} + \frac{61 \;g_{2R}^{5}}{6} + 12 \;g_{2R}^{3} \;g_{3C}^{2} + \frac{3 \;g_{2R}^{3}}{2} \;g_{RR}^{2} + 2 \;g_{2R}^{3} \;g_{LR}^{2}\\
		&+\;g_{LL} \;g_{2R}^{3} \;g_{RL} + \;g_{2R}^{3} \;g_{RR} \;g_{LR} + \frac{3 \;g_{2R}^{3}}{2} \;g_{RL}^{2}
		\\
		(4\pi)^4\beta^{\rm 2loop}_{LL} = 
		&\frac{8 \;g_{LL}^{5}}{3} + 4 \;g_{LL}^{3} \;g_{2L}^{2} + 6 \;g_{LL}^{3} \;g_{2R}^{2} + 4 \;g_{LL}^{3} \;g_{3C}^{2} + \;g_{LL}^{3} \;g_{RR}^{2} + 6 \;g_{LL}^{3} \;g_{RL}^{2} + \frac{5}{6} \;g_{LR}^{2} \;g_{RL}^{3}\\
		&+\frac{14 \;g_{LL}^{4}}{3} \;g_{RL} + \frac{10 \;g_{LL}^{2}}{3} \;g_{RL}^{3} + \frac{8 \;g_{LL}}{3} \;g_{LR}^{4} + \frac{11 \;g_{LL}}{6} \;g_{RL}^{4} + \frac{7}{6} \;g_{LR}^{4} \;g_{RL}\\
		&+\frac{16 \;g_{LL}^{3}}{3} \;g_{LR}^{2} + \;g_{LL}^{2} \;g_{2L}^{2} \;g_{RL} + 3 \;g_{LL} \;g_{RR}^{2} \;g_{LR}^{2} + \frac{\;g_{RL}}{2} \;g_{2L}^{2} \;g_{LR}^{2} + 3 \;g_{RR} \;g_{LR}^{3} \;g_{RL}\\
		&+3 \;g_{LL}^{2} \;g_{2R}^{2} \;g_{RL} + 4 \;g_{LL} \;g_{2L}^{2} \;g_{LR}^{2} + 4 \;g_{LL} \;g_{2L}^{2} \;g_{RL}^{2} + 4 \;g_{LL} \;g_{3C}^{2} \;g_{LR}^{2} + 4 \;g_{LL} \;g_{3C}^{2} \;g_{RL}^{2}\\
		&+6 \;g_{LL} \;g_{2R}^{2} \;g_{LR}^{2} + 4 \;g_{LL} \;g_{LR}^{2} \;g_{RL}^{2} + \frac{3 \;g_{2R}^{2}}{2} \;g_{LR}^{2} \;g_{RL} + \frac{5 \;g_{RR}^{2}}{2} \;g_{LR}^{2} \;g_{RL}\\
		&+\frac{7 \;g_{RR}}{2} \;g_{LL}^{3} \;g_{LR} + \frac{5 \;g_{LL}^{2}}{3} \;g_{RR}^{2} \;g_{RL} + \frac{5 \;g_{LL}}{6} \;g_{RR}^{3} \;g_{LR} + \frac{7 \;g_{LL}}{2} \;g_{RR} \;g_{LR}^{3}\\
		&+\frac{9 \;g_{LL}}{2} \;g_{2R}^{2} \;g_{RL}^{2} + \frac{11 \;g_{LL}}{6} \;g_{RR}^{2} \;g_{RL}^{2} + \frac{11 \;g_{RR}^{3}}{6} \;g_{LR} \;g_{RL} + \frac{11 \;g_{RR}}{6} \;g_{LR} \;g_{RL}^{3}\\
		&+\frac{35 \;g_{LL}^{2}}{6} \;g_{LR}^{2} \;g_{RL} + \frac{\;g_{LL} \;g_{RR}}{2} \;g_{2L}^{2} \;g_{LR} + 4 \;g_{2L}^{2} \;g_{RR} \;g_{LR} \;g_{RL} + 4 \;g_{3C}^{2} \;g_{RR} \;g_{LR} \;g_{RL}\\
		&+7 \;g_{LL}^{2} \;g_{RR} \;g_{LR} \;g_{RL} + \frac{3 \;g_{LL}}{2} \;g_{2R}^{2} \;g_{RR} \;g_{LR} + \frac{25 \;g_{LL}}{6} \;g_{RR} \;g_{LR} \;g_{RL}^{2} + \frac{9 \;g_{RR}}{2} \;g_{2R}^{2} \;g_{LR} \;g_{RL}
		\\
		(4\pi)^4\beta^{\rm 2loop}_{LR} = 
		&\frac{5 \;g_{LL}^{2}}{6} \;g_{RR}^{3} + 4 \;g_{2L}^{2} \;g_{LR}^{3} + 6 \;g_{2R}^{2} \;g_{LR}^{3} + 4 \;g_{3C}^{2} \;g_{LR}^{3} + 6 \;g_{RR}^{2} \;g_{LR}^{3} + \frac{8}{3} \;g_{LR}^{5} + \;g_{LR}^{3} \;g_{RL}^{2}\\
		&+\frac{7 \;g_{RR}}{6} \;g_{LL}^{4} + \frac{8 \;g_{LL}^{4}}{3} \;g_{LR} + \frac{11 \;g_{RR}^{4}}{6} \;g_{LR} + \frac{10 \;g_{RR}^{3}}{3} \;g_{LR}^{2} + \frac{14 \;g_{RR}}{3} \;g_{LR}^{4}\\
		&+3 \;g_{LL}^{3} \;g_{RR} \;g_{RL} + \frac{\;g_{LL}^{2} \;g_{RR}}{2} \;g_{2L}^{2} + \frac{16 \;g_{LL}^{2}}{3} \;g_{LR}^{3} + \;g_{2L}^{2} \;g_{RR} \;g_{LR}^{2} + 3 \;g_{2R}^{2} \;g_{RR} \;g_{LR}^{2}\\
		&+4 \;g_{LL}^{2} \;g_{2L}^{2} \;g_{LR} + 4 \;g_{LL}^{2} \;g_{3C}^{2} \;g_{LR} + 4 \;g_{LL}^{2} \;g_{RR}^{2} \;g_{LR} + 3 \;g_{LL}^{2} \;g_{LR} \;g_{RL}^{2} + 4 \;g_{2L}^{2} \;g_{RR}^{2} \;g_{LR}\\
		&+\frac{3 \;g_{RR}}{2} \;g_{LL}^{2} \;g_{2R}^{2} + 6 \;g_{LL}^{2} \;g_{2R}^{2} \;g_{LR} + \frac{5 \;g_{RR}}{2} \;g_{LL}^{2} \;g_{RL}^{2} + 4 \;g_{3C}^{2} \;g_{RR}^{2} \;g_{LR} + \frac{5 \;g_{RR}}{3} \;g_{LR}^{2} \;g_{RL}^{2}\\
		&+\frac{7 \;g_{LL}^{3}}{2} \;g_{LR} \;g_{RL} + \frac{7 \;g_{LL}}{2} \;g_{LR}^{3} \;g_{RL} + \frac{5 \;g_{LL}}{6} \;g_{LR} \;g_{RL}^{3} + \frac{9 \;g_{2R}^{2}}{2} \;g_{RR}^{2} \;g_{LR}\\
		&+\frac{35 \;g_{RR}}{6} \;g_{LL}^{2} \;g_{LR}^{2} + \frac{11 \;g_{LL}}{6} \;g_{RR}^{3} \;g_{RL} + \frac{11 \;g_{LL}}{6} \;g_{RR} \;g_{RL}^{3} + \frac{11 \;g_{RR}^{2}}{6} \;g_{LR} \;g_{RL}^{2}\\
		&+4 \;g_{LL} \;g_{2L}^{2} \;g_{RR} \;g_{RL} + \frac{\;g_{LL} \;g_{2L}^{2}}{2} \;g_{LR} \;g_{RL} + 4 \;g_{LL} \;g_{3C}^{2} \;g_{RR} \;g_{RL} + 7 \;g_{LL} \;g_{RR} \;g_{LR}^{2} \;g_{RL}\\
		&+\frac{9 \;g_{LL}}{2} \;g_{2R}^{2} \;g_{RR} \;g_{RL} + \frac{3 \;g_{LL}}{2} \;g_{2R}^{2} \;g_{LR} \;g_{RL} + \frac{25 \;g_{LL}}{6} \;g_{RR}^{2} \;g_{LR} \;g_{RL}
		\\
		(4\pi)^4\beta^{\rm 2loop}_{RL} = 
		&6 \;g_{LL}^{2} \;g_{RL}^{3} + \frac{5 \;g_{LL}}{6} \;g_{RR}^{4} + 4 \;g_{2L}^{2} \;g_{RL}^{3} + 4 \;g_{3C}^{2} \;g_{RL}^{3} + \;g_{LR}^{2} \;g_{RL}^{3} + \frac{11}{6} \;g_{RL}^{5}\\
		&+\frac{8 \;g_{LL}^{4}}{3} \;g_{RL} + \frac{7 \;g_{LL}^{3}}{6} \;g_{RR}^{2} + \frac{10 \;g_{LL}}{3} \;g_{RL}^{4} + \frac{9 \;g_{2R}^{2}}{2} \;g_{RL}^{3} + \frac{11 \;g_{RR}^{2}}{3} \;g_{RL}^{3}\\
		&+\frac{14 \;g_{LL}^{3}}{3} \;g_{RL}^{2} + \frac{\;g_{LL} \;g_{2L}^{2}}{2} \;g_{RR}^{2} + \;g_{LL} \;g_{2L}^{2} \;g_{RL}^{2} + 3 \;g_{LL} \;g_{RR}^{3} \;g_{LR} + \frac{11 \;g_{RR}^{4}}{6} \;g_{RL}\\
		&+4 \;g_{LL}^{2} \;g_{2L}^{2} \;g_{RL} + 4 \;g_{LL}^{2} \;g_{3C}^{2} \;g_{RL} + 4 \;g_{LL}^{2} \;g_{RR}^{2} \;g_{RL} + 3 \;g_{LL} \;g_{2R}^{2} \;g_{RL}^{2} + 3 \;g_{RR}^{2} \;g_{LR}^{2} \;g_{RL}\\
		&+6 \;g_{LL}^{2} \;g_{2R}^{2} \;g_{RL} + \frac{3 \;g_{LL}}{2} \;g_{2R}^{2} \;g_{RR}^{2} + 4 \;g_{2L}^{2} \;g_{RR}^{2} \;g_{RL} + 4 \;g_{3C}^{2} \;g_{RR}^{2} \;g_{RL} + \frac{5 \;g_{RR}}{2} \;g_{LR} \;g_{RL}^{3}\\
		&+\frac{7 \;g_{LL}}{2} \;g_{RR}^{2} \;g_{LR}^{2} + \frac{7 \;g_{LL}}{3} \;g_{LR}^{2} \;g_{RL}^{2} + \frac{5 \;g_{RR}^{3}}{2} \;g_{LR} \;g_{RL} + \frac{7 \;g_{RR}}{6} \;g_{LR}^{3} \;g_{RL}\\
		&+\frac{8 \;g_{RR}}{3} \;g_{LL}^{3} \;g_{LR} + \frac{8 \;g_{LL}^{2}}{3} \;g_{LR}^{2} \;g_{RL} + \frac{8 \;g_{LL}}{3} \;g_{RR} \;g_{LR}^{3} + \frac{9 \;g_{2R}^{2}}{2} \;g_{RR}^{2} \;g_{RL}\\
		&+4 \;g_{LL} \;g_{2L}^{2} \;g_{RR} \;g_{LR} + 4 \;g_{LL} \;g_{3C}^{2} \;g_{RR} \;g_{LR} + \frac{25 \;g_{LL}}{6} \;g_{RR}^{2} \;g_{RL}^{2} + \frac{\;g_{2L}^{2} \;g_{RR}}{2} \;g_{LR} \;g_{RL}\\
		&+\frac{35 \;g_{RR}}{6} \;g_{LL}^{2} \;g_{LR} \;g_{RL} + 6 \;g_{LL} \;g_{2R}^{2} \;g_{RR} \;g_{LR} + 7 \;g_{LL} \;g_{RR} \;g_{LR} \;g_{RL}^{2} + \frac{3 \;g_{RR}}{2} \;g_{2R}^{2} \;g_{LR} \;g_{RL}
		\\
		(4\pi)^4\beta^{\rm 2loop}_{RR} = 
		& \;g_{LL}^{2} \;g_{RR}^{3} + 4 \;g_{2L}^{2} \;g_{RR}^{3} + 4 \;g_{3C}^{2} \;g_{RR}^{3} + \frac{11 \;g_{RR}^{5}}{6} + 6 \;g_{RR}^{3} \;g_{LR}^{2} + \frac{5}{6} \;g_{LR} \;g_{RL}^{4}\\
		&+\frac{9 \;g_{2R}^{2}}{2} \;g_{RR}^{3} + \frac{10 \;g_{RR}^{4}}{3} \;g_{LR} + \frac{11 \;g_{RR}^{3}}{3} \;g_{RL}^{2} + \frac{8 \;g_{RR}}{3} \;g_{LR}^{4} + \frac{7}{6} \;g_{LR}^{3} \;g_{RL}^{2}\\
		&+3 \;g_{LL} \;g_{LR} \;g_{RL}^{3} + \;g_{2L}^{2} \;g_{RR}^{2} \;g_{LR} + \frac{\;g_{LR}}{2} \;g_{2L}^{2} \;g_{RL}^{2} + \frac{14 \;g_{RR}^{2}}{3} \;g_{LR}^{3} + \frac{11 \;g_{RR}}{6} \;g_{RL}^{4}\\
		&+3 \;g_{LL}^{2} \;g_{RR} \;g_{RL}^{2} + 4 \;g_{2L}^{2} \;g_{RR} \;g_{LR}^{2} + 4 \;g_{2L}^{2} \;g_{RR} \;g_{RL}^{2} + 3 \;g_{2R}^{2} \;g_{RR}^{2} \;g_{LR} + 4 \;g_{3C}^{2} \;g_{RR} \;g_{LR}^{2}\\
		&+\frac{5 \;g_{LL}}{2} \;g_{RR} \;g_{RL}^{3} + 6 \;g_{2R}^{2} \;g_{RR} \;g_{LR}^{2} + \frac{3 \;g_{2R}^{2}}{2} \;g_{LR} \;g_{RL}^{2} + 4 \;g_{3C}^{2} \;g_{RR} \;g_{RL}^{2} + 4 \;g_{RR} \;g_{LR}^{2} \;g_{RL}^{2}\\
		&+\frac{7 \;g_{RR}}{6} \;g_{LL}^{3} \;g_{RL} + \frac{7 \;g_{LL}^{2}}{3} \;g_{RR}^{2} \;g_{LR} + \frac{7 \;g_{LL}^{2}}{2} \;g_{LR} \;g_{RL}^{2} + \frac{5 \;g_{LL}}{2} \;g_{RR}^{3} \;g_{RL}\\
		&+\frac{8 \;g_{LL}^{3}}{3} \;g_{LR} \;g_{RL} + \frac{8 \;g_{RR}}{3} \;g_{LL}^{2} \;g_{LR}^{2} + \frac{8 \;g_{LL}}{3} \;g_{LR}^{3} \;g_{RL} + \frac{9 \;g_{RR}}{2} \;g_{2R}^{2} \;g_{RL}^{2}\\
		&+\frac{\;g_{LL} \;g_{RR}}{2} \;g_{2L}^{2} \;g_{RL} + 4 \;g_{LL} \;g_{2L}^{2} \;g_{LR} \;g_{RL} + 4 \;g_{LL} \;g_{3C}^{2} \;g_{LR} \;g_{RL} + \frac{25 \;g_{RR}^{2}}{6} \;g_{LR} \;g_{RL}^{2}\\
		&+\frac{3 \;g_{LL}}{2} \;g_{2R}^{2} \;g_{RR} \;g_{RL} + 6 \;g_{LL} \;g_{2R}^{2} \;g_{LR} \;g_{RL} + 7 \;g_{LL} \;g_{RR}^{2} \;g_{LR} \;g_{RL} + \frac{35 \;g_{LL}}{6} \;g_{RR} \;g_{LR}^{2} \;g_{RL}
	\end{align*}
}
\underline{\textsl{SUSY}}
 {\scriptsize
	\begin{align*} 
		(4\pi)^4{\beta^{\rm 2loop}_{LL}}& =8 \;g_{LL}^3 \;g_{3 C}^2+8 \;g_{LR}^2 \;g_{LL} \;g_{3 C}^2+8 \;g_{RL}^2 \;g_{LL} \;g_{3 C}^2+8 \;g_{LR} \;g_{RL} \;g_{RR} \;g_{3 C}^2+7 \;g_{LL}^3 \;g_{2 L}^2+4 \;g_{RL} \;g_{LL}^2 \;g_{2 L}^2\\&+7 \;g_{RL}^2 \;g_{LL} \;g_{2 L}^2+2 \;g_{LR} \;g_{RR} \;g_{LL} \;g_{2 L}^2+2 \;g_{LR}^2 \;g_{RL} \;g_{2 L}^2+7 \;g_{LR} \;g_{RL} \;g_{RR} \;g_{2 L}^2+9 \;g_{LL}^3 \;g_{2 R}^2+6 \;g_{RL} \;g_{LL}^2 \;g_{2 R}^2\\&+9 \;g_{LR}^2 \;g_{LL} \;g_{2 R}^2+\frac{15}{2} \;g_{RL}^2 \;g_{LL} \;g_{2 R}^2+3 \;g_{LR} \;g_{RR} \;g_{LL} \;g_{2 R}^2+3 \;g_{LR}^2 \;g_{RL} \;g_{2 R}^2+\frac{15}{2} \;g_{LR} \;g_{RL} \;g_{RR} \;g_{2 R}^2+\frac{13 \;g_{LL}^5}{3}\\&+8 \;g_{RL} \;g_{LL}^4+\frac{26}{3} \;g_{LR}^2 \;g_{LL}^3+10 \;g_{RL}^2 \;g_{LL}^3+\frac{5}{3} \;g_{RR}^2 \;g_{LL}^3+6 \;g_{LR} \;g_{RR} \;g_{LL}^3+\frac{20}{3} \;g_{RL}^3 \;g_{LL}^2+\frac{10}{3} \;g_{RL} \;g_{RR}^2 \;g_{LL}^2\\&+10 \;g_{LR}^2 \;g_{RL} \;g_{LL}^2+\frac{35}{3} \;g_{LR} \;g_{RL} \;g_{RR} \;g_{LL}^2+\frac{13}{3} \;g_{LR}^4 \;g_{LL}+\frac{7}{2} \;g_{RL}^4 \;g_{LL}+\frac{5}{3} \;g_{LR} \;g_{RR}^3 \;g_{LL}+\frac{20}{3} \;g_{LR}^2 \;g_{RL}^2 \;g_{LL}\\&+5 \;g_{LR}^2 \;g_{RR}^2 \;g_{LL}+\frac{7}{2} \;g_{RL}^2 \;g_{RR}^2 \;g_{LL}+6 \;g_{LR}^3 \;g_{RR} \;g_{LL}+\frac{25}{3} \;g_{LR} \;g_{RL}^2 \;g_{RR} \;g_{LL}+\frac{5}{3} \;g_{LR}^2 \;g_{RL}^3+\frac{7}{2} \;g_{LR} \;g_{RL} \;g_{RR}^3\\&+5 \;g_{LR}^2 \;g_{RL} \;g_{RR}^2+2 \;g_{LR}^4 \;g_{RL}+\frac{7}{2} \;g_{LR} \;g_{RL}^3 \;g_{RR}+5 \;g_{LR}^3 \;g_{RL} \;g_{RR} +7 \;g_{LR}^2 \;g_{LL} \;g_{2 L}^2\\ 
		(4\pi)^4{\beta^{\rm 2loop}_{LR}}& =8 \;g_{LR}^3 \;g_{3 C}^2+8 \;g_{LL}^2 \;g_{LR} \;g_{3 C}^2+8 \;g_{RR}^2 \;g_{LR} \;g_{3 C}^2+8 \;g_{LL} \;g_{RL} \;g_{RR} \;g_{3 C}^2+7 \;g_{LR}^3 \;g_{2 L}^2+4 \;g_{RR} \;g_{LR}^2 \;g_{2 L}^2\\&+7 \;g_{RR}^2 \;g_{LR} \;g_{2 L}^2+2 \;g_{LL} \;g_{RL} \;g_{LR} \;g_{2 L}^2+2 \;g_{LL}^2 \;g_{RR} \;g_{2 L}^2+7 \;g_{LL} \;g_{RL} \;g_{RR} \;g_{2 L}^2+9 \;g_{LR}^3 \;g_{2 R}^2+6 \;g_{RR} \;g_{LR}^2 \;g_{2 R}^2\\&+9 \;g_{LL}^2 \;g_{LR} \;g_{2 R}^2+\frac{15}{2} \;g_{RR}^2 \;g_{LR} \;g_{2 R}^2+3 \;g_{LL} \;g_{RL} \;g_{LR} \;g_{2 R}^2+3 \;g_{LL}^2 \;g_{RR} \;g_{2 R}^2+\frac{15}{2} \;g_{LL} \;g_{RL} \;g_{RR} \;g_{2 R}^2+\frac{13 \;g_{LR}^5}{3}\\&+8 \;g_{RR} \;g_{LR}^4+\frac{26}{3} \;g_{LL}^2 \;g_{LR}^3+\frac{5}{3} \;g_{RL}^2 \;g_{LR}^3+10 \;g_{RR}^2 \;g_{LR}^3+6 \;g_{LL} \;g_{RL} \;g_{LR}^3+\frac{20}{3} \;g_{RR}^3 \;g_{LR}^2+10 \;g_{LL}^2 \;g_{RR} \;g_{LR}^2\\&+\frac{10}{3} \;g_{RL}^2 \;g_{RR} \;g_{LR}^2+\frac{35}{3} \;g_{LL} \;g_{RL} \;g_{RR} \;g_{LR}^2+\frac{13}{3} \;g_{LL}^4 \;g_{LR}+\frac{7}{2} \;g_{RR}^4 \;g_{LR}+\frac{5}{3} \;g_{LL} \;g_{RL}^3 \;g_{LR}+5 \;g_{LL}^2 \;g_{RL}^2 \;g_{LR}\\&+\frac{20}{3} \;g_{LL}^2 \;g_{RR}^2 \;g_{LR}+\frac{7}{2} \;g_{RL}^2 \;g_{RR}^2 \;g_{LR}+\frac{25}{3} \;g_{LL} \;g_{RL} \;g_{RR}^2 \;g_{LR}+6 \;g_{LL}^3 \;g_{RL} \;g_{LR}+\frac{5}{3} \;g_{LL}^2 \;g_{RR}^3\\&+\frac{7}{2} \;g_{LL} \;g_{RL}^3 \;g_{RR}+5 \;g_{LL}^2 \;g_{RL}^2 \;g_{RR}+5 \;g_{LL}^3 \;g_{RL} \;g_{RR}+7 \;g_{LL}^2 \;g_{LR} \;g_{2 L}^2 +\frac{7}{2} \;g_{LL} \;g_{RL} \;g_{RR}^3+2 \;g_{LL}^4 \;g_{RR}\\ 
		(4\pi)^4{\beta^{\rm 2loop}_{RL}}& =8 \;g_{RL}^3 \;g_{3 C}^2+8 \;g_{LL}^2 \;g_{RL} \;g_{3 C}^2+8 \;g_{RR}^2 \;g_{RL} \;g_{3 C}^2+8 \;g_{LL} \;g_{LR} \;g_{RR} \;g_{3 C}^2+7 \;g_{RL}^3 \;g_{2 L}^2+4 \;g_{LL} \;g_{RL}^2 \;g_{2 L}^2\\&+7 \;g_{RR}^2 \;g_{RL} \;g_{2 L}^2+2 \;g_{LR} \;g_{RR} \;g_{RL} \;g_{2 L}^2+2 \;g_{LL} \;g_{RR}^2 \;g_{2 L}^2+7 \;g_{LL} \;g_{LR} \;g_{RR} \;g_{2 L}^2+\frac{15}{2} \;g_{RL}^3 \;g_{2 R}^2+6 \;g_{LL} \;g_{RL}^2 \;g_{2 R}^2\\&+9 \;g_{LL}^2 \;g_{RL} \;g_{2 R}^2+\frac{15}{2} \;g_{RR}^2 \;g_{RL} \;g_{2 R}^2+3 \;g_{LR} \;g_{RR} \;g_{RL} \;g_{2 R}^2+3 \;g_{LL} \;g_{RR}^2 \;g_{2 R}^2+9 \;g_{LL} \;g_{LR} \;g_{RR} \;g_{2 R}^2+\frac{7 \;g_{RL}^5}{2}\\&+10 \;g_{LL}^2 \;g_{RL}^3+\frac{5}{3} \;g_{LR}^2 \;g_{RL}^3+7 \;g_{RR}^2 \;g_{RL}^3+5 \;g_{LR} \;g_{RR} \;g_{RL}^3+8 \;g_{LL}^3 \;g_{RL}^2+4 \;g_{LL} \;g_{LR}^2 \;g_{RL}^2+\frac{25}{3} \;g_{LL} \;g_{RR}^2 \;g_{RL}^2\\&+\frac{35}{3} \;g_{LL} \;g_{LR} \;g_{RR} \;g_{RL}^2+\frac{13}{3} \;g_{LL}^4 \;g_{RL}+\frac{7}{2} \;g_{RR}^4 \;g_{RL}+5 \;g_{LR} \;g_{RR}^3 \;g_{RL}+\frac{13}{3} \;g_{LL}^2 \;g_{LR}^2 \;g_{RL}+\frac{20}{3} \;g_{LL}^2 \;g_{RR}^2 \;g_{RL}\\&+5 \;g_{LR}^2 \;g_{RR}^2 \;g_{RL}+2 \;g_{LR}^3 \;g_{RR} \;g_{RL}+10 \;g_{LL}^2 \;g_{LR} \;g_{RR} \;g_{RL}+\frac{5}{3} \;g_{LL} \;g_{RR}^4+5 \;g_{LL} \;g_{LR} \;g_{RR}^3+2 \;g_{LL}^3 \;g_{RR}^2\\&+\frac{13}{3} \;g_{LL} \;g_{LR}^3 \;g_{RR}+\frac{13}{3} \;g_{LL}^3 \;g_{LR} \;g_{RR}+7 \;g_{LL}^2 \;g_{RL} \;g_{2 L}^2+\frac{20}{3} \;g_{LL} \;g_{RL}^4 +6 \;g_{LL} \;g_{LR}^2 \;g_{RR}^2\\ 
		(4\pi)^4{\beta^{\rm 2loop}_{RR}}& =8 \;g_{RR}^3 \;g_{3 C}^2+8 \;g_{LR}^2 \;g_{RR} \;g_{3 C}^2+8 \;g_{RL}^2 \;g_{RR} \;g_{3 C}^2+8 \;g_{LL} \;g_{LR} \;g_{RL} \;g_{3 C}^2+7 \;g_{RR}^3 \;g_{2 L}^2+4 \;g_{LR} \;g_{RR}^2 \;g_{2 L}^2\\&+7 \;g_{RL}^2 \;g_{RR} \;g_{2 L}^2+2 \;g_{LL} \;g_{RL} \;g_{RR} \;g_{2 L}^2+2 \;g_{LR} \;g_{RL}^2 \;g_{2 L}^2+7 \;g_{LL} \;g_{LR} \;g_{RL} \;g_{2 L}^2+\frac{15}{2} \;g_{RR}^3 \;g_{2 R}^2+6 \;g_{LR} \;g_{RR}^2 \;g_{2 R}^2\\&+9 \;g_{LR}^2 \;g_{RR} \;g_{2 R}^2+\frac{15}{2} \;g_{RL}^2 \;g_{RR} \;g_{2 R}^2+3 \;g_{LL} \;g_{RL} \;g_{RR} \;g_{2 R}^2+3 \;g_{LR} \;g_{RL}^2 \;g_{2 R}^2+9 \;g_{LL} \;g_{LR} \;g_{RL} \;g_{2 R}^2+\frac{7 \;g_{RR}^5}{2}\\&+\frac{5}{3} \;g_{LL}^2 \;g_{RR}^3+10 \;g_{LR}^2 \;g_{RR}^3+7 \;g_{RL}^2 \;g_{RR}^3+5 \;g_{LL} \;g_{RL} \;g_{RR}^3+8 \;g_{LR}^3 \;g_{RR}^2+\frac{25}{3} \;g_{LR} \;g_{RL}^2 \;g_{RR}^2+4 \;g_{LL}^2 \;g_{LR} \;g_{RR}^2\\&+\frac{35}{3} \;g_{LL} \;g_{LR} \;g_{RL} \;g_{RR}^2+\frac{13}{3} \;g_{LR}^4 \;g_{RR}+\frac{7}{2} \;g_{RL}^4 \;g_{RR}+5 \;g_{LL} \;g_{RL}^3 \;g_{RR}+\frac{13}{3} \;g_{LL}^2 \;g_{LR}^2 \;g_{RR}+5 \;g_{LL}^2 \;g_{RL}^2 \;g_{RR}\\&+2 \;g_{LL}^3 \;g_{RL} \;g_{RR}+10 \;g_{LL} \;g_{LR}^2 \;g_{RL} \;g_{RR}+\frac{5}{3} \;g_{LR} \;g_{RL}^4+5 \;g_{LL} \;g_{LR} \;g_{RL}^3+2 \;g_{LR}^3 \;g_{RL}^2+6 \;g_{LL}^2 \;g_{LR} \;g_{RL}^2\\&+\frac{13}{3} \;g_{LL} \;g_{LR}^3 \;g_{RL}+\frac{13}{3} \;g_{LL}^3 \;g_{LR} \;g_{RL} +7 \;g_{LR}^2 \;g_{RR} \;g_{2 L}^2+\frac{20}{3} \;g_{LR} \;g_{RR}^4 +\frac{20}{3} \;g_{LR}^2 \;g_{RL}^2 \;g_{RR} \\ 
		(4\pi)^4{\beta^{\rm 2loop}_{2L}}& =24 \;g_{3 C}^2 \;g_{2 L}^3+3 \;g_{2 L}^3 \;g_{2 R}^2+25 \;g_{2 L}^5+\frac{7}{3} \;g_{LL}^2 \;g_{2 L}^3+\frac{7}{3} \;g_{LR}^2 \;g_{2 L}^3+\frac{7}{3} \;g_{RL}^2 \;g_{2 L}^3\\&+\frac{7}{3} \;g_{RR}^2 \;g_{2 L}^3+\frac{4}{3} \;g_{LL} \;g_{RL} \;g_{2 L}^3+\frac{4}{3} \;g_{LR} \;g_{RR} \;g_{2 L}^3 \\ 
		(4\pi)^4{\beta^{\rm 2loop}_{2R}}& =24 \;g_{3 C}^2 \;g_{2 R}^3+3 \;g_{2 L}^2 \;g_{2 R}^3+\frac{57 \;g_{2 R}^5}{2}+3 \;g_{LL}^2 \;g_{2 R}^3+3 \;g_{LR}^2 \;g_{2 R}^3+\frac{5}{2} \;g_{RL}^2 \;g_{2 R}^3\\&+\frac{5}{2} \;g_{RR}^2 \;g_{2 R}^3+2 \;g_{LL} \;g_{RL} \;g_{2 R}^3+2 \;g_{LR} \;g_{RR} \;g_{2 R}^3 \\ 
		(4\pi)^4{\beta^{\rm 2loop}_{3C}}& =9 \;g_{3 C}^3 \;g_{2 L}^2+9 \;g_{3 C}^3 \;g_{2 R}^2+14 \;g_{3 C}^5+\;g_{LL}^2 \;g_{3 C}^3+\;g_{LR}^2 \;g_{3 C}^3+\;g_{RL}^2 \;g_{3 C}^3+\;g_{RR}^2 \;g_{3 C}^3 \\
	\end{align*}
}
{\bf D-parity conserved}\\

\underline{\textsl{Non-SUSY}}
{\scriptsize
	\begin{align*}
		(4\pi)^4\beta^{\rm 2loop}_{2L} = 
		&2 \;g_{LL}^{2} \;g_{2L}^{3} + \frac{61 \;g_{2L}^{5}}{6} + 3 \;g_{2L}^{3} \;g_{2R}^{2} + 12 \;g_{2L}^{3} \;g_{3C}^{2} + \frac{3 \;g_{2L}^{3}}{2} \;g_{RR}^{2} + 2 \;g_{2L}^{3} \;g_{LR}^{2}\\
		&+\;g_{LL} \;g_{2L}^{3} \;g_{RL} + \;g_{2L}^{3} \;g_{RR} \;g_{LR} + \frac{3 \;g_{2L}^{3}}{2} \;g_{RL}^{2}
		\\
		(4\pi)^4\beta^{\rm 2loop}_{3C} = 
		&\frac{\;g_{LL}^{2} \;g_{3C}^{3}}{2} + \frac{9 \;g_{2L}^{2}}{2} \;g_{3C}^{3} - 26 \;g_{3C}^{5} + \frac{\;g_{3C}^{3} \;g_{RR}^{2}}{2} + \frac{\;g_{3C}^{3}}{2} \;g_{LR}^{2} + \frac{\;g_{3C}^{3}}{2} \;g_{RL}^{2}\\
		&+\frac{9 \;g_{2R}^{2}}{2} \;g_{3C}^{3}
		\\
		(4\pi)^4\beta^{\rm 2loop}_{2R} = 
		&2 \;g_{LL}^{2} \;g_{2R}^{3} + 3 \;g_{2L}^{2} \;g_{2R}^{3} + \frac{61 \;g_{2R}^{5}}{6} + 12 \;g_{2R}^{3} \;g_{3C}^{2} + \frac{3 \;g_{2R}^{3}}{2} \;g_{RR}^{2} + 2 \;g_{2R}^{3} \;g_{LR}^{2}\\
		&+\;g_{LL} \;g_{2R}^{3} \;g_{RL} + \;g_{2R}^{3} \;g_{RR} \;g_{LR} + \frac{3 \;g_{2R}^{3}}{2} \;g_{RL}^{2}
		\\
		(4\pi)^4\beta^{\rm 2loop}_{LL} = 
		&\frac{32 \;g_{LL}^{5}}{9} + 6 \;g_{LL}^{3} \;g_{2L}^{2} + 6 \;g_{LL}^{3} \;g_{2R}^{2} + 4 \;g_{LL}^{3} \;g_{3C}^{2} + \frac{11 \;g_{LL}^{3}}{9} \;g_{RR}^{2} + \frac{17 \;g_{LL}}{9} \;g_{RL}^{4}\\
		&+\frac{22 \;g_{LL}^{3}}{3} \;g_{RL}^{2} + \frac{34 \;g_{LL}^{2}}{9} \;g_{RL}^{3} + \frac{32 \;g_{LL}}{9} \;g_{LR}^{4} + \frac{29}{18} \;g_{LR}^{4} \;g_{RL} + \frac{17}{18} \;g_{LR}^{2} \;g_{RL}^{3}\\
		&+\frac{58 \;g_{LL}^{4}}{9} \;g_{RL} + \frac{64 \;g_{LL}^{3}}{9} \;g_{LR}^{2} + 3 \;g_{LL}^{2} \;g_{2L}^{2} \;g_{RL} + 3 \;g_{LL}^{2} \;g_{2R}^{2} \;g_{RL} + 4 \;g_{LL} \;g_{3C}^{2} \;g_{LR}^{2}\\
		&+6 \;g_{LL} \;g_{2L}^{2} \;g_{LR}^{2} + 6 \;g_{LL} \;g_{2R}^{2} \;g_{LR}^{2} + 4 \;g_{LL} \;g_{3C}^{2} \;g_{RL}^{2} + \frac{3 \;g_{2L}^{2}}{2} \;g_{LR}^{2} \;g_{RL} + \frac{3 \;g_{2R}^{2}}{2} \;g_{LR}^{2} \;g_{RL}\\
		&+\frac{9 \;g_{LL}}{2} \;g_{2L}^{2} \;g_{RL}^{2} + \frac{9 \;g_{LL}}{2} \;g_{2R}^{2} \;g_{RL}^{2} + \frac{11 \;g_{LL}}{3} \;g_{RR}^{2} \;g_{LR}^{2} + \frac{11 \;g_{RR}}{3} \;g_{LR}^{3} \;g_{RL}\\
		&+\frac{17 \;g_{LL}}{9} \;g_{RR}^{2} \;g_{RL}^{2} + \frac{17 \;g_{RR}^{3}}{9} \;g_{LR} \;g_{RL} + \frac{17 \;g_{RR}^{2}}{6} \;g_{LR}^{2} \;g_{RL} + \frac{17 \;g_{RR}}{9} \;g_{LR} \;g_{RL}^{3}\\
		&+\frac{29 \;g_{RR}}{6} \;g_{LL}^{3} \;g_{LR} + \frac{17 \;g_{LL}^{2}}{9} \;g_{RR}^{2} \;g_{RL} + \frac{17 \;g_{LL}}{18} \;g_{RR}^{3} \;g_{LR} + \frac{29 \;g_{LL}}{6} \;g_{RR} \;g_{LR}^{3}\\
		&+\frac{145 \;g_{LL}^{2}}{18} \;g_{LR}^{2} \;g_{RL} + \frac{3 \;g_{LL}}{2} \;g_{2L}^{2} \;g_{RR} \;g_{LR} + \frac{44 \;g_{LL}}{9} \;g_{LR}^{2} \;g_{RL}^{2} + 4 \;g_{3C}^{2} \;g_{RR} \;g_{LR} \;g_{RL}\\
		&+\frac{77 \;g_{RR}}{9} \;g_{LL}^{2} \;g_{LR} \;g_{RL} + \frac{3 \;g_{LL}}{2} \;g_{2R}^{2} \;g_{RR} \;g_{LR} + \frac{9 \;g_{RR}}{2} \;g_{2L}^{2} \;g_{LR} \;g_{RL} + \frac{9 \;g_{RR}}{2} \;g_{2R}^{2} \;g_{LR} \;g_{RL}\\
		&+\frac{85 \;g_{LL}}{18} \;g_{RR} \;g_{LR} \;g_{RL}^{2}
		\\
		(4\pi)^4\beta^{\rm 2loop}_{LR} = 
		& 6 \;g_{2L}^{2} \;g_{LR}^{3} + 6 \;g_{2R}^{2} \;g_{LR}^{3} + 4 \;g_{3C}^{2} \;g_{LR}^{3} + \frac{17 \;g_{RR}^{4}}{9} \;g_{LR} + \frac{32}{9} \;g_{LR}^{5} + \frac{11}{9} \;g_{LR}^{3} \;g_{RL}^{2}\\
		&+\frac{29 \;g_{RR}}{18} \;g_{LL}^{4} + \frac{32 \;g_{LL}^{4}}{9} \;g_{LR} + \frac{17 \;g_{LL}^{2}}{18} \;g_{RR}^{3} + \frac{34 \;g_{RR}^{3}}{9} \;g_{LR}^{2} + \frac{22 \;g_{RR}^{2}}{3} \;g_{LR}^{3}\\
		&+4 \;g_{LL}^{2} \;g_{3C}^{2} \;g_{LR} + \frac{64 \;g_{LL}^{2}}{9} \;g_{LR}^{3} + 3 \;g_{2L}^{2} \;g_{RR} \;g_{LR}^{2} + 3 \;g_{2R}^{2} \;g_{RR} \;g_{LR}^{2} + \frac{58 \;g_{RR}}{9} \;g_{LR}^{4}\\
		&+\frac{3 \;g_{RR}}{2} \;g_{LL}^{2} \;g_{2L}^{2} + 6 \;g_{LL}^{2} \;g_{2L}^{2} \;g_{LR} + \frac{3 \;g_{RR}}{2} \;g_{LL}^{2} \;g_{2R}^{2} + 6 \;g_{LL}^{2} \;g_{2R}^{2} \;g_{LR} + 4 \;g_{3C}^{2} \;g_{RR}^{2} \;g_{LR}\\
		&+\frac{11 \;g_{RR}}{3} \;g_{LL}^{3} \;g_{RL} + \frac{11 \;g_{LL}^{2}}{3} \;g_{LR} \;g_{RL}^{2} + \frac{9 \;g_{2L}^{2}}{2} \;g_{RR}^{2} \;g_{LR} + \frac{9 \;g_{2R}^{2}}{2} \;g_{RR}^{2} \;g_{LR}\\
		&+\frac{17 \;g_{RR}}{6} \;g_{LL}^{2} \;g_{RL}^{2} + \frac{17 \;g_{LL}}{9} \;g_{RR}^{3} \;g_{RL} + \frac{17 \;g_{LL}}{9} \;g_{RR} \;g_{RL}^{3} + \frac{17 \;g_{RR}}{9} \;g_{LR}^{2} \;g_{RL}^{2}\\
		&+\frac{29 \;g_{LL}^{3}}{6} \;g_{LR} \;g_{RL} + \frac{29 \;g_{LL}}{6} \;g_{LR}^{3} \;g_{RL} + \frac{17 \;g_{LL}}{18} \;g_{LR} \;g_{RL}^{3} + \frac{17 \;g_{RR}^{2}}{9} \;g_{LR} \;g_{RL}^{2}\\
		&+\frac{44 \;g_{LL}^{2}}{9} \;g_{RR}^{2} \;g_{LR} + \frac{145 \;g_{RR}}{18} \;g_{LL}^{2} \;g_{LR}^{2} + \frac{3 \;g_{LL}}{2} \;g_{2L}^{2} \;g_{LR} \;g_{RL} + 4 \;g_{LL} \;g_{3C}^{2} \;g_{RR} \;g_{RL}\\
		&+\frac{9 \;g_{LL}}{2} \;g_{2L}^{2} \;g_{RR} \;g_{RL} + \frac{9 \;g_{LL}}{2} \;g_{2R}^{2} \;g_{RR} \;g_{RL} + \frac{3 \;g_{LL}}{2} \;g_{2R}^{2} \;g_{LR} \;g_{RL} + \frac{77 \;g_{LL}}{9} \;g_{RR} \;g_{LR}^{2} \;g_{RL}\\
		&+\frac{85 \;g_{LL}}{18} \;g_{RR}^{2} \;g_{LR} \;g_{RL}
		\\
		(4\pi)^4\beta^{\rm 2loop}_{RL} = 
		& \frac{9 \;g_{2L}^{2}}{2} \;g_{RL}^{3} + \frac{9 \;g_{2R}^{2}}{2} \;g_{RL}^{3} + 4 \;g_{3C}^{2} \;g_{RL}^{3} + \frac{11}{9} \;g_{LR}^{2} \;g_{RL}^{3} + \frac{17}{9} \;g_{RL}^{5}\\
		&+\frac{32 \;g_{LL}^{4}}{9} \;g_{RL} + \frac{29 \;g_{LL}^{3}}{18} \;g_{RR}^{2} + \frac{22 \;g_{LL}^{2}}{3} \;g_{RL}^{3} + \frac{17 \;g_{LL}}{18} \;g_{RR}^{4} + \frac{17 \;g_{RR}^{4}}{9} \;g_{RL}\\
		&+\frac{58 \;g_{LL}^{3}}{9} \;g_{RL}^{2} + 3 \;g_{LL} \;g_{2L}^{2} \;g_{RL}^{2} + 3 \;g_{LL} \;g_{2R}^{2} \;g_{RL}^{2} + \frac{34 \;g_{LL}}{9} \;g_{RL}^{4} + \frac{34 \;g_{RR}^{2}}{9} \;g_{RL}^{3}\\
		&+6 \;g_{LL}^{2} \;g_{2L}^{2} \;g_{RL} + 6 \;g_{LL}^{2} \;g_{2R}^{2} \;g_{RL} + 4 \;g_{LL}^{2} \;g_{3C}^{2} \;g_{RL} + \frac{3 \;g_{LL}}{2} \;g_{2L}^{2} \;g_{RR}^{2} + 4 \;g_{3C}^{2} \;g_{RR}^{2} \;g_{RL}\\
		&+\frac{3 \;g_{LL}}{2} \;g_{2R}^{2} \;g_{RR}^{2} + \frac{11 \;g_{LL}}{3} \;g_{RR}^{3} \;g_{LR} + \frac{9 \;g_{2L}^{2}}{2} \;g_{RR}^{2} \;g_{RL} + \frac{9 \;g_{2R}^{2}}{2} \;g_{RR}^{2} \;g_{RL}\\
		&+\frac{29 \;g_{LL}}{6} \;g_{RR}^{2} \;g_{LR}^{2} + \frac{17 \;g_{RR}^{3}}{6} \;g_{LR} \;g_{RL} + \frac{11 \;g_{RR}^{2}}{3} \;g_{LR}^{2} \;g_{RL} + \frac{17 \;g_{RR}}{6} \;g_{LR} \;g_{RL}^{3}\\
		&+\frac{32 \;g_{RR}}{9} \;g_{LL}^{3} \;g_{LR} + \frac{32 \;g_{LL}}{9} \;g_{RR} \;g_{LR}^{3} + \frac{29 \;g_{LL}}{9} \;g_{LR}^{2} \;g_{RL}^{2} + \frac{29 \;g_{RR}}{18} \;g_{LR}^{3} \;g_{RL}\\
		&+\frac{44 \;g_{LL}^{2}}{9} \;g_{RR}^{2} \;g_{RL} + \frac{32 \;g_{LL}^{2}}{9} \;g_{LR}^{2} \;g_{RL} + 4 \;g_{LL} \;g_{3C}^{2} \;g_{RR} \;g_{LR} + \frac{85 \;g_{LL}}{18} \;g_{RR}^{2} \;g_{RL}^{2}\\
		&+6 \;g_{LL} \;g_{2L}^{2} \;g_{RR} \;g_{LR} + 6 \;g_{LL} \;g_{2R}^{2} \;g_{RR} \;g_{LR} + \frac{3 \;g_{RR}}{2} \;g_{2L}^{2} \;g_{LR} \;g_{RL} + \frac{3 \;g_{RR}}{2} \;g_{2R}^{2} \;g_{LR} \;g_{RL}\\
		&+\frac{145 \;g_{RR}}{18} \;g_{LL}^{2} \;g_{LR} \;g_{RL} + \frac{77 \;g_{LL}}{9} \;g_{RR} \;g_{LR} \;g_{RL}^{2}
		\\
		(4\pi)^4\beta^{\rm 2loop}_{RR} = 
		& \frac{11 \;g_{LL}^{2}}{9} \;g_{RR}^{3} + \frac{9 \;g_{2L}^{2}}{2} \;g_{RR}^{3} + \frac{9 \;g_{2R}^{2}}{2} \;g_{RR}^{3} + 4 \;g_{3C}^{2} \;g_{RR}^{3} + \frac{17 \;g_{RR}^{5}}{9} + \frac{17 \;g_{RR}}{9} \;g_{RL}^{4}\\
		&+\frac{34 \;g_{RR}^{4}}{9} \;g_{LR} + \frac{22 \;g_{RR}^{3}}{3} \;g_{LR}^{2} + \frac{32 \;g_{RR}}{9} \;g_{LR}^{4} + \frac{29}{18} \;g_{LR}^{3} \;g_{RL}^{2} + \frac{17}{18} \;g_{LR} \;g_{RL}^{4}\\
		&+3 \;g_{2L}^{2} \;g_{RR}^{2} \;g_{LR} + 3 \;g_{2R}^{2} \;g_{RR}^{2} \;g_{LR} + 4 \;g_{3C}^{2} \;g_{RR} \;g_{LR}^{2} + \frac{34 \;g_{RR}^{3}}{9} \;g_{RL}^{2} + \frac{58 \;g_{RR}^{2}}{9} \;g_{LR}^{3}\\
		&+6 \;g_{2L}^{2} \;g_{RR} \;g_{LR}^{2} + \frac{3 \;g_{2L}^{2}}{2} \;g_{LR} \;g_{RL}^{2} + 6 \;g_{2R}^{2} \;g_{RR} \;g_{LR}^{2} + \frac{3 \;g_{2R}^{2}}{2} \;g_{LR} \;g_{RL}^{2} + 4 \;g_{3C}^{2} \;g_{RR} \;g_{RL}^{2}\\
		&+\frac{11 \;g_{RR}}{3} \;g_{LL}^{2} \;g_{RL}^{2} + \frac{11 \;g_{LL}}{3} \;g_{LR} \;g_{RL}^{3} + \frac{9 \;g_{RR}}{2} \;g_{2L}^{2} \;g_{RL}^{2} + \frac{9 \;g_{RR}}{2} \;g_{2R}^{2} \;g_{RL}^{2}\\
		&+\frac{29 \;g_{LL}^{2}}{9} \;g_{RR}^{2} \;g_{LR} + \frac{29 \;g_{LL}^{2}}{6} \;g_{LR} \;g_{RL}^{2} + \frac{17 \;g_{LL}}{6} \;g_{RR}^{3} \;g_{RL} + \frac{17 \;g_{LL}}{6} \;g_{RR} \;g_{RL}^{3}\\
		&+\frac{29 \;g_{RR}}{18} \;g_{LL}^{3} \;g_{RL} + \frac{32 \;g_{LL}^{3}}{9} \;g_{LR} \;g_{RL} + \frac{32 \;g_{RR}}{9} \;g_{LL}^{2} \;g_{LR}^{2} + \frac{32 \;g_{LL}}{9} \;g_{LR}^{3} \;g_{RL}\\
		&+6 \;g_{LL} \;g_{2L}^{2} \;g_{LR} \;g_{RL} + 4 \;g_{LL} \;g_{3C}^{2} \;g_{LR} \;g_{RL} + \frac{85 \;g_{RR}^{2}}{18} \;g_{LR} \;g_{RL}^{2} + \frac{44 \;g_{RR}}{9} \;g_{LR}^{2} \;g_{RL}^{2}\\
		&+\frac{3 \;g_{LL}}{2} \;g_{2L}^{2} \;g_{RR} \;g_{RL} + \frac{3 \;g_{LL}}{2} \;g_{2R}^{2} \;g_{RR} \;g_{RL} + 6 \;g_{LL} \;g_{2R}^{2} \;g_{LR} \;g_{RL} + \frac{77 \;g_{LL}}{9} \;g_{RR}^{2} \;g_{LR} \;g_{RL}\\
		&+\frac{145 \;g_{LL}}{18} \;g_{RR} \;g_{LR}^{2} \;g_{RL}
	\end{align*}
}
\underline{\textsl{SUSY}}

{\scriptsize
	\begin{align*} 
		(4\pi)^4{\beta^{\rm 2loop}_{LL}}& =8 \;g_{LL}^3 \;g_{3 C}^2+8 \;g_{LR}^2 \;g_{LL} \;g_{3 C}^2+8 \;g_{RL}^2 \;g_{LL} \;g_{3 C}^2+8 \;g_{LR} \;g_{RL} \;g_{RR} \;g_{3 C}^2+9 \;g_{LL}^3 \;g_{2 L}^2+6 \;g_{RL} \;g_{LL}^2 \;g_{2 L}^2\\&+9 \;g_{LR}^2 \;g_{LL} \;g_{2 L}^2+\frac{15}{2} \;g_{RL}^2 \;g_{LL} \;g_{2 L}^2+3 \;g_{LR} \;g_{RR} \;g_{LL} \;g_{2 L}^2+3 \;g_{LR}^2 \;g_{RL} \;g_{2 L}^2+\frac{15}{2} \;g_{LR} \;g_{RL} \;g_{RR} \;g_{2 L}^2\\&+9 \;g_{LL}^3 \;g_{2 R}^2+6 \;g_{RL} \;g_{LL}^2 \;g_{2 R}^2+9 \;g_{LR}^2 \;g_{LL} \;g_{2 R}^2+\frac{15}{2} \;g_{RL}^2 \;g_{LL} \;g_{2 R}^2+3 \;g_{LR} \;g_{RR} \;g_{LL} \;g_{2 R}^2\\&+3 \;g_{LR}^2 \;g_{RL} \;g_{2 R}^2+\frac{15}{2} \;g_{LR} \;g_{RL} \;g_{RR} \;g_{2 R}^2+\frac{47 \;g_{LL}^5}{9}+\frac{88}{9} \;g_{RL} \;g_{LL}^4+\frac{94}{9} \;g_{LR}^2 \;g_{LL}^3+\frac{34}{3} \;g_{RL}^2 \;g_{LL}^3\\&+\frac{17}{9} \;g_{RR}^2 \;g_{LL}^3+\frac{22}{3} \;g_{LR} \;g_{RR} \;g_{LL}^3+\frac{64}{9} \;g_{RL}^3 \;g_{LL}^2+\frac{32}{9} \;g_{RL} \;g_{RR}^2 \;g_{LL}^2+\frac{110}{9} \;g_{LR}^2 \;g_{RL} \;g_{LL}^2\\&+\frac{119}{9} \;g_{LR} \;g_{RL} \;g_{RR} \;g_{LL}^2+\frac{47}{9} \;g_{LR}^4 \;g_{LL}+\frac{32}{9} \;g_{RL}^4 \;g_{LL}+\frac{16}{9} \;g_{LR} \;g_{RR}^3 \;g_{LL}+\frac{68}{9} \;g_{LR}^2 \;g_{RL}^2 \;g_{LL}\\&+\frac{17}{3} \;g_{LR}^2 \;g_{RR}^2 \;g_{LL}+\frac{32}{9} \;g_{RL}^2 \;g_{RR}^2 \;g_{LL}+\frac{22}{3} \;g_{LR}^3 \;g_{RR} \;g_{LL}+\frac{80}{9} \;g_{LR} \;g_{RL}^2 \;g_{RR} \;g_{LL}+\frac{16}{9} \;g_{LR}^2 \;g_{RL}^3\\&+\frac{32}{9} \;g_{LR} \;g_{RL} \;g_{RR}^3+\frac{16}{3} \;g_{LR}^2 \;g_{RL} \;g_{RR}^2+\frac{22}{9} \;g_{LR}^4 \;g_{RL}+\frac{32}{9} \;g_{LR} \;g_{RL}^3 \;g_{RR}+\frac{17}{3} \;g_{LR}^3 \;g_{RL} \;g_{RR} \\ 
		(4\pi)^4{\beta^{\rm 2loop}_{LR}}& =8 \;g_{LR}^3 \;g_{3 C}^2+8 \;g_{LL}^2 \;g_{LR} \;g_{3 C}^2+8 \;g_{RR}^2 \;g_{LR} \;g_{3 C}^2+8 \;g_{LL} \;g_{RL} \;g_{RR} \;g_{3 C}^2+9 \;g_{LR}^3 \;g_{2 L}^2+6 \;g_{RR} \;g_{LR}^2 \;g_{2 L}^2\\&+9 \;g_{LL}^2 \;g_{LR} \;g_{2 L}^2+\frac{15}{2} \;g_{RR}^2 \;g_{LR} \;g_{2 L}^2+3 \;g_{LL} \;g_{RL} \;g_{LR} \;g_{2 L}^2+3 \;g_{LL}^2 \;g_{RR} \;g_{2 L}^2+\frac{15}{2} \;g_{LL} \;g_{RL} \;g_{RR} \;g_{2 L}^2\\&+9 \;g_{LR}^3 \;g_{2 R}^2+6 \;g_{RR} \;g_{LR}^2 \;g_{2 R}^2+9 \;g_{LL}^2 \;g_{LR} \;g_{2 R}^2+\frac{15}{2} \;g_{RR}^2 \;g_{LR} \;g_{2 R}^2+3 \;g_{LL} \;g_{RL} \;g_{LR} \;g_{2 R}^2\\&+3 \;g_{LL}^2 \;g_{RR} \;g_{2 R}^2+\frac{15}{2} \;g_{LL} \;g_{RL} \;g_{RR} \;g_{2 R}^2+\frac{47 \;g_{LR}^5}{9}+\frac{88}{9} \;g_{RR} \;g_{LR}^4+\frac{94}{9} \;g_{LL}^2 \;g_{LR}^3+\frac{17}{9} \;g_{RL}^2 \;g_{LR}^3\\&+\frac{34}{3} \;g_{RR}^2 \;g_{LR}^3+\frac{22}{3} \;g_{LL} \;g_{RL} \;g_{LR}^3+\frac{64}{9} \;g_{RR}^3 \;g_{LR}^2+\frac{110}{9} \;g_{LL}^2 \;g_{RR} \;g_{LR}^2\\&+\frac{32}{9} \;g_{RL}^2 \;g_{RR} \;g_{LR}^2+\frac{119}{9} \;g_{LL} \;g_{RL} \;g_{RR} \;g_{LR}^2+\frac{47}{9} \;g_{LL}^4 \;g_{LR}+\frac{32}{9} \;g_{RR}^4 \;g_{LR}+\frac{16}{9} \;g_{LL} \;g_{RL}^3 \;g_{LR}\\&+\frac{17}{3} \;g_{LL}^2 \;g_{RL}^2 \;g_{LR}+\frac{68}{9} \;g_{LL}^2 \;g_{RR}^2 \;g_{LR}+\frac{32}{9} \;g_{RL}^2 \;g_{RR}^2 \;g_{LR}+\frac{80}{9} \;g_{LL} \;g_{RL} \;g_{RR}^2 \;g_{LR}+\frac{22}{3} \;g_{LL}^3 \;g_{RL} \;g_{LR}\\&+\frac{32}{9} \;g_{LL} \;g_{RL} \;g_{RR}^3+\frac{22}{9} \;g_{LL}^4 \;g_{RR}+\frac{32}{9} \;g_{LL} \;g_{RL}^3 \;g_{RR}+\frac{16}{3} \;g_{LL}^2 \;g_{RL}^2 \;g_{RR}+\frac{17}{3} \;g_{LL}^3 \;g_{RL} \;g_{RR}\\&+\frac{16}{9} \;g_{LL}^2 \;g_{RR}^3  \\
		(4\pi)^4{\beta^{\rm 2loop}_{RL}}& =8 \;g_{RL}^3 \;g_{3 C}^2+8 \;g_{LL}^2 \;g_{RL} \;g_{3 C}^2+8 \;g_{RR}^2 \;g_{RL} \;g_{3 C}^2+8 \;g_{LL} \;g_{LR} \;g_{RR} \;g_{3 C}^2+\frac{15}{2} \;g_{RL}^3 \;g_{2 L}^2+6 \;g_{LL} \;g_{RL}^2 \;g_{2 L}^2\\&+9 \;g_{LL}^2 \;g_{RL} \;g_{2 L}^2+\frac{15}{2} \;g_{RR}^2 \;g_{RL} \;g_{2 L}^2+3 \;g_{LR} \;g_{RR} \;g_{RL} \;g_{2 L}^2+3 \;g_{LL} \;g_{RR}^2 \;g_{2 L}^2+9 \;g_{LL} \;g_{LR} \;g_{RR} \;g_{2 L}^2\\&+\frac{15}{2} \;g_{RL}^3 \;g_{2 R}^2+6 \;g_{LL} \;g_{RL}^2 \;g_{2 R}^2+9 \;g_{LL}^2 \;g_{RL} \;g_{2 R}^2+\frac{15}{2} \;g_{RR}^2 \;g_{RL} \;g_{2 R}^2+3 \;g_{LR} \;g_{RR} \;g_{RL} \;g_{2 R}^2+3 \;g_{LL} \;g_{RR}^2 \;g_{2 R}^2\\&+9 \;g_{LL} \;g_{LR} \;g_{RR} \;g_{2 R}^2+\frac{32 \;g_{RL}^5}{9}+\frac{64}{9} \;g_{LL} \;g_{RL}^4+\frac{34}{3} \;g_{LL}^2 \;g_{RL}^3+\frac{17}{9} \;g_{LR}^2 \;g_{RL}^3+\frac{64}{9} \;g_{RR}^2 \;g_{RL}^3\\&+\frac{16}{3} \;g_{LR} \;g_{RR} \;g_{RL}^3+\frac{88}{9} \;g_{LL}^3 \;g_{RL}^2+\frac{44}{9} \;g_{LL} \;g_{LR}^2 \;g_{RL}^2+\frac{80}{9} \;g_{LL} \;g_{RR}^2 \;g_{RL}^2+\frac{119}{9} \;g_{LL} \;g_{LR} \;g_{RR} \;g_{RL}^2\\&+\frac{47}{9} \;g_{LL}^4 \;g_{RL}+\frac{32}{9} \;g_{RR}^4 \;g_{RL}+\frac{16}{3} \;g_{LR} \;g_{RR}^3 \;g_{RL}+\frac{47}{9} \;g_{LL}^2 \;g_{LR}^2 \;g_{RL}+\frac{68}{9} \;g_{LL}^2 \;g_{RR}^2 \;g_{RL}\\&+\frac{17}{3} \;g_{LR}^2 \;g_{RR}^2 \;g_{RL}+\frac{22}{9} \;g_{LR}^3 \;g_{RR} \;g_{RL}+\frac{110}{9} \;g_{LL}^2 \;g_{LR} \;g_{RR} \;g_{RL}+\frac{16}{9} \;g_{LL} \;g_{RR}^4+\frac{17}{3} \;g_{LL} \;g_{LR} \;g_{RR}^3\\&+\frac{22}{9} \;g_{LL}^3 \;g_{RR}^2+\frac{22}{3} \;g_{LL} \;g_{LR}^2 \;g_{RR}^2+\frac{47}{9} \;g_{LL} \;g_{LR}^3 \;g_{RR}+\frac{47}{9} \;g_{LL}^3 \;g_{LR} \;g_{RR} \\ 
		(4\pi)^4{\beta^{\rm 2loop}_{RR}}& =8 \;g_{RR}^3 \;g_{3 C}^2+8 \;g_{LR}^2 \;g_{RR} \;g_{3 C}^2+8 \;g_{RL}^2 \;g_{RR} \;g_{3 C}^2+8 \;g_{LL} \;g_{LR} \;g_{RL} \;g_{3 C}^2+\frac{15}{2} \;g_{RR}^3 \;g_{2 L}^2+6 \;g_{LR} \;g_{RR}^2 \;g_{2 L}^2\\&+9 \;g_{LR}^2 \;g_{RR} \;g_{2 L}^2+\frac{15}{2} \;g_{RL}^2 \;g_{RR} \;g_{2 L}^2+3 \;g_{LL} \;g_{RL} \;g_{RR} \;g_{2 L}^2+3 \;g_{LR} \;g_{RL}^2 \;g_{2 L}^2+9 \;g_{LL} \;g_{LR} \;g_{RL} \;g_{2 L}^2\\&+\frac{15}{2} \;g_{RR}^3 \;g_{2 R}^2+6 \;g_{LR} \;g_{RR}^2 \;g_{2 R}^2+9 \;g_{LR}^2 \;g_{RR} \;g_{2 R}^2+\frac{15}{2} \;g_{RL}^2 \;g_{RR} \;g_{2 R}^2+3 \;g_{LL} \;g_{RL} \;g_{RR} \;g_{2 R}^2\\&+3 \;g_{LR} \;g_{RL}^2 \;g_{2 R}^2+9 \;g_{LL} \;g_{LR} \;g_{RL} \;g_{2 R}^2+\frac{32 \;g_{RR}^5}{9}+\frac{64}{9} \;g_{LR} \;g_{RR}^4+\frac{17}{9} \;g_{LL}^2 \;g_{RR}^3+\frac{34}{3} \;g_{LR}^2 \;g_{RR}^3\\&+\frac{64}{9} \;g_{RL}^2 \;g_{RR}^3+\frac{16}{3} \;g_{LL} \;g_{RL} \;g_{RR}^3+\frac{88}{9} \;g_{LR}^3 \;g_{RR}^2+\frac{80}{9} \;g_{LR} \;g_{RL}^2 \;g_{RR}^2+\frac{44}{9} \;g_{LL}^2 \;g_{LR} \;g_{RR}^2\\&+\frac{47}{9} \;g_{LR}^4 \;g_{RR}+\frac{32}{9} \;g_{RL}^4 \;g_{RR}+\frac{16}{3} \;g_{LL} \;g_{RL}^3 \;g_{RR}+\frac{47}{9} \;g_{LL}^2 \;g_{LR}^2 \;g_{RR}+\frac{17}{3} \;g_{LL}^2 \;g_{RL}^2 \;g_{RR}\\&+\frac{68}{9} \;g_{LR}^2 \;g_{RL}^2 \;g_{RR}+\frac{22}{9} \;g_{LL}^3 \;g_{RL} \;g_{RR}+\frac{110}{9} \;g_{LL} \;g_{LR}^2 \;g_{RL} \;g_{RR}+\frac{16}{9} \;g_{LR} \;g_{RL}^4+\frac{17}{3} \;g_{LL} \;g_{LR} \;g_{RL}^3\\&+\frac{22}{9} \;g_{LR}^3 \;g_{RL}^2+\frac{22}{3} \;g_{LL}^2 \;g_{LR} \;g_{RL}^2+\frac{47}{9} \;g_{LL} \;g_{LR}^3 \;g_{RL}+\frac{47}{9} \;g_{LL}^3 \;g_{LR} \;g_{RL}+\frac{119}{9} \;g_{LL} \;g_{LR} \;g_{RL} \;g_{RR}^2 \\ 
		(4\pi)^4{\beta^{\rm 2loop}_{2L}}& =24 \;g_{3 C}^2 \;g_{2 L}^3+3 \;g_{2 L}^3 \;g_{2 R}^2+\frac{57 \;g_{2 L}^5}{2}+3 \;g_{LL}^2 \;g_{2 L}^3+3 \;g_{LR}^2 \;g_{2 L}^3+\frac{5}{2} \;g_{RL}^2 \;g_{2 L}^3+\frac{5}{2} \;g_{RR}^2 \;g_{2 L}^3\\&+2 \;g_{LL} \;g_{RL} \;g_{2 L}^3+2 \;g_{LR} \;g_{RR} \;g_{2 L}^3 \\ 
		(4\pi)^4{\beta^{\rm 2loop}_{2R}}& =24 \;g_{3 C}^2 \;g_{2 R}^3+3 \;g_{2 L}^2 \;g_{2 R}^3+\frac{57 \;g_{2 R}^5}{2}+3 \;g_{LL}^2 \;g_{2 R}^3+3 \;g_{LR}^2 \;g_{2 R}^3+\frac{5}{2} \;g_{RL}^2 \;g_{2 R}^3+\frac{5}{2} \;g_{RR}^2 \;g_{2 R}^3\\&+2 \;g_{LL} \;g_{RL} \;g_{2 R}^3+2 \;g_{LR} \;g_{RR} \;g_{2 R}^3 \\ 
		(4\pi)^4{\beta^{\rm 2loop}_{3C}}& =9 \;g_{3 C}^3 \;g_{2 L}^2+9 \;g_{3 C}^3 \;g_{2 R}^2+14 \;g_{3 C}^5+\;g_{LL}^2 \;g_{3 C}^3+\;g_{LR}^2 \;g_{3 C}^3+\;g_{RL}^2 \;g_{3 C}^3+\;g_{RR}^2 \;g_{3 C}^3 \\
\end{align*}}
\newpage
\subsubsection*{I. Normalizations of abelian charges for embeddings}

\begin{table}[h!]
	\scriptsize
	\centering
	\renewcommand*{\arraystretch}{1.5}
  \begin{tabular}{| c | c | c |}
  \hline
  Symmetry breaking & Branching rule & U(1) normalization \\
    \hline
    $SU(3)\rightarrow SU(2) \otimes U(1)$ & $3= (2,-1)\oplus (1,2)$ & $\frac{1}{2\sqrt{3}}$ \\ \hline
    $SU(4)\rightarrow SU(3) \otimes U(1)$ & $4= (3,-1/3)\oplus (1,1)$   & $\frac{1}{2}\sqrt{\frac{3}{2}}$ \\ \hline
    $SU(5)\rightarrow SU(2) \otimes SU(3)\times U(1)$ & $5= (2,1,3)\oplus (1,3,-2)$   & $\frac{1}{2\sqrt{15}}$ \\ \hline
    $SU(6)\rightarrow SU(5) \otimes U(1)$& $6= (1,-5)\oplus (5,1)$ & $\frac{1}{2\sqrt{15}}$ \\ \hline
    $SU(6)\rightarrow SU(2) \otimes SU(4)\times U(1)$ & $6= (2,1,2)\oplus (1,4,-1)$   & $\frac{1}{2\sqrt{6}}$ \\ \hline
    $SU(6)\rightarrow SU(3) \otimes SU(3)\times U(1)$ & $6= (3,1,1)\oplus (1,3,-1)$   & $\frac{1}{2\sqrt{3}}$ \\ \hline
     $E(6)\rightarrow SO(10) \otimes U(1)$ & $27= (1,4)\oplus (10, -2) \oplus (16, 1)$   & $\frac{1}{2\sqrt{6}}$ \\ \hline
$SO(10)\rightarrow SU(5) \otimes U(1)$ & $10= (5,2)\oplus (\bar{5},2)$   & $\frac{1}{2\sqrt{10}}$ \\ \hline  
  \end{tabular}
\end{table}

\subsubsection*{II.$SO(10),E(6)$ Representations, Dynkin labels and normalizations}
\begin{table}[h!]
	\scriptsize
\centering
\renewcommand*{\arraystretch}{1.2}
\begin{tabular}{ |c|c|c|c| } 
\hline
Group & Representation & Dynkin labels & N(normalization) \\
\hline
\multirow{3}{4em}{SO(10)}& 1 & (00000) & 0 \\ 
& 10 & (10000) & 1\\ 
& $\overline{16}$ & (00010) & 1\\
& 16 & (00001) & 2 \\ 
& 45 & (01000) & 8\\
& 54 & (20000) & 12\\
& 120 & (00100) & 28\\  
& 126 & (00020) & 35\\
& 144 & (10010) & 34\\
\hline
\multirow{3}{4em}{E(6)}& 1 & (000000) & 0 \\ 
& 27 & (100000) & 3\\ 
& $\overline{27}$ & (000010) & 3\\
& 78 & (000001) & 12 \\ 
& 351 & (000100) & 75\\
& $351^\prime$ & (000020) & 84\\
& 650 & (100010) & 150\\ 
& 1728 & (100001) & 480\\ 
& 2430 & (000002) & 810\\
\hline
\end{tabular}
\end{table}


\subsubsection*{III. $SU(N)$ with $N\in [2:6]$ Representations, Dynkin labels and normalizations}

\begin{table}[h!]
	\scriptsize
\centering
\begin{tabular}{ |c|c|c|c| } 
\hline
Group & Representation & Dynkin labels & N(normalization) \\
\hline
\multirow{4}{4em}{SU(2)} & 1 & (0) & 0 \\
 & 2 & (1) & 1/2 \\ 
& 3 & (2) & 2\\ 
& 4 & (3) & 5\\ 
\hline
\multirow{3}{4em}{SU(3)} & 1 & (00) & 0\\ 
& 3 & (10) & 1/2\\ 
& $\bar{3}$ & (01) & 1/2\\
& 6 & (20) & 5/2\\ 
& 8 & (11) & 3\\  
\hline
\multirow{3}{4em}{SU(4)} & 1 & (000) & 0\\ 
& 4 & (100) & 1/2\\ 
& $\bar{4}$ & (001) & 1/2\\
& 6 & (010) & 1\\ 
& 10 & (200) & 3\\
& 15 & (101) & 4\\
& 20 & (011) & 13/2\\ 
& $20^\prime$ & (020) & 8\\ 
\hline
\multirow{3}{4em}{SU(5)}& 1 & (0000) & 0 \\ 
& 5 & (1000) & 1/2\\ 
& $\bar{5}$ & (0001) & 1/2\\
& 10 & (0100) & 3/2\\ 
& 15 & (2000) & 7/2\\
& 24 & (1001) & 5\\
& 35 & (0003) & 14\\  
\hline
\multirow{3}{4em}{SU(6)}& 1 & (00000) & 0 \\ 
& 6 & (10000) & 1/2\\ 
& $\bar{6}$ & (00001) & 1/2\\
& 15 & (01000) & 2 \\ 
& 20 & (00100) & 3\\
& 35 & (10001) & 6\\
& 84 & (01001) & 19\\  
& 105 & (00101) & 26\\
& $105^\prime$ & (00020) & 32\\
& 126 & (00004) & 60\\
\hline
\end{tabular}
\end{table}

\newpage

\subsection*{\underline{Precision correlations among $M_X,M_R,g_U$:}}
\subsubsection*{Non-Supersymmetric scenario}
\begin{figure}[!htbp]
        \centering
	\includegraphics[scale=0.4]{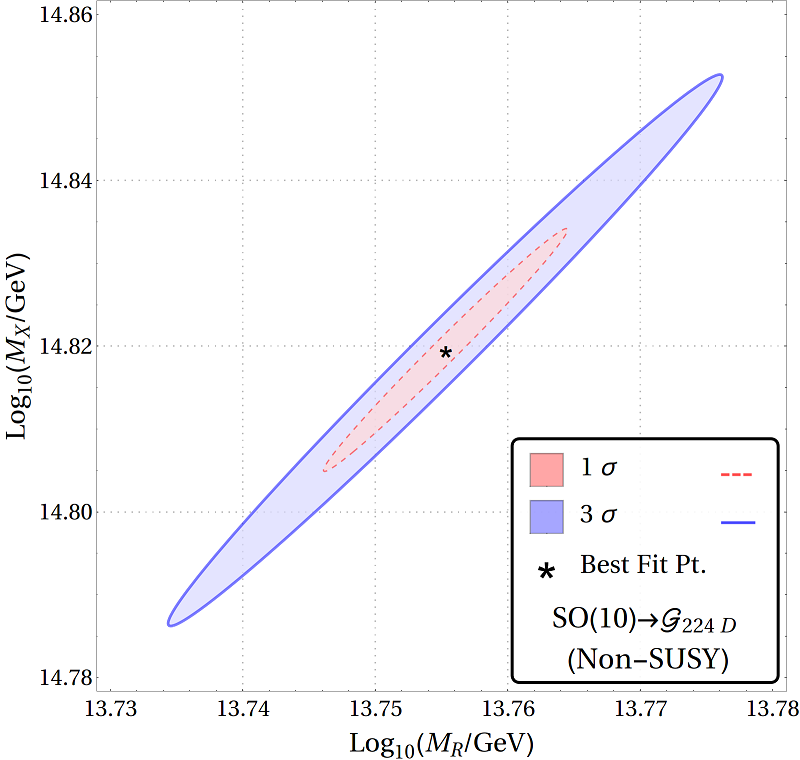}
	\includegraphics[scale=0.4]{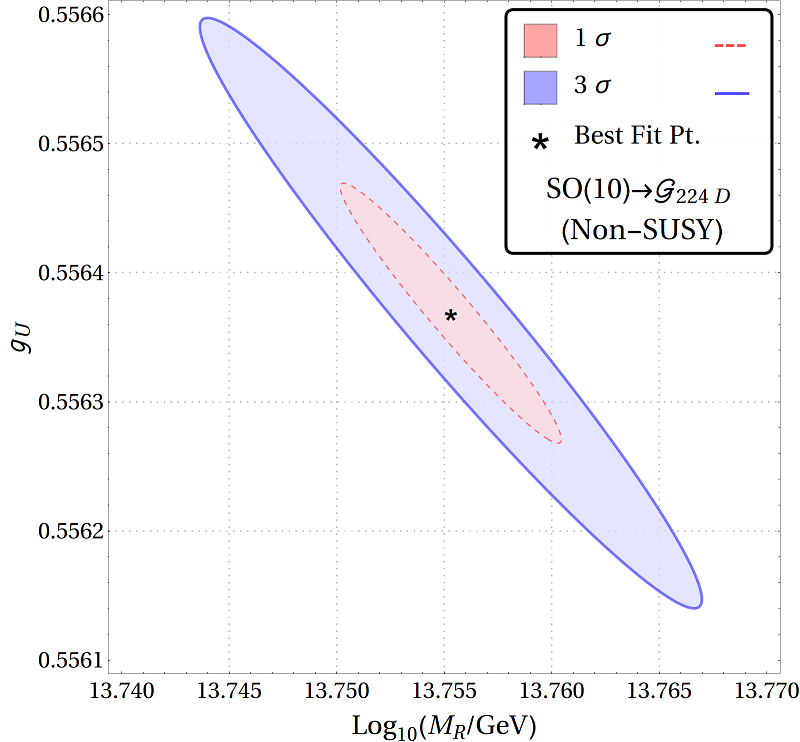}
	\includegraphics[scale=0.4]{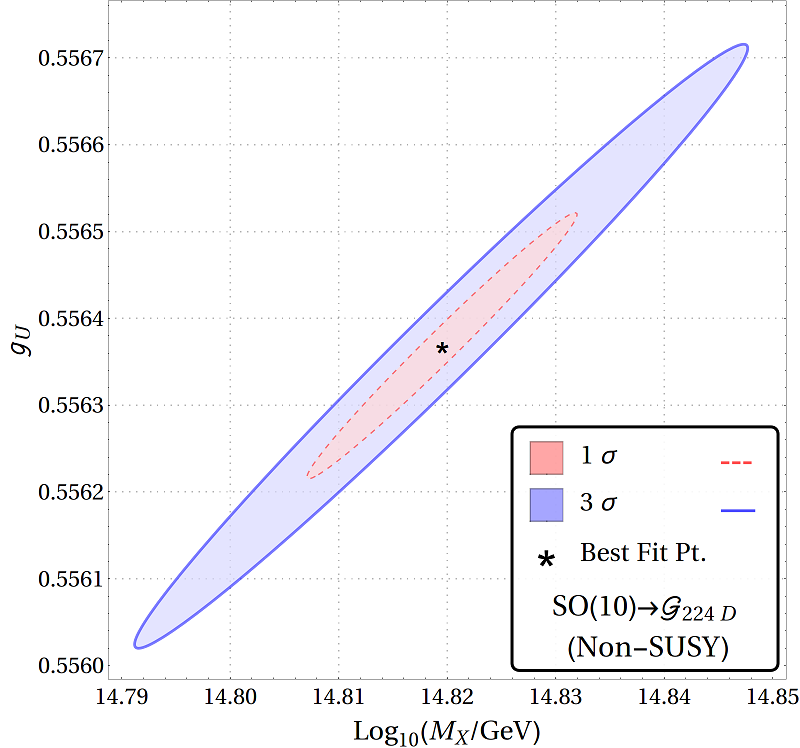}\\
	\includegraphics[scale=0.4]{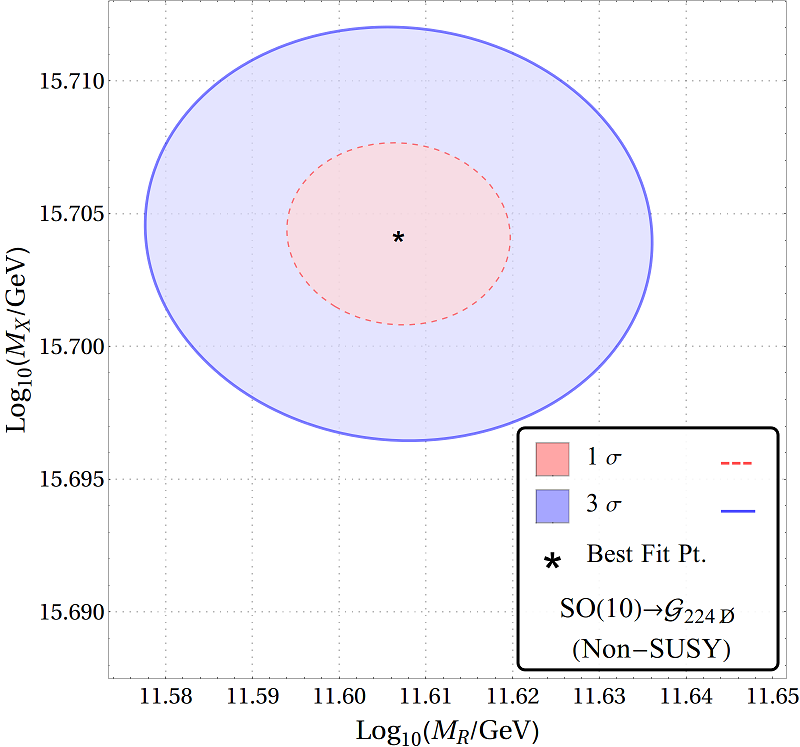}
	\includegraphics[scale=0.4]{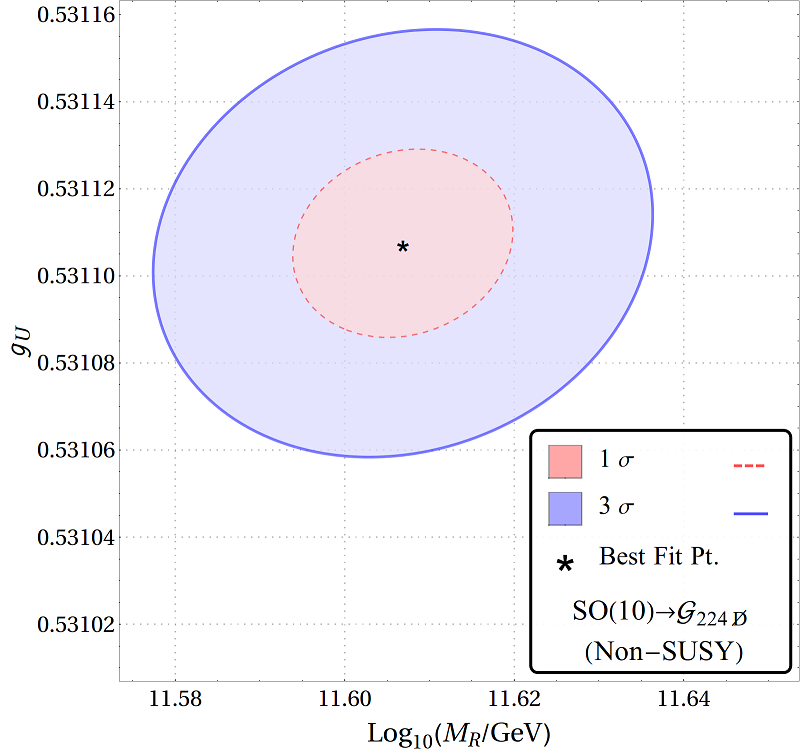}
	\includegraphics[scale=0.4]{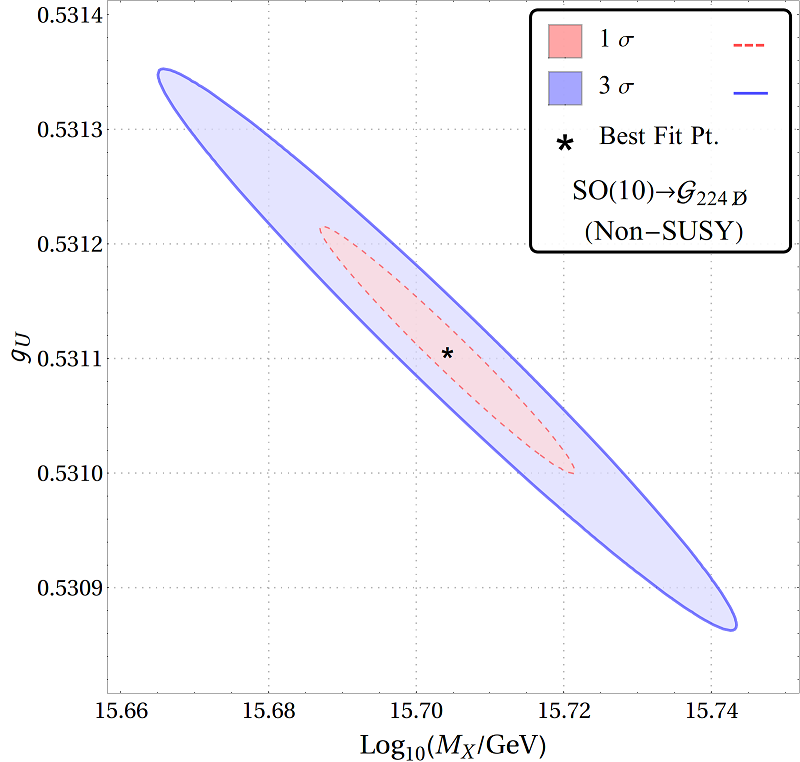}\\
	\includegraphics[scale=0.4]{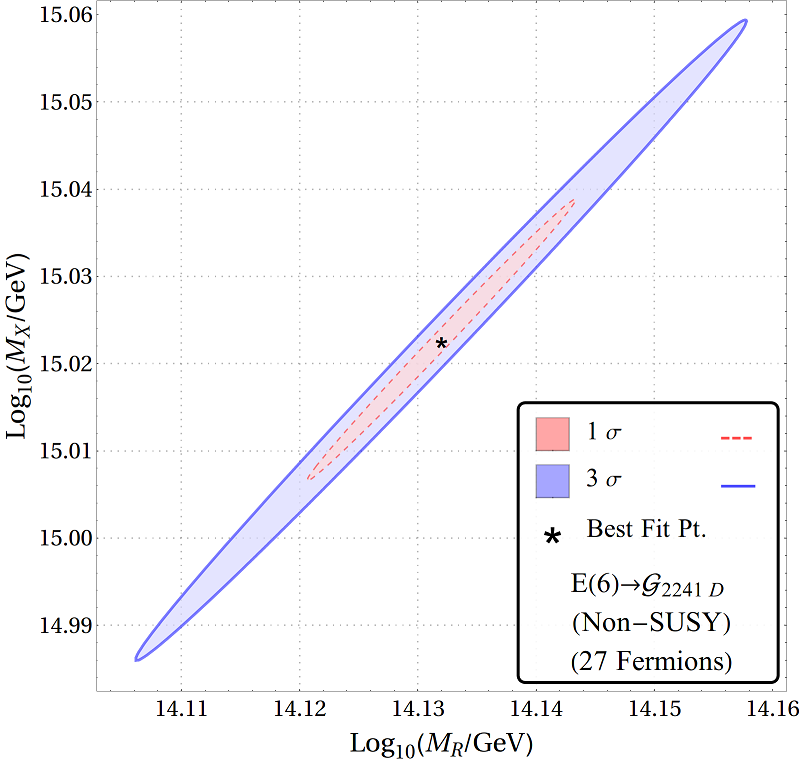}
	\includegraphics[scale=0.4]{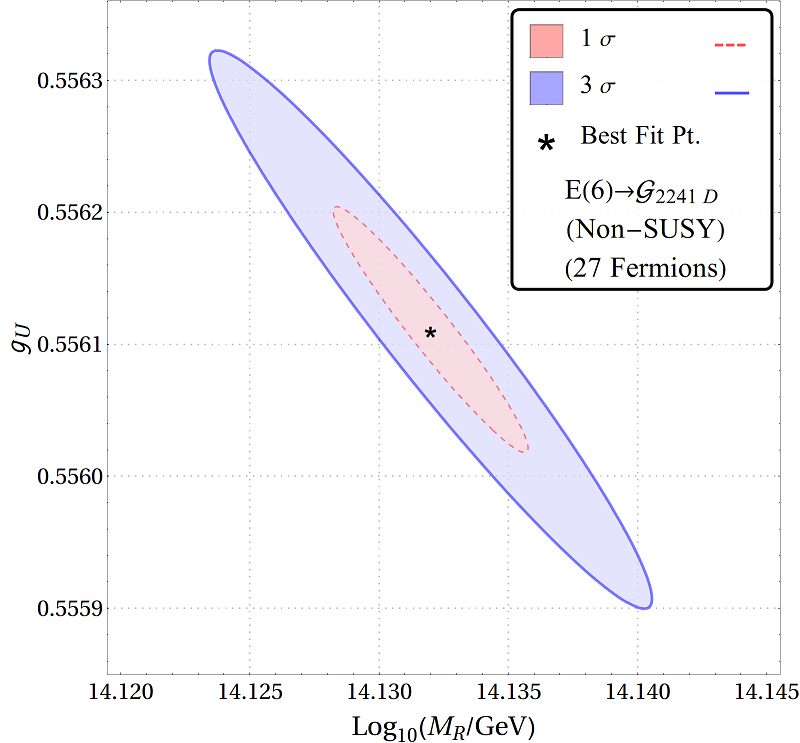}
	\includegraphics[scale=0.4]{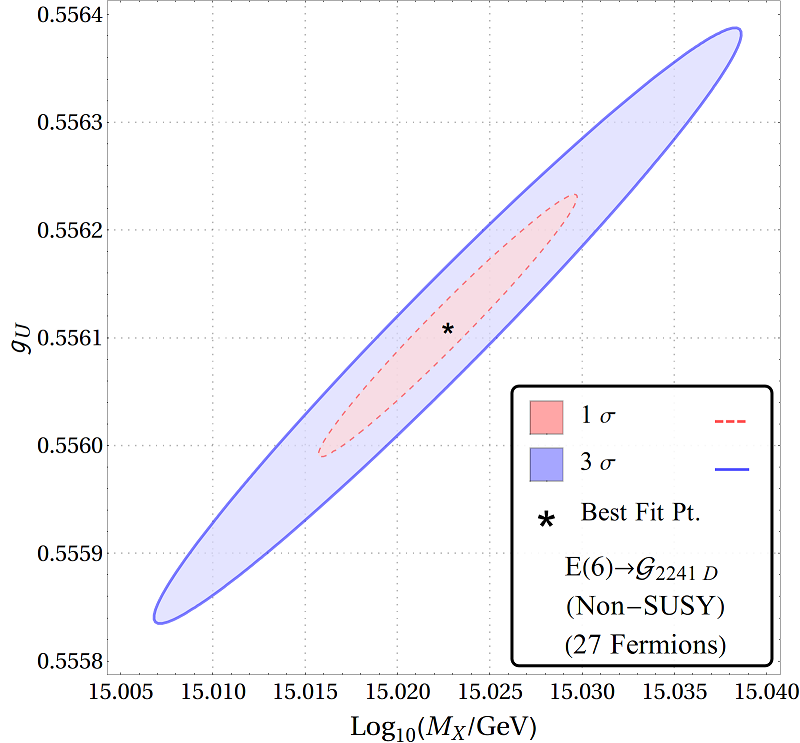}\\
	\includegraphics[scale=0.4]{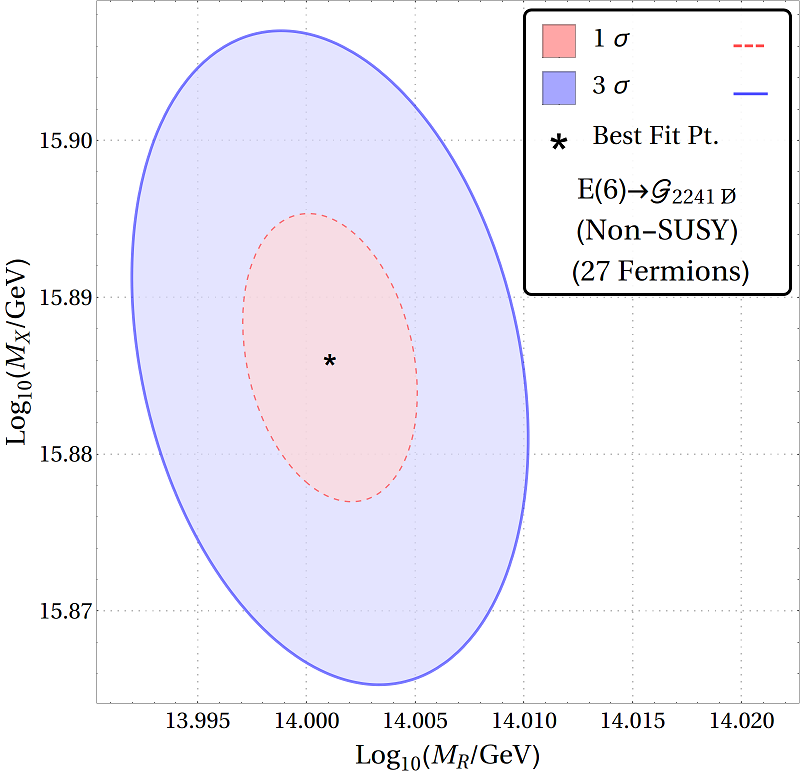}
	\includegraphics[scale=0.4]{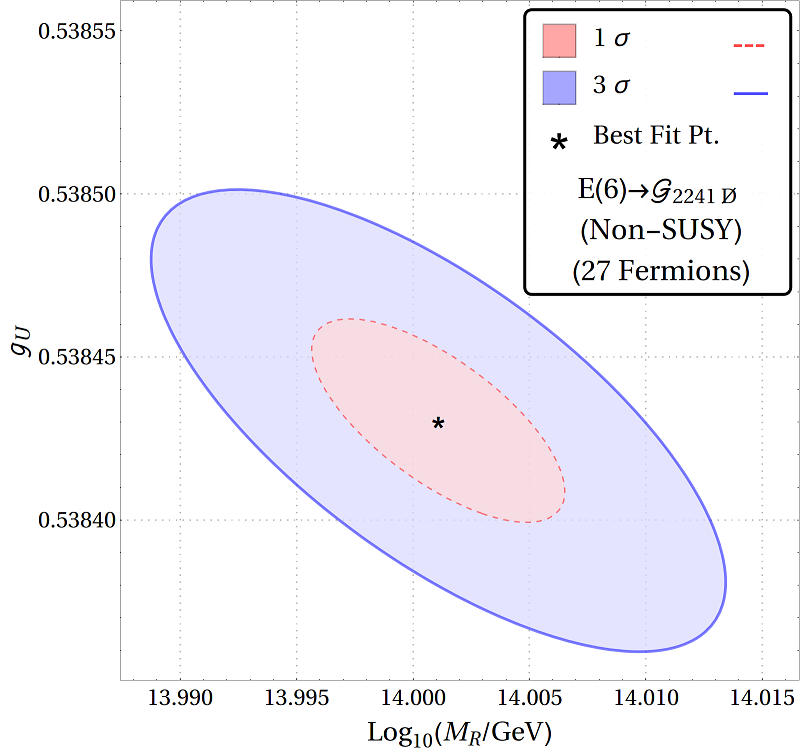}
	\includegraphics[scale=0.4]{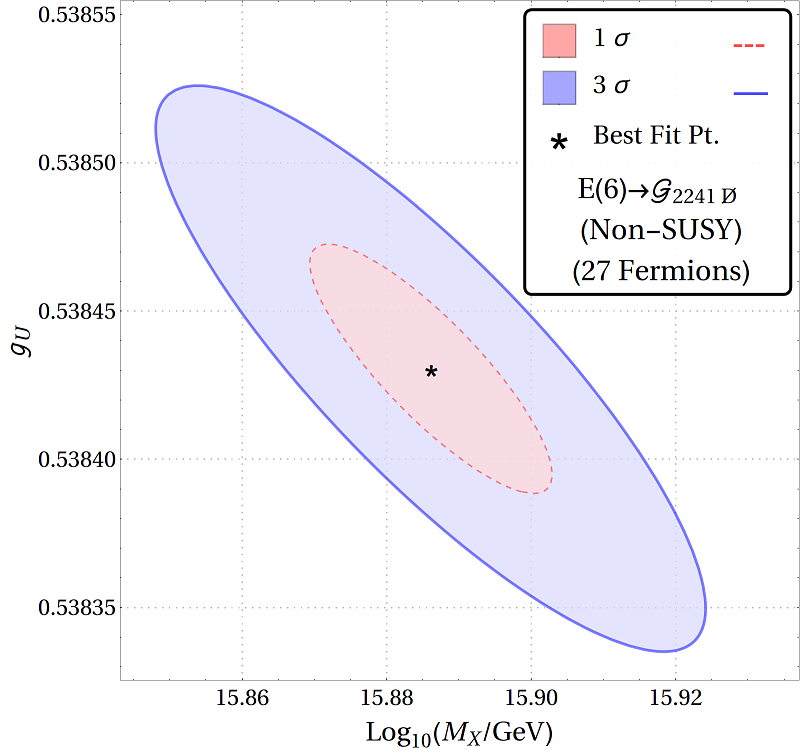}\\
	\includegraphics[scale=0.4]{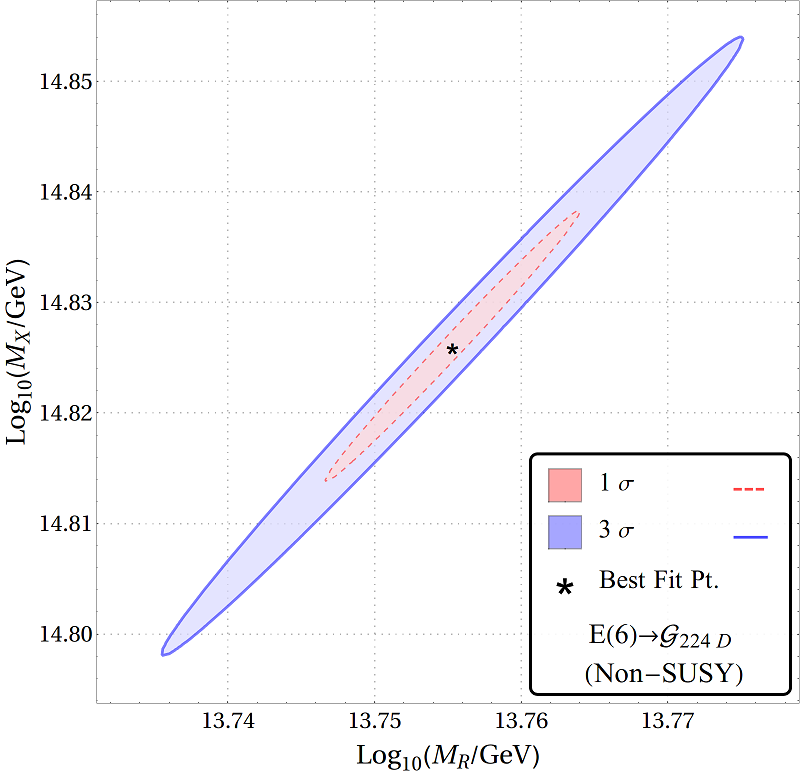}
	\includegraphics[scale=0.4]{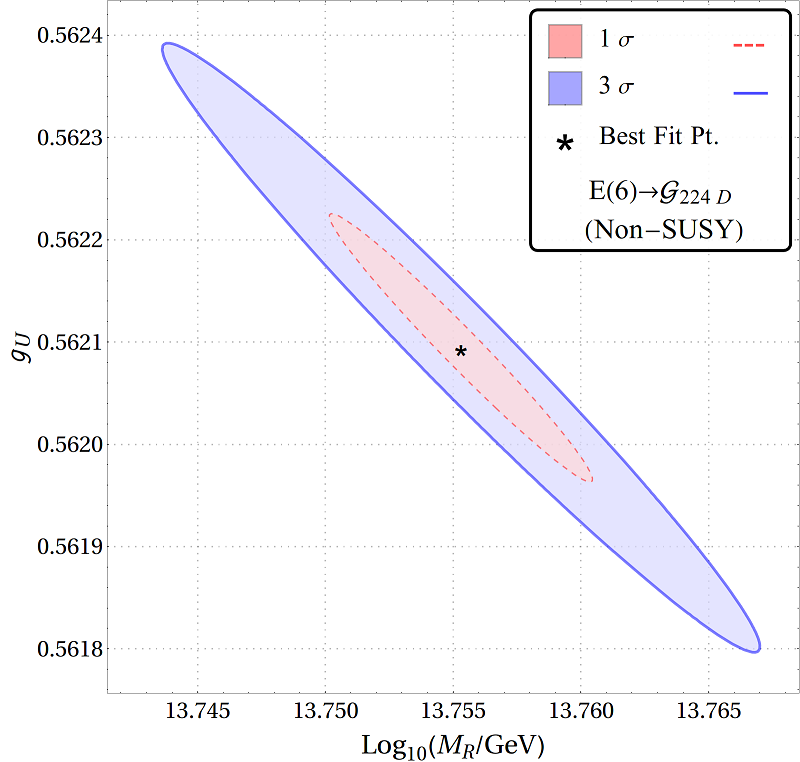}
	\includegraphics[scale=0.4]{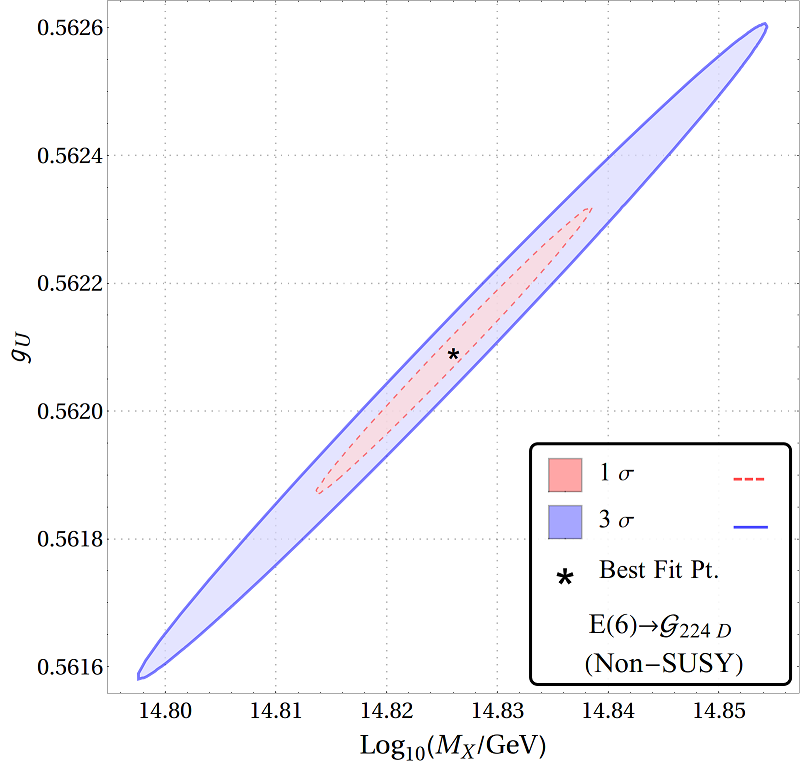}
	\caption{\scriptsize Continued to the next page...}
\end{figure}
\begin{figure}[!htbp]\ContinuedFloat
        \centering
	\includegraphics[scale=0.4]{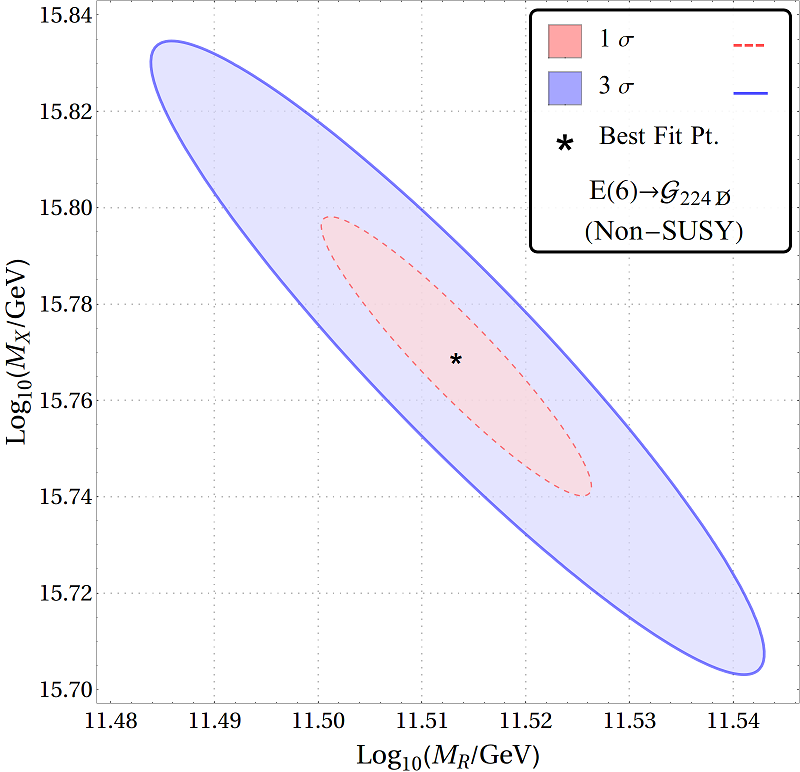}
	\includegraphics[scale=0.4]{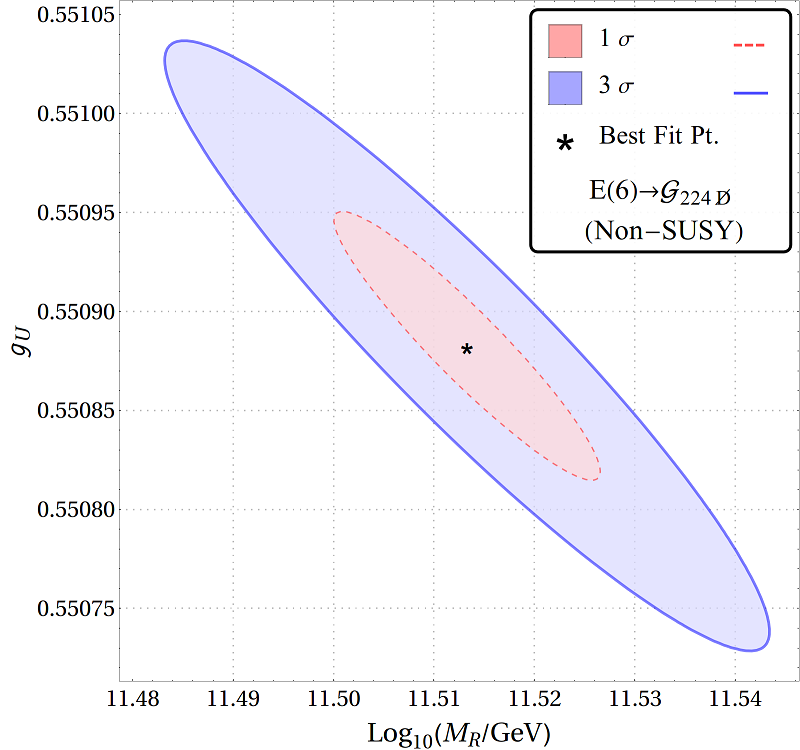}
	\includegraphics[scale=0.4]{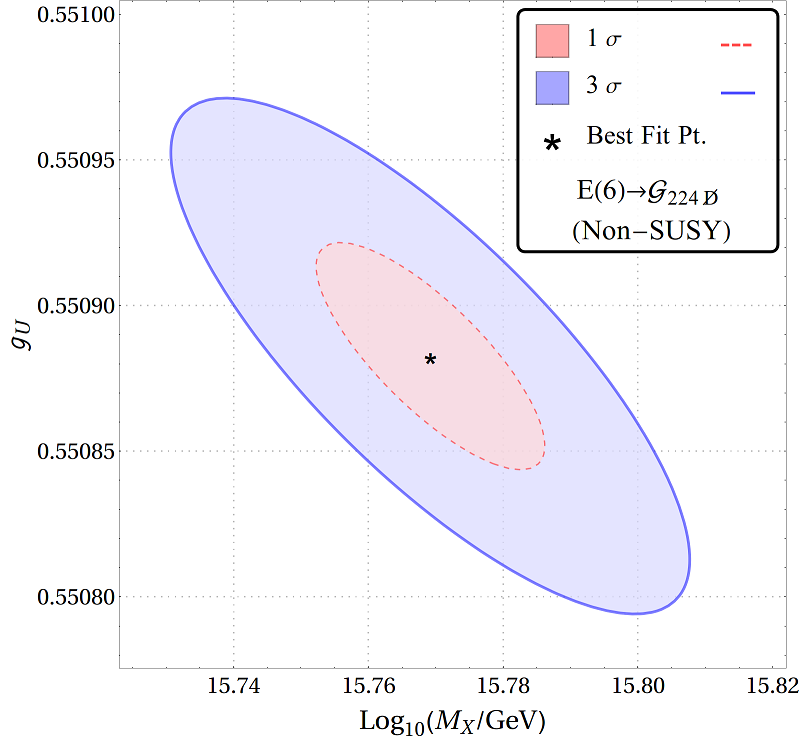}\\
	\includegraphics[scale=0.4]{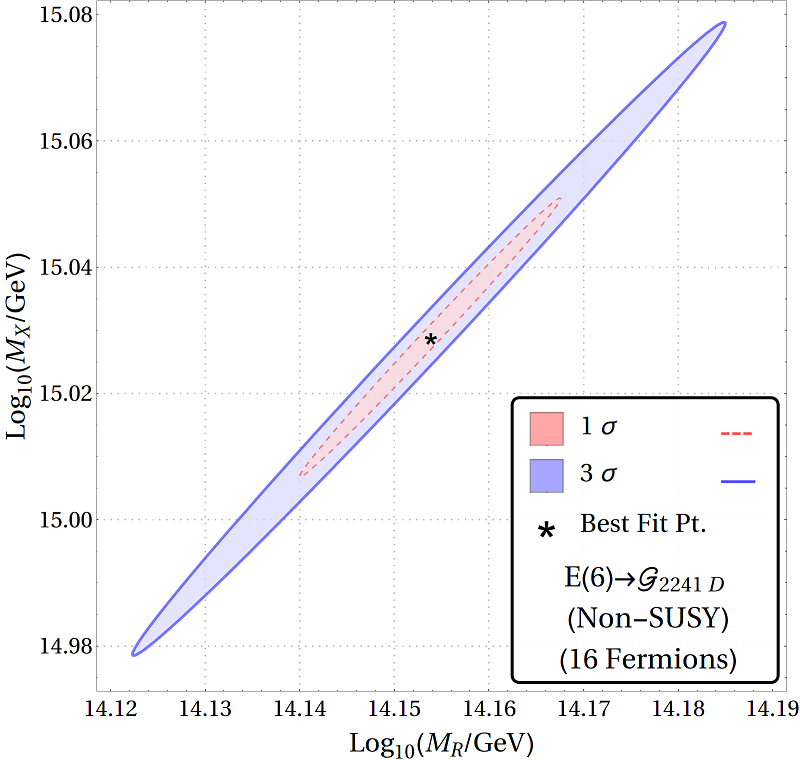}
	\includegraphics[scale=0.4]{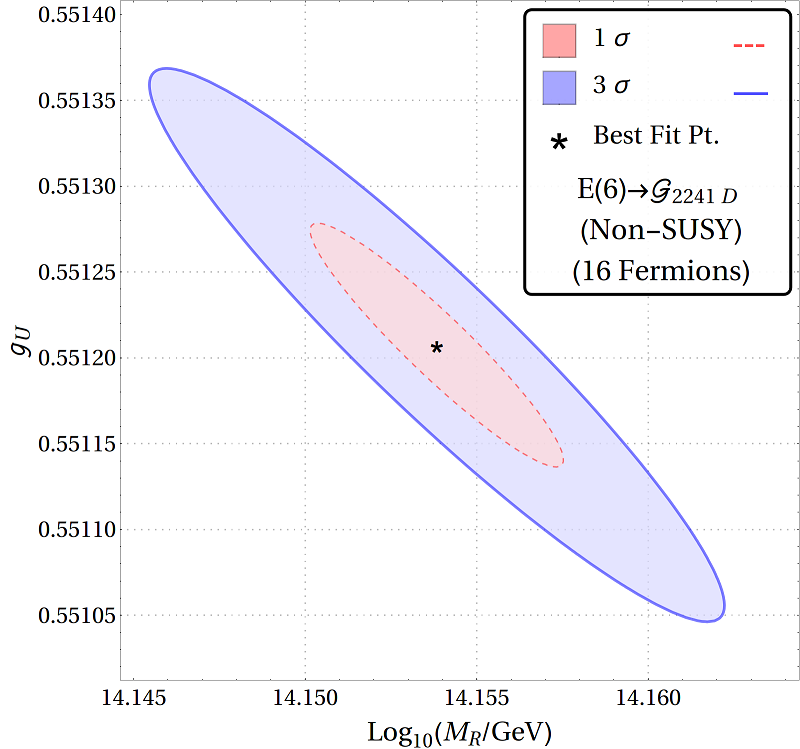}
	\includegraphics[scale=0.4]{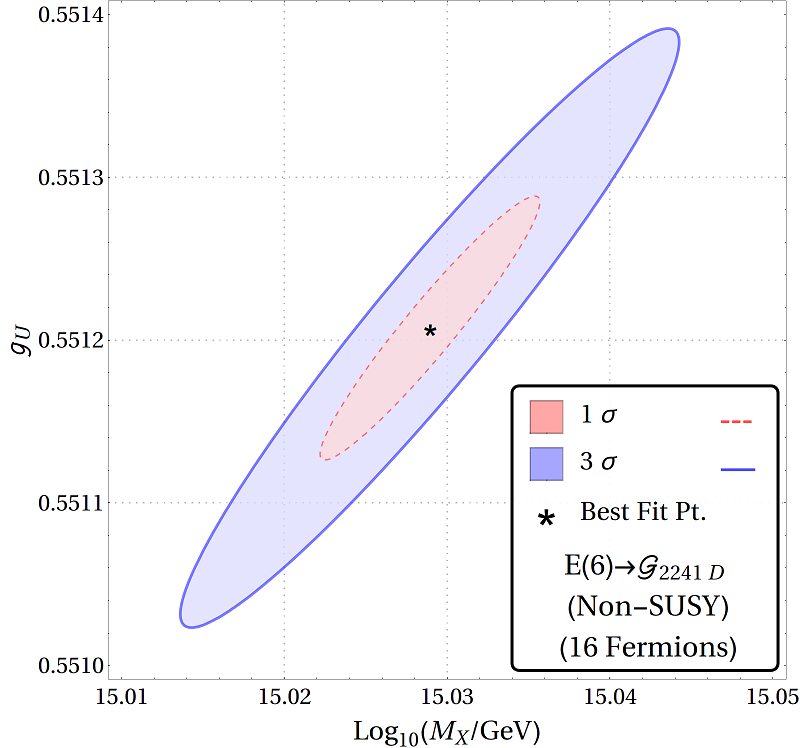}\\
	\includegraphics[scale=0.4]{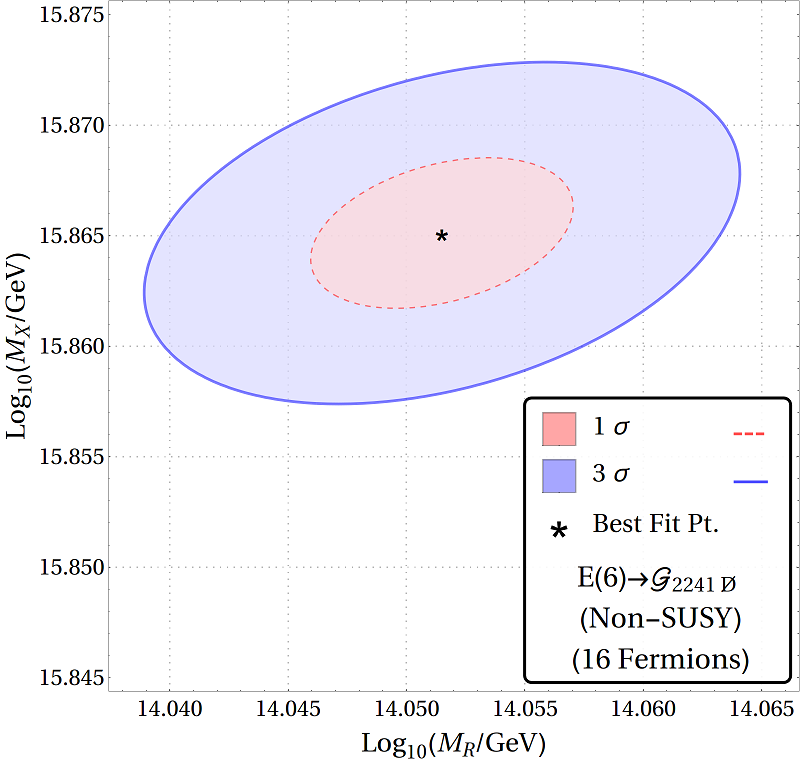}
	\includegraphics[scale=0.4]{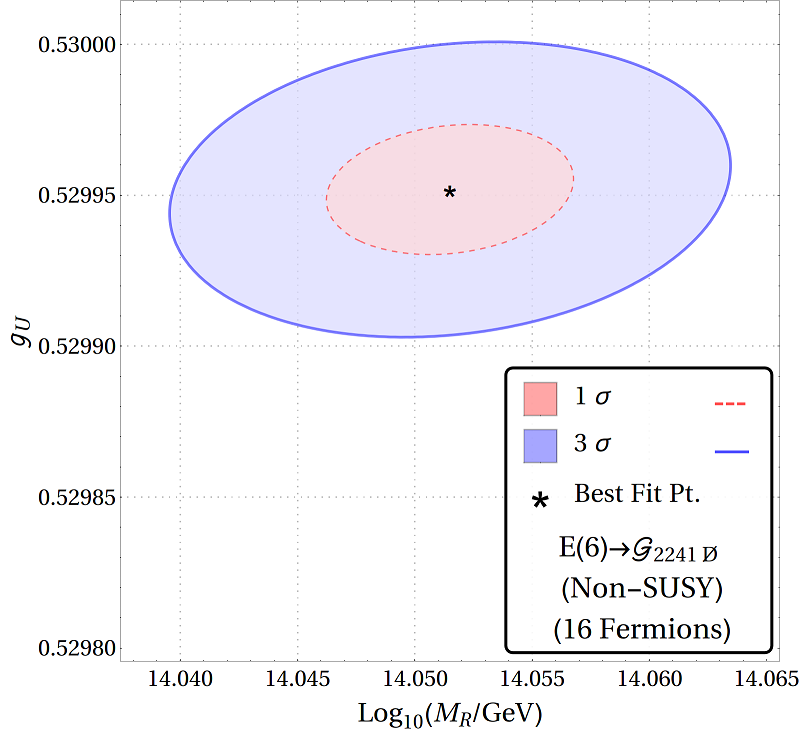}
	\includegraphics[scale=0.4]{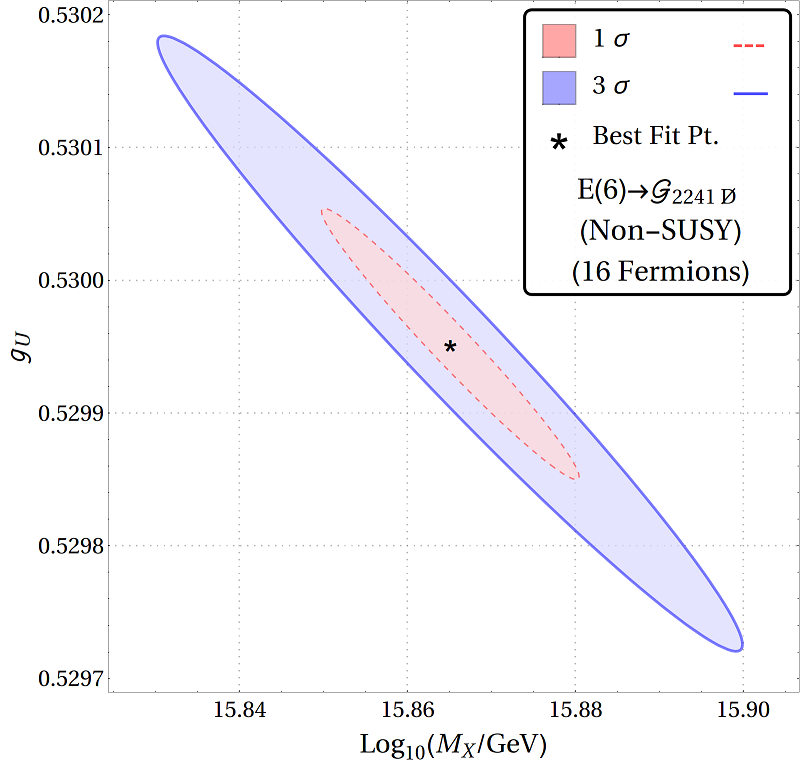}\\
	\includegraphics[scale=0.4]{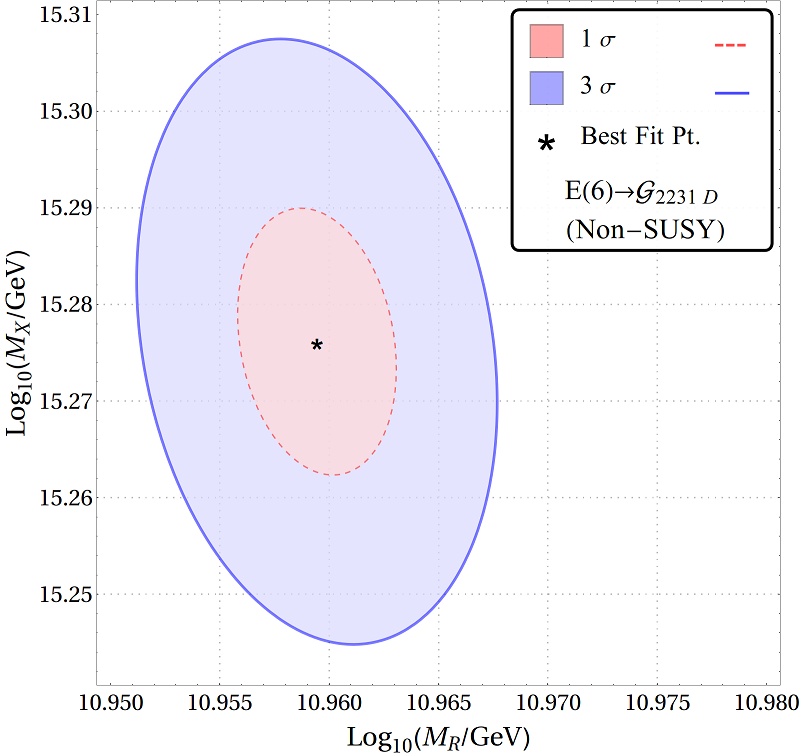}
	\includegraphics[scale=0.4]{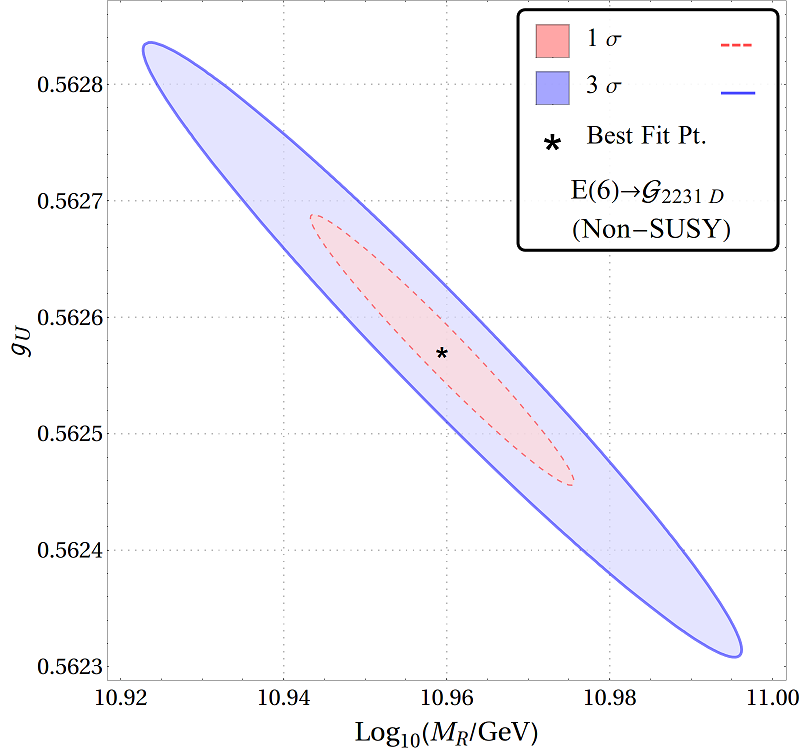}
	\includegraphics[scale=0.4]{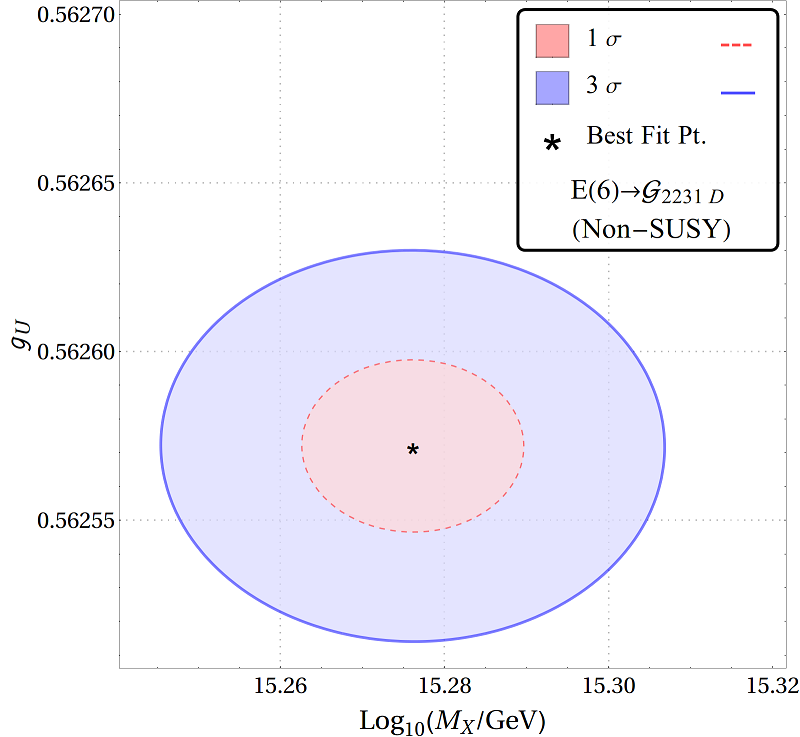}\\
	\includegraphics[scale=0.4]{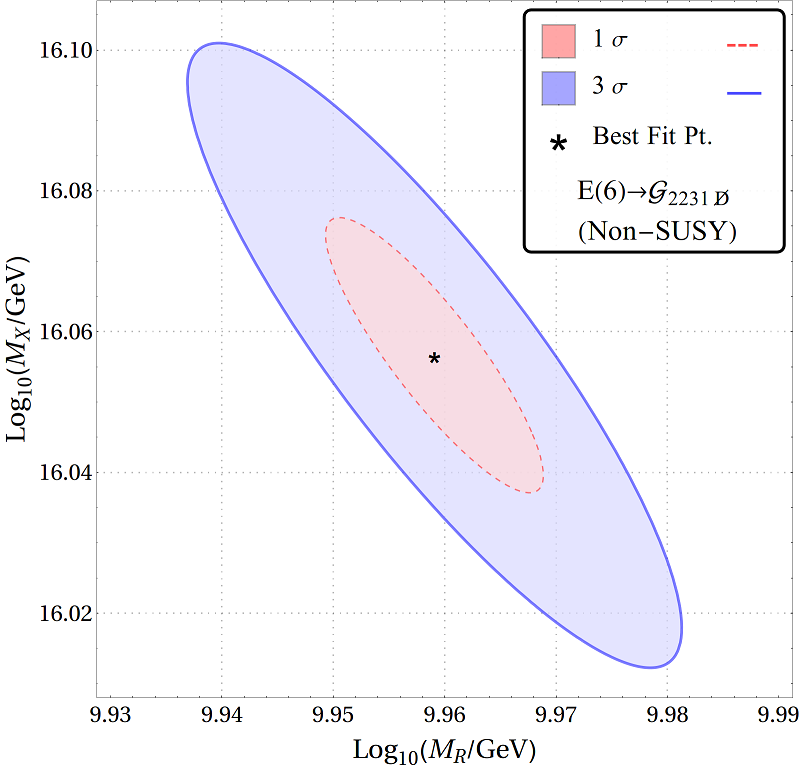}
	\includegraphics[scale=0.4]{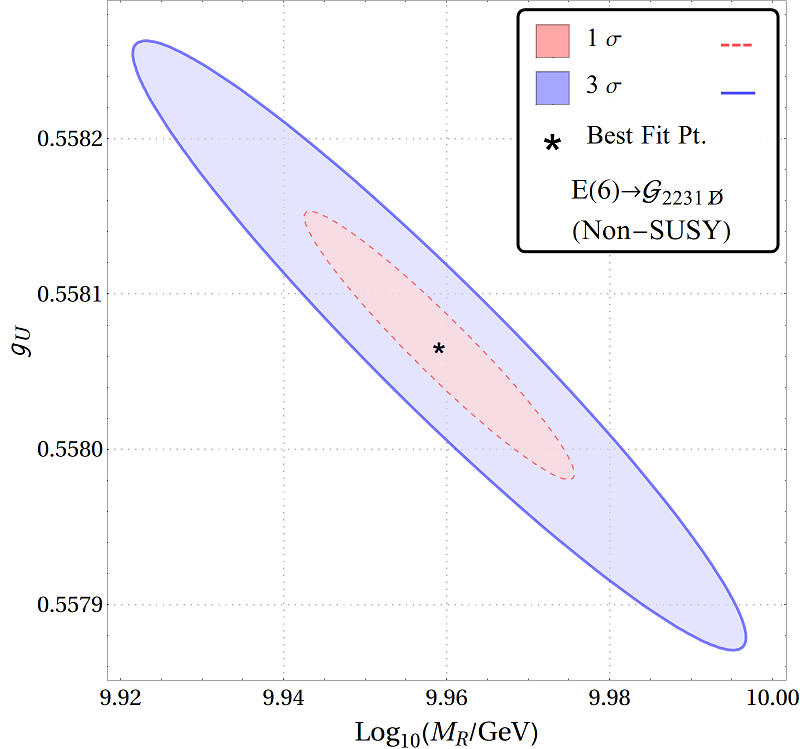}
	\includegraphics[scale=0.4]{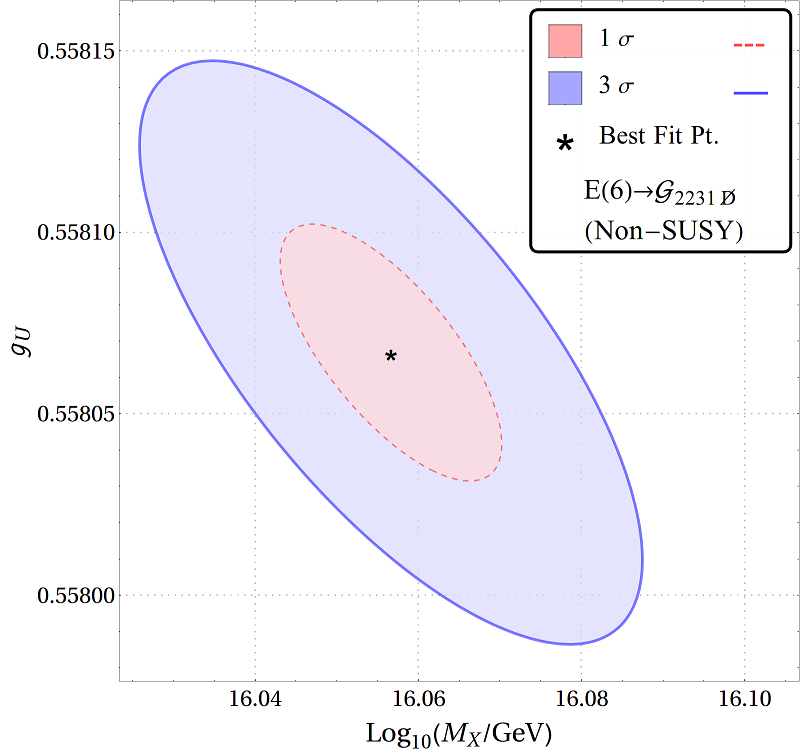}
	\caption{\scriptsize Correlations among $M_R$, $M_X$ and $g_{U}$ satisfying gauge coupling unification for various breaking patterns within Non-supersymmetric scenario.}
	\label{fig:unific-chisq}
\end{figure}

$~$

\subsubsection*{Supersymmetric scenarios}

\begin{figure}[!htbp]
	\centering
	\includegraphics[scale=0.4]{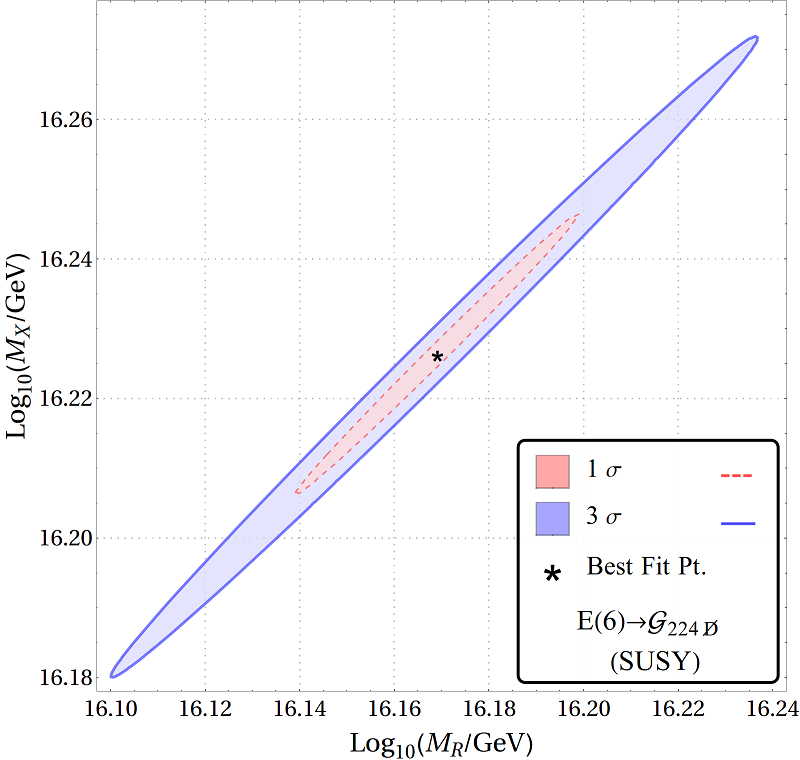}
	\includegraphics[scale=0.4]{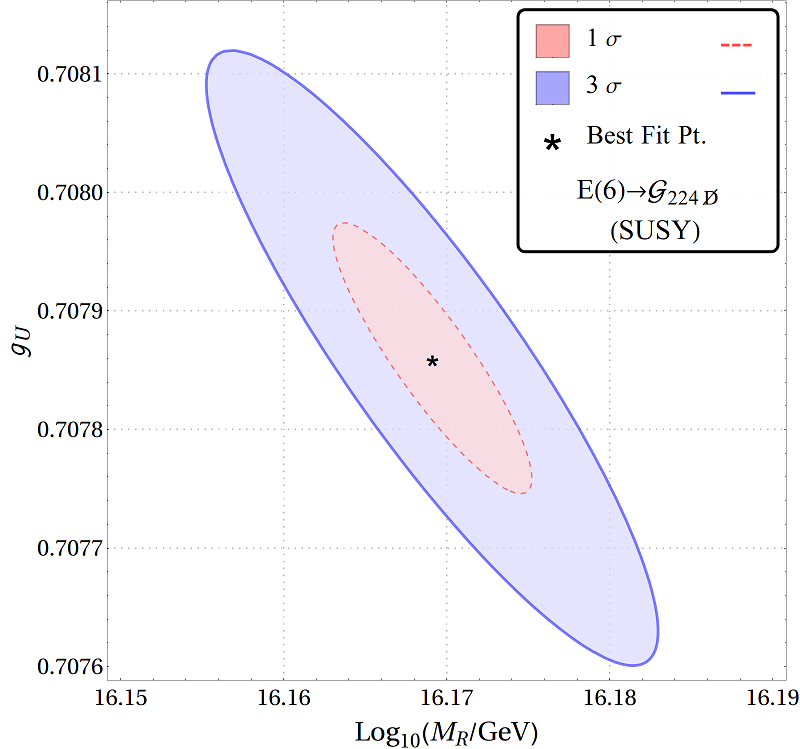}
	\includegraphics[scale=0.4]{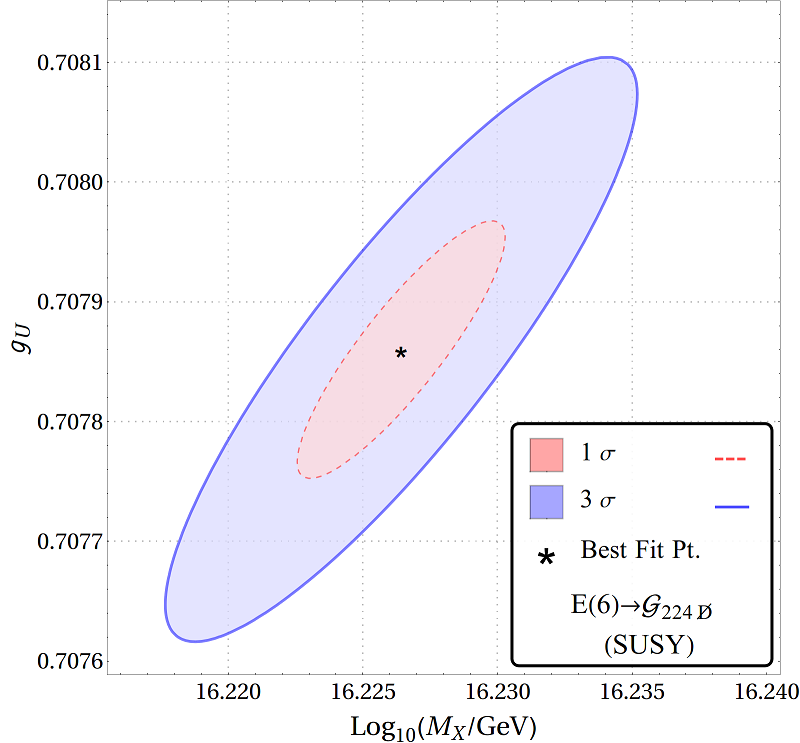}\\
	\includegraphics[scale=0.4]{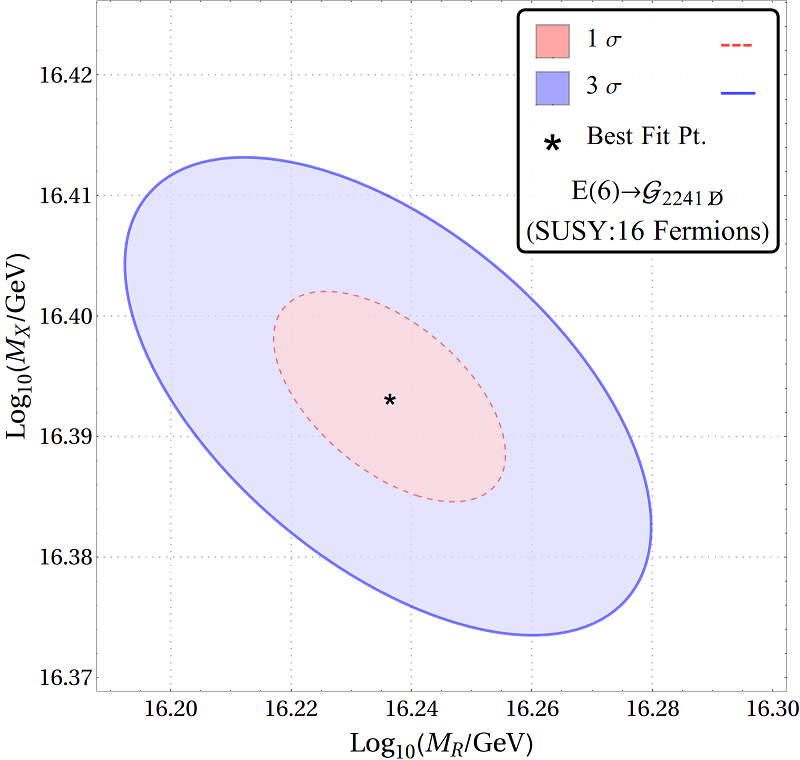}
	\includegraphics[scale=0.4]{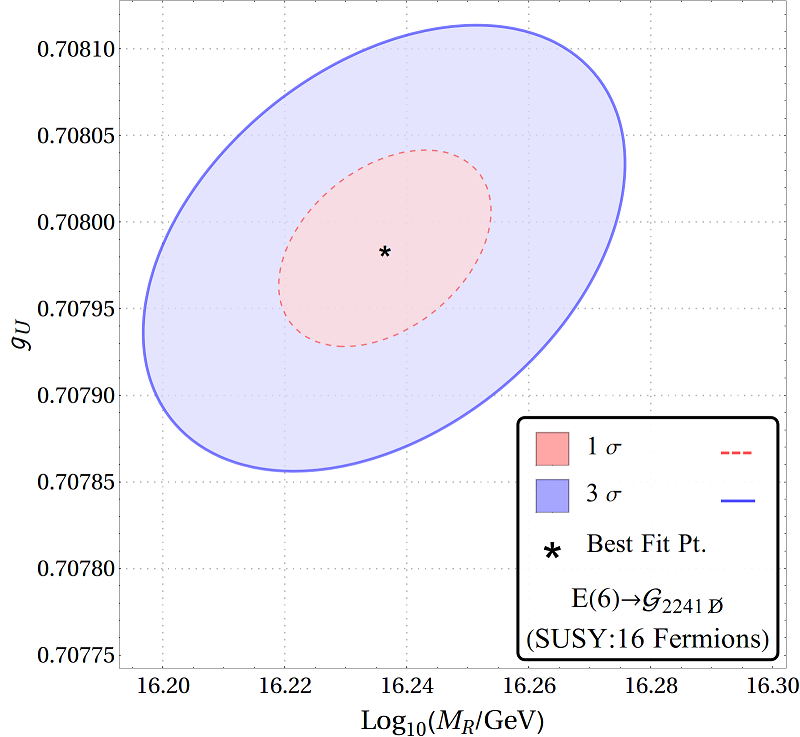}
	\includegraphics[scale=0.4]{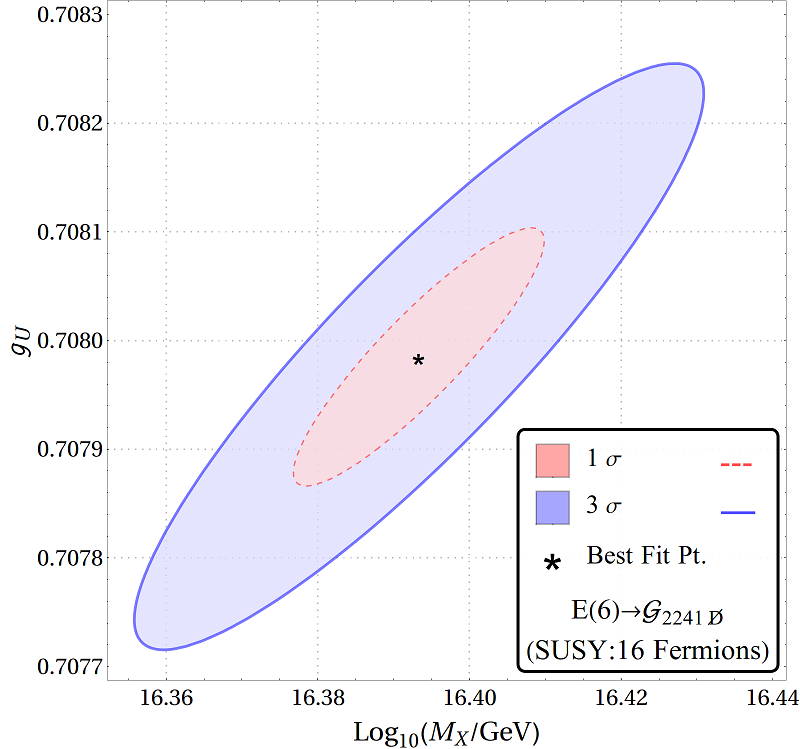}\\
	\includegraphics[scale=0.4]{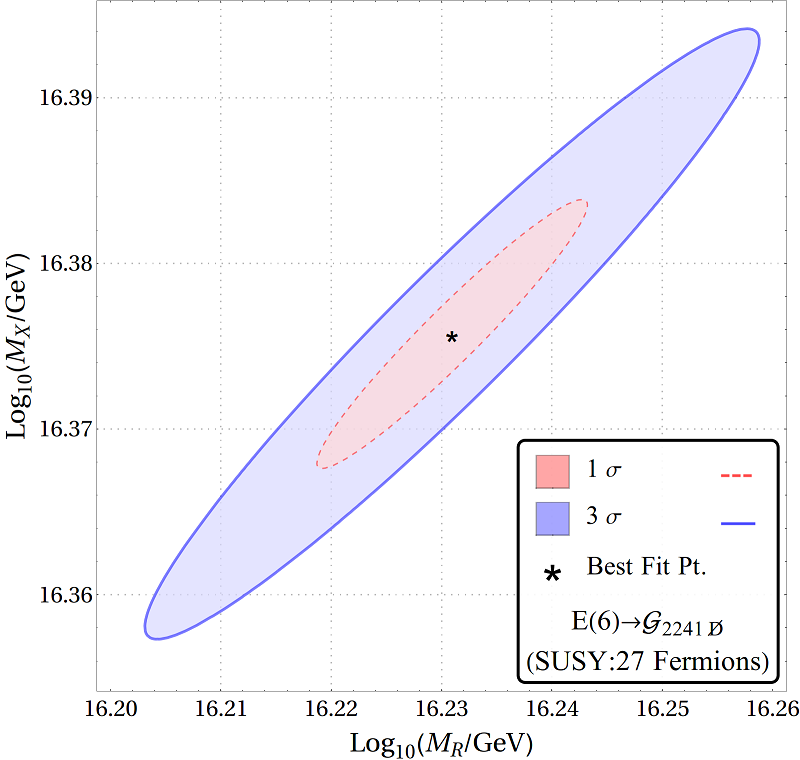}
	\includegraphics[scale=0.4]{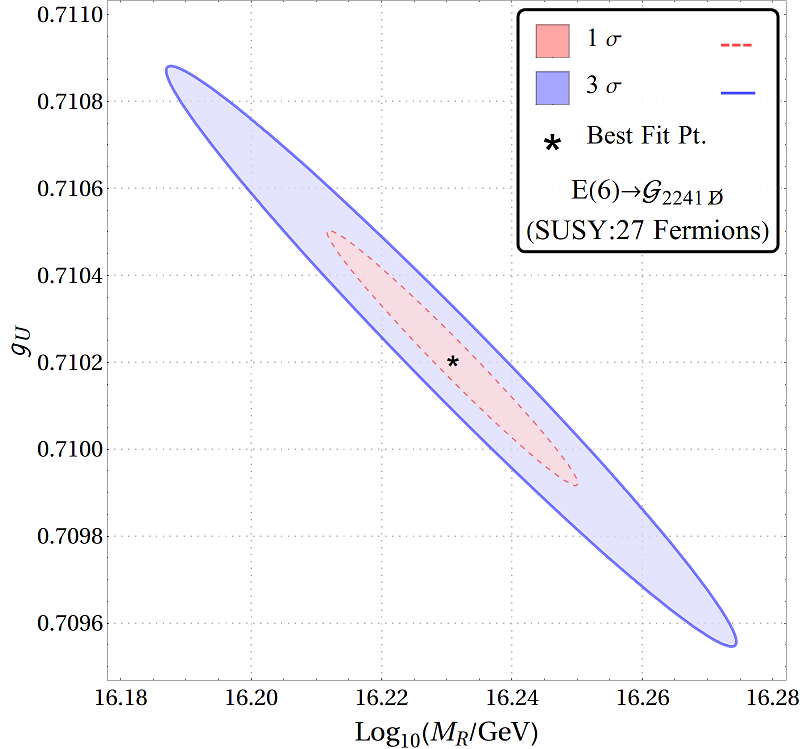}
	\includegraphics[scale=0.4]{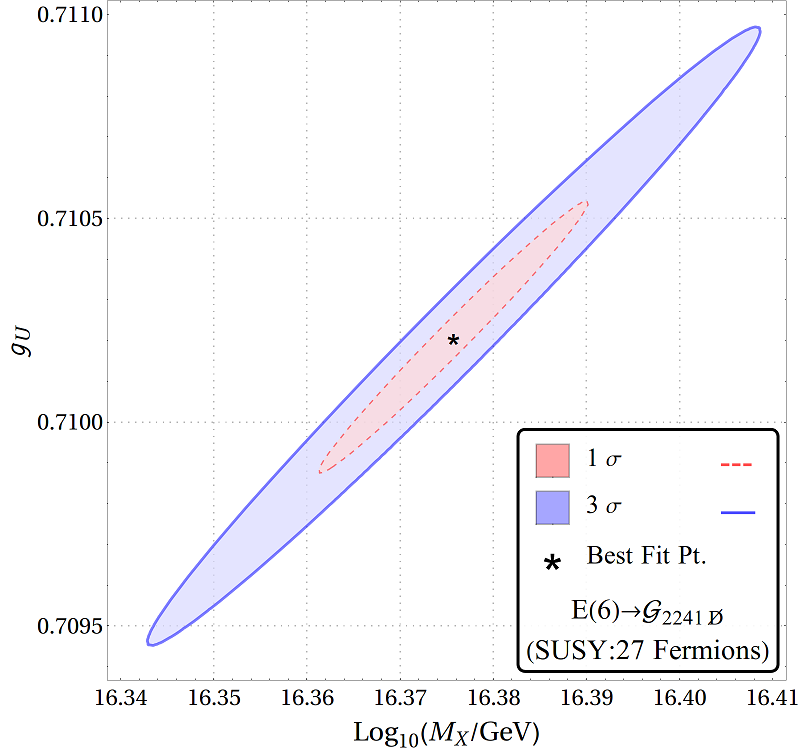}\\
	\includegraphics[scale=0.4]{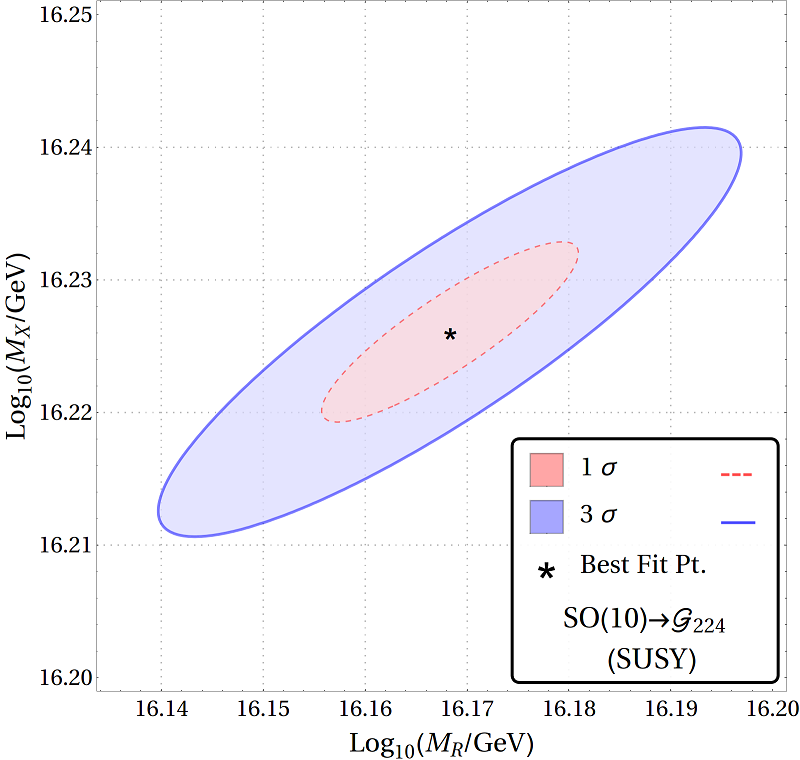}
	\includegraphics[scale=0.4]{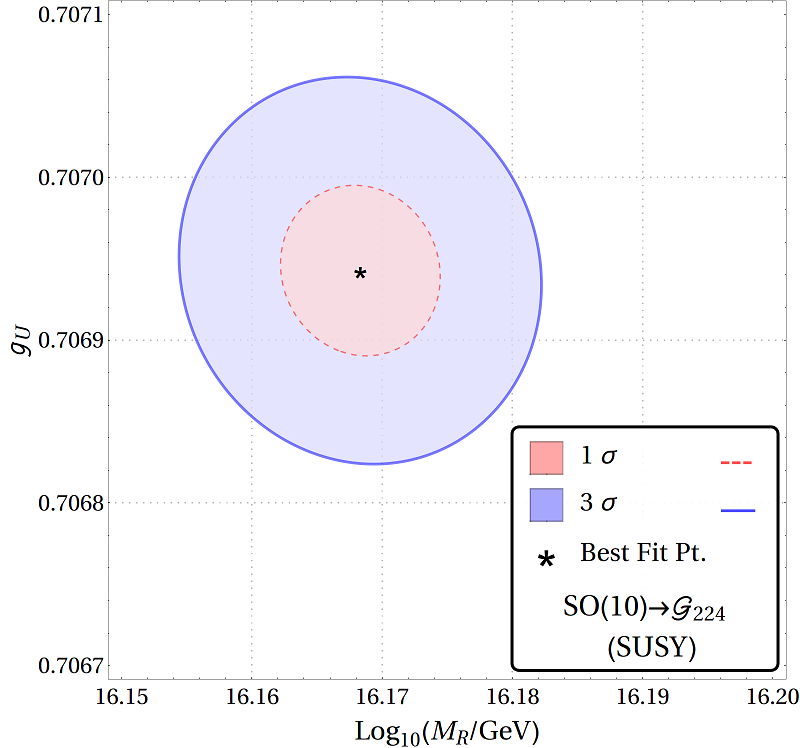}
	\includegraphics[scale=0.4]{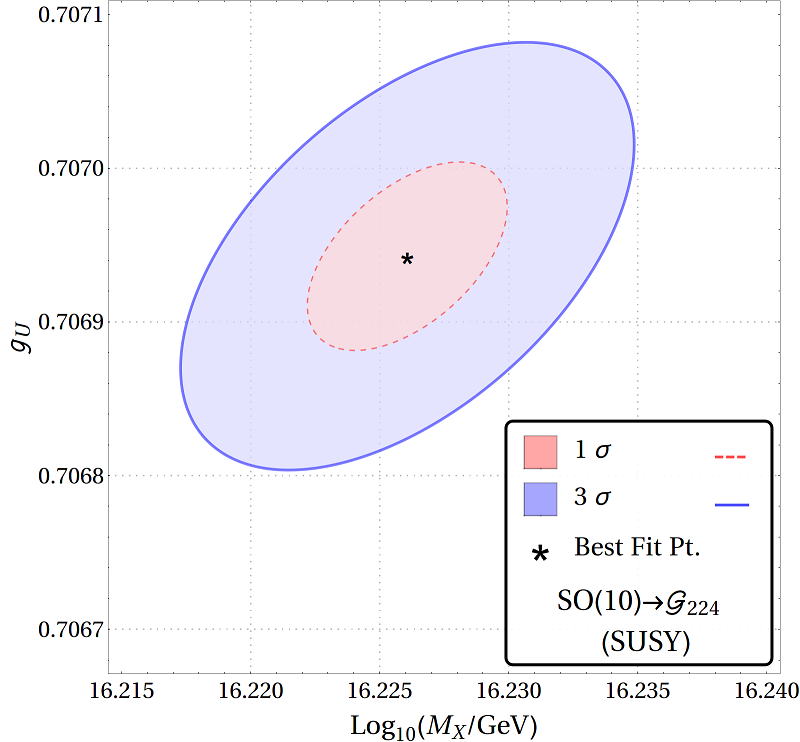}
	\caption{\scriptsize Correlations among $M_R$, $M_X$ and $g_{U}$, for different breaking patterns within supersymmetric scenario, satisfying unification with $M_{SUSY}$ at 1 TeV. Correlations are obtained by fixing $M_{SUSY}$ at 1 TeV.}
	\label{fig:unific-chisq_1tev_S}
\end{figure}

\begin{figure}[!htbp]
	\centering
	\includegraphics[scale=0.4]{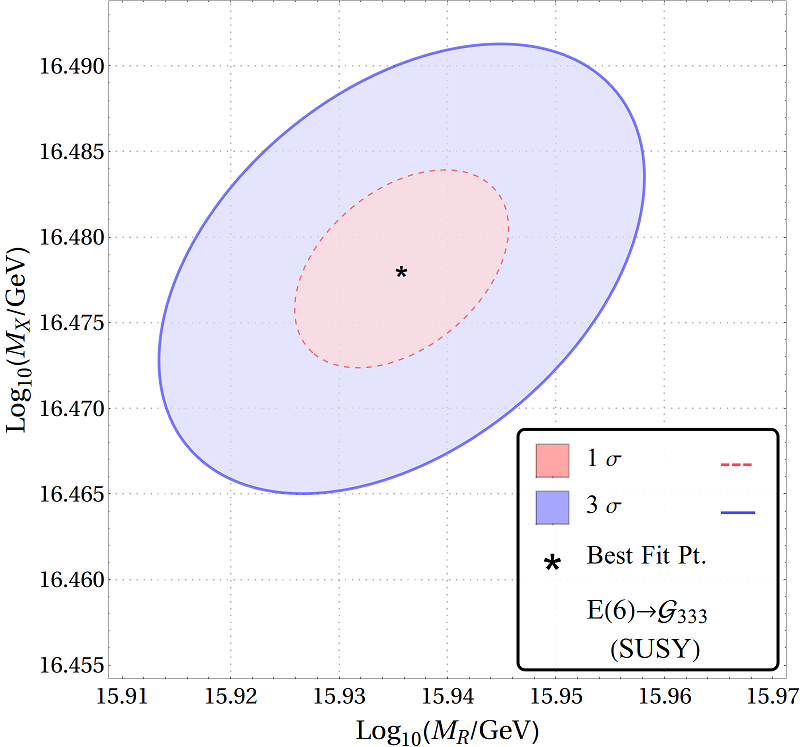}
	\includegraphics[scale=0.4]{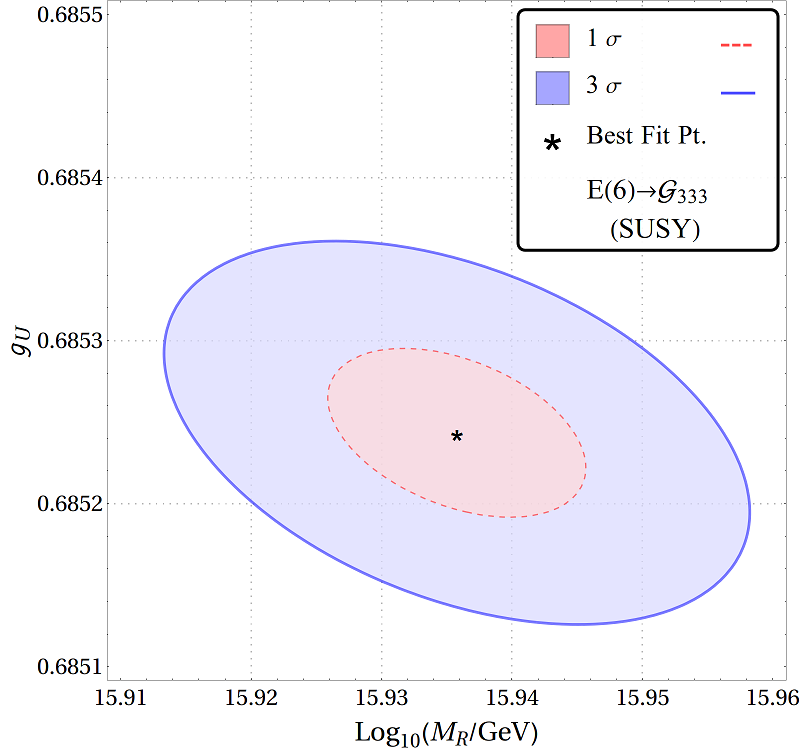}
	\includegraphics[scale=0.4]{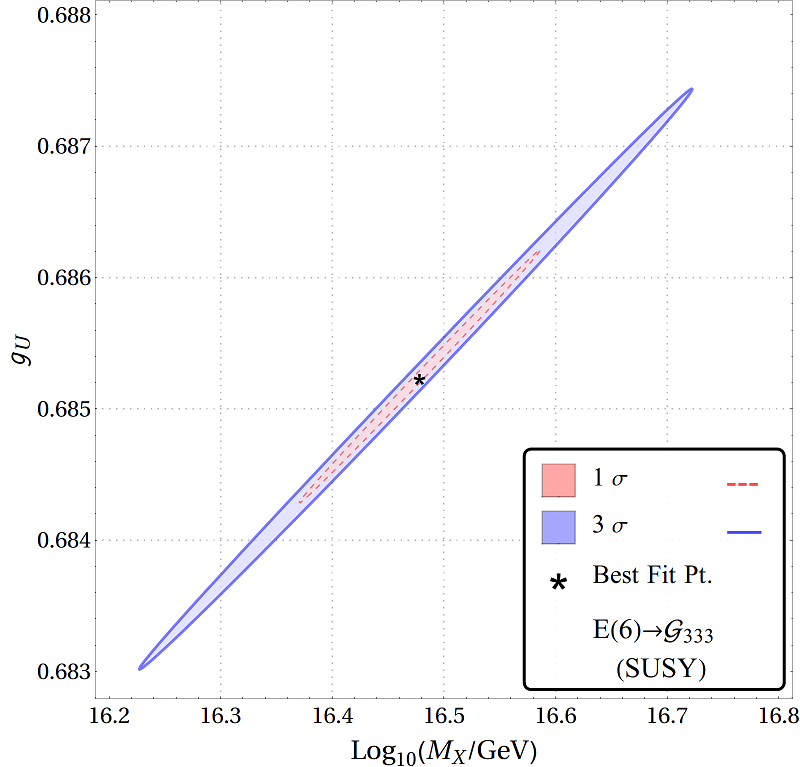}\\
	\includegraphics[scale=0.4]{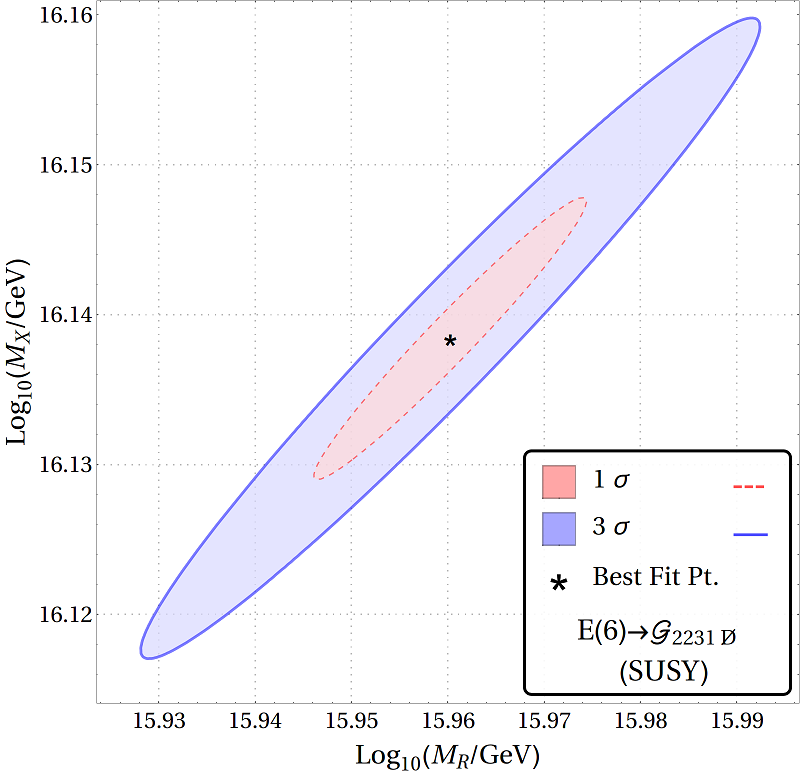}
	\includegraphics[scale=0.4]{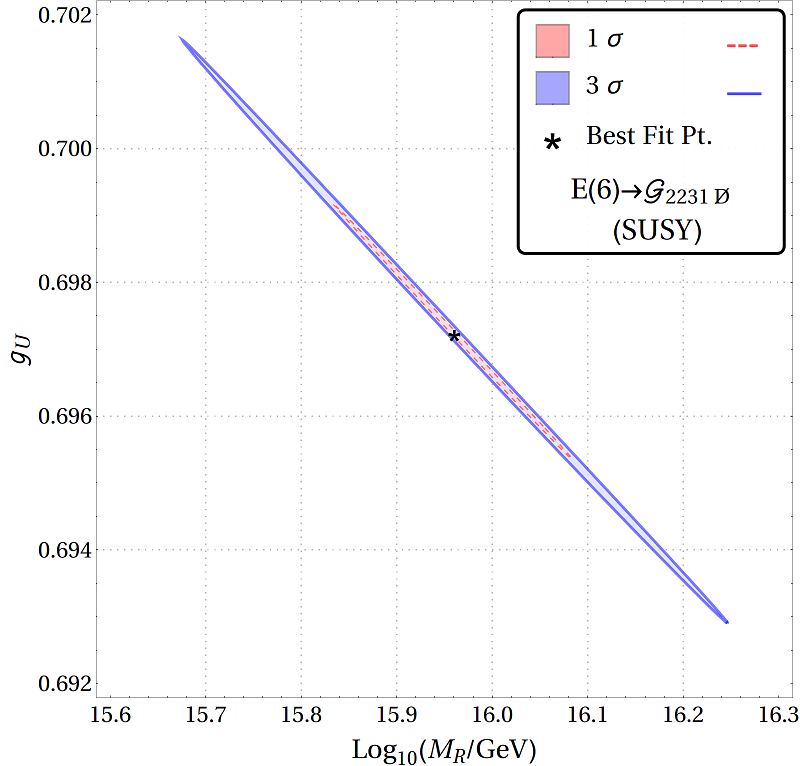}
	\includegraphics[scale=0.4]{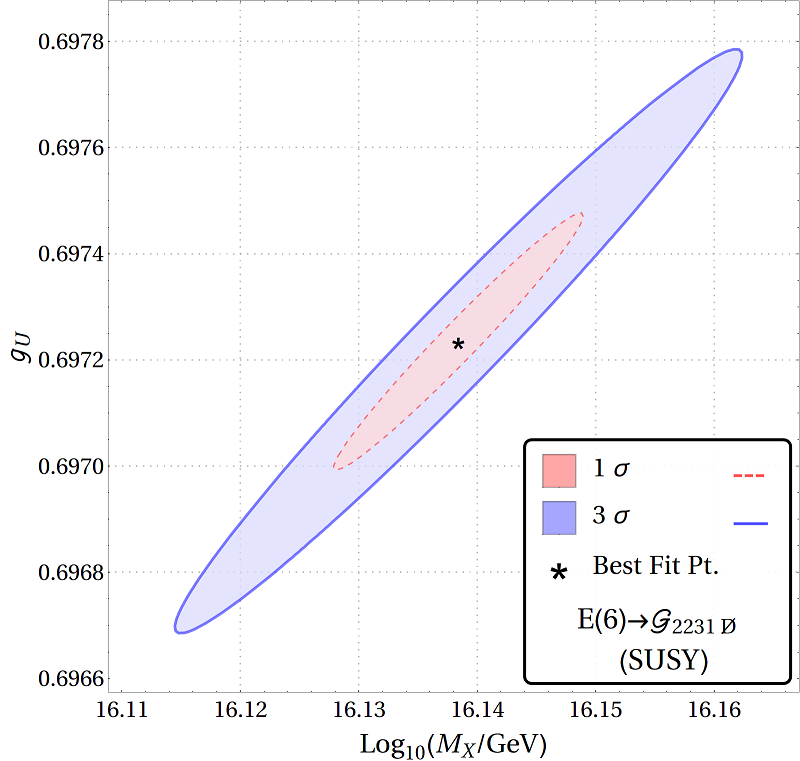}\\
	\includegraphics[scale=0.4]{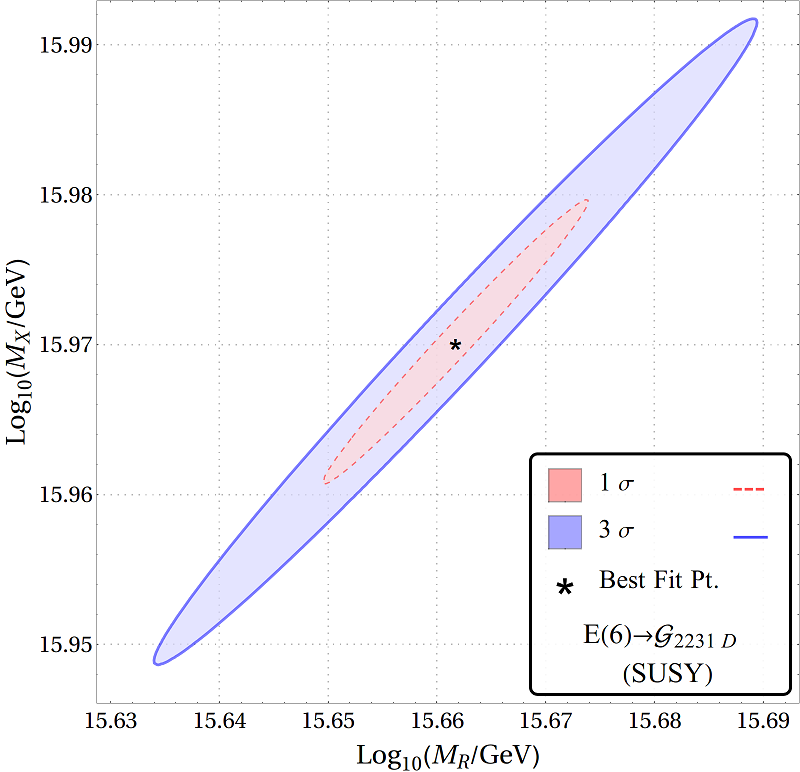}
	\includegraphics[scale=0.4]{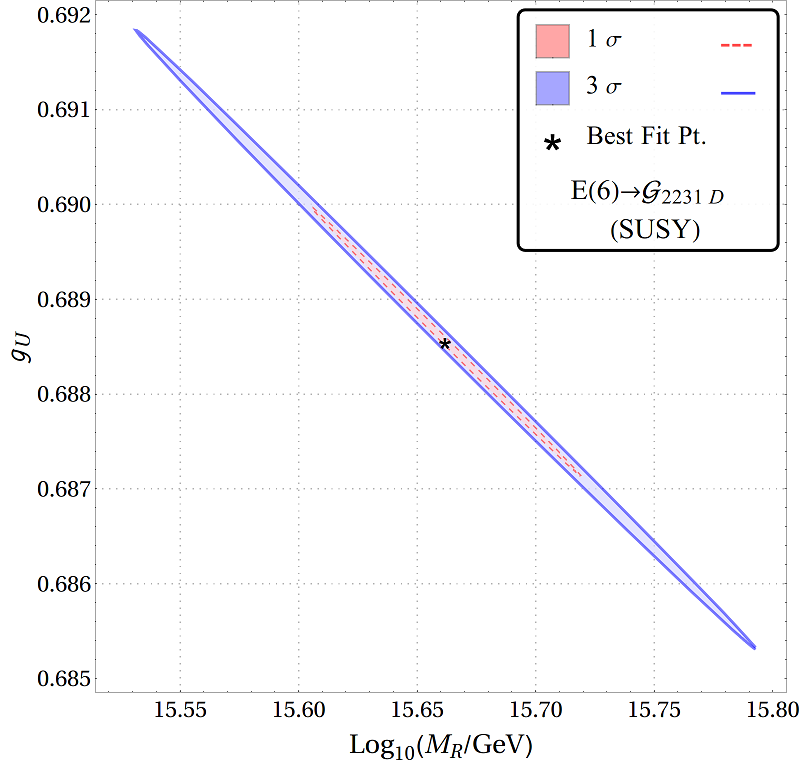}
	\includegraphics[scale=0.4]{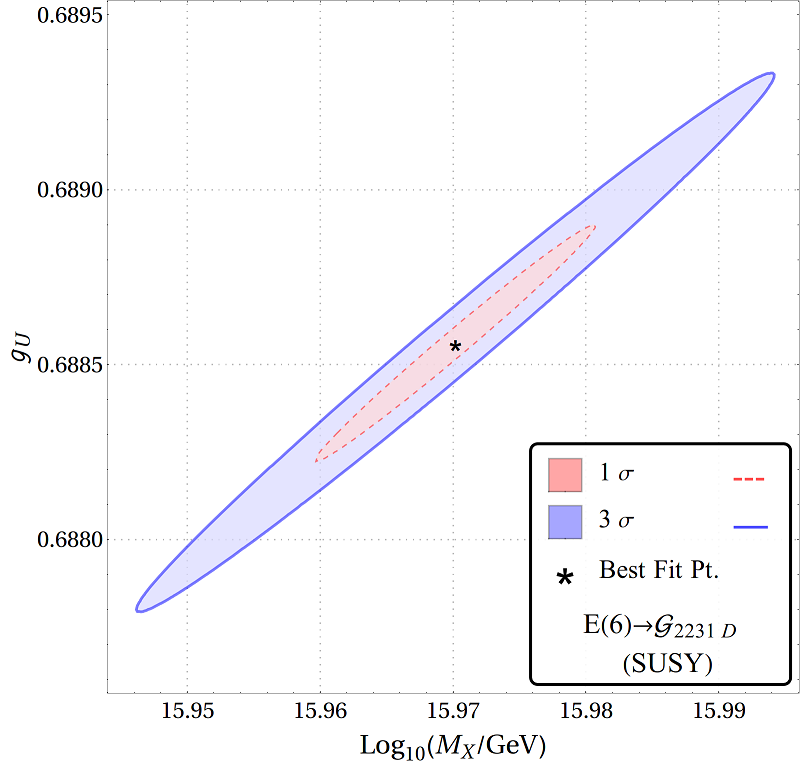}\\
	\includegraphics[scale=0.4]{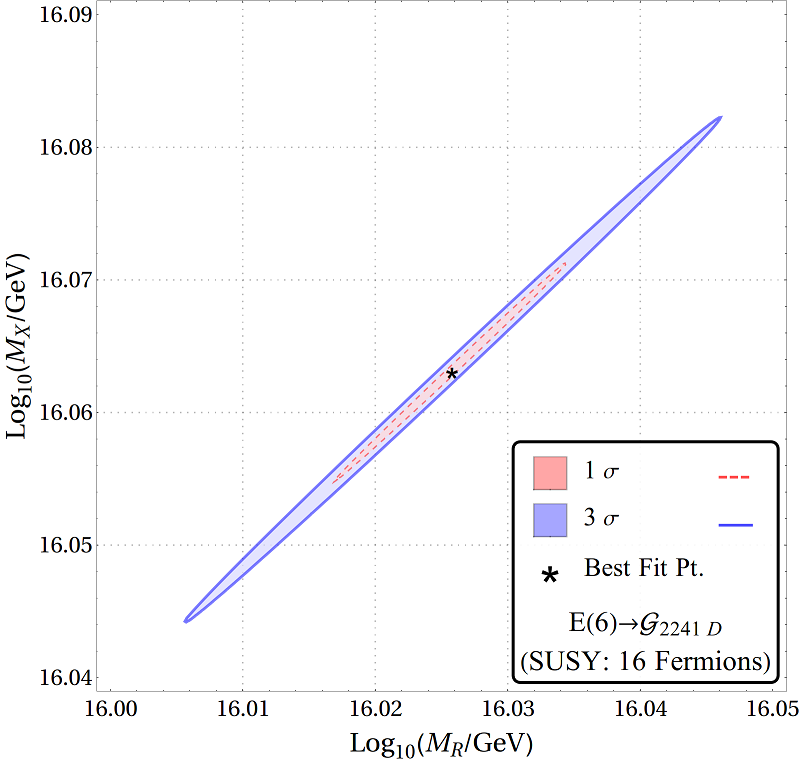}
	\includegraphics[scale=0.4]{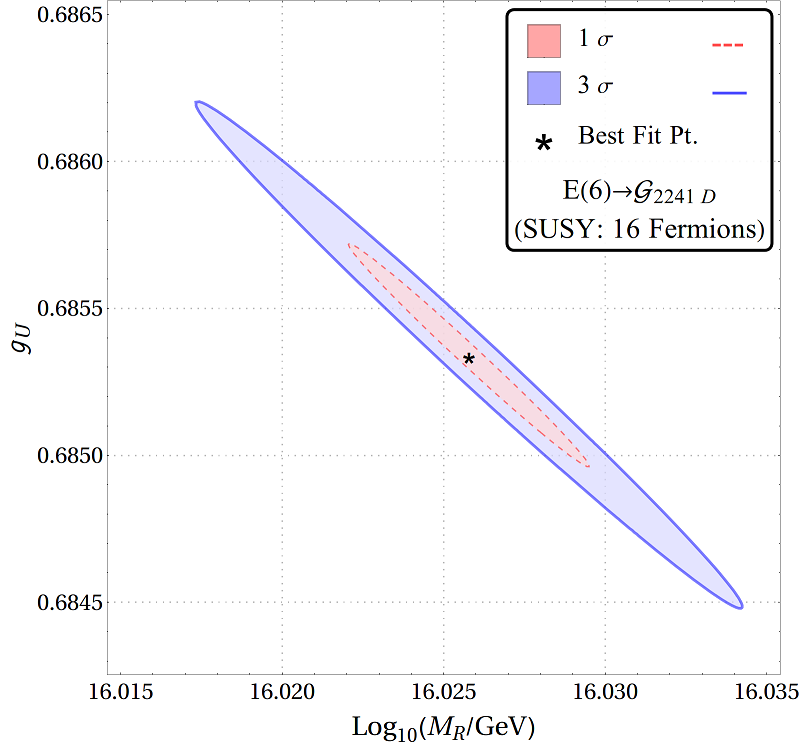}
	\includegraphics[scale=0.4]{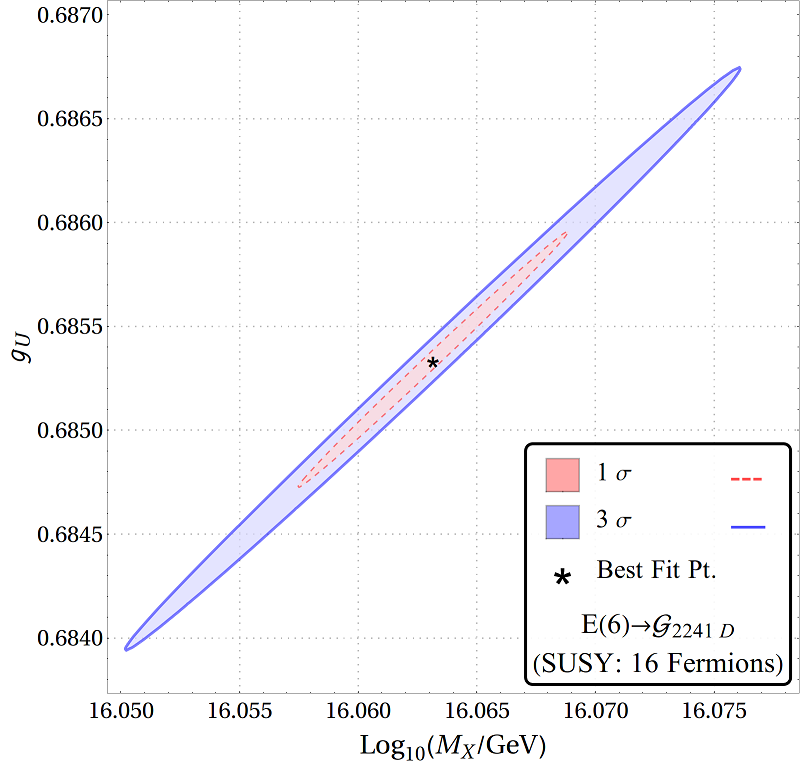}\\
	\includegraphics[scale=0.4]{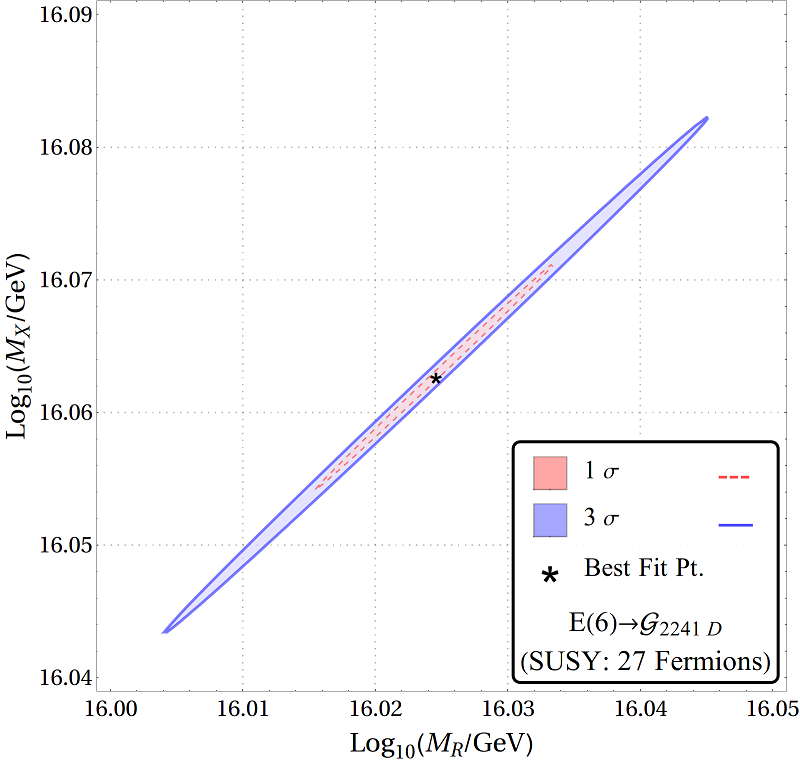}
	\includegraphics[scale=0.4]{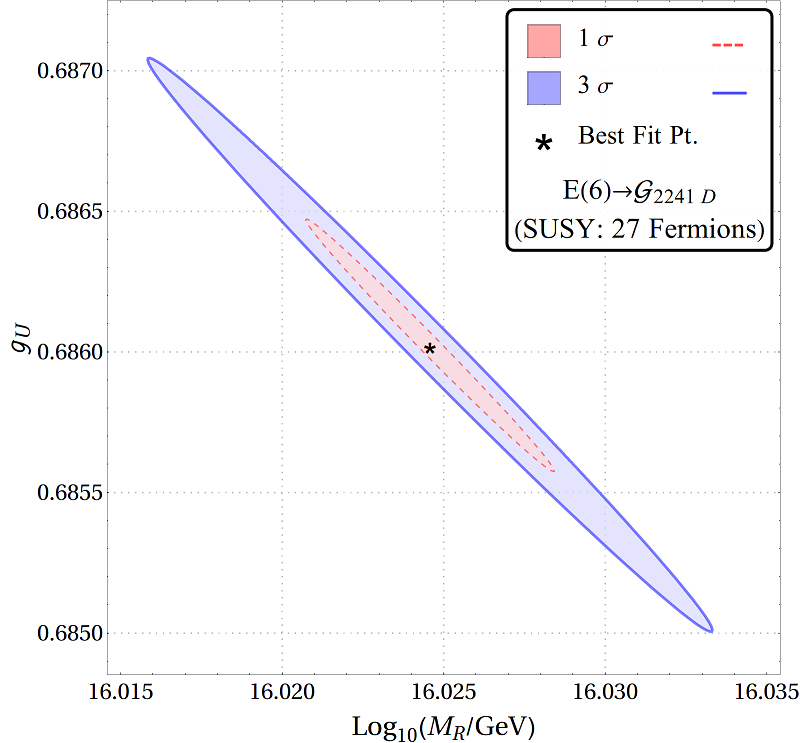}
	\includegraphics[scale=0.4]{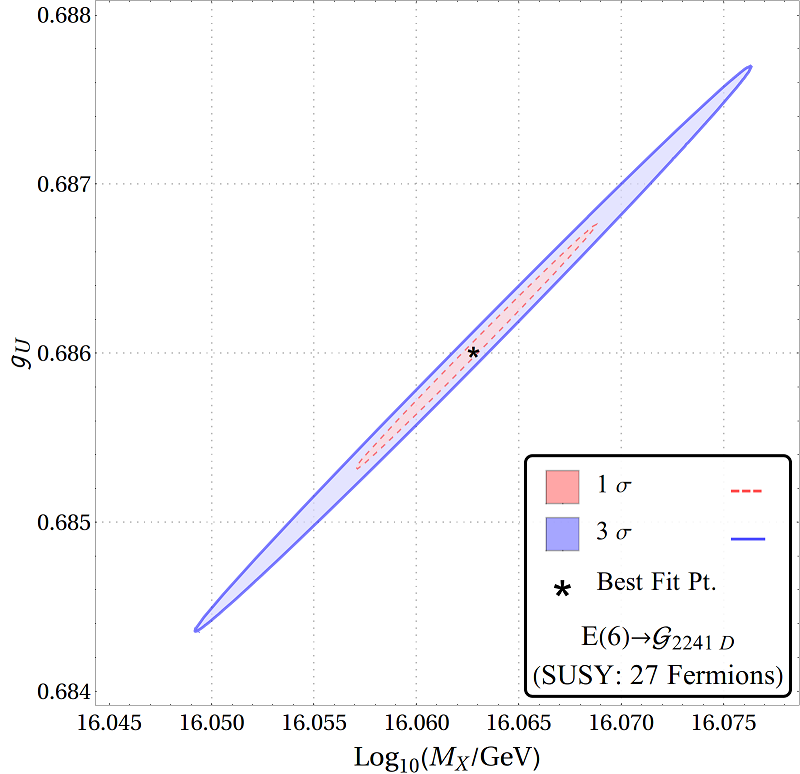}
	\caption{\scriptsize See the caption of next figure...
	}
\end{figure}

\begin{figure}[!hbp]
	\centering
	\includegraphics[scale=0.4]{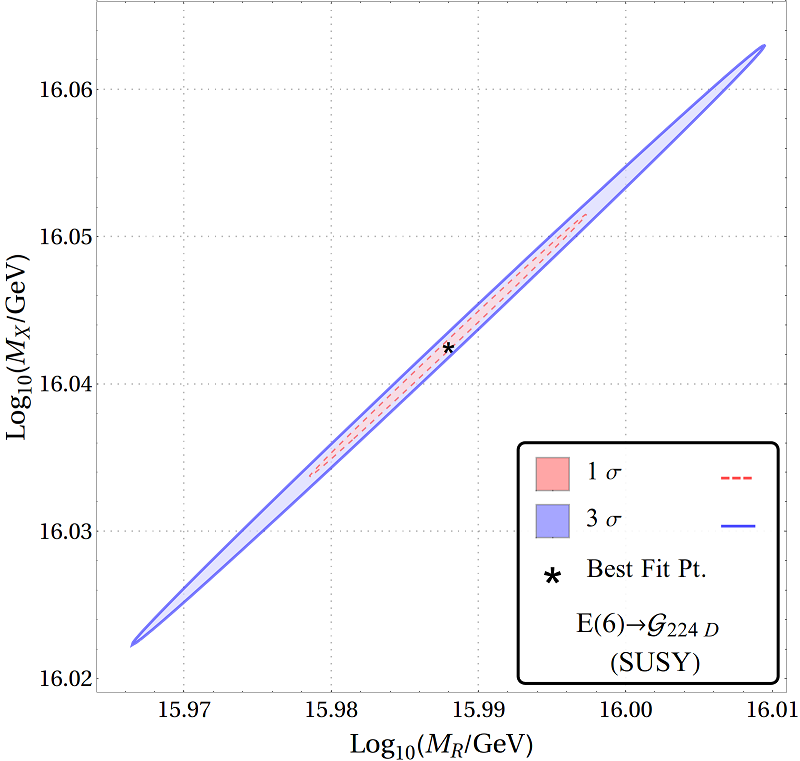}
	\includegraphics[scale=0.4]{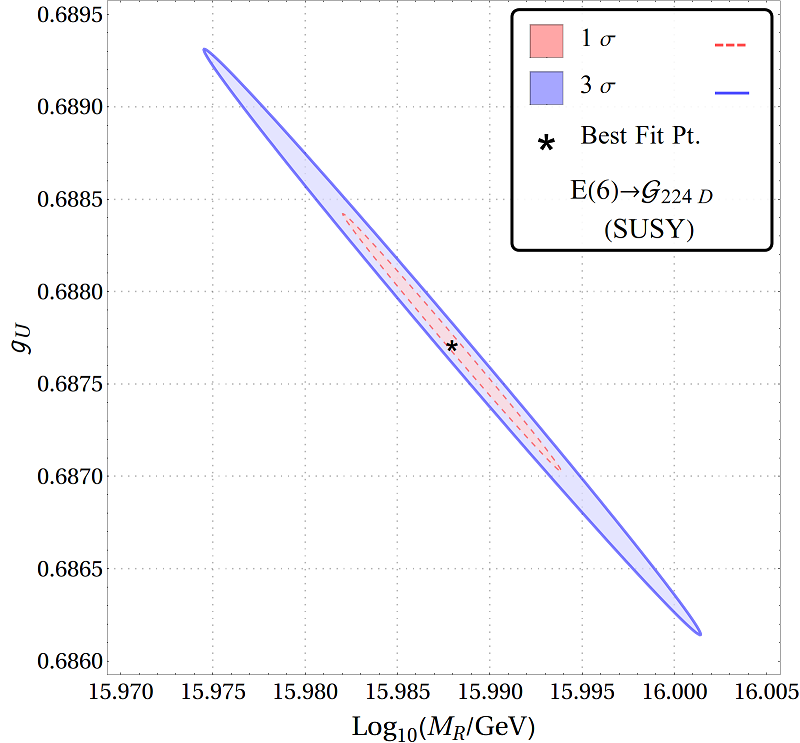}
	\includegraphics[scale=0.4]{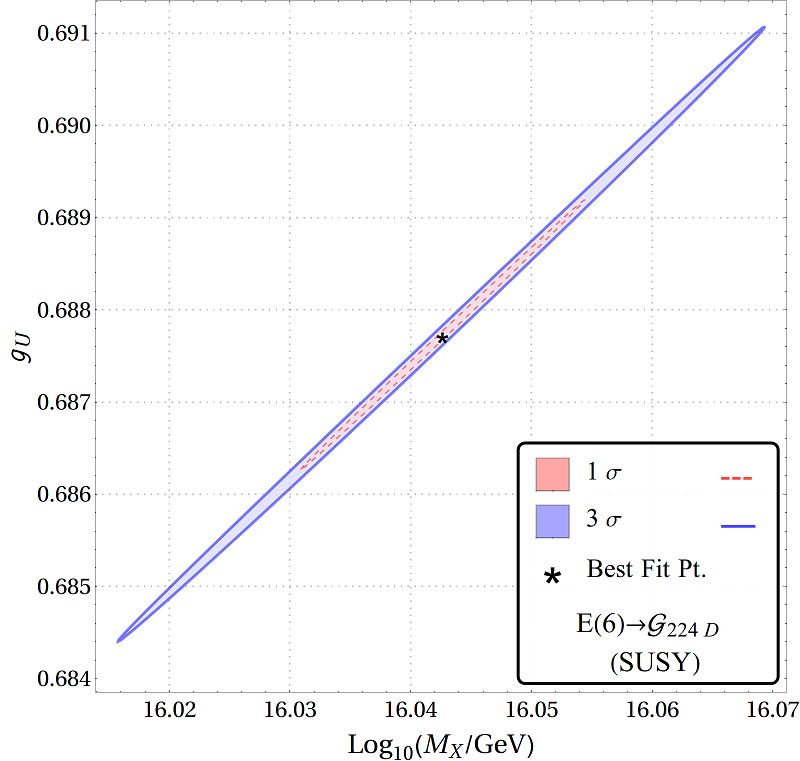}\\
	\includegraphics[scale=0.4]{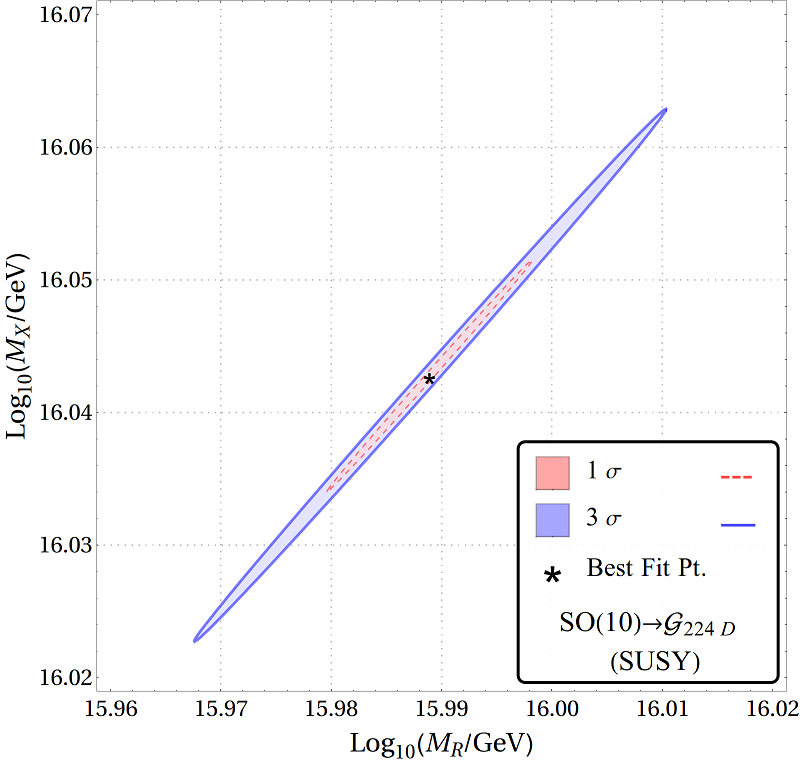}
	\includegraphics[scale=0.4]{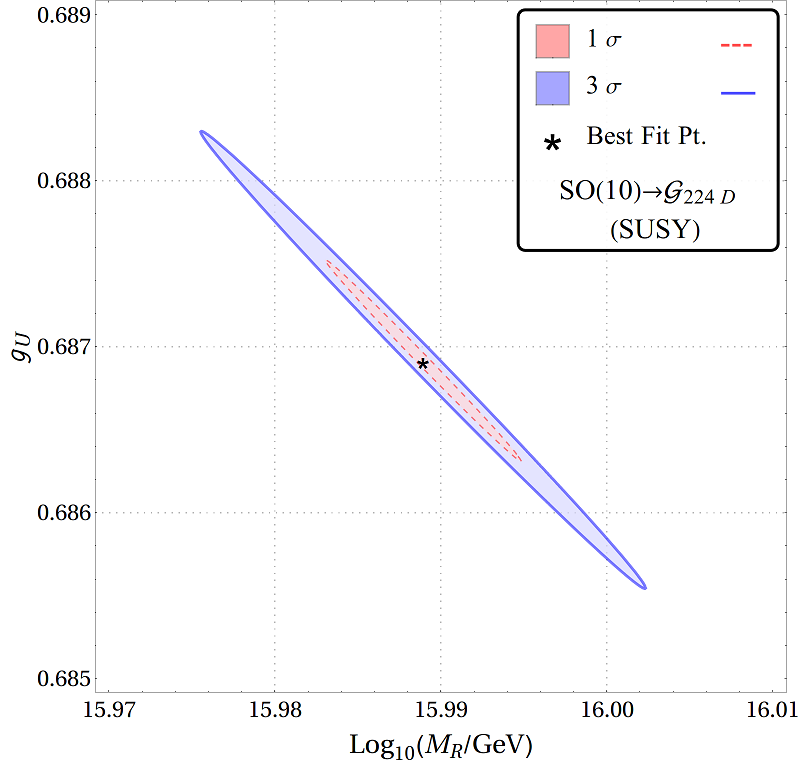}
	\includegraphics[scale=0.4]{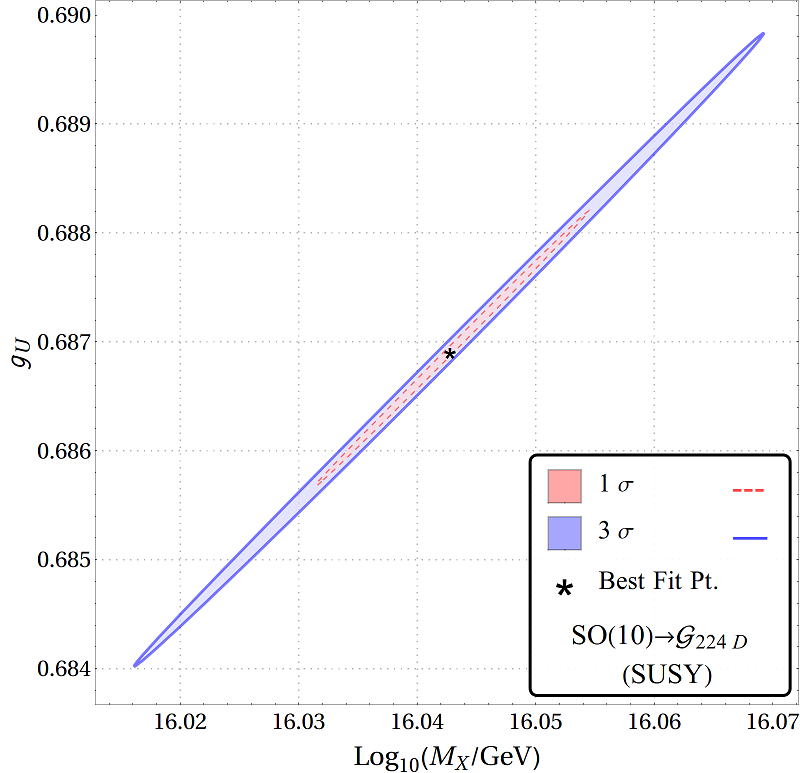}
	\caption{\scriptsize Correlations among $M_R$, $M_X$ and $g_{U}$, for different breaking patterns within supersymmetric scenario, satisfying unification with $M_{SUSY}$ greater than 1 TeV. Correlations are obtained by fixing $M_{SUSY}$ at 15 TeV.}
	\label{fig:unific-chisq_15tev_S_2}
\end{figure}

\pagebreak
\newpage

\section*{}

\providecommand{\href}[2]{#2}
\addcontentsline*{toc}{section}{}
\bibliography{TD-GUT}

\end{document}